\pdfoutput=1 

\documentclass[11pt,a4paper]{mythesis}
\usepackage[left=1.5in,right=1in,top=1in,bottom=1in,includefoot]{geometry}

\usepackage{setspace} 

\usepackage{graphicx}
\usepackage{amsfonts} 
\usepackage{amsmath} 
\usepackage{amssymb}

\newcommand{\Av}{\mbox{\boldmath$A$}}

\newcommand{\Cv}{\mbox{\boldmath$C$}}

\newcommand{\Fv}{\mbox{\boldmath$F$}}

\newcommand{\Iv}{\mbox{\boldmath$I$}}
\newcommand{\Jv}{\mbox{\boldmath$J$}}

\newcommand{\Lv}{\mbox{\boldmath$L$}}

\newcommand{\Nv}{\mbox{\boldmath$N$}}

\newcommand{\Pv}{\mbox{\boldmath$P$}}
\newcommand{\Qv}{\mbox{\boldmath$Q$}}

\newcommand{\Sv}{\mbox{\boldmath$S$}}
\newcommand{\Tv}{\mbox{\boldmath$T$}}

\newcommand{\Yv}{\mbox{\boldmath$Y$}}

\newcommand{\av}{\mbox{\boldmath$a$}}
\newcommand{\bv}{\mbox{\boldmath$b$}}
\newcommand{\cv}{\mbox{\boldmath$c$}}

\newcommand{\ev}{\mbox{\boldmath$e$}}

\newcommand{\gv}{\mbox{\boldmath$g$}}
\newcommand{\hv}{\mbox{\boldmath$h$}}

\newcommand{\kv}{\mbox{\boldmath$k$}}

\newcommand{\nv}{\mbox{\boldmath$n$}}

\newcommand{\rv}{\mbox{\boldmath$r$}}

\newcommand{\vv}{\mbox{\boldmath$v$}}

\newcommand{\xv}{\mbox{\boldmath$x$}}
\newcommand{\yv}{\mbox{\boldmath$y$}}

\newcommand{\muv}{\mbox{\boldmath$\mu$}}
\newcommand{\Gammav}{\mbox{\boldmath$\Gamma$}}
\newcommand{\lambdav}{\mbox{\boldmath$\lambda$}}

\begin{document}

\title{Searching for gravitational waves emitted by binaries with spinning components}
\author{Gareth Jones 
\\
\\
Submitted for the degree of Doctor of Philosophy \\
School of Physics and Astronomy \\
Cardiff University}
\date{January 2008}

\degreedate{Yet to be decided}

\maketitle

\pagenumbering{roman}
\tableofcontents
\listoffigures
\listoftables

\chapter*{Acknowledgements}

Firstly, I would like to thank my supervisor Prof. B. Sathyaprakash 
for advice, support and enthusiasm during my research.
As a member of the LIGO Scientific Collaboration the list of people
from whose efforts I have benefited is enormous.  
I am grateful to them all and especially to the members of the
Compact Binary Coalescence group with whom I have had the pleasure
of working alongside.

I would like to thank Jon Gair, not only for being an
enthusiastic collaborator, but also for accompanying on some
fun nights out in Berlin!

I would like to thank both Sathya and Leonid Grishchuk for
first introducing to the field of gravitational waves during
my undergraduate years and all the 
members of the Cardiff Gravitational Physics group of the
last five years for providing a lively and friendly working environment.
In particular I would like to thank 
Stas Babak for his assistance and support throughout the
early stages of my research, 
Thomas Cokelaer for his useful advice, his friendship and for 
introducing me to an array of films and beers and Craig Robinson 
with whom I have enjoyed
the challenge of the last few years.

I am grateful to Patrick Sutton who has patiently waited for me
to bring this thesis to completion and I am looking forward to
beginning the next stage of my research with him.

I have enjoyed many trips and conferences over the last few years
but I would particularly like to thank the team at the GEO site 
for a very interesting week back in 2004.

I am grateful to have worked with so many people whose hard work and
dedication has been an inspiration and to count many of them
amongst my friends.
I would like also to wish everyone involved in the search for 
gravitational waves the best of luck as we look forward to what are 
going to be some very exciting times!

Finally I would like to thank my family, my fiancee Nicola, my 
friends and bandmates for their interest and support.

\subsubsection{LIGO Scientific Collaboration Acknowledgement}
The second Chapter of this thesis contains analysis of LIGO data 
performed by the author as a member of the LIGO
Scientific Collaboration.

The author gratefully acknowledge the support of the United States
National Science Foundation for the construction and operation of the
LIGO Laboratory and the Science and Technology Facilities Council of the
United Kingdom, the Max-Planck-Society, and the State of
Niedersachsen/Germany for support of the construction and operation of
the GEO600 detector. The authors also gratefully acknowledge the support
of the research by these agencies and by the Australian Research Council,
the Council of Scientific and Industrial Research of India, the Istituto
Nazionale di Fisica Nucleare of Italy, the Spanish Ministerio de
Educaci\'on y Ciencia, the Conselleria d'Economia, Hisenda i Innovaci\'o of
the Govern de les Illes Balears, the Scottish Funding Council, the
Scottish Universities Physics Alliance, The National Aeronautics and
Space Administration, the Carnegie Trust, the Leverhulme Trust, the David
and Lucile Packard Foundation, the Research Corporation, and the Alfred
P. Sloan Foundation.


\chapter*{Conventions}

Masses are quoted in units of solar mass
\begin{equation}
1\,M_{\odot} = 1.98892 \times 10^{30} \mbox{kg}
\end{equation}
Distances are quoted in units of parsecs
\begin{equation}
1\,\mbox{pc} = 3.0856775807 \times 10^{16} \mbox{m}
\end{equation}
The values used here are the same as those used within the LIGO Scientific 
Collaboration Algorithm Library (LAL) and are taken from 
Barnett et al. (1996) \cite{PhysRevD.54.1}.

In mathematical formulae bold face will denote a vector, e.g.~${\bf S}_{1}$ and
overhats to represent unit vectors, e.g.~$\bf{\bf \hat{S}}_{1}$.

All angles will be in radians.

In general Greek indices sum over ($0 \dots 3$) and Latin indices sum over
($1 \dots 3$).

We will denote the inner product as $\left<g,h \right>$ and will use
$||g||$ to mean $\left<g,g \right>$.

There are two possible sign conventions used to define the Fourier
transform. 
Following the conventions used in LAL we shall define the Fourier 
transform $\tilde{g}(t)$  of a time domain function $g(t)$ by
\begin{eqnarray}
\tilde{g}(f) = \int_{-\infty}^{\infty} g(t) e^{-2 \pi i f t} dt
\end{eqnarray}
and the inverse Fourier transform by
\begin{eqnarray}
g(t) = \int_{-\infty}^{\infty} \tilde{g}(f) e^{2 \pi i f t} df.
\end{eqnarray}
Note that in some of the literature referenced in this thesis
the other convention is used.

We will use geometric units (i.e., $c=G=1$) throughout unless we
specify otherwise.



\begin{abstract}
In this thesis we consider the data analysis problem of detecting
gravitational waves emitted by inspiraling binary systems.
Detection of gravitational waves will open a new window on the Universe
enabling direct detection of systems such as binary black holes for
the first time.
In the first Chapter we show how gravitational waves are derived from
Einstein's General theory of Relativity and discuss the emission of
gravitational waves from inspiraling binaries and how this
radiation may be detected using laser interferometers.
Around two thirds of stars inhabit binary systems.
As they orbit each other they will emit both energy and angular momentum
in the form of gravitational waves which will inevitably lead to
their inspiral and eventual merger.
To date, searches for gravitational waves emitted during the inspiral of
binary systems have concentrated on systems with non-spinning components.
In Chapter \ref{ch:spin} we detail the first dedicated search for binaries consisting of spinning
stellar mass compact objects.
We analysed 788 hours of data collected during the third science
run (S3) of the LIGO detectors, 
no detection of gravitational waves was made and we set an upper limit
on the rate of coalescences of stellar mass binaries.
The inspiral of stellar mass compact objects into super massive black holes
will radiate gravitational waves at frequencies detectable by the planned
space-based LISA mission.
In Chapter \ref{ch:emri} we describe the development and testing of a computationally
cheap method to detect the loudest few extreme mass ratio inspiral events that
LISA will be sensitive to.
\end{abstract}

\pagenumbering{arabic}
\doublespacing


\chapter{Introduction}
\label{ch:intro}


Gravitational waves are an inescapable consequence of any theory of gravity
that is consistent with Einstein's
Special Theory of Relativity (1905), in particular its condition that
information cannot propagate at speeds greater than the speed of light
in vacuum, c.
Following Einstein's General Theory of Relativity (1915) we identify 
gravity as a curvature of spacetime and gravitational waves to be
caused by the acceleration of matter.
Gravitational waves carry away both energy and momentum from a 
radiating source and propagate at the speed of light. 

The weak interaction between gravity and matter make the detection of
gravitational waves an exciting but challenging prospect. 
On one hand, their weak interaction with matter means that
gravitational waves will not suffer the scattering and absorption which 
impedes the propagation of electromagnetic radiation through the interstellar medium.
On the other hand, only in the last few decades
has technology advanced to a point where it has been possible to 
construct detectors with good enough sensitivity to observe gravitational waves.
To date, no direct detection of gravitational waves has been made.

The detection of gravitational waves would open a new window on the 
Universe enabling direct observation for the first time
of sources including the inspiral and merger of binary black hole
systems as well as providing deeper insight into known sources such as
x-ray binaries and gamma-ray bursts.
It should not be forgotten that detection of gravitational waves
could provide us with observations of previously unimagined sources. 

The first \emph{indirect} evidence for gravitational waves was identified by 
Hulse and Taylor in 1974 with the observation of a pulsar, now commonly 
referred to as the Hulse-Taylor pulsar \cite{HulseTaylor_1975}. 
Through careful and continuous measurement of the variation in expected arrival 
times of the emitted pulses, Hulse and Taylor concluded that the pulsar was in 
orbit around a common centre of mass with another, as then unobserved, star which 
was later inferred to be a neutron star from its mass.
The system as a whole is known as the Hulse-Taylor binary pulsar or PSR 1913+16. 
In 1983, Taylor and collaborators announced a decrease in the inferred orbital 
period of PSR 1913+16 of $76\,\mu \rm{s}\,\rm{yr}^{-1}$ \cite{WeisbergTaylor_1984}.
With no other explanation it was concluded that the decay of PSR 1913+16's orbit 
was due to the emission of gravitational waves. 
The measured rate of change of the orbital period agrees with the prediction of
General Relativity to within around $0.2\%$ \cite{weisberg-2004}. 
In recognition of their detection of PSR 1913+16, Hulse and Taylor were awarded 
the Nobel Prize for Physics in 1993.
To date, a total of seven binary neutron star systems have been observed
electromagnetically \cite{anderson-2007}
including the first observed double pulsar system, J0737-3039 by 
Burgay et al. (2003) \cite{Burgay2003}. 
As well as providing indirect evidence for gravitational waves these highly 
relativistic systems can be used to test General Relativity 
(see, for example, Will \cite{Will_livingReview}). 

In this Chapter we will begin with the Einstein equations and show
that gravitational waves propagate in flat
spacetime as plane waves at the speed of light and have two independent 
polarizations (Sec.~\ref{sec:planeGWs}).
In Sec.~\ref{sec:sources} we identify binaries consisting of massive compact objects, 
such as neutron stars or black holes, 
as sources of gravitational waves that should be
detectable by current and planned gravitational wave detectors.
In Sec.~\ref{sec:detection} we discuss gravitational wave detectors 
and then move onto describing the optimal method for detecting a signal
with a known form buried in a noisy data stream.

For background reading and guidance with derivations regarding General Relativity I 
have made use of the following material:
Hartle \cite{Hartle}, 
Schutz, \cite{Schutz},
Misner, Thorne and Wheeler \cite{MTW},
Hakim \cite{Hakim},
d'Inverno \cite{inverno}
and lecture notes by Prof. B. Sathyaprakash.
For further reading on data analysis I have made use of:
Whalen \cite{whalen:1971},
Wainstein and Zubakov \cite{wainstein:1962}
Finn (1992) \cite{Finn92}
and Finn and Chernoff (1993) \cite{FC:1993}.

\section{Plane gravitational waves}
\label{sec:planeGWs}

In this Section we will show that a solution to the linearized Einstein 
field equations in vacuum are plane waves propagating at the speed of light.
Furthermore, we will show that by working in a co-ordinate system that 
satisfies some particular gauge conditions the waves can be written 
in terms of two independent polarization states.

\subsection{The vacuum Einstein equations}
We begin by writing the Einstein equations
\begin{equation}
G_{\alpha \beta} = 8 \pi T_{\alpha \beta}
\label{EEgen}
\end{equation}
which relates a measure of the local spacetime curvature $G_{\alpha \beta}$ 
with the distribution of energy-momentum $T_{\alpha \beta}$. 
%
Since both $G_{\alpha \beta}$ and $T_{\alpha \beta}$ are symmetric 
there are $10$ independent equations encoded in Eq.~(\ref{EEgen}). 
These equations are coupled, non-linear partial differential equations. 
Consequently, a general solution to the Einstein equations has not yet been 
derived. Instead we find solutions for the equation under particular conditions.

The Einstein curvature tensor is defined as
\begin{equation}
G_{\alpha \beta} \equiv R_{\alpha \beta} - \frac{1}{2} g_{\alpha \beta} R
\label{curvtensor}
\end{equation}
where $R_{\alpha \beta}$ and $R$ are the Ricci curvature tensor and scalar 
(defined in the next subsection)
and $g_{\alpha \beta}$ is the metric which determines the separation 
between two local events in spacetime.
In a vacuum we see that $T_{\alpha \beta} = 0$ which in turn leads 
to $G_{\alpha \beta} =0$.


\subsection{Linearizing the Einstein equations}
The Einstein equations are non-linear. 
If, however, we consider a region of spacetime whose geometry is almost flat 
we can write a linearized approximation to the Einstein equations for which 
solutions can be found. 
In this Section we will linearize the vacuum Einstein equations.

We write the interval between 2 events in spacetime in 
$(t,x,y,z)$ co-ordinates as
\begin{equation}
\label{spacetimemetric}
ds^2 = g_{\alpha \beta}(\xv) dx^{\alpha} dx^{\beta} 
\end{equation}
where $\gv$ is the {\it metric}, a position dependent second rank tensor which
can be represented by a $4 \times 4$ symmetric matrix. 
For flat spacetime we have $g_{\alpha \beta}$ equals 
the Minkowski metric $\eta_{\alpha \beta}$ defined as diag$(-1,1,1,1)$.
  
When the spacetime is close to being flat we can write the metric as
\begin{equation}
g_{\alpha \beta}(\xv) = \eta_{\alpha \beta} + h_{\alpha \beta}(\xv)
\label{metricg}
\end{equation}
where $\hv$ are small perturbations to the flat metric satisfying 
$|h_{\alpha \beta}| \ll 1$.
We can write the Ricci curvature tensor in terms of the 
Christoffel symbols $\Gamma$ as
\begin{equation}
R_{\alpha \beta} = 
\Gamma_{\alpha \beta, \nu}^{\nu} 
- \Gamma_{\alpha \nu, \beta}^{\nu}  
+ \Gamma_{\alpha \beta}^{\nu} \Gamma_{\nu \delta}^{\delta}
- \Gamma_{\alpha \delta}^{\nu} \Gamma_{\beta \nu}^{\delta}
\label{riccifull}
\end{equation}
where we abbreviated notation for the partial derivative such that
$\partial f(x^{\beta}) / \partial x^{\alpha} = f_{\beta,\alpha}$.
The Christoffel symbols can we written in terms of the metric
\begin{equation}
\Gamma_{\alpha \beta}^{\nu} = \frac{1}{2} 
g^{\nu \delta} 
(
  g_{\delta \alpha, \beta}
+ g_{\delta \beta, \alpha}
- g_{\alpha \beta, \delta}
).
\label{christg}
\end{equation}
Substituting for $g_{\alpha \beta}$ into Eq.~(\ref{christg}), 
neglecting terms beyond first order in $h_{\alpha \beta}$ and 
remembering that $\eta_{\alpha \beta}$ is a constant we find
\begin{equation}
\Gamma_{\alpha \beta}^{\nu} = \frac{1}{2}
\eta^{\nu \delta}
(
  h_{\delta \alpha, \beta}
+ h_{\delta \beta, \alpha}
- h_{\alpha \beta, \delta}
).
\label{christh}
\end{equation}
Substituting $\Gamma_{\alpha \beta}^{\nu}$ back into
Eq.~(\ref{riccifull}) for the Ricci curvature tensor and again neglecting 
terms of $h_{\alpha \beta}$ beyond first order we find
\begin{eqnarray}
R_{\alpha \beta} &=& 
\frac{1}{2} \eta^{\nu \delta}
(
  h_{\delta \beta, \alpha \nu} 
- h_{\alpha \beta, \delta \nu}
- h_{\delta \nu, \alpha \beta}
+ h_{\alpha \nu, \delta \beta} 
) \\
&=&
\frac{1}{2}
(
  h_{\beta \mu, \alpha} ^{\phantom{\beta \mu, \alpha} \mu}
- h_{\alpha \beta, \mu} ^{\phantom{\alpha \beta, \mu} \mu}
- h_{,\alpha \beta}
+ h_{\alpha \mu, \beta} ^{\phantom{\alpha \mu, \beta} \mu} 
)
\label{riccih}
\end{eqnarray}
where we raise the indices of $h_{\alpha \beta}$ using $\eta^{\alpha \beta}$.
We are able approximate $g^{\alpha \beta} = \eta^{\alpha \beta}$ when raising 
indices of $h_{\alpha \beta}$ since the use of the full metric as given 
in Eq.~(\ref{metricg})
would involve terms second order in terms of $h_{\alpha \beta}$.
We have also defined the trace of $h_{\alpha \beta}$ to be 
$h \equiv h^{\mu}_{\phantom{\mu} \mu}$.
By contracting once more we can find the Ricci scalar
\begin{eqnarray}
R &=& \eta^{\alpha \beta} R_{\alpha \beta} = 
\frac{1}{2}
(
h^{\mu \nu}_{\phantom{\mu \nu} ,\mu \nu} - h_{, \mu}^{\phantom{\mu} \mu}
).
\end{eqnarray}
Substituting the expressions for the Ricci curvature tensor $R_{\alpha \beta}$ 
and Ricci scalar $R$ into Eq.~(\ref{curvtensor}) for the 
Einstein 
tensor we find
\begin{eqnarray}
G_{\alpha \beta} &=& \frac{1}{2}
(
 h_{\beta \mu, \alpha} ^{\phantom{\beta \mu, \alpha} \mu}
- h_{\alpha \beta, \mu} ^{\phantom{\alpha \beta, \mu} \mu}
- h_{,\alpha \beta}
+ h_{\alpha \mu, \beta} ^{\phantom{\alpha \mu, \beta} \mu}
- \eta^{\alpha \beta} h^{\mu \nu}_{\phantom{\mu \nu}, \mu \nu} 
+ \eta^{\alpha \beta} h_{, \mu}^{\phantom{\mu} \mu}
).
\end{eqnarray}
We can abbreviate this expression by introducing the `trace reverse' of 
$h_{\alpha \beta}$ which is defined as
\begin{eqnarray}
\bar{h}_{\alpha \beta} \equiv h_{\alpha \beta} - 
\frac{1}{2}\eta_{\alpha \beta} h.
\label{tracereverse}
\end{eqnarray}
It is called the `trace reverse' because $\bar{h} = -h$.
We can then rewrite our expression for the Einstein tensor as
\begin{eqnarray}
\label{einsteinCurvatureFull}
G_{\alpha \beta} &=& -\frac{1}{2}
(
  \bar{h}_{\alpha \beta, \mu} ^{\phantom{\alpha \beta, \mu} \mu} 
+ \eta_{\alpha \beta} \bar{h}_{\mu \nu,}^{\phantom{\mu \nu} \mu \nu}
- \bar{h}_{\alpha \mu, \beta} ^{\phantom{\alpha \mu, \beta} \mu}
- \bar{h}_{\beta \mu, \alpha} ^{\phantom{\beta \mu, \alpha} \mu} 
) 
\end{eqnarray}
We will now go on to show that under a special class of co-ordinate 
transformations we are able to simplify this equation further.
%

\subsection{Gauge transformations}
Through particular small co-ordinate transformations we are able to find 
a co-ordinate system which
\begin{itemize}
\item preserves the form of our nearly-flat metric 
$g_{\alpha \beta}(\xv) = \eta_{\alpha \beta} + h_{\alpha \beta}(\xv)$,
\item keeps the metric perturbations small $|h_{\alpha \beta}| \ll 1$,
\item leaves $\eta = \mathrm{diag}(-1,1,1,1)$ and
\item allows us to modify (and simplify) the functional form of $h_{\alpha \beta}$.
\end{itemize}

We will now derive the form of these co-ordinate transformations.
We will consider a co-ordinate transformation with the standard form
\begin{equation}
x'^{\alpha} = x^{\alpha} + \xi^{\alpha}(\xv)
\label{coordtrans}
\end{equation}
where $\xi$ are of similarly small size as the metric perturbation 
$h_{\alpha \beta}(\xv)$. 
The metric will transform as
\begin{equation}
g'_{\alpha \beta}(\xv') = 
\frac{\partial x^{\nu}}    {\partial x'^{\alpha}} 
\frac{\partial x^{\delta}} {\partial x'^{\beta}} 
g_{\nu \delta}(\xv).
\end{equation}
Considering first order derivatives of our co-ordinates we find
\begin{eqnarray}
\frac{\partial x^{\alpha}}{\partial x'^{\beta}} 
&=& \frac{\partial (x'^{\alpha} - \xi^{\alpha} )}{\partial x'^{\beta}} \\
&=& \delta^{\alpha}_{\beta} - \frac{\partial \xi^{\alpha}}{\partial x'^{\beta}}
\end{eqnarray}
where in first order equations of $\xi$ we can interchange $\xi^{\alpha}(\xv)$
and $\xi^{\alpha}(\xv')$. 
Using this relationship we find the metric transformation becomes
\begin{eqnarray}
g'_{\alpha \beta}(\xv') 
&=&
\bigg(\delta^{\nu}_{\alpha} - \frac{\partial \xi^{\nu}}{\partial x'^{\alpha}} \bigg)
\bigg(\delta^{\delta}_{\beta} - \frac{\partial \xi^{\delta}}{\partial x'^{\beta}} \bigg) 
g_{\nu \delta}(\xv) \\
&=&
\bigg(   
\delta^{\nu}_{\alpha} \delta^{\delta}_{\beta}
- \delta^{\nu}_{\alpha}    \frac{\partial \xi^{\delta}}{\partial x'^{\beta}} 
- \delta^{\delta}_{\beta}  \frac{\partial \xi^{\nu}}{\partial x'^{\alpha}}
\bigg)
g_{\nu \delta}(\xv) \\
&=&
g_{\alpha \beta}(\xv) - 
\bigg(
\delta^{\nu}_{\alpha}    \frac{\partial \xi^{\delta}}{\partial x'^{\beta}}
+
\delta^{\delta}_{\beta}  \frac{\partial \xi^{\nu}}{\partial x'^{\alpha}}
\bigg)
g_{\nu \delta}(\xv).
\end{eqnarray}
where we can neglect terms greater than first order in $\xi^{\alpha}$ or 
of $\xi^{\alpha} h_{\alpha \beta}$.
Substituting in Eq.~(\ref{metricg}) for the metric we obtain
\begin{eqnarray}
h'_{\alpha \beta}(\xv')
&=&
h_{\alpha \beta}(\xv) -
\bigg(
\delta^{\nu}_{\alpha}    \frac{\partial \xi^{\delta}}{\partial x'^{\beta}}
+
\delta^{\delta}_{\beta}  \frac{\partial \xi^{\nu}}{\partial x'^{\alpha}}
\bigg)
\eta_{\nu \delta}(\xv) \\
&=&
h_{\alpha \beta}(\xv) 
- \eta_{\alpha \delta} \frac{\partial \xi^{\delta}}{\partial x'^{\beta}}
- \eta_{\nu \beta}     \frac{\partial \xi^{\nu}}{\partial x'^{\alpha}} \\
&=&
h_{\alpha \beta}(\xv)
- \frac{\partial \xi_{\alpha}}{\partial x'^{\beta}}
- \frac{\partial \xi_{\beta}}{\partial x'^{\alpha}}
\end{eqnarray}
Note that we assume that $\eta_{\alpha \beta}$ is unchanged as we transform 
between co-ordinate systems. 
We have therefore shown that we can apply co-ordinate 
transforms Eq.~(\ref{coordtrans}) whilst maintaining the linearized form 
of the metric Eq.~(\ref{metricg}) and giving rise to metric perturbations 
given by
\begin{equation}
h'_{\alpha \beta}(\xv') = h_{\alpha \beta}(\xv)
- \frac{\partial \xi_{\alpha}}{\partial x'^{\beta}}
- \frac{\partial \xi_{\beta}}{\partial x'^{\alpha}}.
\label{gaugetrans}
\end{equation}
Transformations of this kind are known as {\it gauge transformations}.
We will now find the corresponding co-ordinate transformation in terms of
the `trace reverse' of $h_{\alpha \beta}$. 
From Eq.~(\ref{gaugetrans}) we can show that the trace of $h_{\alpha \beta}$
has gauge transformations
\begin{equation}
h' = h - 2\xi^{\mu}_{\phantom{\mu} ,\mu}.
\label{gaugetranstrace}
\end{equation}
Substituting in Eqs.~(\ref{gaugetrans}) and (\ref{gaugetranstrace})
into the right hand side of our equation for the `trace reverse' 
of $h_{\alpha \beta}$ Eq.~(\ref{tracereverse}) we find that
\begin{equation}
\bar{h}'_{\alpha \beta} = 
\bar{h}_{\alpha \beta} - \xi_{\alpha,\beta} - \xi_{\beta,\alpha} + 
\eta_{\alpha \beta} \xi^{\mu}_{\phantom{\mu} ,\mu}.
\label{gaugetranstr}
\end{equation}

\subsection{Applying the Lorentz gauge condition}
\label{sec:lorentzgauge}
If we make a co-ordinate transformation such that
\begin{eqnarray}
\bar{h}^{\alpha \beta}_{\phantom{\alpha \beta}, \beta} = 0
\ \ \ \ {\rm i.e., \ \ \ \ } 
\bar{h}_{\alpha\beta,}^{\phantom{\alpha\beta,} \beta} = 
\xi_{\alpha,\beta}^{\phantom{\alpha,\beta} \beta}
\label{lorentzgauge}
\end{eqnarray}
we can re-write our previous expression for the Einstein 
tensor 
as
\begin{equation}
G_{\alpha \beta} = -\frac{1}{2} 
\bar{h}_{\alpha \beta, \mu}^{\phantom{\alpha \beta, \mu} \mu} = 
-\frac{1}{2} \Box \bar{h}_{\alpha\beta} = 
0. 
\label{EEBox}
\end{equation}
%
Equation (\ref{lorentzgauge}) is called the {\it Lorentz gauge condition} due to its 
similarity with the Lorentz condition used within electromagnetism.
Recognising that the linearized Einstein equations are wave equations 
suggests solutions of the form
\begin{equation}
\label{hbarwave}
\bar{h}_{\alpha \beta} (\xv) = A_{\alpha \beta} e^{i \kv \cdot \xv}
\end{equation}
where $\kv$ is a four-vector and must be null ($\kv \cdot \kv = 0$) 
in order to satisfy the linearized vacuum Einstein equations
Eq.~(\ref{EEBox}). 
The speed of the waves propagation is given by $|k_{0}|/|\kv|$ where 
$\kv$ is the spatial
components of $\kv$: $k_{1}, k_{2}$ and $k_{3}$.
For a null vector we have $|k_{0}| = |\kv|$ which leads to a wave speed $= 1$ which
is the speed of light.
This means that in vacuum flat spacetime, small perturbations of the 
metric propagate as plane waves at the speed of light.
These propagations of perturbations of the metric are what we call 
{\it gravitational waves}.

Using our equations for the Einstein tensor in vacuum 
Eq.~(\ref{EEBox}) and the Lorentz gauge condition 
Eq.~(\ref{lorentzgauge}) with our plane wave solution 
for $\bar{h}_{\alpha \beta}$, 
it is simple to find the following relations 
\begin{eqnarray}
k^2 A_{\alpha \beta} &=& 0 \\
\label{firstrel}
k^{\alpha} A_{\alpha \beta}  &=& 0
\label{secondrel}
\end{eqnarray}
which we shall use later when finding the number of independent 
components of $\bar{h}_{\alpha \beta}$.

\subsection{Applying the Transverse-Traceless gauge conditions}
\label{sec:applyingTT}
In this Section we show that by applying two more gauge conditions 
we can write the metric perturbation $h_{\alpha \beta}$ using 
only two independent components.

We are able to perform further gauge transformations as long as we ensure
that the Lorentz gauge condition is still satisfied. By substituting our
gauge transformation for $\bar{h}_{\alpha \beta}$ Eq.~(\ref{gaugetranstr}) 
into the Lorentz gauge condition Eq.~(\ref{lorentzgauge}) we find:
\begin{eqnarray}
0 &=& \bar{h}'^{\alpha \beta}_{\phantom{\alpha \beta},\beta} \\
  &=& \bar{h}^{\alpha \beta}_{\phantom{\alpha \beta},\beta} 
-\xi^{\alpha, \beta}_{\phantom{\alpha, \beta} \beta}
-\xi^{\beta, \alpha}_{\phantom{\beta, \alpha} \beta}
+ \eta_{\alpha \beta} \xi^{\mu}_{\phantom{\mu} ,\mu \beta} \\
  &=& \bar{h}^{\alpha \beta}_{\phantom{\alpha \beta},\beta}
-\xi^{\alpha, \beta}_{\phantom{\alpha, \beta} \beta}
-\xi^{\beta, \alpha}_{\phantom{\beta, \alpha} \beta}
+\xi^{\mu, \alpha}_{\phantom{\mu, \alpha} \mu}.
\end{eqnarray}
The third and fourth terms on the right hand side cancel and we know
from the Lorentz gauge condition that 
$\bar{h}^{\alpha \beta}_{\phantom{\alpha \beta},\beta} =0$
which leaves us with
\begin{equation}
0 = \xi^{\alpha, \beta}_{\phantom{\alpha, \beta} \beta} = \Box \xi^{\alpha}.
\end{equation}

We can see immediately that there will be wavelike solutions for our
co-ordinate transformation $\xi$ as we did for $\bar{h}_{\alpha \beta}$.
We can therefore write solutions for the co-ordinate transformation as
\begin{equation}
\xi_{\alpha} (\xv) = i B_{\alpha} e^{i \kv \cdot \xv}.
\label{xiwave}
\end{equation}
We find that by choosing particular values of $B_{\alpha}$ we can choose
a co-ordinate system for which $\bar{h}_{\alpha \beta}$
has a very simple form.
Substituting the wave solutions for 
$\bar{h}_{\alpha \beta}$ Eq.~(\ref{hbarwave})
and
$\xi_{\alpha}$ Eq.~(\ref{xiwave}) 
into Eq.~(\ref{gaugetranstr}) we find
\begin{equation}
A'_{\alpha \beta} =  
A_{\alpha \beta} 
+ k_{\alpha} B_{\beta} 
+ k_{\beta} B_{\alpha}
- \eta_{\alpha \beta} B^{\mu} k_{\mu}.
\end{equation}
It is clear that by judicious choice of $B_{\alpha}$ 
(and therefore $\xi_{\alpha}$) we can impose
further conditions on $A_{\alpha}$ 
(and therefore $\bar{h}_{\alpha \beta}$).
We will now show that by using our gauge transformations it is possible 
to describe the plane wave solution of the Einstein equations in vacuum using 
only two independent components.

We will consider a wave travelling in the $z$-direction. 
We are always able to perform a co-ordinate transformation to make this 
true so the solutions we obtain will be generic. 
Remembering that $\kv$ is null we will have 
\begin{eqnarray}
k_{\alpha} &=& (k,0,0,k) \\ 
k^{\alpha} &=& (-k,0,0,k)  
\end{eqnarray}
From the relation in Eq.~(\ref{secondrel}) we can now show that
\begin{equation}
A_{\mu 0} = A_{\mu 3}.
\end{equation}
Making use of this and the fact that $A_{\alpha \beta}$ is symmetric 
we can write the components of $A_{\alpha \beta}$ as follows
\begin{eqnarray}
A'_{00} &=& A_{00} + k B_{0} + k B_{3} \\ 
A'_{01} &=& A_{01} + k B_{1} \\ 
A'_{02} &=& A_{02} + k B_{2} \\ 
A'_{11} &=& A_{11} + k B_{0} - k B_{3} \\
A'_{12} &=& A_{12} \\
A'_{22} &=& A_{22} + k B_{0} - k B_{3}. 
\end{eqnarray}
By choosing the following values for $B_{\alpha}$ 
\begin{eqnarray}
B_{1} = - \frac{A_{01}}{k} \\
B_{2} = - \frac{A_{02}}{k}  \\ 
\end{eqnarray}
we can set $A'_{01} = A'_{02} = 0$.
By choosing
\begin{eqnarray}
B_{0} &=& - \frac{1}{4k} (2 A_{00} + A_{11} + A_{22}) \\
B_{3} &=&   \frac{1}{4k} (- 2 A_{00} + A_{11} + A_{22})
\end{eqnarray}
we can further set $A'_{00} = 0$ and $A'_{11} + A'_{22} = 0$.
We can then write $A_{\alpha \beta}$ as
\begin{equation}
A_{\alpha \beta}^{TT} = 
\left( \begin{array}{cccc}
0 & 0 & 0 & 0  \\
0 & A_{11} & A_{12} & 0  \\
0 & A_{12} & -A_{11} & 0  \\
0 & 0 & 0 & 0  \\
\end{array} \right)
\end{equation}
The superscript $TT$ refers to the fact that our choice of co-ordinate 
transformation (made here by specifying the components of $B_{\alpha}$) 
lead to a metric perturbation $\bar{h}_{\alpha \beta}$  Eq.~(\ref{hbarwave})
which is traceless and transverse.

We will briefly review the various steps we have used to arrive at our
traceless transverse form of the metric perturbation keeping track of the
number of independent components.
The original (small) metric perturbation $h_{\alpha \beta}$ has 16 components,
due to symmetry only 10 of these are independent.
The Lorentz gauge condition Eq.~(\ref{lorentzgauge}) represents 4 independent
equations which reduces the number of independent components of 
$h_{\alpha \beta}$ to 6. 
Similarly our (4) choices of $B_{\alpha}$ in the wave equation for 
$\xi_{\alpha}$ further reduce the number of independent parameters of
$h_{\alpha \beta}$ to 2.

We write the trace reverse metric in the TT gauge as
\begin{equation}
\label{metricTT}
\bar{h}_{\alpha \beta}^{TT} =
\left( \begin{array}{cccc}
0 & 0 & 0 & 0  \\
0 & A_{+} & A_{\times} & 0  \\
0 & A_{\times} & -A_{+} & 0  \\
0 & 0 & 0 & 0  \\
\end{array} \right)
e^{i \kv \cdot \xv}
\end{equation}
where we have renamed the 2 independent components $A_{+}$ 
and $A_{\times}$.
We find that these two components represent two independent
polarizations of the gravitational waveform which we
call $+$ (``plus'') and $\times$ (``cross'').
The reasons for these names will become clear when we discuss
the effect of a gravitational wave on a ring of freely falling
test masses (see Figs. \ref{test_masses} and \ref{ifo_detection_1}).


Having found that perturbations of the space-time metric can travel
as gravitational waves through vacuum at the speed of light we will now
move on to discuss sources of gravitational waves and methods by which
we should be able to detect them.

\section{Sources of gravitational waves}
\label{sec:sources}

In the previous Section we found a linearized approximation to the Einstein
equations in vacuum:
\begin{eqnarray}
-\frac{1}{2} \Box \bar{h}_{\alpha \beta} = 0.
\end{eqnarray}
We will consider the linearized approximation to the Einstein equations with a 
source:
\begin{eqnarray}
\label{EEsource}
\Box \bar{h}_{\alpha \beta} = -16 \pi T_{\alpha \beta} 
\end{eqnarray}
where $T_{\alpha \beta}$ is the 
energy-momentum-stress tensor (which we will call the
energy-momentum tensor for brevity and is also sometimes
call the stress-energy tensor).
Note that in {\it non}-linearized gravity the Einstein equations with a source
(Eq.~(\ref{EEsource})) would require another term $\tau_{\alpha \beta}$ 
on the right hand side to represent 
the {\it gravitational} (rather than matter) sources
of gravitational curvature and waves. 

In general, wave equations have two solutions of the form
$f(t-r)$ and $f(t+r)$ where $r \equiv |\xv|$.
The first solution describes a wave propagating {\it outward} from
the source {\it after} the event which generated it.
We call this first term the {\it retarded} or {\it causal} part of 
the solution.
The second solution will describe a wave propagating {\it inward} onto
the source {\it before} the event at the source we are considering.
We call this second term the {\it advanced} part of the solution.
We will only consider the causal part of the wave equation's
solution and will neglect the advanced part.

We can find a solution to the linearized approximation to the Einstein
equations Eq.~(\ref{EEsource}) using Green's function for the
d'Alembertian \cite{GreenWave} which will yield 
\begin{eqnarray}
\label{GreenSoln}
\bar{h}_{\alpha \beta}(t,\xv) &=& 
4 
\int 
\frac{ T_{\alpha \beta}(t - |\xv -\xv'|, \xv')}
{|\xv -\xv'|}
d^{3} x'
\end{eqnarray}
where $\xv'$ describes the spatial positions of mass elements
(i.e., $\delta$-function sources) within the source and
$\xv$ is the spatial position of the observer.
We have neglected the advanced part of the solution as previously
discussed.
Assuming that our source is concentrated at the origin and
assuming that the observers distance $D \equiv |\xv|$ 
from the source is large i.e., $|\xv| \gg |\xv'|$ we can make
the approximation that $D \sim |\xv -\xv'|$.
The region far from the source where this approximation can be made is
called the {\it far zone} (sometimes also called the radiation or wave 
zone).
Making this approximation yields
\begin{eqnarray}
\label{barhfarzone}
\bar{h}_{\alpha \beta}(t,\xv) &=&
\frac{4}{D} 
\int T_{\alpha \beta}(t - D,\xv') 
d^{3} x'.
\end{eqnarray}

We only need to consider the spatial components of the 
metric perturbation $\hat{h}_{ij}$ since the TT 
gauge transformation will set $h^{TT}_{0 \alpha} = 0$
(see Sec.~\ref{sec:applyingTT}).
Our metric perturbation must also satisfy the Lorentz gauge condition
Eq.~(\ref{lorentzgauge}). 
We find that the Lorentz gauge condition will be obeyed automatically  
as a consequence of the conservation of energy and momentum in flat 
space which can be written in terms of the energy-momentum tensor
as $T^{\alpha \beta}_{\phantom{\alpha \beta}, \beta} = 0$ \cite{Hartle}.
This conservation law leads to the identities:
\begin{eqnarray}
T^{tt}_{\phantom{tt},t} &=& - T^{tk}_{\phantom{tk},k} \\
T^{kt}_{\phantom{kt},t} &=& - T^{kl}_{\phantom{kl},l} 
\end{eqnarray}
which can be used to show that:
\begin{eqnarray}
T^{tt}_{\phantom{tt},tt} = T^{kl}_{\phantom{kl},lk}
\end{eqnarray}
where superscript $t$ denotes the zeroth, temporal part of a tensor.
It is then possible to show that (see Sec.~\ref{proofTijInt}) 
\begin{eqnarray}
\label{TijInt}
\int T^{ij} d^{3}x = 
\frac{1}{2} \frac{d^{2}}{dt^{2}} \int x^{i} x^{j} T^{tt} d^{3}x.
\end{eqnarray}

We consider a source with only small velocities. This assumption
called the {\it slow motion approximation} will mean that the frequency
$\Omega$ of any oscillations will be small and therefore that the
wavelength $\lambda$ of the gravitational waves emitted will be large
compared to the source, $\lambda \gg R_{\rm{source}}$. 
Consequently, the slow motion approximation is sometimes equivalently
made as the {\it long wavelength approximation}.
%
Under the slow motion approximation we find the energy-momentum tensor 
is dominated by the $T^{tt}$ component which is itself dominated by the 
rest mass density $\mu$.
This property of the slow motion approximation can be observed simply
by considering a pressureless perfect fluid whose energy-momentum
tensor is given by
$T_{\alpha \beta} = \mu u_{\alpha} u_{\beta}$,
where $\mu$ is the rest mass density of some matter and $u_{\alpha}$ is its
four-velocity.
Under the slow motion approximation we are able to neglect the
three spatial terms of our four-velocity since $u_{i} \ll 1$.

We define {\it mass-quadrupole moment} (also known
as the {\it second mass moment}) as
\begin{eqnarray}
\label{quadmassmoment}
I^{ij}(t) \equiv \int \mu x^{i} x^{j}  d^{3}x
\end{eqnarray}
Using this definition we can rewrite Eq.~(\ref{barhfarzone})
for the metric perturbation as
\begin{eqnarray}
\label{metricquad}
\bar{h}_{ij}(t,\xv) = \frac{2}{D} \ddot{I}^{ij}(t-D) 
\end{eqnarray}
where an overdot represents derivation with respect to time.
We have now derived an expression relating the generation of metric
perturbations to the motion of masses. 
In the derivation of this expression we have made the following 
assumptions: 
i) in order to linearize gravity we have assumed
that the spacetime metric is almost flat and the perturbations to
the metric are small, 
ii) in order to simplify our wave equation solution (Eq.~(\ref{GreenSoln}))
we have assumed that the distance from the observer to the source is
much larger than the size of the source and
iii) in order to simplify the derivation of the metric perturbation
in terms of the mass-quadrupole moment we have assumed that the source
has small velocities.

Considering the relationship between the quadrupole moment and the
metric perturbation Eq.~(\ref{metricquad}) we will consider what
might constitute a source of gravitational waves.
The source must have non-stationary (accelerating) distributions of mass  
or time-derivatives of Eq.~(\ref{metricquad}) ensure no gravitational waves 
will be generated.
Furthermore, a spinning source that has an axisymmetric distribution 
of mass about its spin axis will not emit gravitational waves. Although
the source is non-stationary its mass distribution is stationary in time.
We will see shortly that the weak coupling of gravitational waves to 
matter means that we will require very massive, astrophysical events in
order to generate gravitational waves with large enough amplitude to be
detected by current and planned detectors.
Sources that will emit detectable gravitational waves include 
binary star systems, 
non-axisymmetric explosions of stars
and spinning pulsars with ``mountains'' on their surface.


\subsection{Gravitational wave amplitude}
From dimensional analysis (see e.g. Hartle \cite{Hartle} Chapter 23)
we can estimate the amplitude of gravitational waves.
Considering a source with characteristic mass $M$, period of oscillation
$P$ and size $R$ we approximate $\ddot{I}^{ij} \sim M R^{2} / P^{2}$.
For an observer at a distance $r$ from the source we then have 
\begin{eqnarray}
\label{strainform} 
\bar{h} \sim \left( \frac{M}{r} \right)\left(\frac{M}{P}\right)^{2/3}.
\end{eqnarray} 
Assuming some characteristic values 
we find
\begin{eqnarray}
\label{strainvalue}
\bar{h}^{ij} \sim 10^{-22} 
\left( \frac{M} {10 M_{\odot}}     \right)^{5/3} 
\left( \frac{P} {1 \,{\rm{hour}}}  \right)^{-2/3}
\left( \frac{D} {1 {\rm{Mpc}}}     \right)^{-1}.
\end{eqnarray}

We will find that metric perturbations of size 
$\sim 10^{-21} - 10^{-22}$ will cause strains that are just about
measurable using current laser-interferometric detectors. 
We will discuss these more in Sec.~\ref{sec:detection}. 

\subsection{Gravitational waves emitted by a binary system}
\begin{figure}
\begin{center}
\includegraphics[angle=0, width=0.9\textwidth]{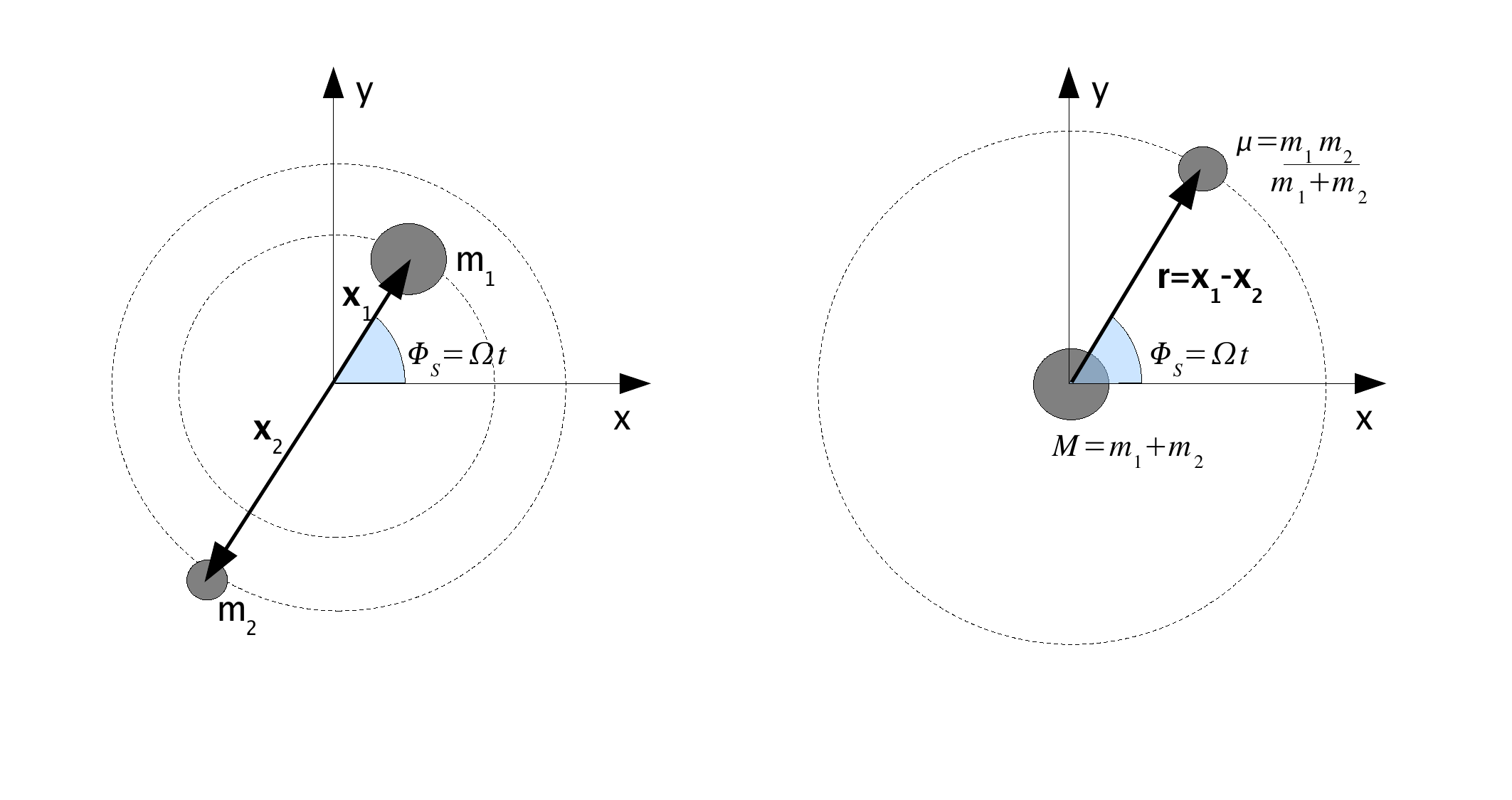}
\caption{The left plot shows a binary with masses $m_{1}$ and $m_{2}$
at positions $\xv_{1}$ and $\xv_{2}$ measured from their common 
centre of mass which we use as the origin. 
The bodies orbit their common centre of mass with orbital frequency
$\Omega$.
The right hand plots shows the equivalent system where we only consider
the motion of a single body with mass equal to the reduced mass 
$\mu  = m_{1} m_{2}/(m_{1} + m_{2}) $ of the binary which orbits the
centre of mass at position $\rv \equiv \xv_{1} - \xv_{2}$ \cite{LL}.}
\label{one_body}
\end{center}
\end{figure}

We will now consider the gravitational waves emitted by a binary
system with bodies of mass $m_{1}$ and $m_{2}$ orbiting their
common centre of mass (which we will take as our origin) 
with position vectors $\xv_{1}$ and $\xv_{2}$.
We will evaluate the mass-quadrupole moment
$I^{ij}$ (Eq.~(\ref{quadmassmoment}))
for the binary by considering the equivalent one body problem.
The equivalent one body problem consists of a body with
mass equal to the reduced mass 
$\mu  = m_{1} m_{2}/(m_{1} + m_{2}) $ of the binary
orbiting the centre of mass at position 
$\rv \equiv \xv_{1} - \xv_{2}$ \cite{LL}.
Figure \ref{one_body} shows this binary and the equivalent
one body system.
By approximating the binary's components as
($\delta-$function) point masses we can simplify the mass-quadrupole
moment and write it as $I^{ij} = \mu r^{i} r^{j}$.

For our binary we will have
\begin{eqnarray}
r^{x}(t) &=& r \cos ( \Omega t ) \\
r^{y}(t) &=& r \sin ( \Omega t ) \\
r^{z}(t) &=& 0
\end{eqnarray}
where $r \equiv |\rv|$.
Taking the time derivative of the mass-quadrupole $I^{ij}$ twice 
and using the centripetal acceleration
$\ddot{r}^{i} = - (|\dot{\rv}^{2}|/|\rv|) \hat{r}^{i}$ we find
\begin{eqnarray}
\bar{h}^{ij} = \frac{4}{D} \mu |\dot{\rv}^{2}| 
\left(
\hat{\dot{r}}^{i} \hat{\dot{r}}^{j} - \hat{r}^{i} \hat{r}^{j}
\right)
\end{eqnarray}
Taking the time derivatives of $\rv$:
\begin{eqnarray}
\dot{r}^{x}(t) &=& - \Omega r \sin ( \Omega t ) \\
\dot{r}^{y}(t) &=&   \Omega r \cos ( \Omega t ) \\
\dot{r}^{z}(t) &=& 0.
\end{eqnarray}
we can then write the metric perturbation as
\begin{eqnarray}
\label{binaryh}
\bar{h}^{ij} = - \frac{4 \Omega^{2} \mu r^{2}}{D}
\left(
\begin{array}{ccc}
\cos[2 \Omega (t-D)] &  \sin[2 \Omega (t-D)] & 0 \\
\sin[2 \Omega (t-D)] & -\cos[2 \Omega (t-D)] & 0 \\
0 & 0 & 0 \\
\end{array}
\right).
\end{eqnarray}

We will briefly discuss the properties of gravitational waves from 
binary systems.
Intuitively we can imagine that as system loses energy
to gravitational waves its orbit will shrink. 
This is referred to as the {\it inspiral} of a binary.
Note that in Newtonian gravity, no gravitational waves would be emitted, 
the system would not lose energy and the inspiral would not occur.
From Kepler's third law, the shrinkage of the binaries orbit 
will cause the period to decrease.
From Eq.~(\ref{strainform}) we see that as the period decreases the 
gravitational wave amplitude will increase.
From Eq.~(\ref{binaryh}) we see that the gravitational wave frequency 
is proportional (twice) to the frequency of sources orbit 
\footnote{Note that this is an approximation. In reality the gravitational
wave will contain many harmonics of the orbital frequency. 
In neglecting the higher harmonics we consider only the {\it restricted}
waveform.}.
Therefore, as period decreases the orbital frequency and therefore
gravitational wave frequency will also increase.
Consequently gravitational waves emitted during the inspiral of
a binary system is described as chirp since they increase in both
amplitude and frequency with time.

\section{Detection of gravitational waves}
\label{sec:detection}
We consider freely-falling test masses (i.e., with no force applied).
The co-ordinate position of the freely-falling test masses will 
remain constant as a gravitational wave passes.
However, since the metric changes we can observe a change
in the {\it proper} distance between two freely falling test masses.
Initially we consider only the $+$ polarization components of the metric
perturbation in Eq.~(\ref{metricTT}). 
Remembering the form of the metric with only small perturbations
$g_{\alpha \beta}(\xv) = \eta_{\alpha \beta} + h_{\alpha \beta}(\xv)$
we can write the proper separation $ds$ in terms of the
co-ordinate separation $dt, dx, dy, dz$ between two events as
\begin{eqnarray}
ds^{2} &=& g_{\alpha \beta}(\xv) dx^{\alpha} dx^{\beta} \\
&=&
- dt^{2}
+ [1 + h_{xx}] dx^{2} 
+ [1 - h_{xx}] dy^{2}
+ dz^{2}
\end{eqnarray}
for a plus polarized gravitational wave propagating in the $z$-direction.

Now we consider a freely-falling test mass initially at a 
co-ordinate distance $L_{x(\rm{co-ord})}$ along the $x$-axis from the origin. 
We evaluate the proper distance between them in the $x$-direction 
(at time $t$ at $z=0$):
\begin{eqnarray}
L_{x}(t) = \int_{0}^{L_{x(\rm{co-ord})}} [1 + h_{xx}(t,0)]^{1/2} dx
\sim 
L_{x(\rm{co-ord})} 
\left[
1 + \frac{1}{2} h_{xx}(t,0)
\right]
\end{eqnarray}
where we have used the binomial expansion to approximate the right hand side.
The time-dependent variation in the proper distance between test masses 
along $x$-axis is given by
\begin{eqnarray}
\label{timedepL}
\delta L_{x}(t) = \frac{1}{2} L_{x(\rm{co-ord})} h_{xx}(t,0). 
\end{eqnarray}
Note that in flat space ($h_{\alpha \beta} = 0$) the co-ordinate 
separation $L_{x(\rm{co-ord})}$ will be equal to the (constant) proper 
distance $L_{x}$ between the particles along the $x$-axis
(since $\eta_{ii} = 1$).
Rewriting Eq.~(\ref{timedepL}) as 
\begin{eqnarray}
\frac{\delta L_{x}(t)}{L_{x(\rm{co-ord})}} = \frac{1}{2} h_{xx}(t,0)
\end{eqnarray}
we identify the left hand side as a dimensionless {\it strain} along
the $x$-axis caused by the passing of the gravitational wave.
We can generalise this to
\begin{eqnarray}
\label{genstrain}
\frac{\delta L(t)}{L} = \frac{1}{2} h_{ij}(t,0) n^{i} n^{j}
\end{eqnarray}
where $\nv$ is a unit vector in the $x-y$ plane and $L$ would be the proper
distance in flat space (equal to the co-ordinate separation).

\begin{figure}
\begin{center}
\includegraphics[angle=0, width=1.1\textwidth]{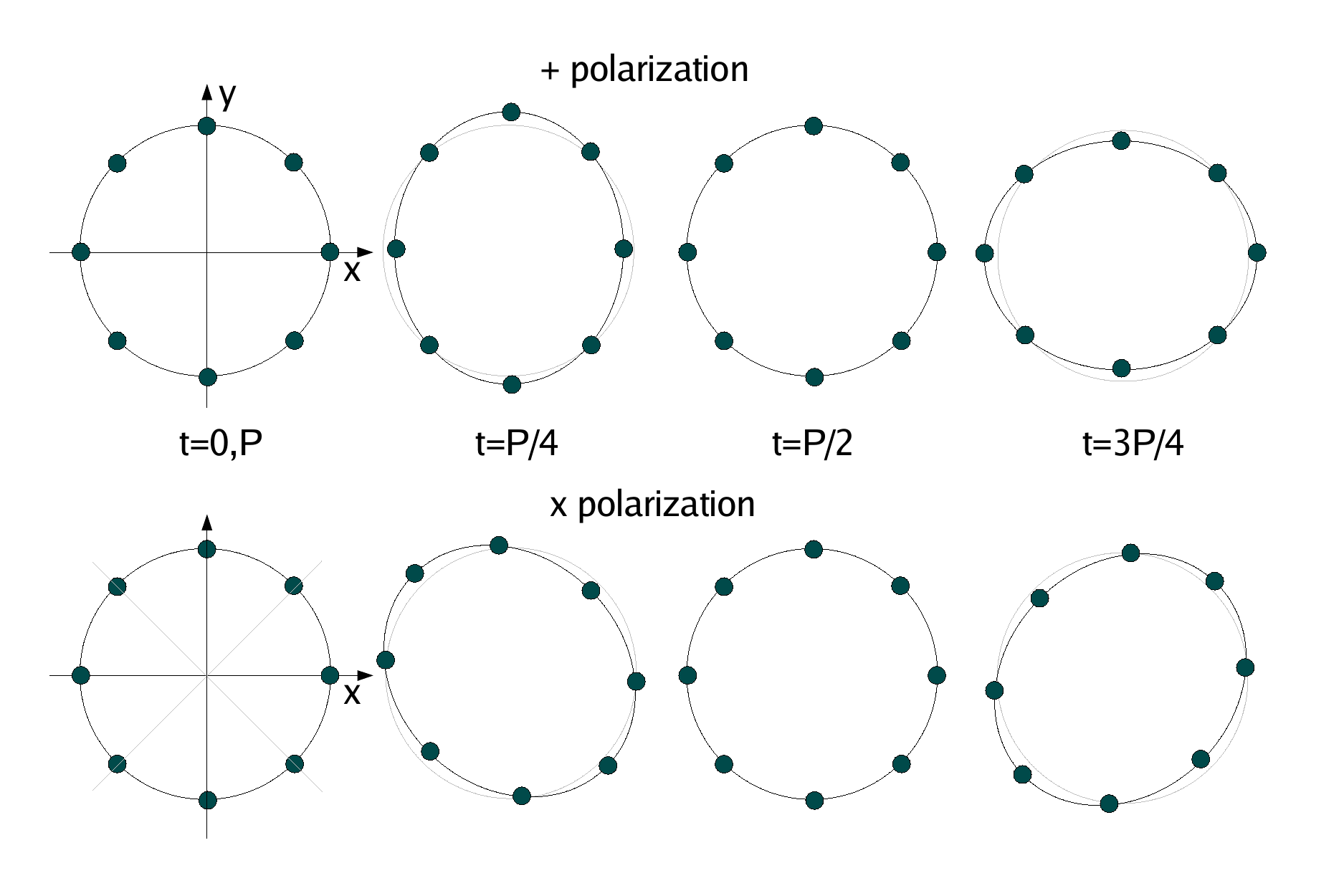}
\caption{Plots showing the change in positions of a ring of test masses 
in the $x-y$ plane as gravitational wave propagates in the $z$-direction. 
The top plot shows the effect of a plus ($+$) polarization gravitational
wave.
Over the course of a single period $P$ of this gravitational wave the ring of 
test masses is contracted in the $x$ direction and simultaneously 
expanded in the $y$ direction (at $P/4$)
direction and then expanded in the $x$ direction and simultaneously
contracted in the $y$ direction (at $3P/4$).
The bottom plot shows the effect of a cross ($\times$) polarization
gravitational wave. 
Its effect on the ring of masses is equivalent to the plus polarization
waveform rotated through $45^{\circ}$.
In this plot the expansion and contraction of the ring of masses has been
exaggerated and is far greater than we would expect from a typical
gravitational wave.}
\label{test_masses}
\end{center}
\end{figure}

The strains caused by the plus polarization part of the gravitational wave 
emitted by a binary system (see Eq.~(\ref{binaryh})) and propagating in the 
$z$-direction are given by:
\begin{eqnarray}
\frac{\delta L_{x}(t)}{L_{x(\rm{co-ord})}} &=& 
- \frac{2 \Omega^{2} \mu r^{2}}{D_{z}} \cos[2 \Omega (t-D_{z})] \\
\frac{\delta L_{y}(t)}{L_{y(\rm{co-ord})}} &=&
+ \frac{2 \Omega^{2} \mu r^{2}}{D_{z}} \cos[2 \Omega (t-D_{z})] \\
\frac{\delta L_{z}(t)}{L_{z(\rm{co-ord})}} &=& 0.
\end{eqnarray}
As expected, since we are are in the Transverse Traceless gauge we have 
no strain in the direction of the waves propagation (i.e., no longitudinal
strain) and we have (sinusoidal) oscillations in the plane transverse to
the waves propagation.  
Note the difference in sign in the strains caused along the
$x$ and $y$ directions. 
This indicates that as the gravitational wave causes proper
distances in the $x$-direction to increase it simultaneously causes
proper distances in the $y$-direction to decrease (and vice versa).
The top plot of Fig.~\ref{test_masses} shows the effect of a plus 
polarized gravitational wave propagating in the $z$-direction on a 
ring of freely falling test masses.

For a cross polarized gravitational wave propagating in the $z$-direction
we can write the proper separation $ds$ in terms of the
co-ordinate separation $dt, dx, dy, dz$ between two events as
\begin{eqnarray}
\label{dscross}
ds^{2} &=& g_{\alpha \beta}(\xv) dx^{\alpha} dx^{\beta} \\
&=&
- dt^{2}
+ dx^{2}
+ dy^{2}
+ 2h_{xy} \, dx \, dy
+ dz^{2}.
\end{eqnarray}
We will now show that a cross polarized gravitational wave will 
have similar effect on a ring of freely falling test masses as a 
plus polarized gravitational wave if we rotate our axes by $45 ^{\circ}$.
Consider rotating the $x$ and $y$ axes through $45 ^{\circ}$ about the
$z$-axis:
\begin{eqnarray}
x \rightarrow x' &=& \frac{1}{\sqrt{2}} (x+y) \\
y \rightarrow y' &=& \frac{1}{\sqrt{2}} (x-y) 
\end{eqnarray}
which lead to the identities 
\begin{eqnarray}
2 dx dy &=& dx'^{2} - dy'^{2} \\
dx^{2} + dy^{2} &=& dx'^{2} + dy'^{2}. 
\end{eqnarray}
Rewriting the proper separation (Eq.~(\ref{dscross})) using these identities
we find it has the same form as the proper separation caused by a plus
polarized gravitational wave in un-rotated axes:
\begin{eqnarray}
ds^{2} &=&
- dt^{2}
+ [1 + h_{xy}] dx'^{2}
+ [1 - h_{xy}] dy'^{2}
+ dz^{2}
\end{eqnarray}
The bottom plot of Fig.~\ref{test_masses} shows the effect of a cross
polarized gravitational wave propagating in the $z$-direction on a
ring of freely falling test masses.

\subsection{Gravitational wave detectors}
\label{sec:gwdetectors}
The search for gravitational waves is dominated by two different types of 
detector, resonant bars and laser-interferometers.
Resonant bar detectors typically consist of a massive metal cylinder which
has been cryogenically cooled. 
A passing gravitational wave will cause stretching and contraction of the
bar which can be measured 
(see Mauceli et al. (1996) \cite{MauceliAllegro} for a description
of the Allegro detector).
These detectors have best sensitivity to gravitational waves with
frequencies in a narrow band about their own resonant frequencies, 
typically $\sim 900$ Hz (see Table 1 of Astone et al. (2003) \cite{PhysRevD.68.022001}).
We will find that many sources of gravitational waves including the inspiral of
binaries will emit across a wide range of frequencies. 
Whereas resonant bar detectors achieve good sensitivity over only a 
relatively narrow band of frequencies, 
laser interferometers have good 
sensitivity over a broad band of frequencies and it is these detectors 
that we shall focus upon.

Despite not being ideal for searches for gravitational waves from the
inspiral of binaries, resonant bars have been used for searches
for gravitational waves with unknown form and/or short duration and bandwidth. 
For a review of gravitational wave searches using resonant bar detectors 
see Astone et al. (2003) \cite{PhysRevD.68.022001}.
Recent searches for gravitational wave stochastic background and 
short duration gravitational wave bursts using both resonant bar and 
laser interferometers are described in
Abbott et al. (2007) \cite{abbottLIGOALLEGRO}
and Baggio et al. (2008) \cite{baggioLIGOAURIGA} (see Fig.~2 of this
paper for a comparison of the sensitivities of these different types
of detector).

\begin{figure}
\begin{center}
\includegraphics[angle=0, width=1.1\textwidth]{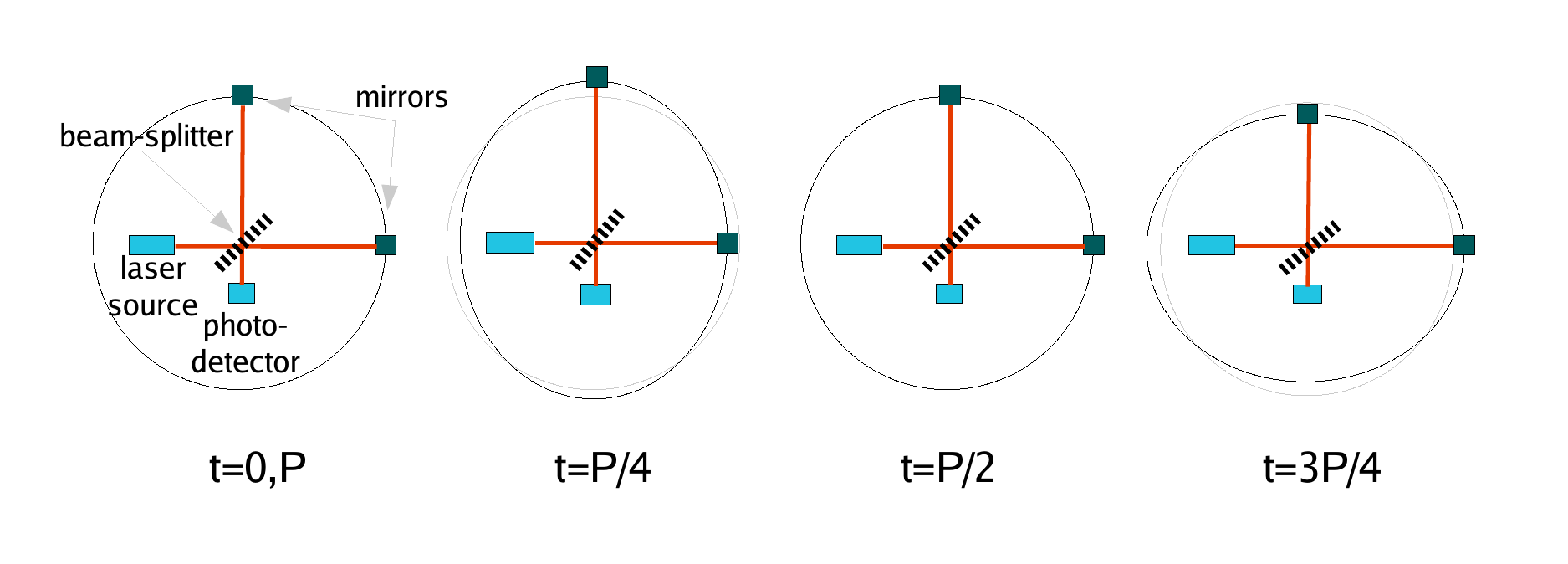}
\caption{Plot showing the effect of a plus polarization gravitational 
wave on a simple Michelson interferometer. 
The gravitational wave causes the interferometers mirrors to move similarly to 
the test masses in the upper plot of Fig.~\ref{test_masses}. 
The interferometer is designed so that when it is in its unperturbed 
configuration the laser beams reflected along the $x$ and $y$ arms will
destructively interfere when recombined at the beam-splitter 
(at $t=0,P/2,P \dots$) and a dark fringe will be measured by
the photodetector.
A passing gravitational wave would cause variation in the proper distance
between the beam-splitter and the mirrors
enabling detection of the gravitational wave through measurement of
the intensity of the recombined laser beam.
}
\label{ifo_detection_1}
\end{center}
\end{figure}

A Michelson interferometer with arms along the $x$ and $y$
directions is shown in the upper plot of Fig.~\ref{ifo_detection_1}.
The interferometer works as follows: the laser source sends
a laser beam to a beam-splitter which splits it into two coherent
beams which then travel at right angles to each other along the
interferometers arms. 
The laser beams are reflected back by mirrors at the end of
each arm and are recombined at the beam-splitter which then
directs the recombined beam to a photodetector which measures its intensity.

The two mirrors and the beam-splitter behave similarly to the
test masses shown in Fig.~\ref{test_masses} and move accordingly
with the passing of a gravitational wave.
We measure the movement of the two mirrors and the beam-splitter 
through the intensity of the recombined laser beam measured at the
photodetector.
The real gravitational wave 
detectors that we will discuss shortly are designed so that when there is no
gravitational wave (i.e, the mirrors have proper distances
$L_{x} = L_{y}$ from the beam-splitter) the laser beams
interfere destructively and we measure a dark fringe at the 
photodetector.

Constructive interference will occur when the difference in the path
travelled by the laser is $\Delta L = n \lambda$ where
$\lambda$ is the wavelength of the laser (assumed to be
monochromatic) and $n = 0,1,2 \dots$.
Destructive interference occurs when
$\Delta L = (n + 1/2) \lambda$.
The path difference between the laser beams travelling along
the $x$ and $y$ arms can be written
\begin{eqnarray}
\Delta L = 2 L_{x} - 2 L_{y} - \frac{\lambda}{2}
\end{eqnarray}
where $L_{x}$ and $L_{y}$ are the proper distances of the
mirrors from the beam-splitter (the prefactor of 2 indicating
that the laser beam makes a return trip) and the subtraction
of $\lambda/2$ ensures we have destructive interference
when $L_{x} = L_{y}$
\footnote{Note that in real ground-based interferometric detectors
such as LIGO (discussed shortly) the optical configuration is
maintained so that the photodetector is kept at a dark fringe.
The feedback signal, known as the error signal, required to
maintain this configuration is what is measured and used to
infer the passing of a gravitational wave. The LIGO and GEO
detectors are detailed in Abbott et al. (2004) \cite{abbott-2004-517}.
}.

From our equations for the strain caused by a passing gravitational
wave (e.g., Eq.~(\ref{genstrain})) we can see that by increasing the length
of the interferometers arms ($L$) we will increase the strain we are seeking
to measure ($\delta L(t)$). 

From Eq.~(\ref{dscross}) for the proper separation caused by a cross 
polarization gravitational wave we can see that the it will
not be detectable by the interferometer we have shown in 
Fig.~\ref{ifo_detection_1}: the strain
it induces will cause the $x$ and $y$ arms of the interferometer to extend
and compress equally and at the same time as each other. Therefore the path
travelled by the laser beams will remain equal 
$L_{x}(t) = L_{y}(t)$ and we would always measure a dark fringe at the
photodetector.
Equally, if we rotated the interferometer in Fig.~\ref{ifo_detection_1}
by $45^{\circ}$ it would be sensitive to only cross polarization gravitational 
waves but not to plus polarization waves.



\subsection{Characterising the detectors}
\begin{figure}
\begin{center}
\includegraphics[angle=0, width=0.8\textwidth]{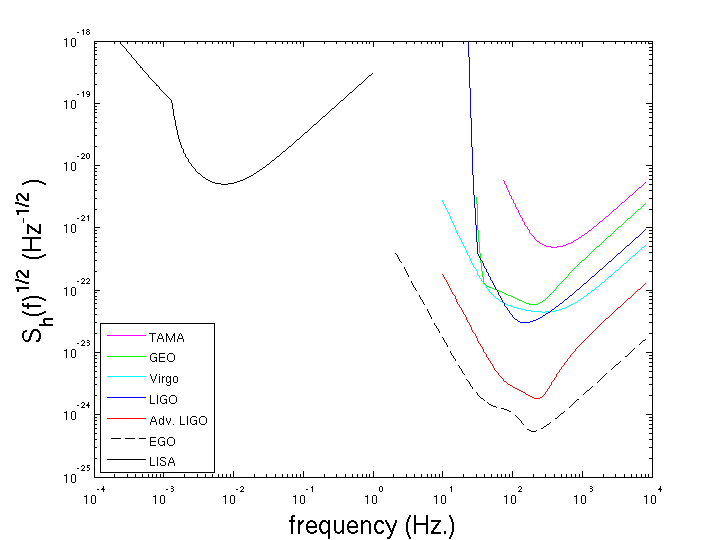}
\caption{Amplitude spectra of current (TAMA, GEO, LIGO, Virgo) and planned 
(Advanced LIGO, EGO, LISA) laser-interferometric gravitational
wave detectors at their design sensitivities.
Fits to the TAMA, GEO and LIGO data were published in 
Damour et al. (2001) \cite{DIS2001}.
The noise curve data was provided by M.-K. Fujimoto (TAMA), 
G.Cagnoli and J. Hough (GEO) and K. Blackburn (LIGO).
The Virgo noise curve data was provided by  J-Y. Vinet 
(available on Virgo home page \cite{VirgoHomePage}).
The Advanced LIGO noise curve data was provided by
Kip Thorne and the fit by B.S.Sathyaprakash.
The EGO noise curve is given by Van Den Broeck and Sengupta (2007) \cite{broeck-2007-24}.
The LISA noise curve is given by Barack and Cutler (2004) \cite{leor04}.
}
\label{PSDcurves_all}
\end{center}
\end{figure}

\begin{figure}
\begin{center}
\includegraphics[angle=0, width=0.8\textwidth]{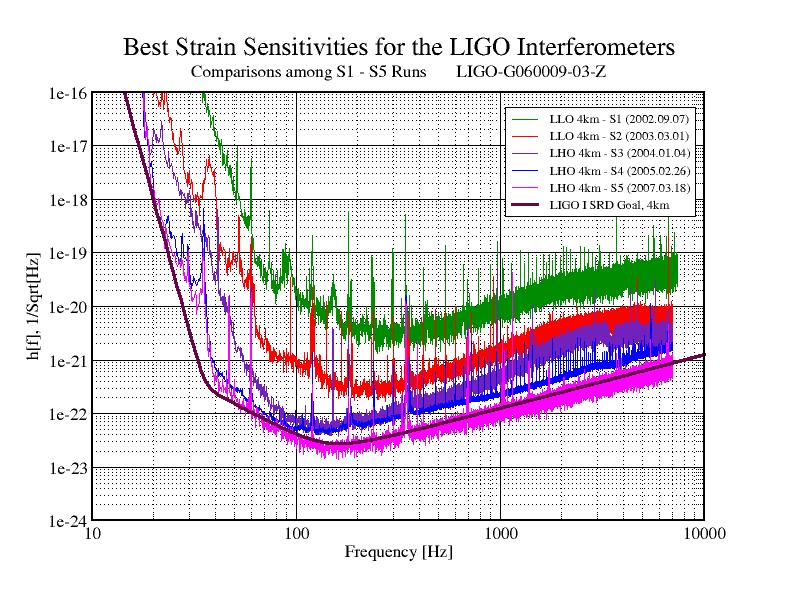}
\caption{Plot showing the best (lowest) amplitude spectra obtained by LIGO during each 
of its first five science runs. The design sensitivity curve is also shown.
We see a steady improvement in LIGO's best sensitivity as we progress through the science runs
and in November 2005, during its fifth science run (S5), 
LIGO achieved its design sensitivity above $\sim 50$ Hz.
In contrast to the smooth shape of the design sensitivity curve, the real spectra
include sharp spikes in which the detector has reduced sensitivity over a narrow
band of frequencies.
These narrow-band spectral lines are caused by vibrations in the wires used to
suspend the interferometer's mirrors (``violin modes''), laser noise and harmonics
to the U.S. power mains frequency of $60$ Hz \cite{cuoco-2001-64}.
Methods for removing these lines are described in Searle et al. (2003) \cite{searle-2003-20}.
This figure was created by the LIGO Laboratory and has been assigned LIGO document 
number LIGO-G06009-03-Z \cite{RefLIGO_sensitivities_S1-S5}.
}
\label{LIGO_sensitivities_S1-S5}
\end{center}
\end{figure}

We characterise gravitational wave detectors by their power or amplitude
spectral density.
$S_{h}(f)$ is the noise power spectral density per $\rm{Hz}$ of a data stream.
$S_{h}^{\rm{one-sided}}(f) =  2 S_{h}^{\rm{two-sided}}(f)$.
The amplitude spectral density $S_{h}(f)^{1/2}$ is the square root of 
the power spectral density and has units $\rm{Hz}^{-1/2}$.
We will discuss the power density in the context of data analysis in 
Sec.~\ref{sec:DA}. 
Figure \ref{PSDcurves_all} shows the amplitude spectral density curves
for a number of current and planned laser interferometric gravitational
wave detectors.
Figure \ref{LIGO_sensitivities_S1-S5} shows the best amplitude spectral density
curves obtained by LIGO during each of its first five science runs.
Lower values of amplitude spectral density indicate sensitivity to smaller
strains and
we shall see that $S_h(f)$ appears in the denominator of our equation 
for signal to noise ratio (see Sec.~\ref{sec:DA}). 

From our equations for the emission of gravitational waves (see e.g.,
Eq.~(\ref{binaryh})) we can see that the amplitude of the strain caused
will be proportional to the inverse of the distance from the source
to the detector. 
Therefore, sensitivity to smaller strain means sensitivity to more
distant sources.
Improvements in sensitivity (i.e., reductions in $S_{h}(f)$) by a factor $x$
would lead to a proportional increase in the distance to which a given source
could be observed with a particular strain and therefore a factor $x^3$
increase in the volume to which we could observe the source.

In this thesis we present results from the analysis of data 
collected by the Laser Interferometer Gravitational-wave observatory
(LIGO) and develop
an algorithm to be used with data collected by 
the 
Laser Interferometer Space Antenna (LISA). 
We will now briefly describe these detectors.


\subsection{LIGO}
\label{subsecLIGO}
The Laser Interferometer Gravitational-wave Observatory (LIGO) consists of 
three detectors located at two sites across the US. 
The LIGO Hanford Observatory (LHO) in Washington state consists of two 
co-located interferometers of arm length 4km and 2km and are known as 
H1 and H2 respectively.
The LIGO Livingston Observatory (LLO) in Louisiana consists of a single 4km
interferometer known as L1.
See Abbott et al. (2004) \cite{abbott-2004-517} for a more detailed
description of the LIGO detectors.

\begin{figure}
\begin{center}
\includegraphics[angle=0, width=0.8\textwidth]{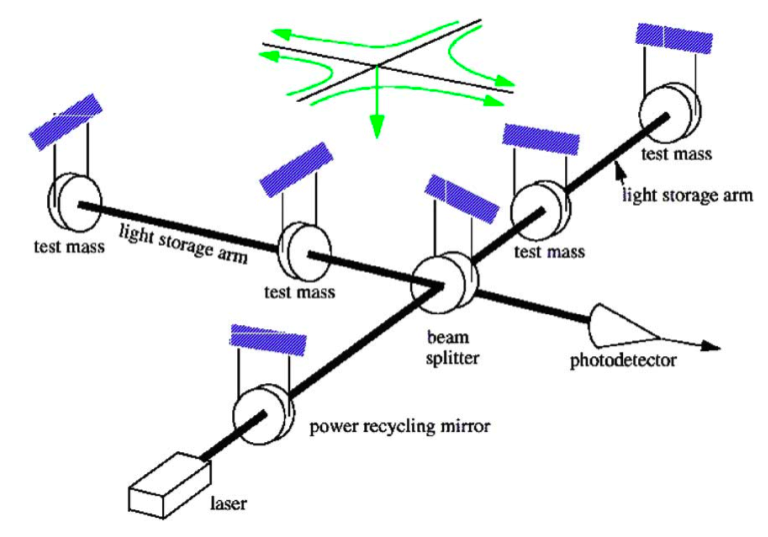}
\caption{
A schematic layout of a LIGO interferometer showing Fabry-Perot optical cavities
and power recycling (see Sec.~\ref{subsecLIGO}).
This figure was reproduced from B. Abbott et al., Nucl. Instrum. Meth.
A 517 (2004) 154-179 \cite{abbott-2004-517} with permission from the authors.
}
\label{fig:LIGOschematic_screenGrab}
\end{center}
\end{figure}

The sensitivity of ground-based laser interferometric detectors is primarily
limited by three different sources of noise, seismic disturbances at low frequencies, 
thermal noise at intermediate frequencies and shot noise, caused by statistical
fluctuations in the laser power, at high
frequencies.
For a detailed breakdown of the various sources of noise which
contribute to LIGO's amplitude spectrum see 
Sigg (2008) \cite{SiggLIGOStatus}. 
Figure \ref{fig:LIGOschematic_screenGrab} shows a schematic layout
of a LIGO interferometer.
The main additions to the LIGO interferometers beyond the simple Michelson
interferometer described in Sec.~\ref{sec:gwdetectors} are 
i) the second set of test mass mirrors along the interferometer arms which form a 
Fabry-Perot optical cavity with the test mass mirrors at the ends of the arms 
and ii) the power recycling mirror between the beam-splitter and the laser source.
The goal of these extra mirrors is to increase the time that the laser beam 
spends in each of the interferometer's arms.
When the interferometer is ``locked'' into resonance, i.e., its mirrors are positioned
correctly, the laser beam will bounce back and forth $\sim 50$ times in the optical
cavity in each arm.  
This effectively increases the arm lengths of the interferometer and therefore improves
its strain sensitivity (see, for example, Eq.~\ref{timedepL}) \cite{LIGOH2FirstLock}.
When the mirrors are not correctly positioned we described the interferometer
as being ``unlocked'' (see Sec.~\ref{sec:dataselection}).
When the interferometer is locked and the arms are not being disturbed by environmental
noise or a passing gravitational wave, almost all of the laser light will return from
the arms to the beam-splitter and back towards the laser source.
The power recycling mirror reflects this laser light back towards the beam-splitter and
into the arms of the interferometer, effectively increasing the laser power by a factor of 
$\sim 40$ \cite{LIGO-CoherentOptics-P070058-A} which will reduce the level of shot noise
\cite{abbott-2004-517}. 

Construction of LIGO began in 1994 and was substantially completed in 2000. 
During October 2002 LIGO and GEO took part in the first science run (S1)
\cite{abbott-2004-517}.
No gravitational waves were observed. 
Although neither detector had achieved their design sensitivities (see
Fig.~\ref{LIGO_sensitivities_S1-S5}), LIGO had sufficiently good sensitivity to be able
to set a better (i.e. lower) upper limit on the rate of coalescences of
binary neutron star inspirals than previous experiments 
\cite{LIGOS1iul}
(the process of setting upper limits on the rate of coalescences in the
event that no gravitational waves were observed is discussed later in 
Sec.~\ref{sub:upperlimit}). 
In November 2005 LIGO achieved its design sensitivity above $\sim 50$ Hz.
In this thesis we will describe a search of LIGO data taken during its
third science run (S3) which took place between October 2003 and January 2004.

\subsection{LISA}
\label{subsec:LISA}
The Laser Interferometer Space Antenna (LISA) will consist of 
three spacecraft in heliocentric Earth-trailing orbits, 
$5$ million kilometres apart at the corners of an 
(approximately) equilateral triangle (see Danzmann K et al. (1998) 
\cite{LISAppa} for a full description of the mission).
Each of LISA's spacecraft house freely falling test masses.
A passing gravitational wave will change the (proper) distance
between these test masses. 
There will be two lasers running between each pair of spacecraft, 
one in each direction, and, similarly to ground-based detectors such as LIGO, 
it is the differences in laser phase between the various light travel paths 
that indicate that gravitational waves are passing through the detector. 

However, unlike ground-based detectors LISA will not suffer from low frequency
noise caused by seismic activity and has been designed to have best sensitivity
in the frequency range $\sim 10^{-4} - 10^{-1} \rm{Hz}$. 
In the raw data, the laser phase difference is totally dominated by laser 
frequency noise. 
However, this can be suppressed without eradicating the gravitational wave signal 
using Time Delay Interferometry (TDI, see for instance Vallisneri (2005) 
\cite{vallis05} and references therein).

LISA is a joint NASA/ESA project and is one of five space-based observatories 
that form NASA's Beyond Einstein programme. 
After the last review (2007) \cite{BEPAC} the LISA Pathfinder mission, a 
precursor mission to LISA designed to test its key technologies, is expected
to be launched in 2009. 
While no firm date has been set for the launch of LISA itself it is hoped to
be within the next decade or so.
Once launched LISA will spend around 13 months getting into its orbit and will
then collect data for between 3 and 5 years. 

In Sec.~\ref{ch:emri} we find that through use of time-frequency data analysis 
techniques LISA will be sensitive to the inspiral of stellar mass objects into
supermassive black holes up to distances of a few Gpc, 
the merger of supermassive black holes at cosmological distances $z \sim 3.5$
and the inspiral of binary white dwarfs in the nearby universe.  

\section{Data analysis}
\label{sec:DA}
In this Section we will describe the data analysis methods used in order
to detect a gravitational wave signal in noisy data.
We will consider a data stream $x(t)$ which may either contain only noise $n(t)$
or noise and a gravitational wave signal $s(t)$. 
We discretely sample the data stream with an interval $\Delta t$ so 
that $x_{j} = x(t_{j})$ where $t_{j} = j \Delta t$.

Our data analysis can be viewed within the framework of a hypothesis test.
We have two hypotheses: 
\begin{itemize}
\item $\mathcal{H}_{0}$: our null hypothesis is that there
is no signal present, $x(t) = n(t)$
\item $\mathcal{H}_{1}$: a signal is present in the data, $x(t) = n(t) + s(t)$ 
\end{itemize}

There are two types of error associated with this test:
\begin{itemize}
\item Type I error: rejecting $\mathcal{H}_{0}$ when it is true. In this case 
our analysis would infer a signal was present when there was no signal
present. We refer to this type of error as a {\it false alarm}.
\item Type II error: accepting $\mathcal{H}_{0}$ when it is false. In this
case our analysis would not infer a signal was present when a signal was present.
We refer to this type of error as a {\it false dismissal}.
\end{itemize}

It is not possible to decrease the probability of false alarm and false dismissal 
simultaneously; decreasing the probability of a false alarm would 
increase the probability of a false dismissal and vice versa.

We can approach the problem of choosing a detection method in two different ways.
When taking the Neyman-Pearson approach, the probability of false dismissal is minimized
having chosen a particular value for the false alarm probability.
When taking the Bayesian approach, the probability of the null hypothesis is estimated
in advance and penalties are assigned to describe the relative
severity of false alarms and false dismissals occurring. These pieces of
information are used to construct the Bayes risk which is subsequently minimized
(see, for example, Whalen (1971) \cite{whalen:1971}).

Significantly both approaches yield a {\it likelihood ratio test} of the form:
\begin{eqnarray}
\mathcal{H}_{0} &&  \; \mathrm{if} \;\; \Lambda < \gamma \\
\mathcal{H}_{1} &&  \; \mathrm{if} \;\; \Lambda \geq \gamma \\
\mathrm{where} && \; \Lambda = \frac{p(x; \mathcal{H}_{1})} {p(x; \mathcal{H}_{0})}
\end{eqnarray}
where $p(x;y)$ is the probability of $x$ occurring given that $y$ is true and where 
$\gamma$ is some thresholding value. The form of this threshold will depend on
whether the Neyman-Pearson or Bayesian approach is taken. The quantity
$\Lambda$ is called the likelihood ratio.

We will now consider the case where the noise $n(t)$ is Gaussian process with a mean of zero,
i.e., $\overline{n(t)} = 0$ where we use an overbar to denote ensemble average. 
The noise can be characterised equivalently by either its autocorrelation $r(t-t')$ or
by its (one-sided) power spectral density $S_{n}(f)$. Indeed, the Wiener--Khinchin theorem 
(also known as the Wiener--Khintchine or Khinchin-–Kolmogorov theorem) shows that for any 
stationary process (i.e., one which can be described at any time by the same probability
distribution) the power spectral density is simply the Fourier transform of its
autocorrelation function.

The real, one-sided noise power spectral density is given by
\begin{equation}
\overline{\tilde{n}(f) \tilde{n}^{*}(f')}
= \frac{1}{2} \delta(f-f') S_{n}(|f|).
\end{equation}
In simple terms, the autocorrelation function simply measures the 
correlation between $n(t)$ at two different times.

The multivariate Gaussian probability density function 
of our data when there is no signal present
(i.e., $x(t) = n(t)$ and $\overline{x(t)} = \overline{n(t)} = 0$)
can be written
\begin{eqnarray}
p(\xv;\mathcal{H}_{0}) =
\frac{1}{ (2 \pi )^{N/2} |\Cv|^{1/2}}
\mathrm{exp} \left[
-\frac{1}{2} (\xv)^{T} \Cv^{-1} (\xv)
\right]
%
\end{eqnarray}
where $\Cv$ is the covariance matrix of the $x_{j}$ and $|\Cv|$ is the
determinant of $\Cv$. 
Following the derivation in Section 2A of Finn (1992) \cite{Finn92} we
find that through use of the Wiener--Khinchin theorem and Parseval
theorem that we can write this probability in the continuum limit
as
\begin{eqnarray}
p(\xv;\mathcal{H}_{0}) = 
\frac{1}{ (2 \pi )^{N/2} |C_{n,ij}|^{1/2}}
\mathrm{exp} [
-\left<x,x \right>)
]
\end{eqnarray}
where we have defined the (symmetric) inner product for {\it any} two 
functions $g$ and $h$ to be
\begin{eqnarray}
\label{inner_any}
\left<g,h \right> \equiv \int_{-\infty}^{\infty}
\left[
\tilde{g}(f) \tilde{h}^{*}(f) + \tilde{h}(f) \tilde{g}^{*}(f)
\right]
\frac{df}{S_{n}(f)}.
\end{eqnarray}
For a real signal $h(t)$ is real we have $\tilde{h}^{*}(f) = \tilde{h}(-f)$
\footnote{To show that $\tilde{h}^{*}(f) = \tilde{h}(-f)$ when $h(t)$ is real
write the (forward) Fourier transform in the form 
$\tilde{h}(f) = 
\int h(t) \cos(2 \pi f t) dt -   
i \int h(t) \sin(2 \pi f t) dt$. 
If $h(t)$ is real, we obtain $\tilde{h}^{*}(f)$ by
inverting the sign of the second term which is wholly imaginary. 
Since $\cos$ is an even function and $\sin$ is an odd function we
can obtain the same expression for $\tilde{h}^{*}(f)$ by replacing
$f$ with $-f$ in our original equation for $\tilde{h}(f)$ and thereby
show that $\tilde{h}^{*}(f) = \tilde{h}(-f)$.
}.
If both $h(t)$ and $g(t)$ are real we can write
\begin{eqnarray}
\label{inner_real_BCV}
\left<g,h \right> &=& 
2 \int_{-\infty}^{\infty}
\tilde{g}^{*}(f) \tilde{h}(f)
\frac{df}{S_{n}(|f|)} \\
&=& 4 \Re
\int_{0}^{\infty}
\tilde{g}^{*}(f) \tilde{h}(f)
\frac{df}{S_{n}(f)}.
\end{eqnarray}
For real functions $h(t)$ and $g(t)$ we can also equivalently write 
\begin{eqnarray}
\label{inner_real_AISS}
\left<g,h \right> \equiv 2 \int_{0}^{\infty} 
\left[
\tilde{g}(f) \tilde{h}^{*}(f) + \tilde{h}(f) \tilde{g}^{*}(f)
\right]
\frac{df}{S_{n}(f)}.
\end{eqnarray}

Since we know that $p(x;\mathcal{H}_{1}) \equiv p(x-s;\mathcal{H}_{0})$
we can write
\begin{eqnarray}
p(\xv;\mathcal{H}_{1}) =
\frac{1}{ (2 \pi )^{N/2} |\Cv|^{1/2}}
\mathrm{exp} \left[
-\frac{1}{2} (\xv)^{T} \Cv^{-1} (\xv)
\right]
%
\end{eqnarray}
Rewriting the inner product
\begin{eqnarray}
\left< x-s,x-s \right>
= \left< x,x \right>
+ \left< s,s \right>
- 2 \left< s,x \right>
\end{eqnarray}
we can find the likelihood ratio 
\begin{eqnarray}
\Lambda = \mathrm{exp}[2 \left< s,x \right> - \left< s,s \right>].
\end{eqnarray}
The inner product of the signal with itself $\left< s,s \right>$ is 
clearly not dependent on the data $x(t)$ and we can choose to rewrite our
statistical test using the likelihood ratio with this term removed. Also
since our expression for the likelihood ratio will then be a monotonic
function of the exponent we can go further and rewrite our test as
\begin{eqnarray}
\mathcal{H}_{0} &&  \; \mathrm{if} \;\; \left< s,x \right> < \gamma_{*} \\
\mathcal{H}_{1} &&  \; \mathrm{if} \;\; \left< s,x \right> \geq \gamma_{*} 
\end{eqnarray}
where $\gamma_{*}$ is some thresholding value.

Matched-filtering is the optimal technique for the detection of a known
signal in stationary, Gaussian noise. 
The optimal filter $q = h / S_{n}(f)$ consists
of an accurate representation of the expected signal, which we
call the template $h$, weighted by the noise spectrum of the detector 
$S_{n}(f)$ so that there are greater contributions to the inner product
$\left<x, h \right>$ when the detector has good sensitivity 
(i.e., when $S_{n}(f)$ is small) \cite{babak:2006}.

\subsection{Properties of the inner product}
\label{matchedfilterresponse}
The mean of $\left< x,h \right>$ for an ensemble of $x$ is given by
\begin{eqnarray}
\overline{ \left< x,h \right>  } = 
2 \int_{-\infty}^{\infty}
\overline{\tilde{x}^{*}(f)}   \tilde{h}(f)
\frac{df}{S_{n}(f)}
\end{eqnarray}
where since the template $h$ is stationary we have 
$\overline{\tilde{h}(f)} = \tilde{h}(f)$.

In the absence of signal $x(t)=n(t)$ we find
\begin{eqnarray}
\overline{ \left< n,h \right>  } =
2 \int_{-\infty}^{\infty}
\overline{\tilde{n}^{*}(f)}  \tilde{h}(f)
\frac{df}{S_{n}(f)}
= 0
\end{eqnarray}
as long as $\overline{n(t)} = 0$. 
The variance of $\left< x,h \right>$ for an ensemble of $x$ is
given by
\begin{eqnarray}
\overline{ 
\left[ \left< x,h \right> - \overline{ \left< x,h \right>  } \right]^{2}}.
\end{eqnarray}
Again, assuming there is no signal we find
\begin{eqnarray}
\overline{
\left[ \left< n,h \right> - \overline{ \left< n,h \right>  } \right]^{2}}
= \overline{\left[ \left< n,h \right> \right]^{2}}
\end{eqnarray}
since we have found previously that $\overline{ \left< n,h \right>} = 0$.
From Eqs.~(\ref{inner_any}) and (\ref{inner_real_AISS}) we can see that 
$\left < a, b \right> \equiv \left < b, a \right>$.
Therefore we can write
\begin{eqnarray}
\overline{\left[ \left< n,h \right> \right]^{2}} &=& 
\overline{\left< n,h \right> \left< h,n \right>} \\
&=& 
4 
\overline{ 
\int_{-\infty}^{\infty}
\int_{-\infty}^{\infty}
\tilde{n}^{*}(f) \tilde{h}(f)
\tilde{h}^{*}(f') \tilde{n}(f')
\frac{df}{S_{n}(f)}
\frac{df'}{S_{n}(f')}} \\
&=&
4
\int_{\infty}^{\infty}
\int_{\infty}^{\infty}
\overline{
\tilde{n}^{*}(f) 
\tilde{n}(f')}
\tilde{h}^{*}(f') 
\tilde{h}(f)
\frac{df}{S_{n}(f)}
\frac{df'}{S_{n}(f')} \\
&=&
4
\int_{\infty}^{\infty}
\int_{\infty}^{\infty}
\frac{1}{2} \delta(f-f') S_{n}(f)
\tilde{h}^{*}(f')
\tilde{h}(f)
\frac{df}{S_{n}(f)}
\frac{df'}{S_{n}(f')} \\
&=& \left< h,h \right>
\end{eqnarray}
If we assume that our template is normalised such that $\left< h,h \right> =1$
we will therefore find that the variance of $\left< n,h \right>$ is unity.

If we perform the same analysis when the detector data consists of signal
and noise, i.e., $x(t)=h(t)+n(t)$ (where we will assume our template $h(t)$
is a perfect description of out signal) we find the mean of the overlap
is given by
\begin{eqnarray} 
\overline{\left[ \left< x,h \right> \right]} &=& \overline{\left[ \left< n+h,h \right> \right]} \\
&=& \overline{\left[ \left< n,h \right> \right]} + \overline{\left[ \left< h,h \right> \right]} \\
&=& 1 
\end{eqnarray} 
and that the variance is given by
\begin{eqnarray}
\overline{\left[ \left< x,h \right> - \overline{ \left< x,h \right>  } \right]^2} 
&=& \overline{\left[ \left< n,h \right>  \right]^2} \\
&=& 1.
\end{eqnarray}

It is also trivial to see how the amplitude of an incoming signal can be measured
immediately from the inner product. Consider a template $h(t)$ and a signal
$s(t) = A h(t)$ where $A$ is a real, dimensionless and time-independent number. 
We find simply that the mean output of
our template with data consisting of signal $A h(t)$ and noise $n(t)$ is 
\begin{eqnarray}
\overline{\left[ \left< x,h \right> \right]} &=& 
\overline{\left[ \left< n,h \right> \right]} 
+ \overline{\left[ \left< h,A h \right> \right]} \\
&=& A.
\end{eqnarray}

\subsection{Definition of signal to noise ratio}

We define the signal to noise ratio (SNR) $\rho$ as the statistic
$\left< x , h \right>$ divided by its standard deviation. 
Using the results from the previous Section we find that when 
our data $x$ contains a signal and stationary, Gaussian noise, 
$x(t) = n(t) + s(t)$
that the expectation value of the SNR is
$\overline{\rho} = \left< s,h \right>$ (assuming that we have normalised
our templates such that $\left< h , h \right> =1 $).
If our data $x$ contains stationary, Gaussian noise but no signal, 
$x(t) = n(t)$ then $\overline{\rho} = 0$.
In practise the detector noise will be neither stationary nor Gaussian.
In order to account for the non-stationarity of the detector noise,
we estimate the noise spectrum $S_{n}(f)$, used
within the calculation of $\left< x,h \right>$, at regular intervals.  
Environmental disturbances and problems with the detector itself
can cause transient artefacts in the detector data meaning that it will
become non-Gaussian. The detector is continuously monitored
allowing data obtained during times of a known environmental disturbance
or problem with the detector to be excluded from subsequent data analysis. 
Details on the methods used to search for gravitational wave signals in real detector
data using matched-filtering is discussed further in Sec.~\ref{sec:pipeline}.
In Sec.~\ref{gaussianresponse} we shown that the linear transformation 
(e.g., the matched-filtering) of a multivariate Gaussian distribution is also a 
multivariate Gaussian distribution. 
We will use this property later when testing our matched-filter algorithm in
Sec.~\ref{testingDTF}.

\chapter{Searching for precessing binary systems in LIGO data}
\label{ch:spin}

Interaction between the spins of the binary's component
bodies and the orbital angular momenta will cause its orbital 
plane to precess during the course of the system's evolution. 
Figures \ref{fig:FOMINJ_plotInj_paper} and \ref{fig:spin_spec_thesis}
compare the waveforms that would be observed from similar binaries,
one consisting of non-spinning components and the other consisting of
spinning components.
It has been found that optimal matched-filter searches should use templates
which take into account the spin modulation of gravitational waves.
In this Chapter we will we summarise how stellar mass binaries (i.e.,
those which LIGO is sensitive to) form and how their components spin up 
(Sec.~\ref{sec:formation}), then move onto modelling their inspiral
orbits and gravitational wave emissions (Sec.~\ref{Sec:TargetModel}).
We then summarise the progress that has been made in building detection
efficient templates to capture these systems (Sec.~\ref{sec:dtf}).
The remainder of the Chapter details the use of the BCV2 detection template
family (Sec.~\ref{sec:BCV2dtf}) to search for signals emitted by
binaries with spinning components in data taken by LIGO
during its third science run. 
No detections were made and in Sec.~\ref{sub:upperlimit} we 
calculate upper limits on the rate of coalescence of neutron star -
black hole binaries with spinning components.

The analysis of LIGO data described in the latter part of this 
Chapter was led by the author
(Gareth Jones) as a member of the LIGO Scientific Collaboration/Virgo 
Compact Binary Coalescence working
group \cite{CBC} and has been previously
published in B.~Abbott et al. (2007) \cite{s3sbbh_arxiv}.

\section{Formation and evolution of stellar mass binary systems}
\label{sec:formation}
We briefly review the current literature regarding the formation and evolution of 
binary systems paying particular attention to the spins of the binary's 
components. The literature focuses upon NS-BH binaries and it turns out that 
the effects of spin are more pronounced in systems with small mass ratio 
(i.e., unequal masses). It is likely that the formation and evolution of 
other stellar mass binaries consisting of compact objects, e.g., BH-BH and 
NS-NS systems will be qualitatively similar and the discussion here 
will be relevant to all these cases.

Stellar mass BHs form either through the collapse of a massive progenitor 
(e.g. a main sequence star that has exhausted the hydrogen in its core) or 
via the accretion-induced collapse of a NS (which itself will have formed 
via collapse of a massive progenitor).  After core collapse, progenitor 
stars with mass $< 1.4 M_{\odot}$ become White Dwarfs, those with mass 
in the range $1.4$ to $\sim 3 M_{\odot}$ become NSs and those with mass 
$\gtrsim 3 M_{\odot}$ become BHs.

As internal densities of a progenitor star collapsing under gravity exceed 
$~10^{10} \rm{kg \, m}^{-3}$ the majority of its protons and electrons will undergo
inverse beta decay to form neutrons (and neutrinos).
In neutron stars it is the repulsive forces  
(arising from degeneracy pressure as described by the Pauli exclusion principle) 
between the neutrons that resist further gravitational collapse.
For progenitor stars with mass $\gtrsim 3 M_{\odot}$ the gravitational forces
exceed the outward degeneracy pressure forces and the star will collapse
further to become a black hole.

A black hole is defined by its event horizon
whose radius will depend on its mass and spin only.
In classical physics anything falling through the event horizon can never return 
from behind it (in quantum physics there are exceptions to this statement such 
as the postulated Bekenstein-Hawking radiation).
Theoretically, black holes are created when {\it any} quantity of matter collapses 
under gravity and becomes smaller than its event horizon. 
In nature there is evidence for stellar mass and supermassive black holes, both
of which are expected to play leading roles in the production of the gravitational
waves we expect to observe with current and planned detectors.
Black holes contain a physical singularity, a point where the curvature of 
spacetime is infinite and physics breaks down (physical singularities are 
different from co-ordinate singularities).
The ``no hair'' theorem states that a black hole can be fully described by
its mass, angular momentum and charge.

The formation of a typical NS-BH binary will begin with two main sequence 
stars in orbit about their common centre of mass. As the more massive of 
these star evolves away from the main sequence it will expand until it 
fills its Roche lobe before transferring mass to its companion.
The Roche lobe is defined as the region of space around an object in a 
binary system within which orbiting material is gravitationally bound to 
that object. If the object expands past its Roche lobe, then the material 
outside of the lobe will fall into the other object in the binary.

The more massive body would eventually undergo core collapse to form a BH, 
and the system as a whole would become a high-mass X-ray binary.
As the second body expands and evolves it would eventually fill its own 
Roche lobe and the binary would then go through a common-envelope phase. This 
common-envelope phase, characterised by unstable mass transfer, would be highly 
dissipative and would probably lead to both contraction and circularization 
of the binary's orbit. Accretion of mass can allow the BH to spin-up.
It has been argued that the common-envelope phase, and associated orbital 
contraction, is essential in the formation of a binary which will coalesce 
within the Hubble time~\cite{Kalogera:2000}. Finally the secondary body would 
undergo supernova to form a NS (or if massive enough, a BH). Prior to the supernova 
of the secondary body we would expect the spin of the BH to be aligned with the 
binary's orbital angular momentum \cite{Kalogera:2000}. However, the ``kick'' 
associated with the supernova of the secondary body could cause the orbital 
angular momentum of the post-supernova binary to become tilted with respect to
the orbital angular momentum of the pre-supernova binary. Since the BH would 
have a small cross-section with respect to the supernova kick we expect any 
change to the direction of its spin angular momentum to be negligible and 
the BH spin to be misaligned with respect to the post-supernova orbital 
angular momentum~\cite{grandclement:102002}. The misalignment between the spin 
and orbital angular momentum is expected to be preserved until the system 
becomes detectable by ground-based interferometers 
\cite{grandclement:102002,ryan-1995-52}.

\subsection{Expected merger rate of compact binaries}
\label{sec:mergerrates}
Estimates of the merger rates of compact binaries consistent with present 
astrophysical understanding are summarised in 
Abbott et al. (2007) \cite{LIGOS3S4all}. 
The rate of merger of NS-NS binaries
can be inferred by the four observed binary systems
containing pulsars which will coalesce within the Hubble time
\cite{Phinney:1991ei,Narayan:1991}.
The current estimate of the merger rate of NS-NS
systems (at $95\%$ confidence) is
$10 - 170 \times 10^{-6} \rm{yr}^{-1} \, \rm{L}_{10}^{-1}$ 
\cite{Kalogera:2004tn,Kalogera:2004nt,kim-2006,kalogera-2006}
where 
$\rm{L}_{10} = 10^{10} \, \rm{L}_{\odot,B}$ is $10^{10}$ times the blue
light luminosity of the Sun 
(for reference, the luminosity of the Milky Way is around
$1.7 \, \rm{L}_{10}$).

Although, we predict that NS-BH and BH-BH systems form 
according to the scenario described previously, there is no direct
astrophysical evidence for these systems.
To predict the merger rate of these systems, 
the authors of Refs.~\cite{OShaughnessy:2005,Oshaughnessy:2006b}
considered various population synthesis models of
compact binary formation which are consistent with the
expected merger rate of NS-NS systems.
They find that the merger rates of binary populations in galactic fields 
are likely to lie (at $95\%$ confidence) in the ranges 
$0.1-15 \times 10^{-6} \rm{yr}^{-1} \, \rm{L}_{10}^{-1}$ and
$0.15-10 \times 10^{-6} \rm{yr}^{-1} \, \rm{L}_{10}^{-1}$ 
for BH-BH and NS-BH binaries respectively.

Compact binary mergers from within dense stellar clusters
or associated with short, hard gamma-ray bursts would increase the 
expected merger rates.
When binary formation in star clusters is taken into account with 
relatively optimistic assumptions, detection rates could be as high 
as a few events per year for initial LIGO 
\cite{PortegiesZwart:2000,imbhlisa-2006,OLeary:2006}.

\subsection{Spin magnitudes}
A compact object can gain spin either during its formation (through the core 
collapse of a massive progenitor or the accretion-induced collapse of a NS) 
or through subsequent accretion episodes.
The dimensionless spin parameter $\chi$ is given by $|\Jv|/M^2$ where $\Jv$ is the 
total angular momentum of the compact object and $M$ is its mass
\footnote{Various conventions exist regarding the symbols used for the various 
spin parameters. Here we will denote the dimensionless spin parameter 
$\chi = |\Jv|/M^2$ and the specific angular momentum $a = |\Jv|/M$ where $\Jv$ is the 
total angular momentum of the compact object and $M$ is its mass.}. 
For a non-spinning body we would have $\chi=0$.

Penrose's hypothesis of cosmic censorship states that physical singularities 
can {\it only} occur behind an event horizon.
In Kerr geometry, used to describe the spacetime surrounding the
event horizon of a spinning black hole, the outermost event horizon 
occurs at $r = M + \sqrt{M^{2} - a^{2}}$
where $r$ is the radial Schwarzschild/Boyer-Lindquist co-ordinate (equal
to the circumference of a circle centred on the central body divided by
$2 \pi$).
For this event horizon to form we require that $a \leq M$ which is
equivalent to $\chi \leq 1$.
For Earth we find $\chi \sim 800$ (found by assuming that the Earth is
a solid sphere and using $|\Jv| = I \omega$ where $I$ is the Earth's
moment of inertia and $\omega$ is the orbital frequency of its spin).

O' Shaughnessy et al. (2005) \cite{ShaughnessyKKB2005} 
consider likely values of 
birth spin and then perform population synthesis in order to model the 
accretion histories of black holes in inspiraling binaries and to place 
bounds on their expected spin. 
Low mass BH birth spins can be estimated by considering the birth spin of 
similar mass NS. Through the observation of radio pulsars NS birth spins have 
been estimated as $\chi = |\Jv| / M^2 \simeq 0.005-0.02$.
However, results from simulations indicate that a large fraction of BHs in 
BH-NS systems were formed by the accretion-induced collapse of a NS that 
has undergone a common envelope phase. We would therefore expect that the 
BH birth spin would be dependent on the spin attained by the NS during the 
poorly understood common-envelope phase. Mildly recycled pulsars in NS-NS 
systems are believed to have been spun up during a common envelope phase 
yet are still measured to have fairly small spins of $\chi \leq 0.01$.
Uncertainties in both the collapse and common-envelope stages of the BH 
evolution lead the authors of \cite{ShaughnessyKKB2005} to place loose 
bounds on BH birth spin of between $\chi = 0$ and $\chi \sim 0.5$, the 
birth spin of a BH forming from the collapse of a maximally spinning NS.
 
Results from the population synthesis performed by O' Shaughnessy et al. 
showed that the evolution of the majority of NS-BH binaries is dominated
by accretion associated with a common-envelope phase rather than by
disk accretion.
Hypercritical accretion occurs when one of the binary's components
spirals through its companion's envelope and rapidly accretes matter at 
super-Eddington (for photons), neutrino-cooled rates 
\cite{ShaughnessyKKB2005}. 
The simulations showed that even with birth 
spin $\chi = 0$ stellar mass BHs ($M < 15 M_{\odot}$) can obtain 
significant spin $\chi \sim 0.8$ through common-envelope phase accretion.
More massive objects are more difficult to spin-up, requiring larger, 
and consequently less likely, transfers of mass. 
For less massive systems,  
$M < 4M_{\odot}$, maximal spin ($\chi = 1$) could easily be obtained through accretion 
alone (i.e., regardless of birth spin). 

In \cite{Thorne:1974ve}, Thorne calculates an upper bound for the spin of a BH.
As matter accretes onto BH its spin will increase, 
radiation emitted by the accretion disk which is subsequently ``swallowed'' 
by the BH causes a counteracting torque which limits the BH's spin to 
$\chi < 0.998$. 
Cook et al. (1994) \cite{Cook:1993qr} consider a variety of NS 
equations of state and calculate the maximum spin the NS could have before
it would break up. 
For NS we find that the maximal spin value is $\chi \sim ~ 0.7$ 
\footnote{This number is obtained by taking $|\Jv|$ and $M$ values from Tables 
6, 7 and 8 in \cite{Cook:1993qr} and calculating $\chi = |\Jv|/ M^2$ with 
$c=G=1$.}.

We can infer the mass of a compact object in a binary system through 
observations of its companion. The mass function is defined as
\begin{equation}
f(m) \equiv \frac{P_{\rm{orb}} K_{2}^{3}} {2 \pi G} = 
\frac{m_{1} \rm{sin}^{3}(\iota)}{(1 + m_{2}/m_{1})^{2}}
\end{equation}
where $m_{1}$ and $m_{2}$ are the masses of the compact object and its 
companion respectively, $P_{\rm{orb}}$ is the orbital period of the 
binary, $K_{2}$ is the velocity amplitude of the companion object and 
$\iota$ is the inclination angle of the binary with respect to the observer.
The mass function $f(m)$ can be calculated for X-ray binaries through 
measurement of the amplitude velocity of the luminous companion and 
the orbital period. By estimating the mass of the companion 
$m_{2}$ and the inclination angle of the binary (e.g., through observation 
of jets) we can obtain a lower limit on the mass of the compact object $m_{1}$.
As of 2006, there are 20 X-ray binaries known to contain a stellar mass BH 
(inferred through dynamical considerations) as well as a further 20 X-ray 
binaries that may contain a stellar mass BH \cite{remillard-2006-44}.

The measurement of BH spin from electromagnetic observations is in progress and 
in Sec.~(8.2) of \cite{remillard-2006-44} four methods are discussed.
Spectral fitting of X-ray continuum data obtained from the Rossi X-ray Timing 
Explorer (RXTE) and the Advanced Satellite for Cosmology and Astrophysics 
(ASCA) was used to place a robust lower bound on the spin of the primary 
component (BH) of the X-ray binary GRS 1915+105. The BH with 
$m_{\rm{BH}} \sim 15M_{\odot}$ was found to have spin 
$\chi > 0.98$ \cite{mcclintock-2006-652}. In Table \ref{tab:xrb} the inferred 
masses and spins of four BH systems, each belonging to an X-ray binary,
are given.

\begin{table}
\caption{Measurements of the masses and spins of four BHs each of which belong 
to an X-ray binary system. The masses were obtained through dynamical 
considerations and the spins were obtained through spectral fitting of X-ray 
continuum data obtained using the RXTE, ASCA and BeppoSAX space telescopes.}
\label{tab:xrb}
\begin{center}
\begin{tabular}{c|c|c}
\hline 
\hline 
System & $m_{\rm{BH}}$ & $\chi_{\rm{BH}}$ \\
\hline 
GRS 1915+105 & $\sim 14.4M_{\odot}$ \cite{harlaftis-2004-414} 
& $> 0.98$ \cite{mcclintock-2006-652} \\
4U 1543-47   & $\sim 9.4 M_{\odot}$ \cite{orosz-2002}         
& $\sim 0.75-0.85$ \cite{shafee-2006-636} \\
GRO J1655-40 & $\sim 6.3 M_{\odot}$ 
\cite{shafee-2006-636, HjellmingRupen1995}  
& $\sim 0.65-0.75$ \cite{shafee-2006-636} \\
LMC X-3      & $\sim 7 M_{\odot}$   \cite{soria-2000}         
& $< 0.26$ \cite{davis-2006-647, mcclintock-2006-652}\\ 
\hline
\end{tabular}
\end{center}
\end{table}

\subsubsection{Effect of spin on kick velocities}
Campanelli et al. \cite{campanelli-2007-659} (2007) use numerical relativity 
simulations to investigate the evolution of a generic binary 
(e.g., unequal mass, misaligned spins). Their results show that spin of the 
binary's components can increase the kick velocity of the post-merger remnant. 
They predict kick velocities of nearly $4000 \rm{\, km \, s}^{-1}$ for some systems 
(anti-aligned maximal spins lying in the orbital plane) which would allow 
these systems to become ejected from their host galaxies (escape velocities 
for giant elliptical and spiral galaxy bulges are in the range 
$450 - 2000 \rm{\, km \, s}^{-1}$ and are smaller for dwarf galaxies). 

\section{Target model}
\label{Sec:TargetModel}
In this Section we describe the target model that we use as a fiducial
representation of the gravitational wave signals expected from precessing
binaries consisting of spinning compact objects.
We will describe the target model that was used by Buonanno, Chen and
Vallisneri in \cite{BCV2} (known as BCV2) in the development of their
detection template family.

\subsection{The adiabatic approximation and circularization of the binary's 
orbit}
\label{sec:aacirc}
For simplicity, the target model waveforms are assumed to be generated by
the inspiral of the binary in the {\it adiabatic limit}.
The part of inspiral observable by ground-based detectors occurs towards 
the end of a long period of adiabatic dynamics throughout which 
the timescale of orbital shrinkage (due to the emission of gravitational
waves) is far larger than the period of a single orbit, i.e., 
$T_{\rm {Orbital-shrinkage}} \gg T_{\rm {Orbit}}$. 
Working under the adiabatic approximation allows us to write the
energy balance equation $dE/dt = - \mathcal{F}$, where $E$ is the
binding energy of the binary (i.e., the energy required 
to disassemble the binary) and $\mathcal{F}$ is the gravitational
wave flux, which in turn simplifies the time evolution of the binary
(see, for example, Sec.~I of Ajith et al. (2005) \cite{Ajith:2005} and
Damour et al. (2001) \cite{DIS2001}).
Under the adiabatic approximation, we can assume our binary to have
instantaneously circular orbits which are i) shrinking due to
the emission of gravitational waves and ii) precessing due to the
effects of spin.

There is evidence that binary's orbit will have circularized
through the emission of gravitational waves before it will be observable
by current detectors.
Eq.~(5.12) of Peters (1964) \cite{Peters:1964} gives the semi-major axis 
of the binary's orbit $a$ as a function of its eccentricity $e$:
\begin{eqnarray}
a(e) \propto \frac{e^{12/19}}{1-e^2}.
\label{eq:peters}
\end{eqnarray}
For small eccentricity we can write $a \propto e^{12/19}$ and through Kepler's
third law, $a \propto P^{2/3}$, where $P$ is the binary's orbital period, we
can write $e \propto P^{19/18}$. Considering the evolution of $e$ and $a$
with the decrease in the binary's period $P$ we see that eccentricity decreases
more rapidly the than orbital separation.
Since the binary will undergo only its final few tens or hundreds of orbits in
the detector's band of good sensitivity we can assume that the binary's orbit 
will have become essentially circular by the time we observe it with ground
based detectors.
Indeed, from Eq.~(\ref{eq:peters}) we can show that a low mass binary system
(e.g., neutron star - neutron star) with high eccentricity in the LISA band
of good sensitivity,
e.g., $e=0.9$ at $f \sim 10^{-3}$ Hz will have completely circularized before 
it enters the LIGO band of good sensitivity ($\sim 40$ Hz).
In Belczynski et al. (2002) \cite{Belczynski:2002} the authors use 
population synthesis to analyse the evolution of binary systems. In their
Figure 5 they show the circularization of binary systems between formation and
when they enter LIGO's band of good sensitivity.
Orbital eccentricity cannot be neglected when discussing extreme mass
ratio inspiral systems that are a relevant source for LISA
(see Chapter \ref{ch:emri}).

\subsection{Equations used to calculate a precessing binary's
time evolution}
\label{sec:EvolEqns}
This target model uses post-Newtonian (PN) equations for the time-evolution
of the instantaneous orbital frequency $\omega$, the spins of the binary's
components $\Sv_{1}$, $\Sv_{2}$ and the orbital angular momentum of the
binary $\Lv_{N}$. 

The first (time) derivative of the orbital angular frequency $\omega$ is
given to 3.5PN order \cite{DTIWW:1995, BDI:1995, WW:1996, BIWW:1996, 
Blanchet:1996, BFIJ:2002, BIJ:064005, DIS2001, DIS:2002} 
with spin effects at 1.5 and 2PN order 
\cite{KidderWW:1993, DTIWW:1995, BDI:1995, WW:1996, kidder:821}.
We quote the waveform as given in Buonanno et al. 
(henceforth PBCV2) \cite{PBCV2} 
\footnote{The expansion of $\dot{\omega}/\omega^2$ given in PBCV2 \cite{PBCV2}
(Eqs.~(1-7))
is equivalent to the expansion given in BCV2 \cite{BCV2} 
(Eq.~(1)) but has been written in a fashion which has made the identification 
of the different PN terms clearer.} 
but have corrected some of the 2.5PN and 3.5PN coefficients for an error
in the contribution of tails to the gravitational wave flux (details of this
follow the equations):
\begin{eqnarray}
\label{omegadot}
\frac{\dot{\omega}}{\omega^2} &=& \frac{96}{5} \eta (M \omega)^{5/3}
(1 + 
\mathcal{P}_{\rm{1PN}} +
\mathcal{P}_{\rm{1.5PN}} +
\mathcal{P}_{\rm{2PN}} +
\mathcal{P}_{\rm{2.5PN}} \nonumber \\ 
&+&
\mathcal{P}_{\rm{3PN}} +
\mathcal{P}_{\rm{3.5PN}}
),
\end{eqnarray}
where
\begin{eqnarray}
\label{omegadot1PN}
\mathcal{P}_{\rm{1PN}} &=& 
- \frac
{743 + 924 \eta}
{336}
(M \omega)^{2/3}, \\
\label{omegadot1.5PN}
\mathcal{P}_{\rm{1.5PN}} &=& 
-\Bigg\{
\frac{1}{12} \sum_{i=1,2}
\left[
\chi_{i} ( \hat{\Lv}_{N} \cdot \hat{\Sv}_{i} )
\left(
113 \frac{m_{i}^{2}}{M^2} + 75 \eta
\right)
\right] 
 \nonumber \\  
&-& 
4 \pi
\Bigg\} (M \omega), \\
\label{omegadot2PN}
\mathcal{P}_{\rm{2PN}} &=& 
\Bigg\{
\left(
\frac{34103}{18144} + \frac{13661}{2016}\eta + \frac{59}{18}\eta^{2}
\right)
-\frac{\eta \chi_{1} \chi_{2}}{48} 
[
247 (\hat{\Sv}_{1} \cdot \hat{\Sv}_{2} ) 
\nonumber \\  
&-& 721 
(\hat{\Lv}_{N} \cdot \hat{\Sv}_{1} ) (\hat{\Lv}_{N} \cdot \hat{\Sv}_{2} )
]
\Bigg\}(M \omega)^{4/3}, \\
\label{omegadot2.5PN}
\mathcal{P}_{\rm{2.5PN}} &=&
- \frac{1}{672}
( 4159 + 15876 \eta ) \pi (M \omega)^{5/3}, \\
\label{omegadot3PN}
\mathcal{P}_{\rm{3PN}} &=&
\left [
\left(
\frac{16447322263}{139708800} 
-\frac{1712}{105} \gamma_{E} 
+ \frac{16}{3} \pi^{2}
\right) \right.  \nonumber \\
&+& \left. \left(
- \frac{273811877}{1088640}
+ \frac{451}{48} \pi^{2} 
- \frac{88}{3} \hat{\theta}
\right) \eta
+ \frac{541}{896} \eta^{2}
- \frac{5605}{2592} \eta^{3} \right. \nonumber \\
&-& \left. \frac{856}{105} {\rm{log}}(16 ( M \omega)^{2/3} )
\right] (M \omega)^{2}, \\
\label{omegadot3.5PN}
\mathcal{P}_{\rm{3.5PN}} &=&
\left(
- \frac{4415}{4032} 
+ \frac{717350}{12096} \eta
+ \frac{182990}{3024} \eta^{2}
\right) \pi (M \omega)^{7/3}
\end{eqnarray}
where $\Lv_{N} = \mu \rv \times \vv$ 
(where $\mu = m_{1} m_{2} / M$ is the reduced mass )
is the Newtonian angular momentum and
$\hat{\Lv}_{N} = \Lv_{N} / |\Lv_{N}|$,
$\gamma_{E} = 0.577 \dots$ is Euler's constant,
$\hat{\theta} = 1039/4620$ was determined in Blanchet et al. (2004) 
\cite{blanchet:091101}.  
We define the accumulated orbital phase 
\begin{eqnarray}
\label{accorbphase}
\Psi \equiv \int \omega dt = \int \frac{\omega}{\dot{\omega}} d\omega. 
\end{eqnarray}

In L. Blanchet (2005) \cite{blanchet:129904} and 
L. Blanchet et al. (2005) \cite{blanchet:129903}
an error in the contribution 
of tails to the gravitational wave flux was identified in the calculations 
presented in L. Blanchet (1996) \cite{Blanchet:1996} and in
L. Blanchet et al. (2002) \cite{BIJ:064005}.
The subsequent correction of this error led to changes in some coefficients
at 2.5PN and 3.5PN order in the expansion of $\dot{\omega}/\omega^{2}$, 
(i.e., Eqs.~(\ref{omegadot2.5PN}) and (\ref{omegadot3.5PN}))
since the publication
of BCV2 \cite{BCV2}.
In the 2.5PN term, $15876$ replaces the value $14532$ 
and in the 3.5PN term $717350$ replaces the value $661775$
and $182990$ replaces the value $149789$. 
These new coefficients can be derived simply using the expansion of 
$(dF/dt)^{3.5PN}$ given in Arun et al. (2005) \cite{arun:069903}.

The equations for the precession of the spins $\Sv_{1}$ and $\Sv_{2}$
are given by (see, for example, 
Eqs.~(4.17b,c) of Kidder (1995) \cite{kidder:821} 
or Eqs.~(11b,c) of ACST (1994) \cite{ACST}):
\begin{eqnarray}
\label{S1PN}
\dot{\Sv}_{1} &=& 
\frac{(M \omega)^{2}}{2 M} 
\left\{
\eta (M \omega)^{-1/3} 
\left(
4 + 3 \frac{m_{2}}{m_{1}} 
\right) \hat{\Lv}_{N} \right. \nonumber \\
&+& \left.  \frac{1}{M^{2}}
[
\Sv_{2} - 3 (\Sv_{2} \cdot \hat{\Lv}_{N}) \hat{\Lv}_{N}
]
\right\} \times \Sv_{1}, \\
\label{S2PN}
\dot{\Sv}_{2} &=&
\frac{(M \omega)^{2}}{2 M}
\left\{
\eta (M \omega)^{-1/3}
\left(
4 + 3 \frac{m_{1}}{m_{2}}
\right) \hat{\Lv}_{N} \right. \nonumber \\
&+& \left. \frac{1}{M^{2}}
[
\Sv_{1} - 3 (\Sv_{1} \cdot \hat{\Lv}_{N}) \hat{\Lv}_{N}
]
\right\} \times \Sv_{2}
\end{eqnarray}
where we have followed BCV2 \cite{BCV2} by using
Kepler's third law ($r = (M/\omega^{2})^{1/3}$)
and the Newtonian expression of the magnitude of the
orbital angular momentum, 
\begin{eqnarray}
|\Lv_{N}| = \mu r^2 \omega = \eta M^{5/3} \omega^{1/3},
\label{LNmag}
\end{eqnarray}
to substitute for $r$ when writing these expressions.   

The equation for the precession of $\hat{\Lv}_{N}$ is 
(see, for example Eq.~(4.17a) of Kidder (1995) \cite{kidder:821}  
or Eq.~(11a) of ACST (1994) \cite{ACST}):
\begin{eqnarray}
\label{LNPN}
\dot{\hat{\Lv}}_{N} &=& \frac{\omega^{2}}{2 M}
\left\{
\left[
\left(
4 + 3 \frac{m_{2}}{m_{1}} 
\right)
\Sv_{1}
+
\left(
4 + 3 \frac{m_{1}}{m_{2}} 
\right)
\Sv_{2}
\right] \times \hat{\Lv}_{N} \right. \nonumber \\
& & \left. 
- \frac{3 \, \omega^{1/3}}{\eta M^{5/3}}
[
(\Sv_{2} \cdot \hat{\Lv}_{N}) \Sv_{1}
+ 
(\Sv_{1} \cdot \hat{\Lv}_{N}) \Sv_{2}
] \times \hat{\Lv}_{N}
\right\}.
\end{eqnarray}

In writing these equations we have assumed that the component
bodies are sufficiently axisymmetric that we are able to neglect their
own gravitational wave emission and therefore assume that
the magnitude of the spin remains constant during the course of the
inspiral, i.e., $d|\Sv_{i}|/dt = 0$.
Therefore, the loss of total angular momentum experienced by the system as it
inspirals is caused by loss of orbital, rather than spin, angular momentum.
Therefore, defining total angular momentum to be $\Jv = \Lv + \Sv$ we have
$d|\Jv|/dt = d|\Lv|/dt$.

Eqs.~(\ref{omegadot}), (\ref{S1PN}), (\ref{S2PN}) and (\ref{LNPN}) form a 
set of coupled differential equations. 
To follow the evolution of a precessing binary we 
numerically integrate these equations
until the adiabatic approximation is no longer valid. 
This occurs either when 
the binary reaches its
Minimum Energy Circular Orbit (MECO, also known as the innermost
circular orbit for non-spinning binaries in Blanchet (2002) 
\cite{BlanchetICO:2002}) after which the system plunges or if
the orbital angular frequency stops evolving i.e., $\dot{\omega} = 0$
(see Sec.~IIB of BCV2 \cite{BCV2}).

\subsection{Response of a detector to gravitational waves from a precessing,
inspiraling binary}
The response of a ground-based interferometric detector to a gravitational
wave emitted by a compact binary has the form
\begin{eqnarray}
\label{hresp}
h_{\rm{resp}} = \frac{\mu}{D} \frac{M}{r} Q^{ij} P_{ij}
\end{eqnarray}
where we have the reduced mass $\mu = m_{1} m_{2}/M$, $D$ is the distance
from the gravitational wave source to the detector and $r$ is the (absolute)
separation of the binary's components.
The tensor $\Qv$ is proportional to the second time derivative of the
mass-quadrupole of the binary
and the tensor $\Pv$ projects this moment onto the detector.

In order to calculate $h_{\rm{resp}}$ we will first find $\Qv$ which can be 
given as
\begin{eqnarray}
\label{Qij}
Q^{ij} = 2 \left[ \lambda^{i} \lambda^{j} - n^{i} n^{j} \right]
\end{eqnarray}
where $n^{i}$ is the unit vector along the separation vector of the binary's
components $\rv$ and $\lambda^{i}$ is the unit vector along the
component's relative velocity $\vv$
\footnote{
In order to obtain $Q^{ij}$ in the form shown above, 
we can evaluate the mass-quadrupole moment 
$I^{ij}$ (Eq.~(\ref{quadmassmoment})) 
for the binary by considering the equivalent one body problem.
The equivalent one body problem consists of a body with
mass equal to the reduced mass $\mu$ of the binary
orbiting the centre of mass (which we will take as our origin)
at position $\rv \equiv \xv_{1} - \xv_{2}$ where
$\xv_{1}$ and $\xv_{2}$ are the position vectors of the original
bodies $m_{1}$ and $m_{2}$ \cite{LL}.
By approximating the binary's components as
($\delta-$function) point masses we can simplify the mass-quadrupole 
moment and write it as $I^{ij} = \mu r^{i} r^{j}$.
Taking the time derivative twice and using the centripetal acceleration
$\ddot{r}^{i} = - (|\dot{\rv}|/|\rv|) \hat{r}^{i}$ it is trivial to
obtain Eq.~(\ref{Qij}) for $Q^{ij}$ modulo a factor of $\mu |\dot{\rv}|^2$.
In the following analysis we not use the one body approach since
we wish to identify the spin associated with each body separately.
}.

In order to find $\lambdav$ and $\nv$ (and therefore $\Qv$) we must
follow the evolution of the binary within a chosen coordinate system.
There are various coordinate systems that can be used and we shall see later
that through expedient choice of the coordinate system we can usefully
isolate the effects of spin upon the gravitational wave that will be
observed at the detector.
Following BCV2 \cite{BCV2} we shall first describe the binary using
a generalisation of the Finn-Chernoff (FC) convention described
in Finn and Chernoff (1993) \cite{FC:1993} (see Sec.~IIIA).
Using the FC convention we specify a fixed source frame defined
by a set of orthogonal basis vectors 
$\{\ev_{x}^{S}, \ev_{y}^{S},\ev_{z}^{S}\}$.
In the analysis presented in Ref.~\cite{FC:1993}, 
Finn and Chernoff considered only binaries with non-spinning
components. For these binaries there would be no spin-induced
precession of the orbital plane and it made sense to specify
$\ev_{z}^{S} = \hat{\Lv}_N$ so that $\{\ev_{x}^{S}, \ev_{y}^{S}\}$
would form a (permanent) orthonormal basis for the orbital plane. 

\begin{figure}
\begin{center}
\includegraphics[angle=0, width=1.1\textwidth]{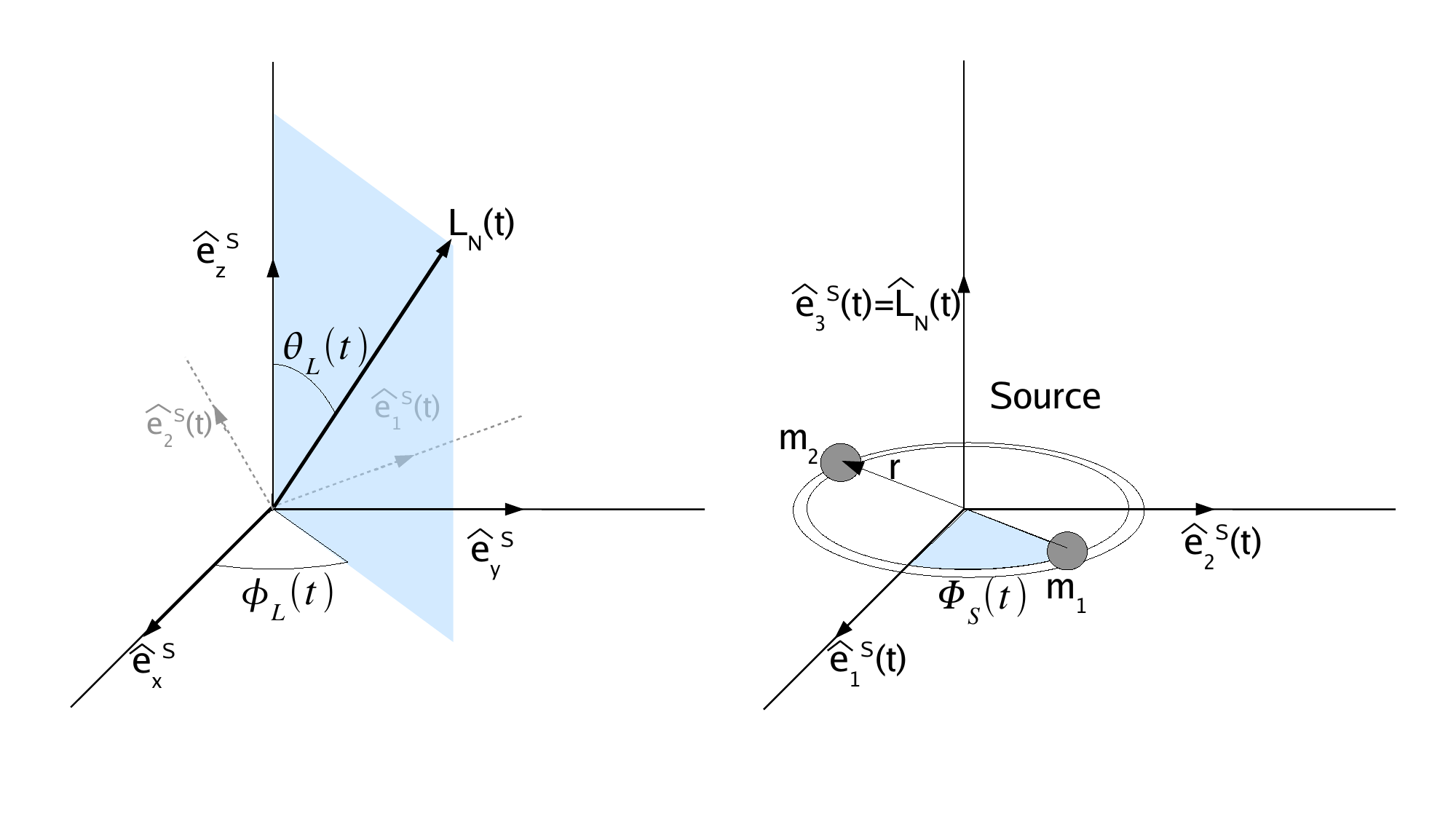}
\caption{The binary in the source frame. The left hand plot shows
the binary's orbital angular momentum $\Lv_{N}$  within the fixed
source frame $\{\ev_{x}^{S}, \ev_{y}^{S},\ev_{z}^{S}\}$.
We also show the orthonormal basis for the instantaneous orbital
plane $\{\ev_{1}^{S}, \ev_{2}^{S}\}$.
The right hand plot shows the binary within the orthonormal basis
$\{\ev_{1}^{S}, \ev_{2}^{S}\}$. The separation vector of the binary's
components $\rv$ and the orbital phase $\Phi_{S}$ are marked
on this plot.
}
\label{fig:source_frame_1}
\end{center}
\end{figure}

However, for binaries consisting of spinning components, the
orbital plane will precess and we specify a (time-dependent)
orthonormal basis for the instantaneous orbital plane 
$\{\ev_{1}^{S}(t), \ev_{2}^{S}(t)\}$ relative to the 
(arbitrarily) fixed $\ev_{z}^{S}$ basis vector:
\begin{eqnarray}
\ev_{1}^{S}(t) = \frac{\ev_{z}^{S} \times \hat{\Lv}_{N}(t)}{\sin \theta_L(t)},
\;
\ev_{2}^{S}(t) = \frac{\ev_{z}^{S} - \hat{\Lv}_{N}(t) \cos \theta_L(t)}
{\sin \theta_L(t)}
\end{eqnarray}
and $\ev_{3}^{S}(t) = \hat{\Lv}_N(t)$
where we have temporarily made explicit the time-dependent quantities.
These co-ordinate frames are shown in Fig.~\ref{fig:source_frame_1}.

We measure the orbital phase of the binary's components 
$\Phi_{S}$ from $\ev_{1}^{S}$.
To aid visualisation of this system it might be useful to note that as 
the orbital angular momentum $\hat{\Lv}_{N}$ precesses, $\ev_{1}^{S}$ 
will remain in the $x-y$ plane of the fixed source frame.
Note that $\Phi_{S}$ is defined as an angle measured in a particular
frame whereas the previously defined accumulated orbital phase $\Psi$
is simply a function (an integral) of the instantaneous angular
orbital frequency $\omega$ (see Eq.~(\ref{accorbphase})).
In general, $\Phi_{S}(t) \neq \Psi$. The relationship between these
phases will be discussed more later (see Sec.~\ref{BCV2analysis}).

Having defined $\{\ev_{1}^{S}, \ev_{2}^{S}\}$ we are able to define 
the polarization tensors of the instantaneous orbital plane
$\{\ev_{+}^{S}, \ev_{\times}^{S}\}$:
\begin{eqnarray}
\ev_{+}^{S} \equiv  \ev_{1}^{S} \otimes \ev_{1}^{S} - 
                    \ev_{2}^{S} \otimes \ev_{2}^{S}, 
\; \, \, \, 
\ev_{\times}^{S}  \equiv  \ev_{1}^{S} \otimes \ev_{2}^{S} +
                          \ev_{2}^{S} \otimes \ev_{1}^{S}
\end{eqnarray}
where $\otimes$ represents the {\it tensor} or {\it outer product}.
The tensor product is defined such that a tensor $\av$ defined as the 
tensor product of two vectors $\bv$, $\cv$ (i.e., $\av = \bv \otimes \cv$) 
will have elements $a^{ij} = b^{i} \times c^{j}$.

We can write the unit vectors of the binary separation and relative
velocity as
\begin{eqnarray}  
\hat{\nv} = \ev_{1}^{S} \cos \Phi_{S} + \ev_{2}^{S} \sin \Phi_{S},
\;
\hat{\lambdav} = -\ev_{1}^{S} \sin \Phi_{S} + \ev_{2}^{S} \cos \Phi_{S}
\end{eqnarray}  
and from this the mass-quadrupole moment as
\begin{eqnarray}
Q^{ij} = -2 \left( 
\left[
\ev_{+}^{S}
\right]^{ij} \cos 2 \Phi_{S}
+
\left[
\ev_{\times}^{S}
\right]^{ij} \sin 2 \Phi_{S}
\right).
\end{eqnarray}

\begin{figure}
\begin{center}
\includegraphics[angle=0, width=1.1\textwidth]{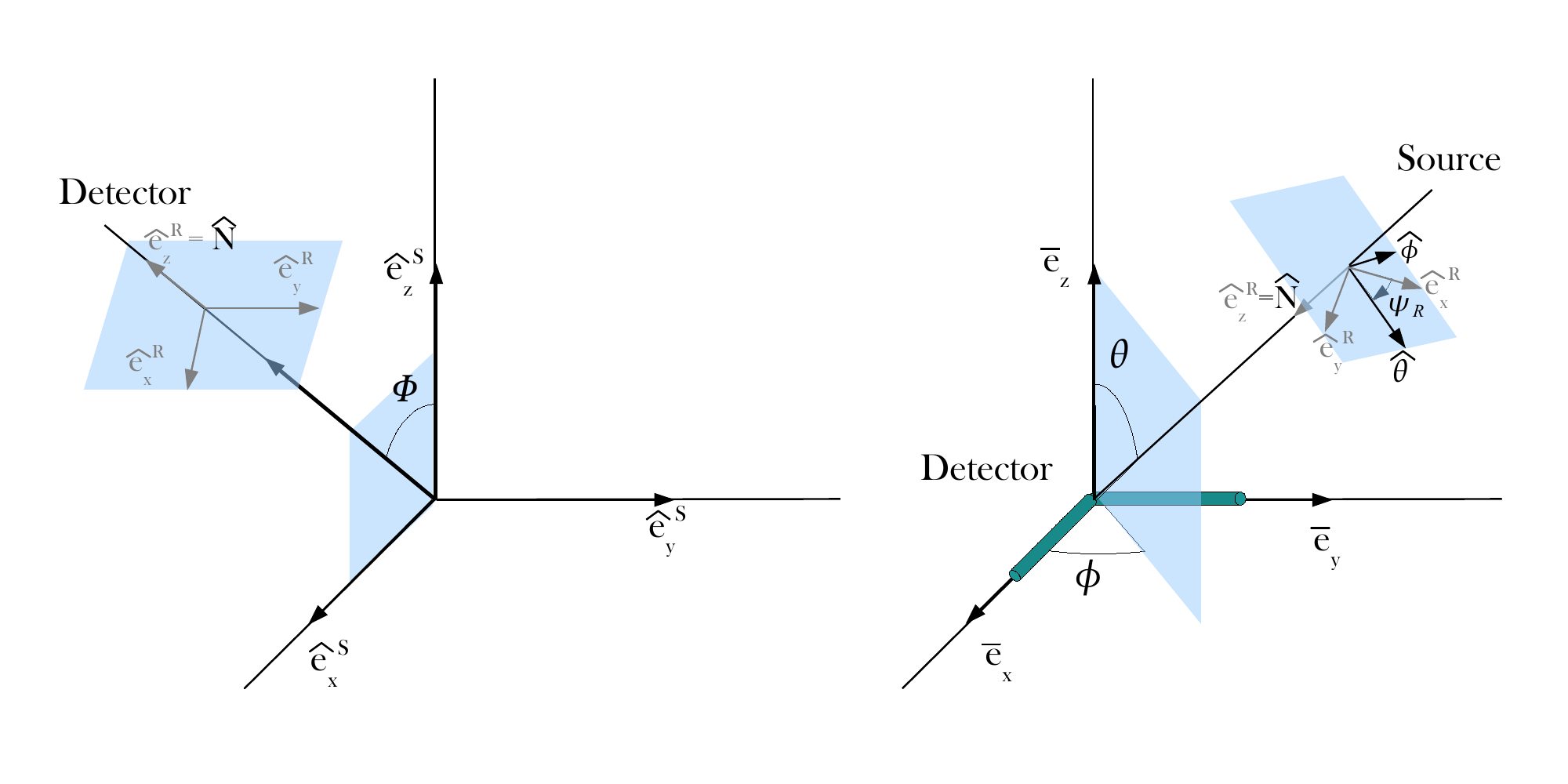}
\caption{The radiation and detector frames. 
The left hand plot shows the fixed radiation frame 
$\{\ev_{x}^{R}, \ev_{y}^{R},\ev_{z}^{R}\}$ and the fixed
source frame $\{\ev_{x}^{S}, \ev_{y}^{S},\ev_{z}^{S}\}$.
We choose $\ev_{z}^{R}$ to lie along the vector $\Nv$ which
points from the source to the detector.
The right hand plot shows the detector in the frame 
$\{\overline{\ev}_{x}, \overline{\ev}_{y},\overline{\ev}_{z}\}$
chosen so that the detector's arms lie along $\overline{\ev}_{x}$
and $\overline{\ev}_{y}$.
}
\label{fig:detector_radiation_frame_1}
\end{center}
\end{figure}

In order to project the quadrupolar moment $\Qv$ of the system onto the
detector we use the tensor $\Pv$ as shown in Eq.~(\ref{hresp}).
We will define the (fixed) radiation source frame relative to our previously
defined fixed source frame:
\begin{eqnarray} 
\ev_{x}^{R} &=& \ev_{x}^{S} \cos \Phi - \ev_{z}^{S} \sin \Phi \\
\ev_{y}^{R} &=& \ev_{y}^{S} \\
\ev_{z}^{R} &=& \ev_{x}^{S} \sin \Phi + \ev_{z}^{S} \cos \Phi
\end{eqnarray} 
where the $\Phi$ is the angle between the vector $\Nv$ which points from the
source to the detector and $\ev_{z}^{S}$. Similarly to how we defined 
$\{ \ev_{+}^{S}, \ev_{\times}^{S} \}$ we also define polarization tensors
of the radiation frame (following the notation of BCV2 \cite{BCV2}):
\begin{eqnarray}
\Tv_{+}       \equiv  \ev_{x}^{R} \otimes \ev_{x}^{R} -
                      \ev_{y}^{R} \otimes \ev_{y}^{R},
\; \, \, \,
\Tv_{\times}  \equiv  \ev_{x}^{R} \otimes \ev_{y}^{R} +
                      \ev_{y}^{R} \otimes \ev_{x}^{R}.
\end{eqnarray}
We also define the detector frame $\{\overline{\ev}_{x}, \overline{\ev}_{y},
\overline{\ev}_{z}\}$ so that the detector's arms lie along
$\overline{\ev}_{x}$ and $\overline{\ev}_{y}$.
The radiation and detection frames are shown in 
Fig.~\ref{fig:detector_radiation_frame_1}.

\begin{figure}
\begin{center}
\includegraphics[angle=0, width=1.1\textwidth]{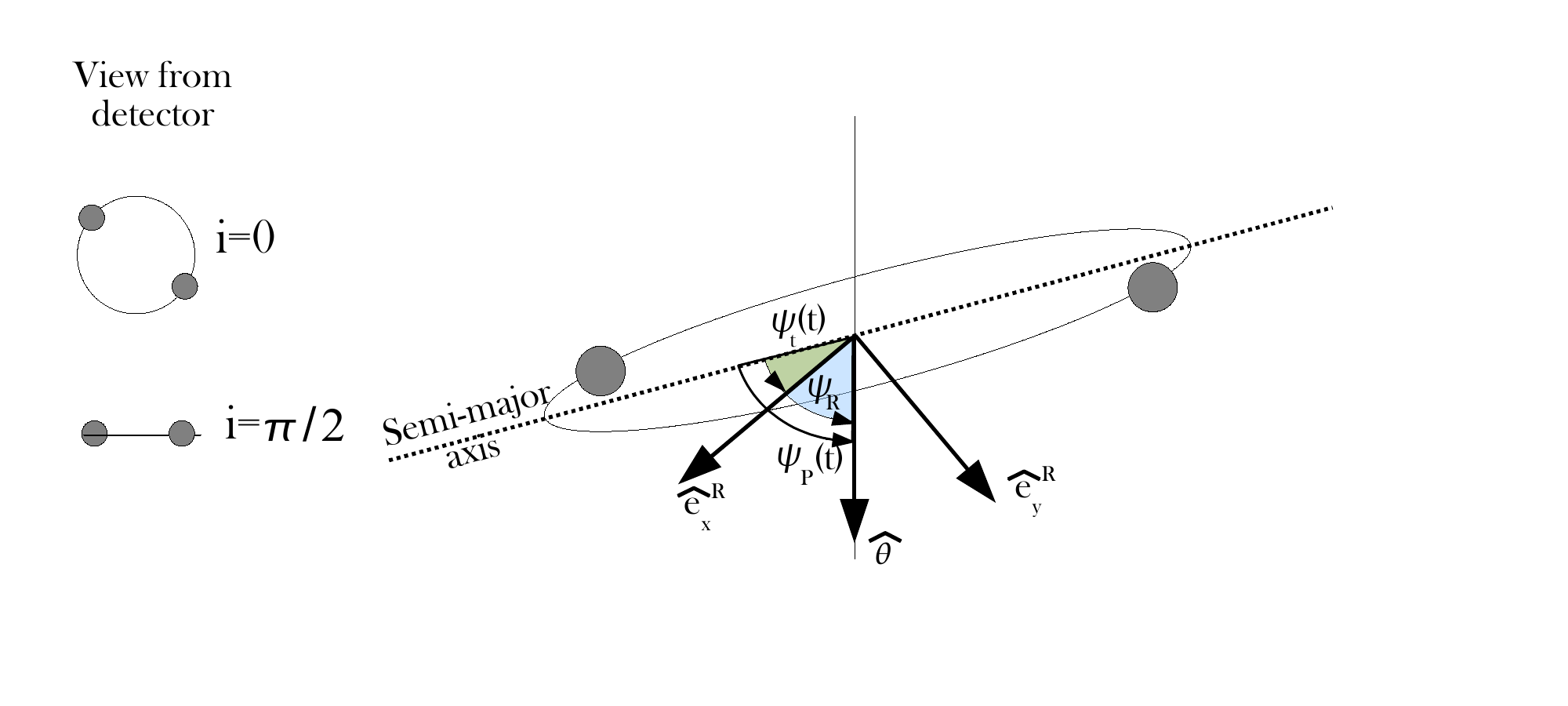}
\caption{The two small diagrams on the left show projections of a circular orbit 
onto the plane of the sky with inclination angle $\iota = 0$ and $\iota = \pi/2$. 
The diagram on the right shows the polarization angle $\psi_{P}(t)$ measured
anti-clockwise from the semi-major axis of the ellipse made by
projecting the binary's orbit onto the plane of the sky to a line of
constant azimuth $\hat{\theta}$ 
(i.e., a vertical line from the detector's horizon).
We see that $\psi_{P}(t)$ is the sum of the angles $\psi_{t}(t)$ measured
between the semi-major axis and $\ev_{x}^{R}$ and 
$\psi_{R}$ measured between $\ev_{x}^{R}$ and $\hat{\theta}$.
Since the radiation frame is fixed, $\psi_{R}$ remains constant with time.
As the binary precesses $\psi_{t}(t)$, and therefore $\psi_{P}(t)$, will evolve. 
}
\label{fig:polarization_FC}
\end{center}
\end{figure}

\begin{figure}
\begin{center}
\includegraphics[angle=0, width=0.9\textwidth]{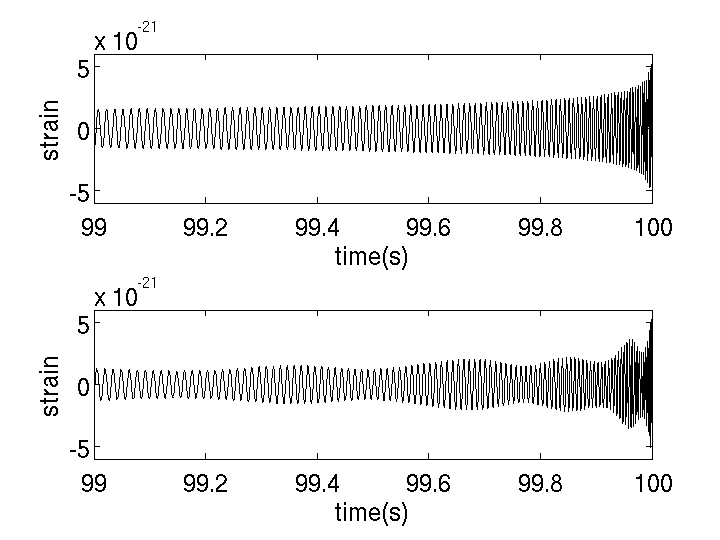}
\caption{
The gravitational waveforms we expect to observe from the late
inspiral phase of two different neutron star - black hole systems,
one consisting of non-spinning bodies (upper plot) and the other consisting of
maximally spinning bodies (lower plot).
Both systems are identical apart from the spin of their component bodies.
The spin-induced precession of the binary's orbital plane causes modulation of
the gravitational wave signal and can be clearly seen
in the lower plot.
}
\label{fig:FOMINJ_plotInj_paper}
\end{center}
\end{figure}

\begin{figure}
\begin{center}
\includegraphics[angle=0, width=0.6\textwidth]{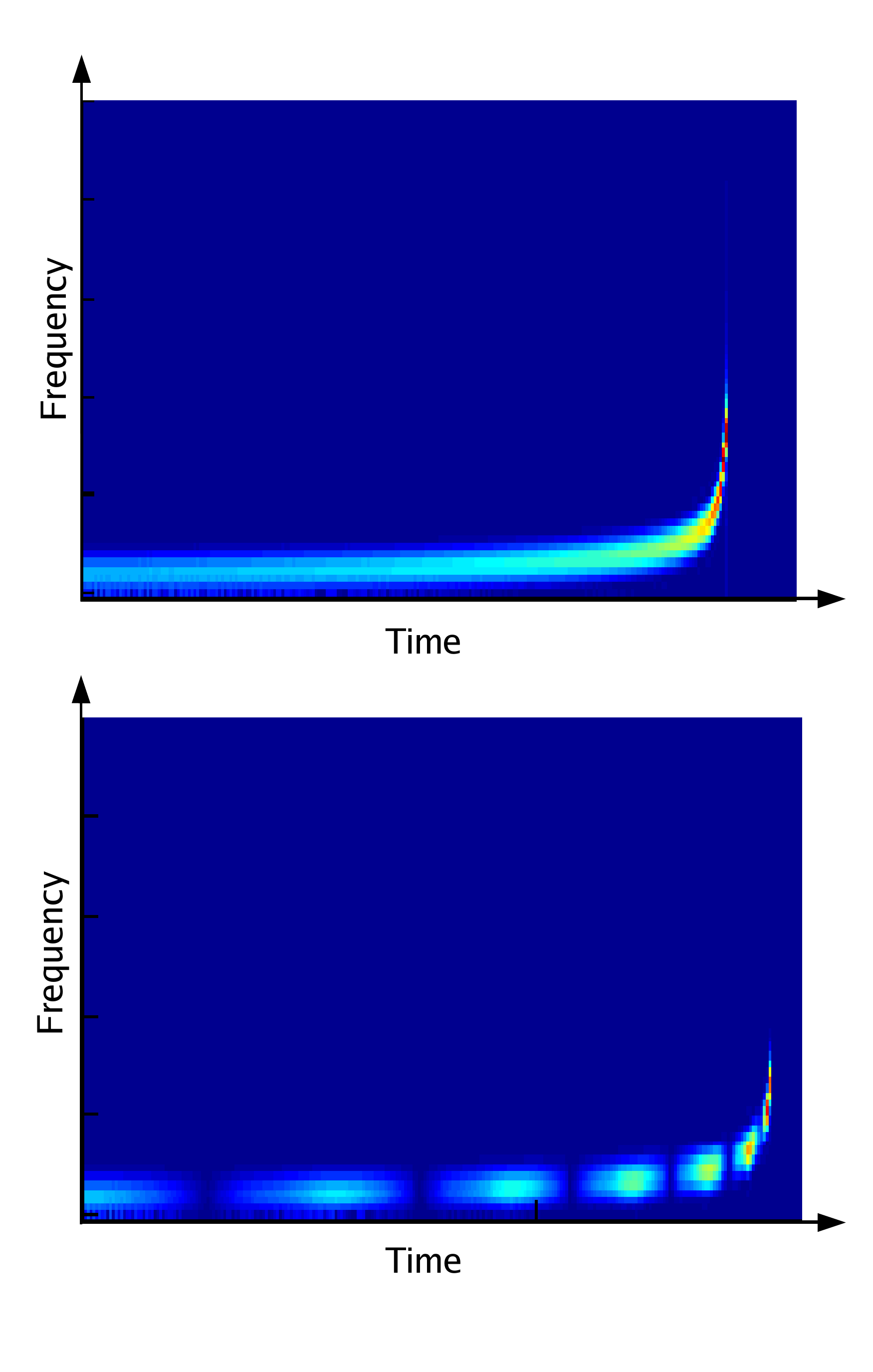}
\caption{
Spectrograms showing the gravitational waveforms we expect to observe from the late
inspiral phase of two different binary systems,
one consisting of non-spinning bodies (upper plot) and the other consisting of
maximally spinning bodies (lower plot).
Both systems are in quasi-circular orbits (i.e., not eccentric, although the
binary with spinning components will precess) and 
are identical apart from the spin of their component bodies.
The spin-induced precession of the binary's orbital plane causes modulation of
the gravitational wave signal and can be clearly seen
in the lower spectrogram.
The motion of LISA will cause similar modulations in the gravitational
waves it will observe.
}
\label{fig:spin_spec_thesis}
\end{center}
\end{figure}

The tensor $\Pv$ will depend upon the sky position $(\theta, \phi)$ and 
polarization angle $\psi_{P}$ of the source in relation
to the detector.
The inclination angle $\iota$ of a binary system is the angle between the
vector $\Nv$ joining the binary and detector, and the binary's 
orbital angular momentum $\Lv$,
\begin{eqnarray}
\iota = \cos^{-1} \hat{\Lv} \cdot \hat{\Nv}.
\end{eqnarray}
A circular orbit with inclination angle $\iota \neq 0, \pi$ will make an ellipse  
on the plane of the sky 
(i.e., the plane containing $\ev_{x}^{R}$ and $\ev_{y}^{R}$ ).
The orientation of this ellipse is described by the polarization angle
$\psi_{P}$. 
For a binary consisting of spinning components, both
inclination $\iota$ and the polarization angle $\psi_{P}$ will be functions of
time due to the precession of the orbital plane.
Using the FC style convention, the polarization angle $\psi_{P}$ is
measured anti-clockwise from the semi-major axis of the ellipse made by
projecting the binary's orbit onto the plane of the sky to a line of
constant azimuth $\hat{\theta}$ 
(i.e., a vertical line from the detector's horizon).
This is shown in Fig.~\ref{fig:polarization_FC}.
Note that there are two parts of the polarization angle shown on this figure;
i) $\psi_{R}$ is the (constant) angle between the x-axis of the radiation frame
$\ev_{x}^{R}$ and $\hat{\theta}$ and 
ii) $\psi_{t}(t)$ which is the angle between the semi-major axis of the ellipse
made by projecting the binary's orbit onto the plane of the sky and
$\ev_{x}^{R}$ which will evolve as the binary precesses.
%

Note that during the relatively short duration of the inspiral we can
make the approximation that the sky position $(\theta, \phi)$ of the 
source is constant. For sources that emit for longer duration in the detectors
band of good sensitivity, such as pulsars that will
be observed by LIGO or inspiral events that will be observed by LISA, it is
necessary to include the time-dependence of the source's sky position when
calculating the detector's response.

The antenna patterns $F_{+}$ and $F_{\times}$  encode the detector's 
directional sensitivity to 
plus ($+$) and cross ($\times$) polarization gravitational waves 
(see, for example Eqs.~(4a,b) of ACST \cite{ACST} or 
Eqs.~(29) and (30) of BCV2 \cite{BCV2}) and are given by
\begin{eqnarray} 
\label{Fpluscross} 
F_{+}(t) = \frac{1}{2}
\left( 1 + \cos^{2} \theta \right) \cos 2 \phi \cos 2 \psi_{R} 
- \cos \theta \sin \phi \sin 2 \psi_{R}, \\
F_{\times}(t) = \frac{1}{2} \left( 1 + \cos^{2} \theta \right) 
\cos 2 \phi \cos 2 \psi_{R}
+ \cos \theta \sin \phi \sin 2 \psi_{R}.
\end{eqnarray}  

The final form for the detector response is
\begin{eqnarray}
\label{hrespfinal}
h_{\rm{resp}} = \frac{\mu}{D} \frac{M}{r}
\underbrace{
-2
\left(
\left[
\ev_{+}^{S}
\right]^{ij} \cos 2 \Phi_{S}
+
\left[
\ev_{\times}^{S}
\right]^{ij} \sin 2 \Phi_{S}
\right)
}_{\Qv}
\underbrace{-2
\left(
\left[\Tv_{+}\right]_{ij} F_{+} +
\left[\Tv_{\times}\right]_{ij} F_{\times}
\right)
}_{\Pv}.
\end{eqnarray}
Note that $\Pv$ does not vary with time and that the time evolution of the 
binary is encoded within $\Qv$.

\subsection{Parameters of the binary}
17 physical parameters are required to fully describe a generic spinning binary 
system relative to a particular observer.
These parameters are 
the masses of the binary's components, $m_{1}$ and $m_{2}$ (2);
the spins of the binary's components, $\Sv_{1}(t)$ and $\Sv_{2}(t)$ (6),
the orbital angular momentum of the system, $\Lv_{N}(t)$ (3),
and the orbital phase $\Phi_{S}(t)$ (1) 
at a particular time $t$;
the eccentricity $e$ and the point of perihelion (or aphelion) (2)
and the distance and direction of the observer from the system $\Nv$ (3). 
Note that in this analysis we assume that the emission
of gravitational waves has circularized the binary's orbit before
it is observable (see Sec.~\ref{sec:aacirc}). 

The set of parameters listed here is not unique since various parameters can be
recoded in terms of other parameters with no loss of information. For 
instance specifying both component masses $m_{1}$ and $m_{2}$ is obviously 
equivalent to specifying both total mass $M = m_{1} + m_{2}$ and the symmetric 
mass ratio $\eta = m_{1} m_{2} / M^{2}$ or the 
reduced mass $\mu =  m_{1} m_{2}/M$.
The absolute separation of the binary's components can be found
using $r = (M/\omega^{2})^{1/3}$ (from Kepler's third law in geometric
units) where $\omega$ is the orbital frequency.
The direction of the orbital angular momentum
relative to the detector can be specified by the inclination angle 
$\iota$ and polarization angle $\psi_{P}$
and its magnitude is given by Eq.~(\ref{LNmag}).
We can write the spins as $\Sv_{i} = \chi_{i} m_{i}^{2} \hat{\Sv}_{i}$,
where $\chi_{i}$ is a dimensionless parameter such that $0< \chi_{i} < 1$
for compact objects.

The parameters used to describe the system relative to an observer can be 
classified into two groups: {\it intrinsic} and {\it extrinsic} parameters. 
Intrinsic parameters describe the system itself and include its masses and 
spins whereas the extrinsic parameters describe the system's distance and 
orientation to an observer. This distinction proves significant in the design 
of the detection template families we use to search for spinning systems as it
turns out that, in general, to determine intrinsic parameters we need to include
them in the templates we use to matched-filter our detector data whereas 
extrinsic parameters can be found automatically by maximising over the
matched-filter output. 





\section{Development of detection template families to capture 
gravitational waves from spinning systems}
\label{sec:dtf}

In the introduction (Sec.~\ref{sec:DA}) we showed that the optimal method 
to detect a known signal in a noisy data stream is to perform 
matched-filtering using templates that accurately represent the signal weighted 
(in the frequency domain) by the power spectrum of the detector noise.
We cannot use the target model waveform 
(described in Sec.~\ref{Sec:TargetModel}) as a detection template since 
the large number of parameters needed to describe the waveform 
(i.e., 17, or 15 if we assume circular orbits) 
mean 
that we would require an intractably huge number of templates to cover the 
parameter space 
(i.e., the range of masses, spin magnitudes and orientations) 
we wish to search. 

\subsubsection{Detection template families}
Instead of using the target model, we will make use of a 
{\it detection template family} (DTF) that is designed to capture the essential
features of the true gravitational wave signal 
(as approximated by the target model)
but which depends on a smaller number of parameters.
Detection templates might be parameterised by either physical parameters
of the source or, as in the case of the DTF we will use, by non-physical
or {\it phenomenological} parameters that describe the properties
of the observed waveform rather than the source itself. 

At their best, DTFs can reduce the computational requirements of a gravitational
wave search while achieving essentially the same detection performance
as exact templates (i.e., as generated using the target model). 
However, DTFs can include non-physical signal shapes that may increase the number 
of spurious triggers caused by noise (i.e., false alarms) which will in turn
require us to set larger SNR thresholds and will affect the calculation of upper 
limits (see Sec.~\ref{sub:upperlimit}). 
Detection template families are also less adequate for parameter estimation, 
since the mapping between the detection template parameters and those of the
binary are not one-to-one, this is why they are called {\it detection} template
families.

Apostolatos (1995) \cite{Apostolatos:1995} introduces 
the fitting factor ($FF$) as a quantitative measure of how well a given 
family of templates can ``fit'' a predicted gravitational waveform.  
The value of the fitting factor gives the reduction in signal to noise ratio (SNR) 
caused by using a given template family rather than the true signal, this is
described in more detail in the next few Sections.
In the terminology of Damour et al. (2001) \cite{DIS2001}) we would say that
DTFs are {\it effectual} (good fitting factor with target model) if not
particularly {\it faithful} (i.e., poor estimation of parameters of target model). 

The distance-range of a search for gravitational wave signals emitted by 
astrophysical systems is limited by the lowest SNR for which a true signal 
can be distinguished from noise.
Using a detection template family with a $FF=0.9$ would result in a $10\%$ 
drop in distance-range and a corresponding $(1-0.9^3 \approx) 27\%$ drop 
in detectable event rate when compared with using ``perfect'' templates with $FF=1$. 
Apostolatos measures low fitting factors when using non-modulated PN templates 
to search for spin-modulated gravitational wave signals 
(Sec.~VIII of Apostolatos (1995) \cite{Apostolatos:1995}. 
Results from this analysis will be discussed later in this Section). 
These results clearly motivate the development of a detection template family 
which can accurately model the spin-induced modulation of the gravitational 
wave signal.
We will now review the analysis of the effects of precessing, 
inspiraling binary systems and see how this has informed the development of a new
detection template family designed to capture their gravitational wave
emission.

\subsection{Previous analysis on the effect of spin on gravitational waves}
In ACST \cite{ACST} the authors consider a simplified form of the 
target model which neglects other post-Newtonian corrections in order 
to emphasise the effects of spin upon upon the system's dynamics and 
gravitational wave emission.
The authors concentrate their analysis on two special binary 
configurations; 
i) $m_{1} \simeq m_{2}$ which could represent a NS-NS system
or a symmetric BH-BH system and
ii) $\Sv_{2} = 0$ which could represent a very
asymmetric system ($m_{1} \gg m_{2}$) for which the spin of the 
lower mass component could be neglected.
For case i) the authors make the additional assumption that spin-spin 
effects can be ignored since they occur at a higher 
post-Newtonian order (2PN), and are therefore typically smaller 
than the leading spin-orbit term (1.5PN). Spin-spin effects
are not present for a system with only one spinning component
as in case ii). 
Making the assumptions described the authors were able to write the
equations governing the system's precession as
\begin{eqnarray}
\label{S1ACST}
\dot{\hat{\Sv}}_{1} &=&
\left(
2 + \frac{3 m_{2}}{2 m_{1}}
\right) \frac{\Jv}{r^{3}}  \times \hat{\Sv}_{1} \\
\label{LNACST}
\dot{\hat{\Lv}}_{N} &=& 
\left(
2 + \frac{3 m_{2}}{2 m_{1}} 
\right)
\frac{\Jv}{r^{3}}
\times \hat{\Lv}_{N}
\end{eqnarray}
where $\Jv = \Lv + \Sv$ and $\Sv = \Sv_{1} + \Sv_{2}$
\footnote{To derive these simplified precession equations take the precession 
equations given in Eq.~(11) of ACST \cite{ACST} and neglect all
spin-spin and higher order terms. 
We use the result that $\Jv \times \Lv 
= (\Lv+\Sv) \times \Lv 
= (\Lv \times \Lv) + (\Sv \times \Lv)$
which reduces to $\Jv \times \Lv = \Sv \times \Lv$
(and similarly $\Jv \times \Sv = \Lv \times \Sv$) to write the
right hand side of the simplified precession equations in terms of
$\Jv$.}.
For these {\it ACST configurations} the authors constructed approximate
solutions to the precession and inspiral equations and were able to gain
insight into the dynamics of these binaries during their inspiral.

The authors identify two distinct evolutionary behaviours of the binary: 
i) {\it simple precession} occurs when total angular momentum $\Jv>0$ and 
the orbital angular momentum $\Lv_{N}$ and the spin angular momentum $\Sv$ 
precess about a near constant $\Jv$,
ii) {\it transitional precession} occurs when $\Lv$ and $\Sv$ are 
anti-aligned and of the same approximate magnitude such that $\Jv \sim 0$ 
and the system temporarily ``loses its gyroscopic bearings and tumbles in 
space''.
As discussed previously in Sec.~\ref{sec:EvolEqns}, 
$|\Sv|$ will remain almost constant 
during the course of inspiral 
while $|\Lv|$ will decay with time. 
Therefore, for 
transitional precession to occur we require that 
initially $|\Lv| > |\Sv|$ and that $\Lv$ and $\Sv$ be
approximately anti-aligned with each other.

\begin{figure}
\begin{center}
\includegraphics[angle=0, width=0.7\textwidth]{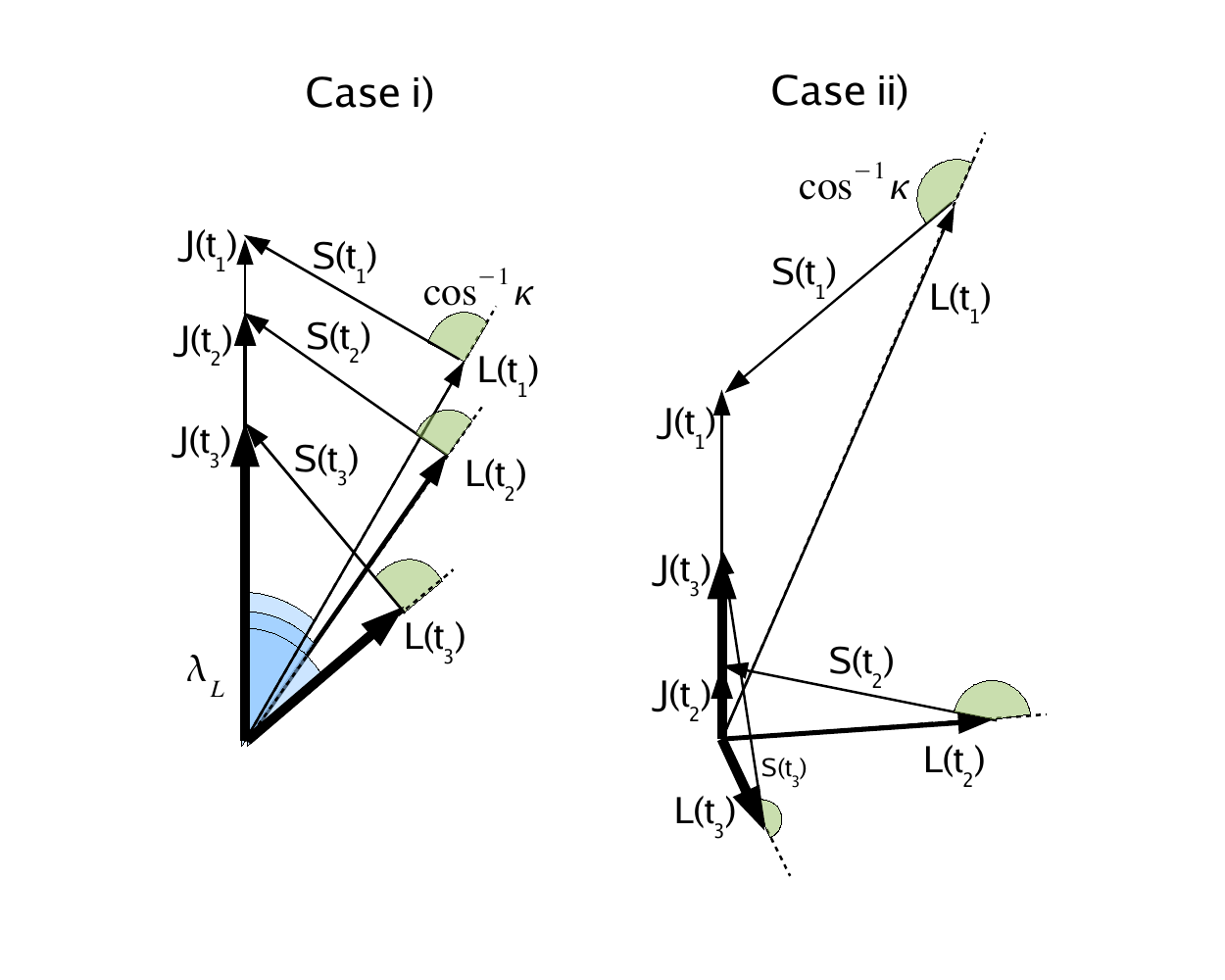}
\caption{The evolution of orbital angular momentum $\Lv$, 
spin angular momentum $\Sv$ and total angular momentum $\Jv$ 
during simple and transitional precession.
In case i) only simple precession occurs as the total angular momentum $\Jv$ 
remains relatively large and roughly constant in direction while $\Lv$ 
and $\Sv$ precess about it.
In case ii) the evolution undergoes simple precession at early times ($t_{1}$)
until at around $t_{2}$, $\Lv$ has become anti-aligned with
and similar in magnitude to $\Sv$ 
so that $\Jv = \Lv + \Sv \sim 0$. The system will undergo a
period of transitional precession, during which the system
will tumble randomly in space, until $|\Lv| < |\Sv|$ and simple
precession is resumed ($t_{3}$).
This figure is based upon Fig.~2 of ACST \cite{ACST}.
}
\label{fig:simple_transitional}
\end{center}
\end{figure}

The evolution of orbital $\Lv$, spin $\Sv$ and total angular momentum
$\Jv$ during simple and transitional precessions is shown in
Fig.~\ref{fig:simple_transitional}.
Considering the simplified precession equations (Eqs.~(\ref{S1ACST}) and
(\ref{LNACST})) we can show that $\Lv$ and $\Sv$ maintain fixed directions
relative to each other as they precess about $\Jv$. 
We can write
\begin{eqnarray}
\label{kappa}
\kappa \equiv \hat{\Sv} \cdot \hat{\Lv}
\end{eqnarray}
where $\kappa$, 
and therefore the opening angle $\rm{cos}^{-1} \kappa$ between $\Lv$ and $\Sv$, 
will remain constant throughout the inspiral
\footnote{To prove that $\equiv \hat{\Sv} \cdot \hat{\Lv}$ is a constant of
the motion consider the time derivative 
$d(\hat{\Sv} \cdot \hat{\Lv})/dt =
\hat{\Sv} \cdot \dot{\hat{\Lv}} + \hat{\Lv} \cdot \dot{\hat{\Sv}}$.
When evaluating the two terms on the right hand side both will contain
cross products of a vector with itself and therefore be equal to zero.}.
The decay of $|\Lv|$ and $|\Jv|$ during the inspiral as $|\Sv|$ remains 
approximately constant will cause the opening angle $\lambda_{L}$ 
between $\Lv$ and $\Jv$ to increase during the inspiral.
This is shown in Fig.~\ref{fig:simple_transitional}. 
The random nature of the motion of $\Jv$ during transitional precession
makes the accurate prediction of the resulting waveform practically
impossible and it is therefore fortunate that most inspiral evolutions
do not exhibit transitional precession 
(this is discussed further in Sec.~\ref{BCV2analysis}).

\begin{figure}
\begin{center}
\includegraphics[angle=0, width=0.7\textwidth]{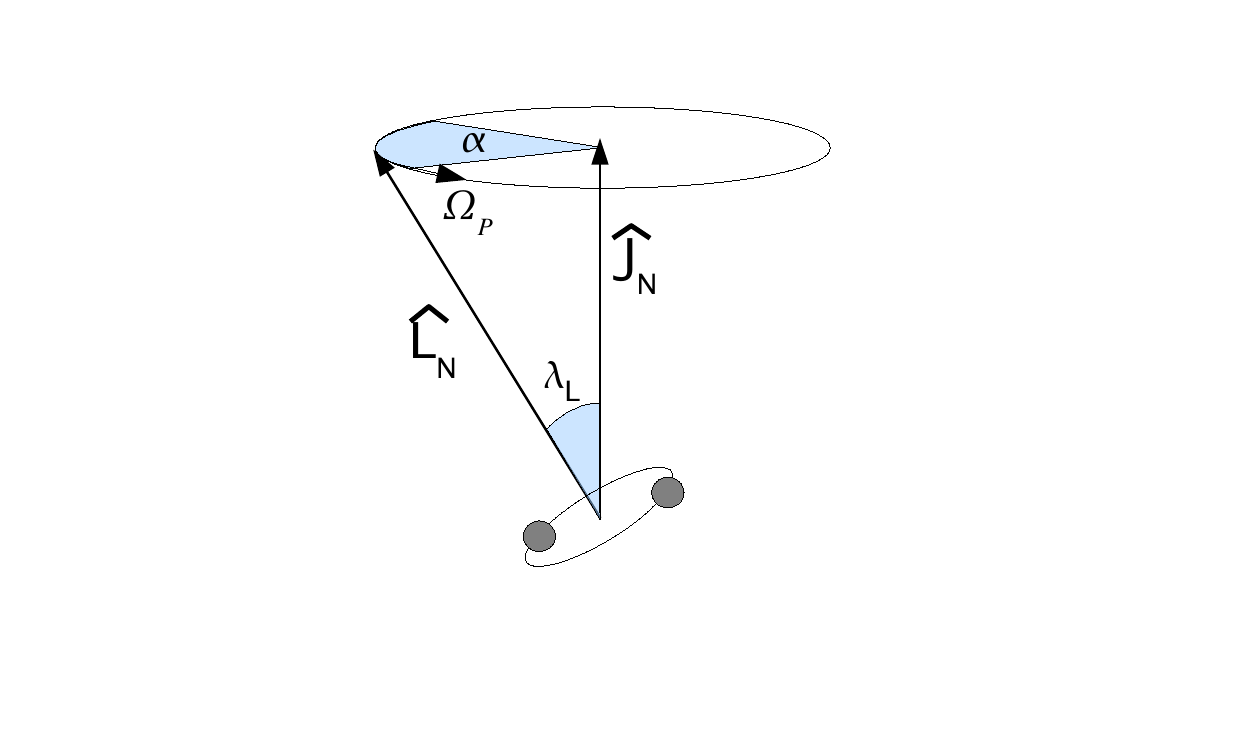}
\caption{During simple precession the orbital angular momentum $\Lv$ 
of the (ACST configuration) binary will precess about the total angular
momentum $\Jv$ with frequency $\Omega_{p}$. The opening angle $\lambda_{L}$
and the precession angle $\alpha$ are also identified.
This figure is based upon Fig.~4 of ACST \cite{ACST}.
} 
\label{fig:orbangmmtm_cone}
\end{center}
\end{figure}

From the simplified precession equations (Eqs.~(\ref{S1ACST}) and 
(\ref{LNACST})) we see that during simple precession 
$\Lv$ and $\Sv$ will precess about $\Jv$ with angular frequency 
\begin{eqnarray}
\Omega_{p} = 
\left(
2 + \frac{3 m_{2}}{2 m_{1}} 
\right) 
\frac{|\Jv|}{r^{3}}
= \frac{d \alpha}{dt}
\end{eqnarray}
where we have also defined the precession angle $\alpha$ 
(see Fig.~\ref{fig:orbangmmtm_cone}). 
The authors of ACST \cite{ACST} considered cases where 
$|\Lv| \gg |\Sv|$ and where $|\Sv| \gg |\Lv|$ and found that the
evolution of the precession angle could be approximated by
a power law in orbital frequency $f = \omega/ 2\pi$:
\footnote{Please note that an error occurs in the first bracketed terms of
the right hand side of Eq.~(45) of \cite{ACST}. The term $1+3M_{1}/4M_{2}$
should in fact read $1+3M_{2}/4M_{1}$ and appears correctly in Eq.~(29)
of \cite{Apostolatos:1995}.}
\begin{equation}
\frac{\alpha(f)}{2 \pi} \simeq
\left\{
\begin{array}{ll}
11 
\left(    
1 + \frac{3 m_{2}}{4 m_{1}} 
\right)
\frac{10 M_{\odot}}{M} \frac{10 {\rm{Hz}}}{f}
& \rm{for} \, |\Lv| \gg |\Sv| 
\\
1.9 
\left(
1 + \frac{3 m_{2}}{4 m_{1}} 
\right)
\frac{m_{1}}{m_{2}} \frac{S}{m_{1}^{2}} 
\left( \frac{10 M_{\odot}}{M} \frac{10 {\rm{Hz}}}{f}
\right)^{2/3}
& \rm{for} \, |\Sv| \gg |\Lv|. 
\end{array} 
\right.
\label{alphaf}
\end{equation}

Apostolatos (1995) \cite{Apostolatos:1995} introduces 
the fitting factor (FF) as a quantitative measure of the reduction in SNR 
caused by using a particular family of templates in order 
to capture a predicted gravitational
waveform.
The fitting factor is given as
\begin{eqnarray}
{\rm{FF}} = 
{\rm{max}}_{\lambda_{1}, \lambda_{2} \dots} 
\frac{\left< h, T_{\lambda_{1}, \lambda_{2} \dots} \right>}
{\sqrt{  
\left< h, h  \right> 
\left< T_{\lambda_{1}, \lambda_{2} \dots}, 
T_{\lambda_{1}, \lambda_{2} \dots} \right>
}}
\end{eqnarray}
where $h$ is our best prediction of the gravitational waveform that will
be observed and $T$ is a template designed to capture $h$ and which is
parameterised by $\lambda_{1}, \lambda_{2} \dots$.
The denominator ensures that $0 \leq {\rm{FF}} \leq 1$.
Apostolatos (1995) \cite{Apostolatos:1995} 
writes the detector response to a precessing,
inspiraling binary as
\begin{eqnarray}
\tilde{h}_{\rm{resp}}(f) \simeq 
\frac{1}{2} \tilde{h}_{C}(f) \times {\rm{AM}} \times {\rm{PM}}
\end{eqnarray}
where $\tilde{h}_{C}(f)$ is a non-modulated carrier signal and
AM and PM represent amplitude and phase modulations caused by
spin-induced precession (see Eq.~(17) of Apostolatos (1995) 
\cite{Apostolatos:1995} and note that the final multiplicative
factor is approximately unity).
Apostolatos investigates the relative influence of phase and 
amplitude modulations upon the fitting factor
when the templates used to detect spin-modulated gravitational
waves do not themselves include the effects of spin.
It was found that
at worst, amplitude modulation alone can only account for
fitting factors dropping to $\sim 0.9$ whereas phase modulation
can cause the fitting factor to drop below $0.6$.

Apostolatos also investigated the effect of the opening angle between
spin and orbital angular momentum (i.e., $\cos^{-1} \kappa$) 
on the fitting factors.
When considering a maximally spinning $10M_{\odot}$ BH and a non-spinning
NS, $FF < 0.9$ were measured for around a quarter of systems with a
$\cos^{-1} \kappa = 30^\circ$.
When the opening angle is increased to $\cos^{-1} \kappa = 140^\circ$ 
then $FF < 0.9$ are measured for nearly all systems.

Building on work in ACST \cite{ACST}, Apostolatos suggests in 
Refs.~\cite{Apostolatos:1995, Apostolatos:1996} that the spin-induced modulational 
effects of the gravitational wave signal's phase could be captured by adding 
modulational terms to the standard non-modulational (NM) frequency domain 
phasing of templates used to search for the inspiral of binaries with 
non-spinning components (see Eq.~(12) of Apostolatos (1996)) 
\cite{Apostolatos:1996}):
\begin{equation}
\label{apostansatz}
\psi_{\rm{Spin-Modulated}}(f) \rightarrow \psi_{\rm{Non-Modulated}}(f) + 
\mathcal{C} \cos(\delta + \mathcal{B} f^{-2/3}).
\end{equation}
This is the Apostolatos ansatz.
It makes sense that the modulational term occurs at $f^{-2/3}$ since this 
corresponds to the power law evolution of the precession angle when 
$|\Sv| \gg |\Lv|$ (see Eq.~(\ref{alphaf})). 
An implementation of this detection template family 
(which we shall refer to as the Apostolatos family) was tested in
Grandcl\'ement et al. (2003) \cite{Grandclement2003}.
Although the fitting factor increased by around $15 - 30\%$ compared
to using templates with no spin-modulation included, the fitting
factors were still only $\sim 0.7$ which would lead to a drop in
expected event rate of up to $80\%$.
We will now describe the work of Buonanno, Chen and Vallisneri which
led to the development of a detection template family
which captures spin-modulated gravitational waves with $FF > 0.9$
and which we shall use to search for
gravitational waveforms in real LIGO data.

\subsection{BCV2 analysis of spinning binary systems}
\label{BCV2analysis}
In BCV2 the authors used the target model described in 
Sec.~\ref{Sec:TargetModel} to further investigate the effects of spin
upon the observed gravitational waveform which led to the development
of a new detection family (which we shall refer to as the
BCV2 DTF).
The BCV2 analysis considers a wider range of systems than ACST and do
not limit themselves the ACST configurations previously discussed
(i.e., either $m_{1} \simeq m_{2}$ or $\Sv_{2} = 0$).
In BCV2 the authors consider BH-BH systems with masses
$(20,10)M_{\odot}$,
$(15,15)M_{\odot}$,
$(20,5)M_{\odot}$,
$(10,10)M_{\odot}$ and
$(7,5)M_{\odot}$
consisting of maximally spinning BHs
and NS-BH systems with masses 
$(1.4,10)M_{\odot}$ consisting of a maximally spinning BH and a
non-spinning NS (Sec VIB and VIC of BCV2 \cite{BCV2})  . 
We shall refer to these as {\it BCV2 configurations}.
The choices of spin are not based upon astrophysical results
(most of the spin measurements summarised earlier were published after
BCV2 \cite{BCV2}) but to emphasise the effects of spins upon the
evolution and emission of these systems.
We summarise the findings of their analysis here.

When ignoring spin-spin coupling (but still considering binaries
consisting of two spinning bodies) the authors of BCV2 \cite{BCV2}
find a generalisation of Eq.~(\ref{kappa}) for the opening angle
between the orbital angular momentum and the spins:
\begin{eqnarray} 
\kappa_{\rm{eff}} \equiv 
\frac{\hat{\Lv}_{N} \cdot \Sv_{{\rm{eff}}}}
{M^{2}}
\end{eqnarray}
where we have defined an {\it effective spin}
\begin{eqnarray}
\Sv_{{\rm{eff}}} \equiv 
\left(1 + \frac{3 m_{2}}{4 m_{1}} \right) \Sv_{1} +
\left(1 + \frac{3 m_{1}}{4 m_{2}}   \right) \Sv_{2}. 
\end{eqnarray}

The authors of BCV2 \cite{BCV2} investigate the regularity that 
transitional precession occurs.
For transitional precession to be observed, we require
that 
$|\Lv_{N}| = |\Sv| \leq |\Sv_{1}| + |\Sv_{2}|$
before the system plunges, i.e., $f_{\rm{trans}}^{\rm{min}} < f_{\rm{Schw}}$
where we assume that plunge occurs at the Schwarzschild radius.
For transitional precession to be observed, they find that
the symmetric mass ratio 
must be less than some limiting value
$\eta \lesssim 0.22$, 
see Sec.~IIIE of BCV2 \cite{BCV2}
\footnote{There appears to be an error in
Eq.~(59) of BCV2 \cite{BCV2} in which both inequalities
should be reversed. 
In the first case we should demand that the minimum frequency 
$f_{\rm{trans}}^{\rm{min}}$
for transitional precession to occur 
be less than the Schwarzschild frequency $f_{\rm{Schw}}$ which would
lead to $f_{\rm{trans}}^{\rm{min}} / f_{\rm{Schw}} \lesssim 1$ and therefore
$\eta \lesssim 0.22.$}.  
Of the BCV2 configuration binaries considered, only the 
$(20,10)M_{\odot}$, $(20,5)M_{\odot}$
and $(10,1.4)M_{\odot}$ binaries satisfy the
condition on $\eta$.
The authors of BCV2 \cite{BCV2} considered $\geq 200$ initial configurations
of each of these binaries and observed no transitional precession
of the $(20,10)M_{\odot}$ and $(10,1.4)M_{\odot}$ binaries and only a few
cases of transitional precession of the $(20,5)M_{\odot}$ binary.  
Indeed, for the $(10,1.4)M_{\odot}$ binary, consisting of 
a maximally spinning BH and a non-spinning
NS, the magnitude of the spin angular momentum was always greater than
the magnitude of the orbital angular momentum meaning that transitional
precession could never occur for any configuration of the binaries 
spin and orbital angular momentum.

The authors of BCV2 \cite{BCV2} also investigated the effects of the spin
terms on the evolution of the orbital angular frequency $\omega$ (Eqs.~(\ref{omegadot}), 
(\ref{omegadot1.5PN}) and (\ref{omegadot2PN})) 
and 
on the accumulated orbital phase $\Psi$ (Eq.~\ref{accorbphase}).
They find that the effects of spin on the accumulated orbital phase $\Psi$
would be largely non-modulational and could be well captured by the 
phasing used to describe binaries with non-spinning components.
It is important to acknowledge that although the accumulated orbital
phase $\Psi$ is not modulated by the effects of spin, the phase (and 
amplitude) of the gravitational waveform observed at the detector
will be modulated by the spin-induced precession of the orbital plane
and that these effects should not be neglected. 
The phase $\Phi_{S}(t)$ which enters the general expression for detector
response $h_{\rm{resp}}$ (see Eq.~(\ref{hrespfinal})) is measured with respect 
to basis vector $\ev_{1}^{S}$ which is always in the instantaneous orbital plane 
(i.e., $\Lv_{N}(t) \cdot \ev_{1}^{S}(t) = 0$, see Fig.~\ref{fig:source_frame_1}). 
In general, $\Phi(t) \neq \Psi(t)$ since $\ev_{1}^{S}$ (and also $\ev_{2}^{S}$)
can have arbitrary rotation about $\Lv_{N}$.
In BCV2 \cite{BCV2} the authors define a new {\it precessing convention}
for the basis $\{ \ev_{1}^{S}, \ev_{2}^{S} \}$ 
such that $\Phi(t) = \Psi(t)$ which allows the use
of $\Psi(f)$ when we write down the detector response $h_{\rm{resp}}$. From
their earlier observations we know that the non-modulational phase 
$\psi_{\rm{NM}}(f)$ used to describe the phasing of binaries with non-spinning
components is a good approximation to $\Psi$.

\section{The BCV2 detection template family for spinning systems}
\label{sec:BCV2dtf}
Following their analysis the authors of BCV2 \cite{BCV2}
proposed a detection template family representing a generalisation of 
the Apostolatos ansatz 
designed to capture gravitational waveforms from precessing, inspiraling
binaries in the adiabatic limit.
Significantly we will find in Sec.~\ref{sec:maximisation} that
the majority of the parameters of this DTF are extrinsic parameters
that can be found in a computationally cheap manner by 
maximisation of the measured SNR.
From BCV2 \cite{BCV2} Eq.~(86) we write the form if the DTF:
\begin{eqnarray}
\label{BCV2DTFa}
h( \dots; f) =
\left[ \sum_{k=1}^n (\alpha_k + i \alpha_{k+n}) 
\mathcal{A}_k (f) \right]
 e^{2 \pi i f t_{0}} e^{i \psi_{\rm{NM}} (f)}
\end{eqnarray}
for $f > 0$ and $h(f) = h^{*}(-f)$ for $f<0$. 
The $\alpha$'s represent the global phase, the strength of the amplitude 
modulation due to spin-induced precession, the relative phase of these 
modulations to the leading order amplitude ($f^{-7/6}$) and the internal 
complex phase of the modulation \cite{BCV2}.
The (real) amplitude functions $\mathcal{A}_k$ depend on the precise form
of the template chosen.
The function $\psi_{\rm{NM}}$ is the non-modulated phasing of a non-spinning
binary and is given as power series of gravitational wave frequency $f$:
\begin{eqnarray} 
\label{psiNMexp}
\psi_{\rm{NM}}(f) = f^{-5/3} \left (   
\psi_{0} 
+ \psi_{1} f^{1/3}  
+ \psi_{2} f^{2/3}  
+ \psi_{3} f  
\dots
\right).
\end{eqnarray} 
We have discussed in Sec.~\ref{BCV2analysis} that the non-modulated phase
$\psi_{\rm{NM}}$ used to describe a binary with non-spinning components 
has been to capture well the accumulated orbital phase of binaries with
spinning components. 
In practice we find that
the phasing of the gravitational wave can be captured well using
only the $\psi_{0}$ and $\psi_{3}$ terms 
and we will neglect the other terms in this expansion 
\footnote{
Here we follow convention set by data analysts and have multiplied the
subscript labels of the $\psi$ values used by Buonanno, Chen and Vallisneri 
by two.
Hence, $\psi_{3}$ here is completely equivalent to the $\psi_{3/2}$
used in BCV2 \cite{BCV2} and similarly for the other terms
in the expansion of $\psi_{\rm{NM}}$.
}.

In BCV2 \cite{BCV2} the authors suggest and test three forms of the
detection template family before recommending the third family which they
refer to as $(\psi_{0}, \psi_{3}, \beta)_{6}$ (Eq.~(90) of  BCV2 \cite{BCV2}):
\begin{eqnarray}
\label{BCV2DTF}
(\psi_{0}  \psi_{3} \beta)_{6}: && \nonumber \\
h(\dots; f) &=& f^{-7/6} 
\big[ 
  (\alpha_{1} + i\alpha_{2})
+ (\alpha_{3} + i\alpha_{4}) \cos(\beta f^{-2/3}) \nonumber\\
&+& (\alpha_{5} + i\alpha_{6)}) \sin(\beta f^{-2/3})
\big] \nonumber  \\
&&\theta(f_{\rm{cut}} -f) 
e^{2 \pi i f t_{0}} \,
\rm{exp} \, i \big[ 
\psi_{0} f^{-5/3} + \psi_{3} f^{-2/3}
\big].
\end{eqnarray}
Rewriting Eq.~(\ref{BCV2DTF}) similarly to Eq.~(\ref{BCV2DTFa}) we find 
the three real 
amplitude functions, $\mathcal{A}_k(f_{\rm{cut}},\beta;f)$
of $(\psi_{0}, \psi_{3}, \beta)_{6}$ to be
\begin{eqnarray}
\label{amp}
\mathcal{A}_1(f_{\rm{cut}},\beta;f) &=& f^{-7/6} 
\theta (f_{\rm{cut}}-f) \nonumber\\
\mathcal{A}_2(f_{\rm{cut}},\beta;f) &=& f^{-7/6} 
\cos(\beta f^{-2/3}) \theta (f_{\rm{cut}}-f) \nonumber\\
\mathcal{A}_3(f_{\rm{cut}},\beta;f) &=& f^{-7/6} 
\sin(\beta f^{-2/3}) \theta (f_{\rm{cut}}-f).
\end{eqnarray}
The $\beta$ parameter varies to capture the effects of spin modulation.
We see that in the Apostolatos ansatz Eq.~(\ref{apostansatz}), the 
term $\cos(\beta f^{-2/3})$ is an approximation for the 
precession angle $\alpha$ when $|\Sv| \gg |\Lv|$ (see Eq.~(\ref{alphaf})). 
The parameter $\beta$ takes a similar role in this DTF. 
In Buonanno et al. (2005) \cite{PBCVT} (known as PBCVT) the authors
provide some physical interpretation of the $\beta$ parameter identifying
it as representing the rate of change of the precession angle $\alpha$, i.e.,
$\Omega_{p} = d \alpha / dt$ at the frequency band of good detector
sensitivity. 

The function $\theta$ is the Heaviside step function which is defined as
\begin{displaymath}
\theta(x) = \left\{
\begin{array}{cl}
 0  & x < 0 \\
 1  & x \geq 0
\end{array} \right.
\end{displaymath}
The parameter $f_{\rm{cut}}$ is used to terminate the template when we no
longer have confidence that the template will provide a good match to
the signal (i.e., at the late stages of inspiral when the adiabatic 
approximation is no longer valid).
For gravitational wave frequency $f \leq f_{\rm{cut}}$ 
then $\theta(f_{\rm{cut}}-f) =1$. 
The choice of $f_{\rm{cut}}$ for our templates is discussed in 
Sec.~\ref{sec:fcutoff}.

Buonanno, Chen and Vallisneri measured the fitting factor of this 
detection template family and their results are presented in 
Sec.~VIC of BCV2 \cite{BCV2} (see Fig. 11 in particular)
\footnote{The authors of BCV2 \cite{BCV2} use a
downhill simplex method called {\tt AMOEBA}\cite{NumericalRecipesInC} in
order to obtain the best possible matches between the DTF and the target
waveforms.
This method works well for signals with high SNR but would not be effective
in searching for weak signals in real detector data.}.
The BCV2 DTF described here $(\psi_{0}, \psi_{3}, \beta)_{6}$
outperforms the other variants of the DTF they considered,
$(\psi_{0}, \psi_{3}, \beta)_{4}$
and $(\psi_{0}, \psi_{3})_{2}$, which have fewer $\alpha$ terms
and therefore less degrees of freedom with which to maximise their
overlap with a given target waveform.
The BCV2 DTF also outperformed the standard (physically parameterised) 
stationary phase approximation templates. 
Average fitting factors of $\simeq 0.93$ were measured for the
NS-BH binaries and even higher $\geq 0.97$ for the BH-BH binaries
considered (i.e., the BCV2 configurations discussed in 
Sec.~\ref{BCV2analysis}).
Lower fitting factors for the asymmetric systems (e.g., 
$(1.4, 10) M_{\odot}$ NS-BH binaries) is unsurprising since we expect
spin modulation to have most effect on these systems thereby
making their waveforms more complicated and thus harder to capture
accurately.

\subsection{Maximisation of overlap over extrinsic parameters}
\label{sec:maximisation}
When listing the parameters used to describe a binary system consisting 
of spinning components we divided the parameters into two categories:
{\it intrinsic} parameters which describe properties of the system itself 
(e.g., masses, spins) and {\it extrinsic} parameters which describe the 
observers relation to the system (e.g., amplitude of observed emission, 
inclination and polarization angle).
Now considering the problem of finding the template $h$ within our DTF 
(as given in Eq.~(\ref{BCV2DTFa})) which yields the highest overlap with 
a given target signal $s$, we find that we can usefully separate the 
parameters used to describe our templates into these categories.
For the extrinsic parameters used to describe the templates (e.g.,
$\psi_{0}$, $\psi_{3}$, $\beta$ and $f_{\rm{cut}}$) we must construct
templates corresponding to each set of these parameters we wish to 
search for.
Conversely, we are able to search automatically through the range of our 
intrinsic parameters (e.g., $t_{0}$ and $\alpha_{1 \dots 6}$) for the 
values which yield the best overlap.

To begin with we will consider the maximisation of the overlap over time.
The overlap between a time-shifted template $h(t-t_{0})$ and a signal $s$
is given by
\begin{eqnarray}
\left< s,h(t-t_{0}) \right> = 
4 \Re
\int_{0}^{\infty}
\tilde{s}^{*}(f) \tilde{h}(f) e^{2 \pi i f t_{0}}
\frac{df}{S_{n}(f)}.
\end{eqnarray}
Note that in the case of no time-shift ($t_{0} = 0$) we re-obtain
the formulae for the inner product given in Eq.~(\ref{inner_real_BCV}).
Rather than evaluate the overlap separately for every value of $t_{0}$ 
(in reality our time-series will be discretized so there will be a finite
number of values) we can employ the inverse Fourier transform to evaluate
all values of $t_{0}$ automatically. Finding the value of $t_{0}$ which 
maximises the overlap is simply a case of noting the time at which the maxima
in the resulting overlap time series occurs. 
We use the computationally efficient Fast Fourier transform (FFT) to carry 
out forward and inverse Fourier transforms 
(see Sec.~\ref{app:matchedfilter} and
e.g., Chapter 12 of Ref.~\cite{NumericalRecipesInC} for documentation of 
FFTs).

Now we shall consider the maximisation of the overlap over the $\alpha$
parameters.
Consider a template characterised only by its extrinsic parameters
$h(t_{0},\alpha_{k})$ which has been normalised such that the inner
product $\left<h,h\right>=1$. The overlap between the template $h$ and the 
signal $s$ is
\begin{eqnarray}
\max_{t_{0},\alpha_{k}} \left< s, h(t_{0},\alpha_{k}) \right>.
\end{eqnarray}
We find it expedient to orthonormalise our amplitude functions 
$\left< \widehat{\mathcal{A}}_{i}, 
\widehat{\mathcal{A}}_{j} \right> = \delta_{ij}$
and then define our {\it basis templates} as
\begin{eqnarray}
\label{basisTemplates}
\hat{h}_{k}(t_{0};f) &=& \widehat{\mathcal{A}}_{k} (f) 
e^{2 \pi i f t_{0}} e^{i \psi_{\rm{NM}}} \nonumber \\
\hat{h}_{k+n}(t_{0};f) &=& i \widehat{\mathcal{A}}_{k} (f) 
e^{2 \pi i f t_{0}} e^{i \psi_{\rm{NM}}}.
\end{eqnarray}
The orthonormalisation of the amplitude functions, 
$\mathcal{A} \rightarrow \widehat{\mathcal{A}}$ is a lengthy
procedure and is described in the Appendix, Sec.~\ref{app:orthonorm}.
We are able to write our original template $h$ in terms of our
basis templates $\hat{h}_{k}$:
\begin{eqnarray}
h(t_{0},\alpha_{k};f) = \sum_{k=1}^{2n} \hat{\alpha}_{k} \hat{h}_{k}. 
\end{eqnarray}
The overlap between the template $h$ and a signal $s$ would be:
\begin{eqnarray}
\label{overlapBasis}
\max_{t_{0},\alpha_{k}} \left< s, h(t_{0},\alpha_{k}) \right> =
\max_{t_{0}} \max_{\hat{\alpha}_{k}} \sum_{k=1}^{2n} 
\hat{\alpha}_{k} \left< s, \hat{h}_{k}(t_{0}) \right>.
\end{eqnarray}
We will require that our templates be normalised and find that this
will lead to a constraint on the $\hat{\alpha}_{k}$ values 
\begin{eqnarray}
\left<h,h\right> &=& 1 \nonumber \\ 
&=& \sum_{k=1}^{2n} 
\left< 
\hat{\alpha}_{k} \hat{h}_{k},  
\hat{\alpha}_{k} \hat{h}_{k}  
\right> \nonumber \\
&=& \sum_{k=1}^{2n} \hat{\alpha}_{k}^{2} 
\left< \hat{h}_{k},\hat{h}_{k} \right> \nonumber \\
&=& \sum_{k=1}^{2n} \hat{\alpha}_{k}^{2}
\end{eqnarray}
where we have made use of the fact that since the amplitude functions
$\widehat{\mathcal{A}}_{k}$ are orthonormal, the basis templates are
each orthonormalised,  $\left< \hat{h}_{i},\hat{h}_{j} \right> = \delta_{ij}$. 
We can find $\hat{\alpha}_{k}$ that maximise the overlap by employing the method
of Lagrange multipliers
\begin{eqnarray}
\label{maxrhosqlagrange}
\Lambda = 
\sum_{k=1}^{2n} \hat{\alpha}_{k} \left< s, \hat{h}_{k}(t_{0}) \right>
 - \lambda \left[ \sum_{k=1}^{2n} \hat{\alpha}_{k}^{2} - 1 \right] 
\end{eqnarray}
which leads to
\begin{eqnarray}
\label{alphaHatMax}
\hat{\alpha}_{k} = 
\frac{ \left< s, \hat{h}_{k}(t_{0}) \right> }
{\sqrt{ \sum_{j=1}^{2n} \left< s, \hat{h}_{j}(t_{0}) \right>^{2} }}.
\end{eqnarray}
Substituting Eq.~(\ref{alphaHatMax}) into Eq.~(\ref{overlapBasis}) we find
the overlap maximised over $\hat{\alpha}_{k}$ 
\begin{eqnarray}
\max_{t_{0},\alpha_{k}} \left< s, h(t_{0},\alpha_{k}) \right> =
\sqrt{ \sum_{j=1}^{2n} \left< s, \hat{h}_{j}(t_{0}) \right>^{2} }.
\end{eqnarray}
In the case where the data to be filtered $x(t)$ contains both signal $s(t)$
and noise $n(t)$, i.e., $x(t) = n(t)+s(t)$ we can define the SNR as
\begin{eqnarray}
\label{BCV2SNR}
\rho = \max_{t_{0},\alpha_{k}} \left< x, h(t_{0},\alpha_{k}) \right> =
\sqrt{ \sum_{j=1}^{2n} \left< x, \hat{h}_{j}(t_{0}) \right>^{2} }
\end{eqnarray}
The implementation of the SNR calculation including the maximisation
over time and $\alpha$ parameters is described in the Appendix, 
Sec.~\ref{app:matchedfilter}.

\subsection{Testing the detection template family}
\label{testingDTF}
We have shown in Secs.~\ref{gaussianresponse}
and \ref{matchedfilterresponse} that when filtering 
data $x$ with a single template $h$:
\begin{itemize}
\item The output of filtering $\left<x,h \right>$ will be a Gaussian 
distributed variable if $x(t)$ is a Gaussian distributed variable. 
\item The expectation value of $\left<n,h \right>$ will be zero if 
$\overline{n(t)} =0$.
\item The variance of $\left<n,h \right>$ will be unity if we use a 
normalised template such that $\left< h,h \right> =1$.
\end{itemize}

Therefore, for a Gaussian distributed variable $x$ with mean $0$ and variance
$1$, i.e. $x \sim N(0,1)$ we expect that $\left<x,h \right> \sim N(0,1)$.
Also, for a Gaussian distributed variable $x$ with mean $\mu$ and variance 
$\sigma^{2}$, i.e. $x \sim N(\mu,\sigma^2)$ we have
\begin{eqnarray}
\chi_{n}^{2} = \sum_{i=1}^{n} \frac{x_{i}^{2} - \mu_{i}} {\sigma_{i}^{2}}.
\end{eqnarray}
where $n$ is the number of degrees of freedom of the $\chi^2$ distribution.
The $\chi_{n}^{2}$ has mean $n$ and variance $2n$.

From Eq.~(\ref{BCV2SNR}) we see that 
\begin{eqnarray}
\rho^2 = \max_{t_{0},\alpha_{k}} \left< x, h(t_{0},\alpha_{k}) \right>^2 =
 \sum_{j=1}^{2n} \left< x, \hat{h}_{j}(t_{0}) \right>^{2}. 
\end{eqnarray}
For $x \sim N(0,1)$ we expect 
$\left< x, \hat{h}_{j}(t_{0}) \right> \sim N(0,1)$
and therefore that $\rho^2 \sim \chi_{2n}^2$ 
(using the range of $n$ used in the summation in this equation).

\begin{figure}
\begin{center}
\includegraphics[angle=0, width=0.45\textwidth]{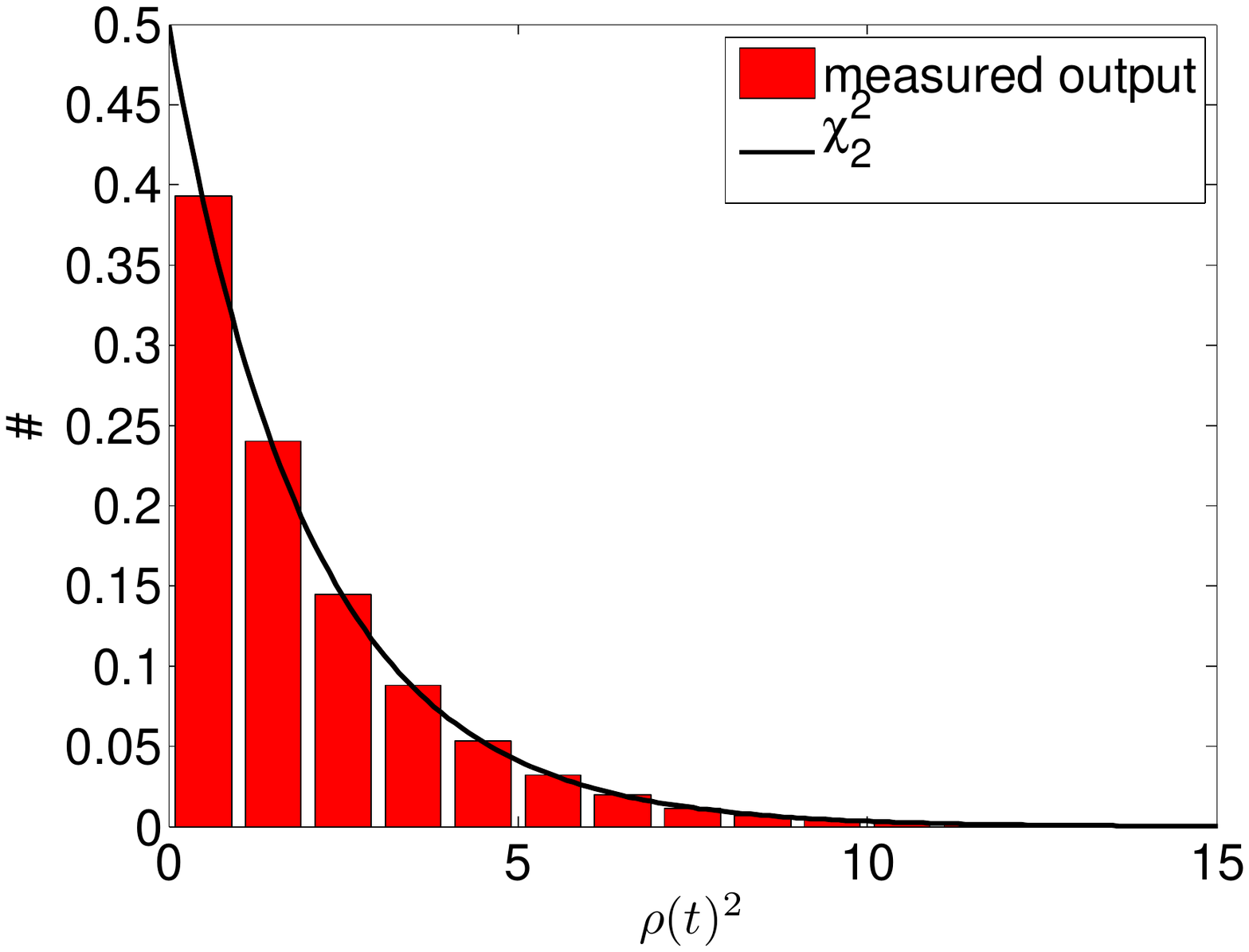}
\includegraphics[angle=0, width=0.45\textwidth]{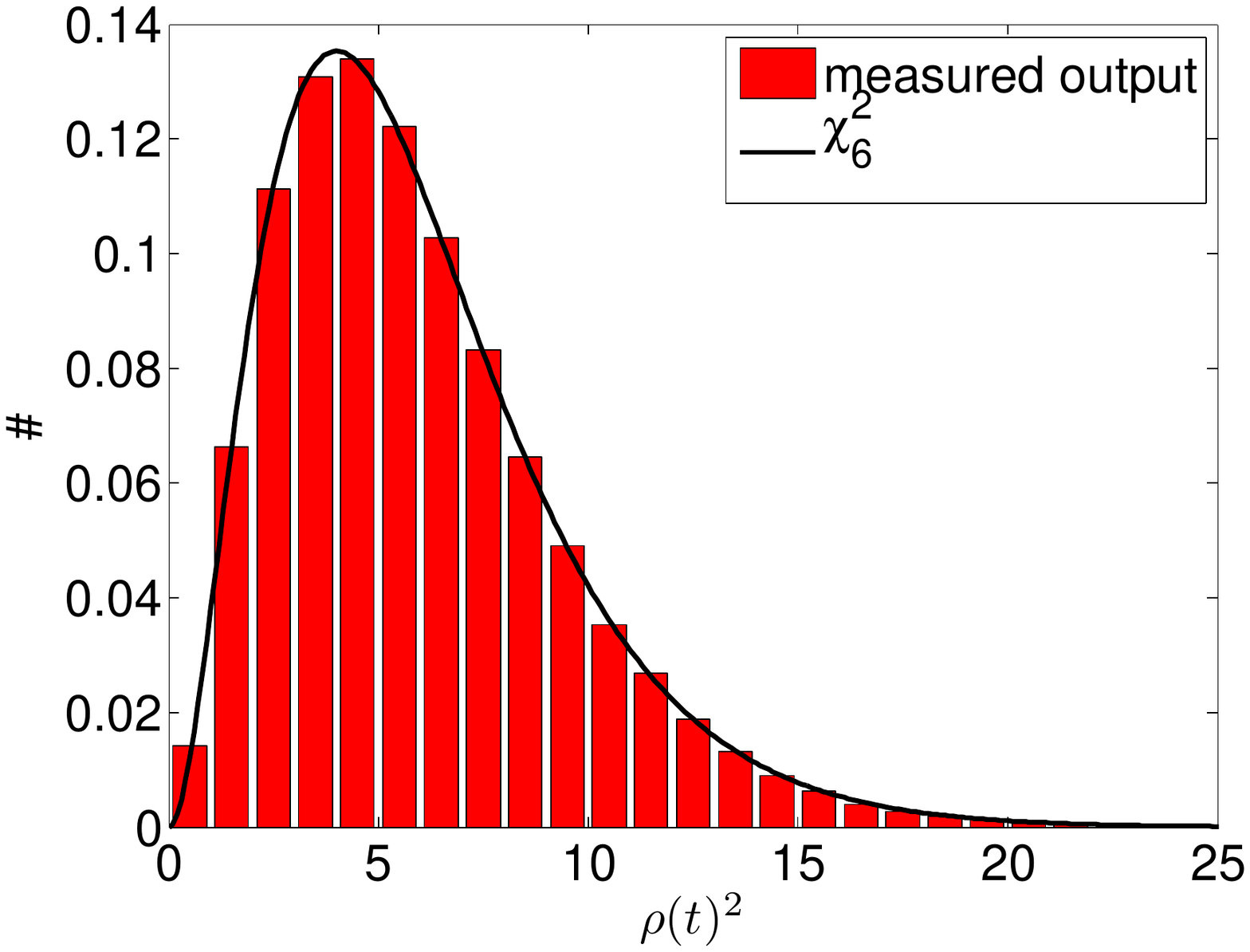}
\caption{Histograms of $\rho^{2}$ when $\beta = 0$ (left hand plot) and 
when $\beta \neq 0$ (right hand plot). As expected $\rho^2$ is distributed 
as a $\chi_{n}^{2}$ with the number of degrees of freedom $n$ equal to the number
of non-zero basis templates $\hat{h}_{i}$ used to calculate $\rho^{2}$.}
\label{fig:rhosq_hist}
\end{center}
\end{figure}

In general, when $\beta \neq 0$ we have 6 (non-zero) basis templates 
$\hat{h}_{j}$ and we would therefore expect $\rho^2 \sim \chi_{6}^2$.
When $\beta = 0$ we find that the (orthonormalised) amplitude functions
$\widehat{\mathcal{A}}_{2}$ and $\widehat{\mathcal{A}}_{3}$ become
zero at all frequencies
(see Eqs.~(\ref{app_moments}), (\ref{app_A2hat}) and (\ref{app_A3hat})).
Consequently we find that 4 of the 6 basis templates $\hat{h}_{j}$ 
defined in Eq.~(\ref{basisTemplates}) become zero at all frequencies.
Therefore, we will expect that when $\beta = 0$, $\rho^2 \sim \chi_{2}^2$.
Figure \ref{fig:rhosq_hist} shows histograms of $\rho^2$ measured
using our detection template family (with $\beta = 0$ and $\beta \neq 0$)
when filtering Gaussian white noise with zero mean and unit standard
deviation. The response of our templates is as we would expect.

\begin{figure}
\begin{center}
\includegraphics[angle=0, width=0.9\textwidth]{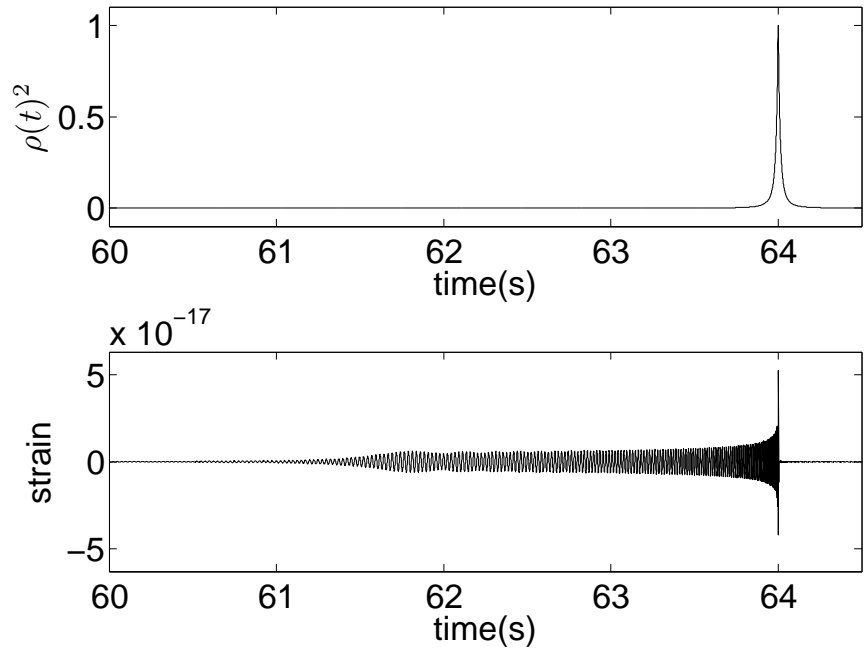}
\caption{Plot showing the overlap ($\rho^2$) measured when filtering 
a template $h$ with itself. As expected we measure an overlap of unity
at the end time of the waveform.}
\label{fig:wave_snr}
\end{center}
\end{figure}

We also expect that for a normalised template $h$ that we would obtain an overlap
of unity if we were to use a template as our input data i.e., $x = h$. 
Figure \ref{fig:wave_snr} shows the overlap measured (top plot) when we
perform this test. As expected an overlap of unity was measured.

\subsection{Estimating non-physical parameters in terms of physical 
parameters}
We are able to construct approximate relationships between the physical 
parameters $M$, $\eta$ and $\chi$ of our spinning binary system 
and the non-physical parameters used in the detection template family 
$\psi_0$, $\psi_3$ and $\beta$.
In Eq.~(3.3) and (3.4) of Arun et al. (2005) \cite{arun:084008} 
the phase of a gravitational 
wave inspiral is given to 3.5PN in the Fourier domain:
\begin{eqnarray}
\label{phase3.5PN}
\psi(f) &\equiv& \Psi_{f}(t_{f}) - \frac{\pi}{4} \nonumber \\
& = & 2 \pi f t_{c} - \phi_{c} - \frac{\pi}{4} + \frac{3}{128 \eta \nu^{5}} 
\sum_{k=0}^{N} \alpha_{k} \nu^{k}
\end{eqnarray}   
where $\nu = (\pi M f)^{1/3}$, $M=m_{1}+m_{2}$ and $\eta=m_{1} m_{2} / M^{2}$.
In BCV1 \cite{BCV1} (see Sec.~VI) it was noted that 
the phasing of the target model could be captured well
using only the $\psi_{0}$ and $\psi_{3}$ terms in the expansion of 
the non-modulational phase of the templates (see Eq.~(\ref{psiNMexp}))
and setting the other $\psi$ coefficients equal to zero.
The values of the $\alpha$ coefficients in Eq.~(\ref{phase3.5PN}) 
corresponding to the same order in frequency as $\psi_{0}$ and 
$\psi_{3}$ are 
$\alpha_{0} = 1$ and $\alpha_{3} = -16 \pi$
respectively.
Equating the terms of Eq.~(\ref{phase3.5PN}) with the
$\psi_{0}$ and $\psi_{3}$ terms of Eq.~(\ref{psiNMexp}) 
corresponding to the same order in frequency we find
\begin{eqnarray}
\label{psi0Meta}
\psi_{0} &=& \frac{3}{128} \frac{1}{\eta (\pi M)^{5/3}} \\
\label{psi3Meta}
\psi_{3} &=& -\frac{3}{8}  \frac{\pi^{1/3}}{\eta M^{2/3}}.
\end{eqnarray}

From ACST \cite{ACST} we find that the evolution of the precessional angle 
$\alpha_{p}$  can be approximated by power laws of $f$ in 2 extreme 
cases; $|\Lv| \gg |\Sv|$ and $|\Sv| \gg |\Lv|$.
The first case, when $|\Lv| \gg |\Sv|$, corresponds to a binary with either 
comparable masses or which is at the early stages of inspiral 
(i.e., large separation).
The second case, $|\Sv| \gg |\Lv|$, corresponds to a binary with small 
mass ratio (i.e., large mass asymmetry) or which is at the late stages of 
inspiral (i.e., small separation) \cite{PBCVT}. 
In Fig.~(3) of \cite{kidder:821}, Kidder shows the evolution of $|\Lv|$ 
and $|\Sv|$ with separation $r$ for both an equal
mass binary and a binary with a small mass ratio. 
This figure effectively illustrates the regimes during which 
the two extreme cases $|\Lv| \gg |\Sv|$ and $|\Sv| \gg |\Lv|$  are relevant.

When $|\Sv| \gg |\Lv|$ we find \footnote{Please note that a small error 
appears in Eq.~(45) of \cite{ACST}. 
The factors $1+ 3m_{1}/ 4m_{2}$ should in fact read $1+ 3m_{2}/ 4m_{1}$. 
Eq.~(42) of \cite{ACST} shows the correct factor as does \cite{PBCVT}.}.
\begin{eqnarray}
\alpha_{p}(f) \sim 3.8 \pi \left(1 + \frac{3}{4} \frac{m_{2}}{m_{1}}\right) 
\frac{m_{1}}{m_{2}} \chi 
\left(\frac{10 M_{\odot}}{M} \frac{10 Hz}{f}\right)^{2/3}.
\end{eqnarray}
We use the parameter $\beta$ as the coefficient in the power law, 
$\alpha_p(f) = \beta f^{-2/3}$ and thus find
that
\begin{eqnarray}
\beta \propto \left(1 + \frac{3}{4} \frac{m_{2}}{m_{1}}\right) \frac{m_{1}}{m_{2}} \chi
\left(\frac{10 M_{\odot}}{M} \right)^{2/3}.
\end{eqnarray}
In the analysis of \cite{ACST} the authors assume either that $m_{1} = m_{2}$ 
(meaning that spin-spin coupling can be ignored) 
or that $\Sv_{2} = 0$ which would correspond to systems with small 
mass ratio, e.g., the inspiral of a NS into a 
spinning BH.
Asymmetric mass systems $m_{1} \gg m_{2}$  can be modelled as systems 
with only a single body spinning since
even if both systems were spinning maximally $|\Sv_{1}| \gg |\Sv_{2}|$.
Maximal value of $\beta$ occurs when $\chi = 1$.

Using the $f^{-2/3}$ power law approximation of $\alpha_{p}(f)$ is only expected 
to perform well matching binaries with $|\Sv| \gg |\Lv|$ but is shown to match well 
with systems with $|\Lv| \gg |\Sv|$ \cite{BCV2}.

\section{Creating template banks}
\label{sec:templatebank}
\begin{figure}
\begin{center}
\includegraphics[angle=0, width=0.9\textwidth]{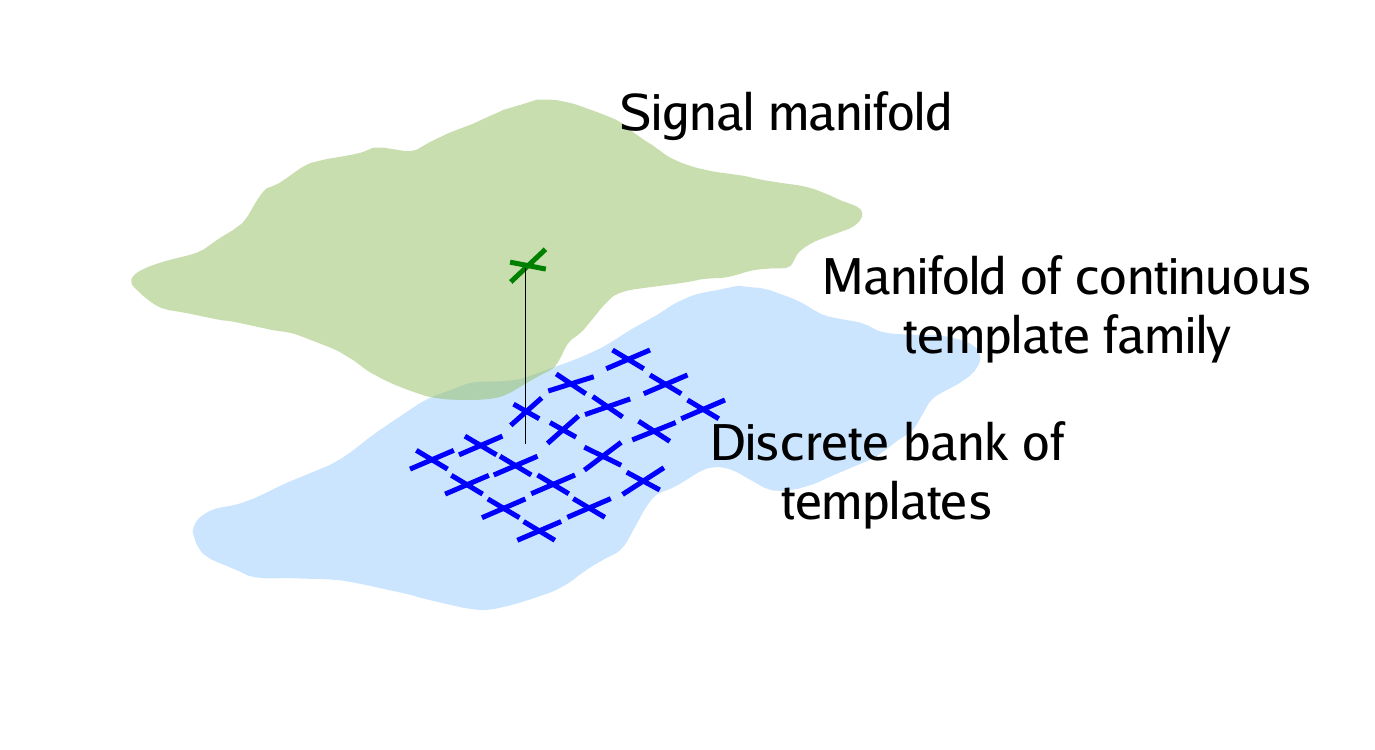}
\caption{Schematic representation of the signal and continuous detection
template family manifolds as 2 dimensional surfaces.
The signal manifold (green) contains all the possible gravitational waveforms 
we might observe from a binary with spinning components.
The continuous detection template family manifold (blue) contains 
all the waveforms that can be represented by our 
detection template family Eq.~(\ref{BCV2DTF}).
We use the metric (of the intrinsic parameters, Eq.~(\ref{tmpltbankmetric})) 
on the continuous detection
template family in order to choose a finite set of templates (blue crosses)
which we refer to as our {\it template bank}.
We place the templates of our bank such that for any point chosen 
within the region  of the continuous detection template family manifold 
we wish to cover, an overlap (or match) greater than the specified 
minimal match will be obtained with one of the templates in the bank. 
The fitting factor (previously discussed) describes the separation of
the signal and template manifolds. If the fitting factor was unity for
a region of parameter space the manifolds would be appear to touch in
that region. 
}
\label{fig:manifolds}
\end{center}
\end{figure}

The detection template formula Eq.~(\ref{BCV2DTF}) describes a continuous
multi-dimensional manifold containing every possible waveform that this
family can generate.
We can also imagine another manifold containing every possible gravitational
waveform we might observe from a binary with spinning components.
The parameters used to describe the detection templates/signals 
act as the co-ordinates on these manifolds.
Figure \ref{fig:manifolds} shows the continuous manifolds of signals and
templates.
We must now select a finite, discretely spaced subset of points on
the continuous detection template manifold which will form the bank of 
templates we shall use to search for gravitational wave signals from 
binaries with spinning components.
There are two important decisions to be made in choosing the templates
for our bank: 
i) we must choose the parameter space we would like our template bank to
cover (i.e., the range of masses and spins of our target waveforms we are
interested in capturing)
and ii) the spacing of the templates within this region of parameter space
which will in turn directly influence the number of templates we will
have in our bank and the computational cost of the search.
We wish to space our templates as sparsely as possible (in order to minimize
the computational cost of the search) while also ensuring that we will
achieve good matches for any point on the continuous detection template 
manifold with one of the templates in the bank.
We will consider the question of template spacing and placement first.

\subsection{Calculating the metric on a continuous detection template 
manifold}
Following the geometric formalism introduced by
Balasubramanian et al. (1996) \cite{Bala_Sathya_Dhurandhar_1996} and
Owen (1996) \cite{Owen:1996}
we will find a metric on the manifold of continuous detection template family
which will inform our choice of template spacing.
Our templates are parameterised by extrinsic parameters $\muv$ and intrinsic
parameters $\lambdav$, i.e., $h(\muv, \lambdav^{a})$ where $a$ is an index
which ranges through all the different intrinsic parameters $\lambdav$ 
(i.e., $a$ is not an exponent).

We consider two templates with slightly different parameters, 
$h(\muv, \lambdav)$ and $h(\muv + \Delta \muv, \lambdav + \Delta \lambdav)$
and calculate the match between them 
(from Owen (1996) \cite{Owen:1996} Eq.~(2.10)):
\begin{eqnarray}
M (\lambdav, \Delta \lambdav) \equiv 
{\rm{max}}_{\muv,\Delta \muv}
\left<
h(\muv, \lambdav),
h(\muv + \Delta \muv, \lambdav + \Delta \lambdav)
\right>.
\end{eqnarray}
We automatically maximise over the extrinsic parameters. 
Therefore, the match describes the proportion of the optimal match (unity
for normalised templates) measured when using templates with intrinsic
parameters differing by $\Delta \lambdav$.
Expanding the match $M$ as a (Taylor) power series about 
$\Delta \lambdav = 0$ we find
\begin{eqnarray}
M (\lambdav, \Delta \lambdav) \simeq 
1 
+ \underbrace{\frac{\partial M}
             {\partial \Delta \lambda^{a}} \Delta \lambda^{a}}_{\sim 0} 
+ \frac{1}{2}
  \frac{\partial^{2} M}
  {\partial \Delta \lambda^{a} \partial \Delta \lambda^{b}} 
  \Delta \lambda^{a} \Delta \lambda^{b}
+ \dots
\end{eqnarray}
where we neglect the first derivative in this expansion since it will tend 
to zero around the maxima of $M$ at $\Delta \lambdav = 0$ and neglect terms 
beyond the second derivative. 
Here the indices $a$ and $b$ both range through all the different intrinsic
parameters (i.e., they are not limited to $0,1,2,3$).
We define the metric on the intrinsic parameter space of the manifold of the 
continuous detection template family 
(from Owen (1996) \cite{Owen:1996} Eq.~(2.12)):
\begin{eqnarray}
\label{tmpltbankmetric}
g_{ab} (\lambdav) = - \frac{1}{2} \frac{\partial^{2} M}
  {\partial \Delta \lambda^{a} \partial \Delta \lambda^{b}}
\end{eqnarray}
for $\Delta \lambdav \sim 0$ which allows us to write the {\it mismatch} as
(from Owen (1996) \cite{Owen:1996} Eq.~(2.13)):
\begin{eqnarray}
\label{mismatch}
1 - M \simeq g_{ij} \Delta \lambda^{a} \Delta \lambda^{b}
\end{eqnarray}
We can use the metric $\gv$ to choose the largest spacings 
$\Delta \lambdav$ of our intrinsic parameters and still obtain matches 
greater than $M$ which we call the {\it minimal match}.

In Balasubramanian et al. (1996) \cite{Bala_Sathya_Dhurandhar_1996}
the authors define the metric
in an alternative but usefully intuitive manner (see their
Eqs.(3.9) and (3.10)). 
The (proper) distance between two nearby templates 
separated by intrinsic parameters $\Delta \lambdav$
on the manifold is given by 
\begin{eqnarray} 
\left< h(\lambdav + \Delta \lambdav) - h(\lambdav), 
       h(\lambdav + \Delta \lambdav) - h(\lambdav) \right>.   
\end{eqnarray} 
Expanding the terms of the inner product we find that the (proper)
distance between these templates is
\begin{eqnarray}
\left< 
\frac{\partial h(\lambdav)} {\partial \Delta \lambda^{a}},
\frac{\partial h(\lambdav)} {\partial \Delta \lambda^{b}}       
\right>
\Delta \lambda^{a}
\Delta \lambda^{b}
\end{eqnarray}
from which, recalling Eq.~(\ref{spacetimemetric}) for the spacetime metric,  
it is natural to define the metric on the continuous detection template 
family as 
\begin{eqnarray}
g_{ab} (\lambdav) \equiv \left<
\frac{\partial h(\lambdav)} {\partial \Delta \lambda^{a}},
\frac{\partial h(\lambdav)} {\partial \Delta \lambda^{b}}
\right>.
\end{eqnarray}
These two definitions of the metric $g_{ab}$ are equivalent.
We will now describe the calculation of the metric for our detection
template and then the placement of templates using this metric.
This work was performed by Dr. Benjamin Owen and Chad Hanna for the 
Inspiral/CBC working group of the LSC.
The testing of the resulting template bank's coverage was
performed by myself.

\subsection{Calculating the metric on the BCV2 continuous detection 
template manifold}

In this search we used a metric based on the {\it strong modulation}
approximation.
The rationale is that binary systems with waveforms only weakly modulated by
spin-induced precession should be detectable with high efficiency by a
search whose matched-filter templates do not model the effects of spin, 
e.g.,~\cite{LIGOS3S4all}.
We therefore concentrate on designing a bank that will capture systems whose
waveforms will be strongly modulated.
The metric calculation and template placement algorithms
become much simpler in the strong modulation limit.
Recently, more precise treatments of the full metric on the BCV2
detection template family parameter space have become 
available,
see Pan et al. (2004) \cite{Pan:2003qt} and 
Buonanno et al. (2005) \cite{Buonanno:2005pt},
and work is in progress to incorporate them into future searches.

In the strong modulation approximation, the orbital plane is assumed to precess
many times as the gravitational wave sweeps through the LIGO band of good 
sensitivity.
Also the opening angle between the orbital and spin angular momentum
($\cos^{-1} \kappa$, see Fig.~\ref{fig:simple_transitional})
is assumed to be large, corresponding to large amplitude modulations of
the signal.
Mathematically this corresponds to the statement that the precession phase
$\mathcal{B}=\beta f^{-2/3}$ sweeps through many times $2\pi$ which means that 
the basis-templates $h_j$ are nearly orthonormal (without requiring the 
Gram-Schmidt procedure).
Below we shall see that this assumption places a condition on the precession 
parameter $\beta$, which for the initial LIGO design noise power spectral 
density \cite{Abramovici:1992ah} corresponds to 
$\beta \gtrsim 200\rm{Hz}^{2/3}$.

\begin{figure}
\begin{center}
\includegraphics[width=0.9\textwidth]{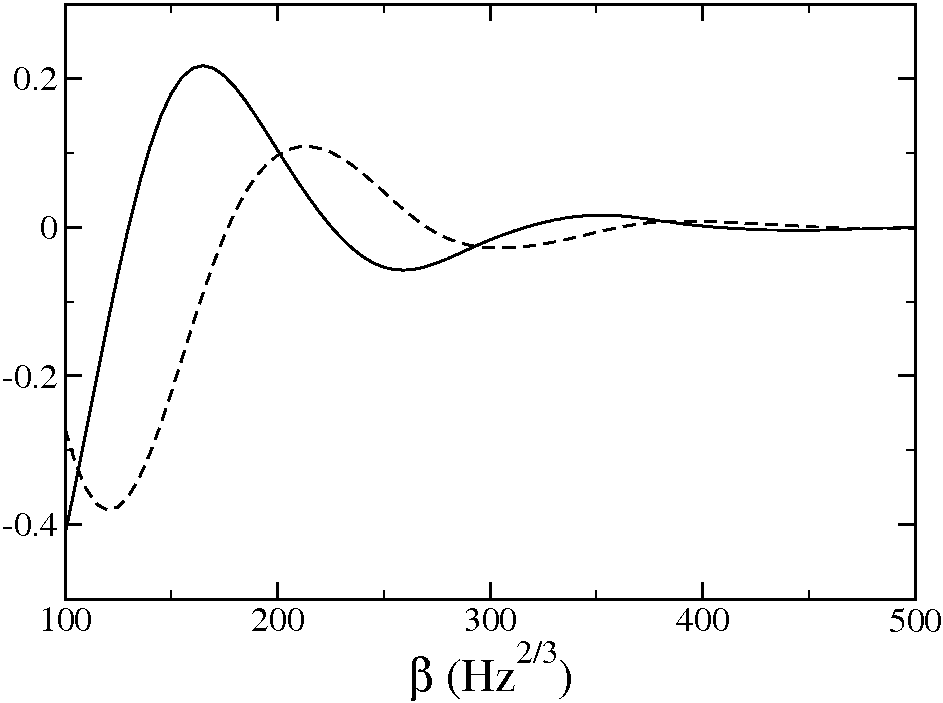}
\end{center}
\caption{Plot of $C_{7}$ (solid line) and $S_{7}$ (dashed-line) that are 
defined in Eq.~(\ref{C7S7}).
For values of $\beta \gtrsim 200\rm{Hz}^{2/3}$ the value of both functions
drop below $0.1$.
For these high values of $\beta$, i.e., the regime of strong
modulation, we find that the basis templates Eq.~(\ref{basisTemplates2})
will be orthogonal to each other without need for the Gram-Schmidt 
procedure. 
This in turn will simplify the calculation of the metric.
This figure was originally produced by Dr. Benjamin Owen.
}
\label{fig:moments}
\end{figure}

The basis templates are written as
\begin{eqnarray}
\label{basisTemplates2}
h_{1}(f) = -i h_{4}(f) &\propto& f^{-7/6} e^{i \psi_{\rm{NM}}} \nonumber \\
h_{2}(f) = -i h_{5}(f)  &\propto& f^{-7/6} \cos(\beta f^{-2/3})  e^{i \psi_{\rm{NM}}} \nonumber \\
h_{3}(f) = -i h_{6}(f)  &\propto& f^{-7/6} \sin(\beta f^{-2/3})  e^{i \psi_{\rm{NM}}} 
\end{eqnarray}
where we have proportionality (rather than equality) since we will require 
our basis templates to be normalized such that 
$\left< h_{j}, h_{j} \right> = 1$.
We can write our detection template as 
\begin{eqnarray}
h = \sum_{j=1}^{2n} \alpha_{j} h_{j}. 
\end{eqnarray}
The overlaps between the various basis templates can be written as
\begin{eqnarray} 
\label{C7S7}
\left< h_{1}, h_{2} \right> = - \left< h_{4}, h_{5}  \right> &=&
4 \Re \underbrace{  \int_{0}^{\infty} f^{-7/3} \cos (\beta f^{-2/3}) 
\frac{df}{S_{n}(f)} }_{= C_{7}(\beta)} \nonumber\\ 
\left< h_{1}, h_{3} \right> = - \left< h_{4}, h_{6}  \right> &=&
4 \Re \underbrace{  \int_{0}^{\infty} f^{-7/3} \sin (\beta f^{-2/3}) 
\frac{df}{S_{n}(f)} }_{= S_{7}(\beta)} \\ 
\left< h_{2}, h_{3} \right> = - \left< h_{5}, h_{6}  \right> &=&
4 \Re \underbrace{  \int_{0}^{\infty} f^{-7/3} \cos (\beta f^{-2/3}) \sin (\beta f^{-2/3}) 
\frac{df}{S_{n}(f)} }_{= \frac{1}{2} S_{7}(2 \beta)} \nonumber 
\end{eqnarray} 
where we have identified the functions $S_{7}(\beta)$ and $C_{7}(\beta)$
which are plotted in Fig.~\ref{fig:moments}.
For values of $\beta \gtrsim 200\rm{Hz}^{2/3}$, i.e., when the waveform
is strongly modulated, both functions have values less than $0.1$. 
The overlaps between different basis templates given above will approach zero
and we can write $\left< h_{i}, h_{j} \right> = \delta_{ij}$.

Therefore, by making the strong modulation approximation we can write 
\begin{eqnarray}
\label{rhosqstrongmod}
\rho^2 = \sum_{j=1}^{2n} \left< x, \alpha_{j} h_{j} \right>^{2}
\end{eqnarray}
similarly to how we construct $\rho^{2}$ for our  basis templates
$\hat{h}_{j}$ which were orthonormalised by the Gram-Schmidt procedure.
We can write the detection template with intrinsic parameters 
$(\psi_{\rm{NM}} + d \psi_{\rm{NM}}, \beta + d \beta)$ as
\begin{eqnarray}
h(f) &=& 
(\alpha_1 + i\alpha_2) e^{i(\psi_{\mathrm{NM}} + d\psi_{\mathrm{NM}})} 
\nonumber\\
&& + (\alpha_3 + i\alpha_4) \cos([\beta + d\beta]f^{-2/3} ) 
e^{i(\psi_{\mathrm{NM}} + d\psi_{\mathrm{NM}})} 
\nonumber\\
&&+ (\alpha_5 + i\alpha_6) \sin([\beta + d\beta]f^{-2/3} ) 
e^{i(\psi_{\mathrm{NM}} + d\psi_{\mathrm{NM}})}
\end{eqnarray}
Expanding the intrinsic parameters up to their second derivatives we
find
\begin{eqnarray}
h(f) &\approx&
\left( 1 + i \, d\psi_{\mathrm{NM}} - \frac{1}{2} d\psi_{\mathrm{NM}}^{2} \right)
\Bigg\{ (\alpha_{1} + i\alpha_{2}) {h}_{1}   \nonumber\\
&&  + (\alpha_{3} + i\alpha_{4}) 
\left[ \left( 1 - \frac{1}{2} d\mathcal{B}^{2} \right) 
{h}_{2} - d\mathcal{B} {h}_{3} \right]  \nonumber\\
&&  + (\alpha_{5} + i\alpha_{6}) 
\left[ \left( 1 - \frac{1}{2} d\mathcal{B}^{2} \right) 
{h}_{3} + d\mathcal{B} {h}_{2} \right] \Bigg\}.
\end{eqnarray}
where we have used $d\mathcal{B} = d\beta f^{-2/3}$.
We define the functional $F$ (originally defined in Owen (1996) 
\cite{Owen:1996} as $\mathcal{J}$) as
\begin{equation}
\label{funF}
F(a) = \frac{1}{I_{7}} \int_{f_{\rm min}/f_{0}}^{f_{\rm max}/f_{0}} dx
\frac
{x^{-7/3}}
{S_{h}(x f_{0})} a(x)
\end{equation}
and the noise moment $I$ is itself defined as
\begin{equation}
\label{funI}
I_{q} \equiv \int_{f_{\rm min}/f_{0}}^{f_{\rm max}/f_{0}} dx
\frac
{x^{-q/3}}
{S_{h}(x f_{0})}
\end{equation}
where $f_{\rm min}$ and $f_{\rm max}$ define the range of frequencies we 
integrate over. 
Since we have shown that our basis templates Eq.~(\ref{basisTemplates2})
are orthogonal when the waveforms are strongly modulated we are able to
write the overlaps between the detection template $h$ and its constituent
basis templates $h_{j}$ as
\begin{eqnarray}
\label{xhrels}
\left < h,h_1 \right > &=& \alpha_1 \left[ 1 - \frac{1}{2} 
F\left(d\psi_{\mathrm{NM}}^2 \right) \right] 
- \alpha_2 F(d\psi_{\mathrm{NM}}), 
\\
\left < h,h_4 \right > &=& \alpha_2 \left[ 1 - \frac{1}{2} 
F\left(d\psi_{\mathrm{NM}}^2 \right) \right] 
+ \alpha_1 F(d\psi_{\mathrm{NM}}), 
\nonumber\\
\left <h,h_2 \right > &=& \alpha_3 \left[ 1 - \frac{1}{2}
F\left(d\psi_{\mathrm{NM}}^2 \right) 
- \frac{1}{2} F\left( d\mathcal{B}^2 \right) \right]
\nonumber\\
& & \, - \alpha_4 F(d\psi_{\mathrm{NM}})
+ \alpha_5 F(d\mathcal{B}) - \alpha_6 F(d\psi_{\mathrm{NM}} \, d\mathcal{B}), 
\nonumber\\
\left < h,h_5 \right > &=& \alpha_4 \left[ 1 - \frac{1}{2} 
F\left( d\psi_{\mathrm{NM}}^2 \right) -
\frac{1}{2} F\left( d\mathcal{B}^2 \right) \right] 
\nonumber\\
& & \,
+ \alpha_3 F(d\psi_{\mathrm{NM}})
+ \alpha_6 F(d\mathcal{B}) + \alpha_5 F(d\psi_{\mathrm{NM}}\, d\mathcal{B}), 
\nonumber\\
\left < h,h_3 \right > &=& \alpha_5 \left[ 1 - \frac{1}{2} 
F\left( d\psi_{\mathrm{NM}}^2 \right) -
\frac{1}{2} F\left( d\mathcal{B}^2 \right) \right] 
\nonumber\\
& & \,
- \alpha_6 F(d\psi_{\mathrm{NM}})
- \alpha_3 F(d\mathcal{B}) + \alpha_4 
F(d\psi_{\mathrm{NM}}\, d\mathcal{B}), 
\nonumber\\
\left < h,h_6 \right > &=& 
\alpha_6 \left[ 1 - \frac{1}{2} F\left( d\psi_{\mathrm{NM}}^2 \right) -
\frac{1}{2} F\left( d\mathcal{B}^2 \right) \right] 
\nonumber\\
& & \,
+ \alpha_5 F(d\psi_{\mathrm{NM}})
- \alpha_4 F(d\mathcal{B}) - \alpha_3 F(d\psi_{\mathrm{NM}}\, d\mathcal{B})
\nonumber.
\end{eqnarray}
Using Eq.~(\ref{rhosqstrongmod}) we can write the square of the overlap 
when filtering a detection template $h(\psi_{\mathrm{NM}}, \beta)$
with another detection template 
$h(\psi_{\mathrm{NM}} + d \psi_{\mathrm{NM}} , \beta + d \beta)$ as
\begin{eqnarray}
\rho^{2} &=& \sum_{j=1}^6 \left< h, h_j \right>^2 \nonumber\\
&=& 
\sum_{j=1}^6 \alpha_j^2 
\left[ 1 - F(d\psi_{\mathrm{NM}}^2) + F(d\psi_{\mathrm{NM}})^2 \right] 
- \sum_{j=3}^6 \alpha_j^2 \left[ F(d\mathcal{B}^2) - F(d\mathcal{B})^2
\right] 
\nonumber\\
& & - \bigg[ 2\left( \alpha_3 \alpha_6 \right. - \left. \alpha_4 \alpha_5 \right)
\times \left[ F(d\psi_{\mathrm{NM}}\, d\mathcal{B}) -
F(d\psi_{\mathrm{NM}}) F(d\mathcal{B}) \right] \bigg].
\end{eqnarray}
We maximise $\rho^{2}$ subject to the constraint 
$\sum_{j=1}^{2n} \alpha_{j}^{2} = 1$ using the method of Lagrange multipliers
(see Eq.~(\ref{maxrhosqlagrange})).
We find $\alpha_{1} = \alpha_{2} = 0$, $\alpha_{3} = -\alpha_{6}$, and 
$\alpha_{4} = \alpha_{5}$, which leads to
\begin{eqnarray}
\label{rhosqF}
\rho^{2} &=& 1 - F(d\psi_{\mathrm{NM}}^2) + F(d\psi_{\mathrm{NM}})^2 
- F(d\mathcal{B}^2) 
\nonumber\\
&& + F(d\mathcal{B})^2 + F(d\psi_{\mathrm{NM}}\, d\mathcal{B}) 
- F(d\psi_{\mathrm{NM}}) F(d\mathcal{B}).
\end{eqnarray}

As well as maximising $\rho^{2}$ with respect to the $\alpha_{j}$
parameters we should also maximise with respect to the time of the
sources coalescence $t_{c}$.
During matched-filtering we maximise over time using an FFT 
(see Sec.~\ref{sec:maximisation}).
Here we will consider a signal perfectly described by a template 
$h(\psi_{\rm{NM}} + d \psi_{\rm{NM}}, \beta + d \beta, t_{c} + d t_{c})$.
We can incorporate the time dependence of the template into our phase
by writing $\psi = \psi_{\rm{NM}} + 2 \pi f t_{c}$.
To calculate the dependence on time of the match between two signals
$h(\psi_{\rm{NM}}, \beta,  t_{c})$ and
$h(\psi_{\rm{NM}} + d \psi_{\rm{NM}}, \beta + d \beta, t_{c} + d t_{c})$
we will replace $d \psi_{\rm{NM}}$ with 
$d \psi = \psi_{0} f^{-5/3} + d\psi_{3} f^{-2/3} + 2 \pi f d t_{c}$
in Eq.~(\ref{rhosqF}).

Using Eq.~(\ref{mismatch}) for the mismatch we can write
\begin{eqnarray}
\rho^{2} = 1 - 2 g_{ab} d\lambda^{a} d\lambda^{b}
\end{eqnarray}
(where we have used match $M = \rho$ for the case of normalized templates
without noise) 
allowing us to identify the metric's components (which we will now call
$\gamma_{ab}$) as
\begin{eqnarray}
\label{gnoproj}
\gamma_{t_c t_c} &=& 2 \pi^{2} \left( J_{1} - J_{4}^{\phantom{4}2} \right), 
\nonumber\\
\gamma_{t_c \psi_0} &=&  \pi \left( J_{9} - J_{4} J_{12} \right), 
\nonumber\\
\gamma_{t_c \psi_3} &=&  \pi \left( J_{6} - J_{4} J_{9} \right), 
\nonumber\\
\gamma_{t_c \beta} &=& \frac{\pi}{2} \left( J_{6} - J_{4} J_{9} \right), 
\nonumber\\
\gamma_{\psi_{0} \psi_{0}} &=& \frac{1}{2} \left(  J_{17} - J_{12}^{\phantom{12} 2} \right) ,  
\nonumber\\
\gamma_{\psi_{0} \psi_{3}} &=&  \frac{1}{2} \left(  J_{14} - J_{9} J_{12} \right) , 
\nonumber\\
\gamma_{\psi_{0} \beta} &=& \frac{1}{4} \left( J_{14} - J_9J_{12} \right), 
\nonumber\\
\gamma_{\psi_{3} \psi_{3}} &=& \frac{1}{2} \left( J_{11} - J_{9}^{\phantom{9} 2} \right) ,  
\nonumber\\
\gamma_{\psi_{3} \beta} &=& \frac{1}{4} \left( J_{11} - J_{9}^{\phantom{9} 2} \right), 
\nonumber\\
\gamma_{\beta \beta} &=& \frac{1}{2} \left( J_{11} - J_{9}^{\phantom{9} 2} \right)
\end{eqnarray}
where $J$ represent the normalized noise moments given by 
Poisson and Will (1995) \cite{Poisson:1995ef} 
\begin{eqnarray}
J_{q} \equiv \frac{I_{q}} {I_{7}}
\end{eqnarray}
and the noise moment $I$ was defined in Eq.~(\ref{funI}).

We are interested in placing templates in the 
($\psi_{0}$, $\psi_{3}$, $\beta$) space so we will project out the time
dependence of the metric using
(from Owen (1996) \cite{Owen:1996} Eq.~(2.28)):
\begin{eqnarray}
g_{ij} = \gamma_{ij} - \frac{\gamma_{ti} \gamma_{tj}}{\gamma_{tt}}
\end{eqnarray}
where the indices $i$ and $j$ range over all the intrinsic parameters
($\psi_{0}$, $\psi_{3}$, $\beta$).
As well as projecting out the time dependence we also neglect the
$\psi_{0} \beta$ and $\psi_{3} \beta$ cross terms which will simplify
the template placement and only result in a small over-coverage of the 
parameter space (neglecting these terms will only decrease the
volume of parameter space a given template achieves match greater than
minimal match by $\sim 3\%$). 
The metric we finally obtain has components
\begin{eqnarray}
\label{metricComps}
g_{\psi_{0} \psi_{0}} &=& 
\frac{1}{2} \left( J_{17} - J_{12}^{\phantom{12} 2} \right)  - 
\frac{ \left( J_{9} - J_{4} J_{12} \right)^2}
{ 2 \left(  J_{1} - J_{4}^{\phantom{4} 2} \right) }, 
\nonumber\\
g_{\psi_{0} \psi_{3}} &=& \frac{1}{2} \left(  J_{14} - J_{9} J_{12} \right)  - 
\frac{(J_{9} - J_{4} J_{12})(J_{6} - J_{4} J_{9}  ) }
{ 2 \left( J_{1} - J_{4}^{\phantom{4} 2} \right) }, 
\nonumber\\
g_{\psi_{0} \beta} &=& 0, 
\nonumber\\
g_{\psi_{3} \psi_{3}} &=& 
\frac
{1}{2} \left(J_{11} - J_{9}^{\phantom{9} 2} \right)  - 
\frac{ (J_{6} - J_{4} J_{9} )^2 }
{ 2 \left(  J_{1} - J_{4}^{\phantom{4} 2} \right) }, 
\nonumber\\
g_{\psi_{3} \beta} &=& 0, 
\nonumber\\
g_{\beta \beta} &=& \frac{1}{2} \left( J_{11} - J_{9}^{\phantom{9} 2} \right) 
- \frac{(J_{6} - J_{4} J_{9})^2}{ 8 \left( J_1 - J_{4}^{\phantom{4} 2}\right)}.
\end{eqnarray}

\subsubsection{Choosing the upper frequency cutoff}
\label{sec:fcutoff}
In practice, our upper bound on frequency is the Nyquist frequency
$f_{\rm{Nyquist}} = 2048 \rm{Hz}$ 
which is defined as half of the sampling frequency 
$f_{s} = 4096 \rm{Hz}$ 
\footnote{LIGO data is sampled at
16384 Hz and then downsampled. We will find that an upper frequency
of 2048 Hz is sufficient since at large frequencies the ground-based 
detectors sensitivity is poor due to the effects of shot noise
and there is only negligible benefits in integrating to higher limits.}. 
Ideally, we would perform the integrals to find the moment functions 
(see Eqs.(\ref{funF}) and (\ref{funI})) required in our
calculation of the template placement metric up to the
upper frequency cutoff $f_{\rm{cut}}$ of our detection templates.
For simplicity we use $f_{\rm{Nyquist}}$ as the upper frequency
cutoff in these integrals.
We find later that despite this approximation our template bank 
achieves high matches for a range of simulated signals
(see Sec.\ref{sub:testingbank}). 

However, we still must provide an estimate of $f_{\rm{cut}}$ in
order to limit our detection templates to the frequencies in which 
we believe they adequately describe true gravitational waveforms that
would be observed (i.e., the adiabatic limit).
We know that after the last stable orbit (LSO, similar to 
the minimum energy circular orbit which is the termination
point of the target waveform) the binary's components
will ``plunge'' and the bodies will merge over a time scale of 
only very few inspiral orbits.
Clearly, the binary is no longer in the adiabatic regime during
its plunge and we choose to set $f_{\rm{cut}}$ to the
frequency of the last stable orbit $f_{\rm{LSO}}$.

Determining $f_{\rm{LSO}}$ is complicated except in the extreme
asymmetric mass ratio limit ($\eta \rightarrow 0$).
We approximate the gravitational wave frequency of the 
last stable orbit (LSO) as the highest gravitational wave
frequency that would be emitted by test mass in Schwarzschild
geometry:
\begin{eqnarray}
\label{fLSO}
f_{\rm{LSO}} = \frac{1}{6^{3/2} \pi M}.
\end{eqnarray}
In practice we estimate $M$ to be the total mass of our binary from
the non-physical parameters of our template bank $\psi_{0}$ and $\psi_{3}$ 
using Eqs.~(\ref{psi0Meta}) and (\ref{psi3Meta}).
For reference, a binary with total mass $M = 1 M_{\odot}$ would have 
$f_{\rm{LSO}} \simeq 4400 \rm{Hz}$ and a binary with total
mass $M = 40 M_{\odot}$ would have $f_{\rm{LSO}} \simeq 110 \rm{Hz}$.
The $f_{\rm{LSO}}$ calculated is very much an upper limit
on the extent of the inspiral stage of the binary's evolution and it is likely
that the evolution will have become non-adiabatic 
(Allen et al. (2005) \cite{allen:2005}).
Despite these limitations, using $f_{\rm{cut}}$ rather than $f_{\rm{Nyquist}}$
will improve the matches obtained by our templates with expected inspiral 
signals.

We know from the studies presented in BCV2 \cite{BCV2} (see Figs. 5 and 6
and surrounding text) that the optimal value of $f_{\rm{cut}}$ depends on 
$\kappa$ (related to the opening angle of the spin and orbital angular 
momenta, see Eq.~(\ref{kappa})).  
Future searches could benefit from allowing multiple values of $f_{\rm{cut}}$
to be specified for each $(\psi_{0}, \psi_{3}, \beta)$ combination present in 
the template bank.
The choice of the lower frequency is dependent on the noise spectrum of the
detectors and is discussed in Sec.~\ref{subsub:lowfreqcut}.

\subsection{Template placement algorithm} 
The spacing of our templates (i.e., the density of our template bank)
is determined by our choice of the {\it minimal match} ($MM$) which is
defined as the lowest match that can be obtained between a signal
and the nearest template, see Owen (1996) \cite{Owen:1996}
and Sathyaprakash and Dhurandhar (1991) \cite{Sathya_Dhurandhar_1991}.
A template bank with minimal match $MM=0.95$ would, therefore,
suffer no more than a $1-MM = 5\%$ loss in SNR due to mismatch between
the parameters of a signal and the best possible template in the bank
(assuming that the signal and templates are from the same family, i.e.,
a fitting factor of unity).

The metric components shown in Eq.~(\ref{metricComps}) are 
not dependent on the intrinsic parameters 
($\psi_{0}$, $\psi_{3}$, $\beta$)
which makes the placement of templates simple.
The templates are placed on the vertices of a body-centred cubic (BCC)
lattice which is the most efficient template placement in three dimensions 
(i.e., it leads to the smallest number of templates to cover a given
region of parameter space).

The metric $\gv$ (whose components are given in Eq.~(\ref{metricComps}))
is diagonalized to form $\gv^{\prime}$ which will only have components
$g_{\psi_{0} \psi_{0}}^{\prime}$,
$g_{\psi_{3} \psi_{3}}^{\prime}$ and
$g_{\beta \beta}$ (the $\beta \beta$ component is unaffected by
the diagonalization).

The spacings of the template banks 
(in a single direction and ignoring the others) 
which yield matches of at least the 
minimal match $MM$ is given by 
$ds_{i}^{\phantom{i} 2} = 2(1-MM) = g_{ii} \Delta (\lambda^{i})^{2}$.
The factor of 2 is so that the point where the match is at its worst (i.e., $MM$)
is equidistant between two templates in the $\lambda^{i}$ direction.
This can be rearranged to find the co-ordinate spacing $\Delta \lambdav$ of
our intrinsic parameters for a given minimal match.
For body centred cubic placement we require
\begin{eqnarray}
\Delta \psi_{0}^{\prime} &=& \frac{4}{3} 
\sqrt{ \frac{2(1-MM)}{g_{\psi_{0} \psi_{0}}^{\prime}}}, 
\label{psi0placement}
\\
\Delta \psi_{3}^{\prime} &=& \frac{4}{3} 
\sqrt{ \frac{2(1-MM)}{g_{\psi_{3} \psi_{3}}^{\prime}}}, 
\label{psi3placement}
\\
\Delta \beta             &=& 
\frac{2}{3} \sqrt{ \frac{2(1-MM)}{g_{\beta \beta}}}. 
\label{betaplacement}
\end{eqnarray}

We will place templates in order to capture systems with asymmetric masses
for which the spin angular momenta is generally larger than the orbital
angular momentum leading to more pronounced spin effects.
We estimate the range of $\psi_{0}$ and $\psi_{3}$ values needed to
cover the physical mass range $1.0 M_{\odot} < m_{1} < 3.0 M_{\odot}$
and $6.0 M_{\odot} < m_{2} < 12.0 M_{\odot}$ using the relationships
given by Eqs.~(\ref{psi0Meta}) and (\ref{psi3Meta}).
This choice of mass region allows us to concentrate on asymmetric mass
ratio binaries with total mass low enough that we can use $f_{\rm{Nyquist}}$
as the upper frequency when evaluating moments for the template bank metric 
calculation.
Due to the approximate nature of these relationships we find that the
range of masses the template bank achieves best matches for is slightly
different and this is discussed in Sec.~\ref{sub:testingbank}.
These choices lead to placing templates with 
$\psi_{0}$ in the range $\sim 2 - 8 \times 10^{5} \rm{Hz}^{5/3}$ and
$\psi_{3}$ in the range $\sim -2 - -5 \times 10^{3} \rm{Hz}^{2/3}$.
We place templates with $\beta$ in the range $0 - \beta_{\rm{max}}$
where
\begin{eqnarray}
\beta_{\rm{max}} = 3.8 \pi 
\left(
1 + \frac{3}{4} \frac{m_{2,{\rm{max}}}}{m_{1,{\rm{min}}}}
\right)
\frac{m_{1,{\rm{max}}}}{m_{2,{\rm{min}}}}
\left(
\frac{10 M_{\odot}}{m_{1,\rm{min}} + m_{2,\rm{min}}}
\frac{10 \rm{Hz}}{150 \rm{Hz}}
\right)^{2/3}
\end{eqnarray}
is chosen to give the highest value of $\beta$ possible given
the mass range we are seeking to cover and the peak sensitivity
of the detector occurring at roughly $f=150 \rm{Hz}$.
By placing templates with small values of $\beta$ we will be
sensitive to weakly spin-modulated binaries as well as the strongly
modulated binaries the template bank was designed to cover.

Starting at the lowest values of 
$\psi_{0}^{\prime}$, 
$\psi_{3}^{\prime}$ and
$\beta$
we place templates in a grid in the plane of constant $\beta = 0$ using the
co-ordinate spacings for $\psi_{0}^{\prime}$ (Eq.~(\ref{psi0placement})) 
and $\psi_{3}^{\prime}$ (Eq.~(\ref{psi3placement})).
We then move to the next layer of $\beta$ using Eq.~(\ref{betaplacement})
and then place another grid of templates.  
Neighbouring layers of templates will have their 
$\psi_{0}^{\prime},\psi_{3}^{\prime}$ co-ordinate values offset from
each other by $\Delta \psi_{0}^{\prime}/2, \Delta\psi_{3}^{\prime}/2$
in order to create the body-centred cubic spacing. 

\begin{figure}
\begin{center}
\includegraphics[width=0.9\textwidth]{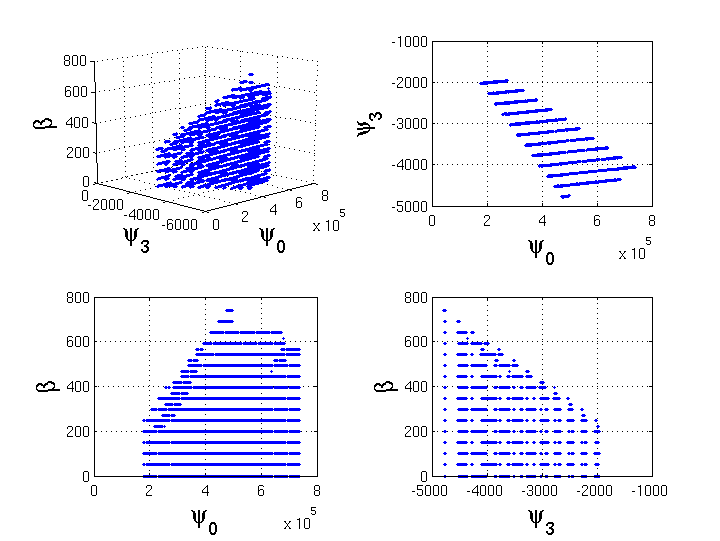}
\end{center}
\caption{A template bank generated with minimal match $= 0.95$
using 2048 seconds of H1 data taken during S3.
The crosses show the positions of individual templates in the
$(\psi_0,\psi_3,\beta)$ parameter space. For each template a value
for the cutoff frequency $f_{\rm cut}$ is estimated using Eq.~(\ref{fLSO}).
This bank requires a 3-dimensional template placement scheme in order
to place templates in the $(\psi_0,\psi_3,\beta)$ parameter space. Previous
searches for non-spinning systems have used 2-dimensional placement schemes.}
\label{fig:templatebank}
\end{figure}

We only place templates within the range of $\psi_{0}, \psi_{3}, \beta$ 
described above.
This can lead to the template bank having ragged edges and some under-coverage
of the targeted region of parameter space near the boundary of the template 
bank.
Owen and Hanna developed a scheme to solve this problem:  
if the next template to be placed using co-ordinate spacing $\Delta x$ 
(where $x$ is $\psi_{0}^{\prime}$, $\psi_{3}^{\prime}$ or $\beta$) 
according to Eqs.(\ref{psi0placement}) to (\ref{betaplacement}) would 
be beyond the boundary of the template bank they assess whether a template
placed using spacing $\Delta x/2$ would be within the boundary of
the template bank. 
If so, this template is included in the template bank. 
Figure \ref{fig:templatebank} shows an example of a template bank 
generated during the search of S3 LIGO data.

\subsection{Testing the template bank}
\label{sub:testingbank}
\begin{figure}
\begin{center}
\includegraphics[width=0.4\textwidth]{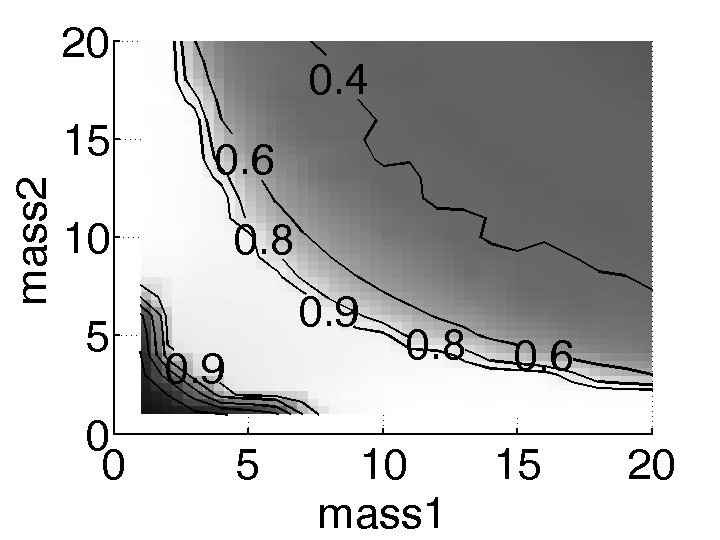}
\includegraphics[width=0.4\textwidth]{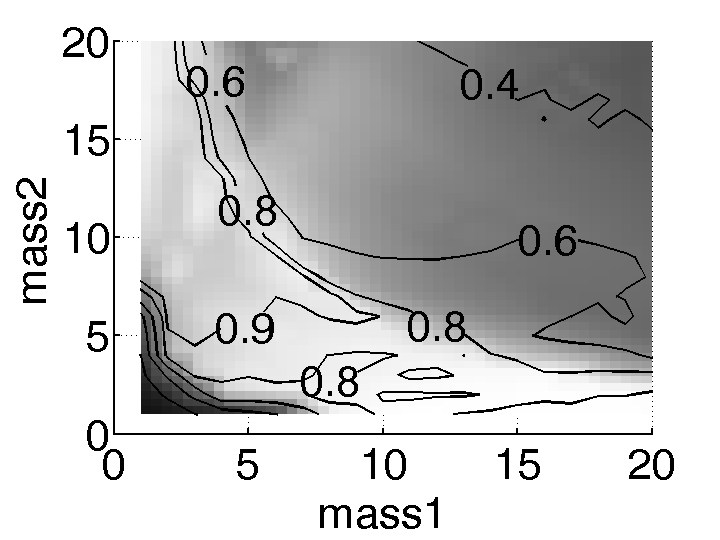} \\
\includegraphics[width=0.4\textwidth]{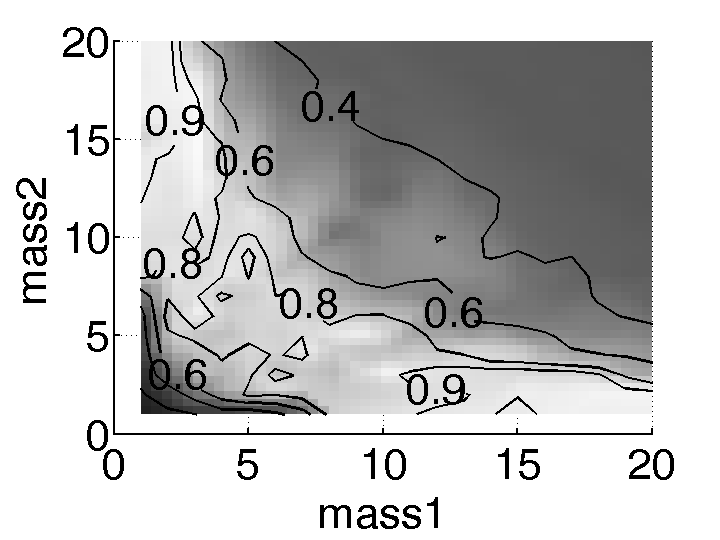} 
\includegraphics[width=0.4\textwidth]{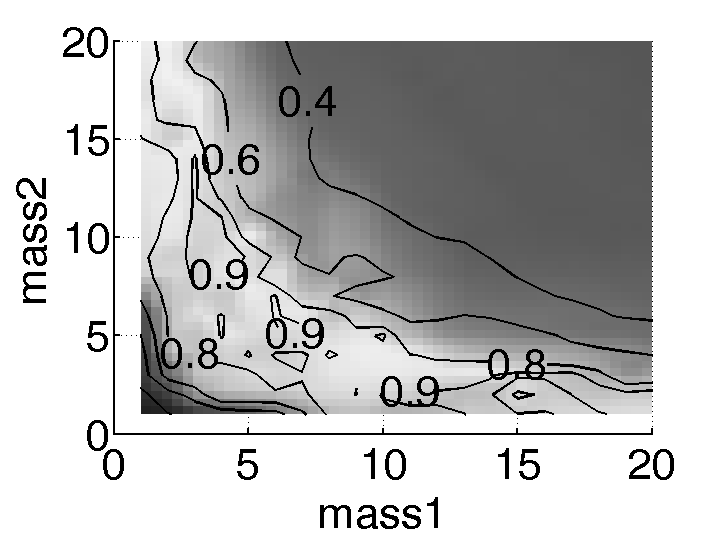}
\end{center}
\caption{Plots showing the best match achieved by filtering a series of 
simulated signals through the template bank described in this Section.
The values on the $x$ and $y$ axes correspond to the component masses of the 
binary source to which the simulated signal corresponds.
The colour of the plots shows the best match achieved for a given 
simulated signal.
The four subplots correspond to four different spin-configurations of the 
binary source. The top-left subplot shows results for a non-spinning 
binary system.  
The top-right subplot shows results for a system consisting of one 
non-spinning object and one maximally spinning object with its spin 
slightly misaligned with the orbital angular momentum. 
We would expect this system to precess.
The bottom two subplots show results for two generic precessing systems 
consisting of two maximally spinning bodies with spins and orbital 
angular momentum all misaligned from each other. 
We see that the region of the mass plane for which we obtain matches $> 0.9$ 
is largest for the non-spinning system and tends to be concentrated in the 
asymmetric mass region loosely bounded by 
$1.0~M_{\odot} < m_{1} < 3.0~M_{\odot}$ and
$12.0~M_{\odot} < m_{2} < 20.0~M_{\odot}$.}
\label{fig:banksims}
\end{figure}

The template bank was tested using a series of simulated signals
constructed using the equations of the target waveforms 
described in Sec.~\ref{Sec:TargetModel}.
We considered a variety of spin configurations including systems
where neither, one or both bodies were spinning.
We also considered masses outside the range we expected the template
bank to have good coverage in order to fully evaluate the range of masses
for which it could be used.
For each spin configuration we created a series of signals for
every combination of (integer) masses in the range:
$1.0~M_{\odot} < m_{1}, m_{2} < 20.0~M_{\odot}$.
Using the initial LIGO design sensitivity we then measured the best
match that could be obtained for each signal using our template bank.
Figure~\ref{fig:banksims} shows a sample of the results from the tests of 
the template bank.
As expected we found that our template bank achieved the highest matches
for non-spinning (and therefore non-precessing) binaries. 
Performance degrades as spin-precessional effects become more pronounced 
i.e., when both bodies are spinning maximally with spins misaligned from 
the orbital angular momenta.
The template achieved matches $>0.9$ for a mass range 
$1.0~M_{\odot} < m_{1} < 3.0~M_{\odot}$ and
$12.0~M_{\odot} < m_{2} < 20.0~M_{\odot}$ 
(and equivalent systems with $m_{1}$ and $m_{2}$ swapped).
The detection template family (described in Sec.~\ref{sec:BCV2dtf}) is capable 
of obtaining high matches for comparable mass systems, the lower matches 
obtained for comparable mass systems are a result of targeting our template 
bank on asymmetric mass ratio systems 
(which are more susceptible to spin effects and conform to the strong 
modulation approximation). 

Matches below the specified minimal match of $0.95$ in the bank's region of 
good coverage are a consequence of (small) differences between the DTF and the 
target waveforms meaning that the DTF {\it cannot} perfectly match the target 
waveforms (see discussion of the fitting factor of the DTF in 
Sec.~\ref{sec:BCV2dtf}).

The fitting factor (see Sec.~\ref{sec:dtf})
measures the reduction of SNR due to differences 
between the DTF and the target waveform \cite{Apostolatos:1995} 
and should not be confused with the minimal match which measures the loss of 
SNR due to discreteness of the template bank \cite{Owen:1996}.  
The DTF performance was evaluated and its fitting factor was measured 
in Sec.~VI of Ref.~\cite{BCV2}, 
for NS-BH systems an average FF of $\approx 0.93$ was measured. 

\section{The data analysis pipeline}
\label{sec:pipeline}
\begin{figure}
\begin{center}
\includegraphics[angle=0, width=0.9\textwidth]{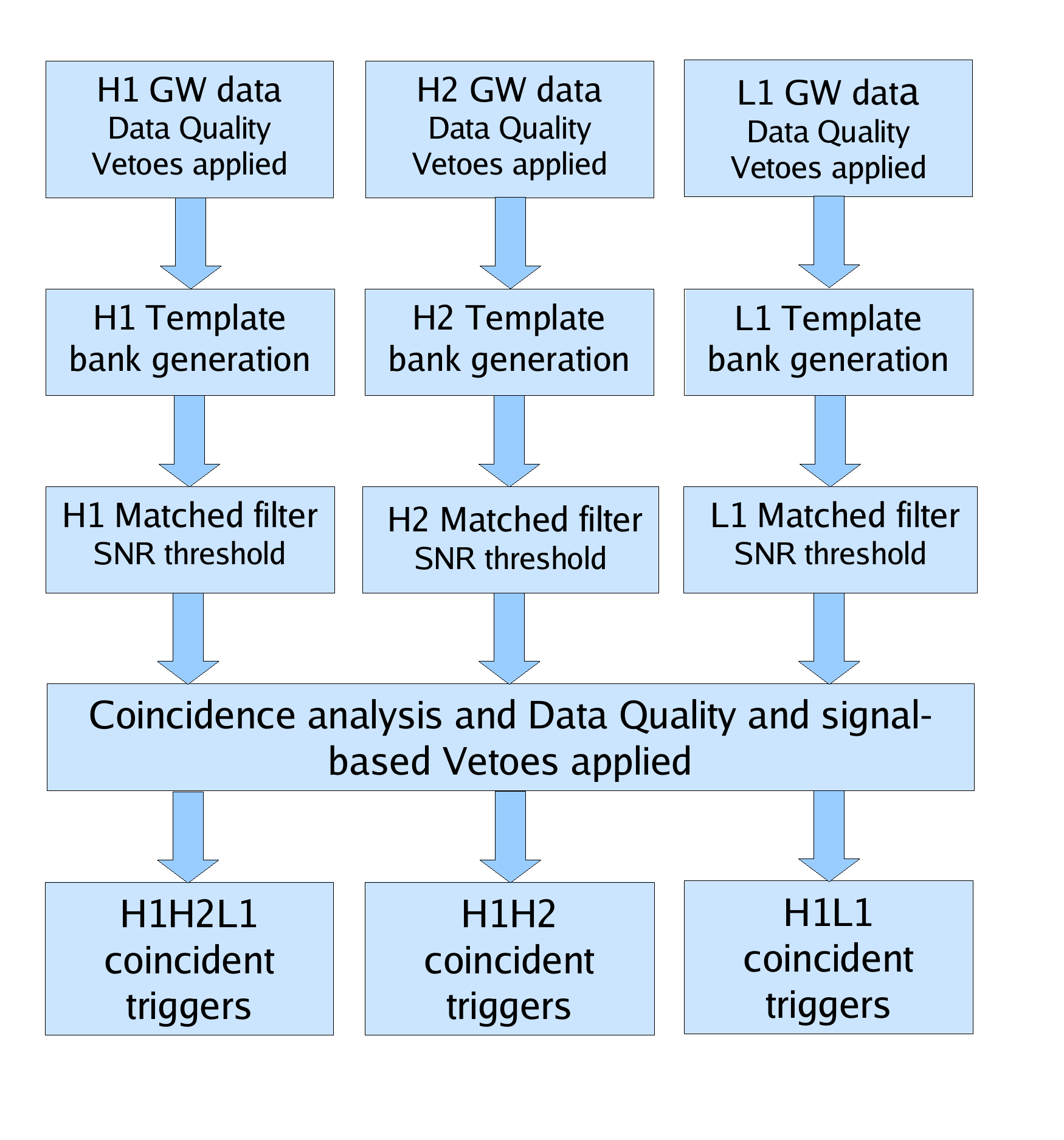}
\caption{Flowchart showing the various stages of the data analysis pipeline.
For each of the LIGO detectors, H1, H2 and L1 (see Sec.~\ref{subsecLIGO}
for a description) we begin our analysis by discarding data taken during
times when there are known environmental disturbances or problems with the
detector (Secs.~\ref{sec:dataselection} to \ref{subsec:S3data}).
We generate a template bank for each detector (Sec.~\ref{subsec:tmpltbank})
and then subsequently matched-filter the data (Sec.~\ref{mfdetdata})
constructing a list of triggers with signal to noise ratio exceeding our
predetermined threshold. 
Triggers occurring within a small time window, with similar parameters
consistent with those expected to be caused by true gravitational wave
signals in two or more detectors are identified (Secs.~\ref{sub:coincAnalysis} to
\ref{sub:vetoes}) as {\it coincident} triggers.
Coincident triggers are then investigated to see if they are consistent 
with our predicted background and whether they could be confidently
claimed as evidence for a gravitational wave ({Sec.~\ref{sub:results})}.
}
\label{fig:pipeline}
\end{center}
\end{figure}

The analysis of real detector data can be divided into a series of well 
defined processes that are collectively known as the 
{\it data analysis pipeline}.
Figure \ref{fig:pipeline} shows the data analysis pipeline
that was used in the analysis of LIGO data taken during its
third science run (S3, see Sec.~\ref{subsec:S3data}).
The pipeline used for this search is the same as was used in searches
for non-spinning binaries in S3 LIGO data. 
The searches for primordial black holes, binary neutron star and
binary black holes with non-spinning components using S3 and S4
LIGO data are described in Ref.~\cite{LIGOS3S4all}.

\begin{figure}
\begin{center}
\includegraphics[angle=0, width=0.9\textwidth]{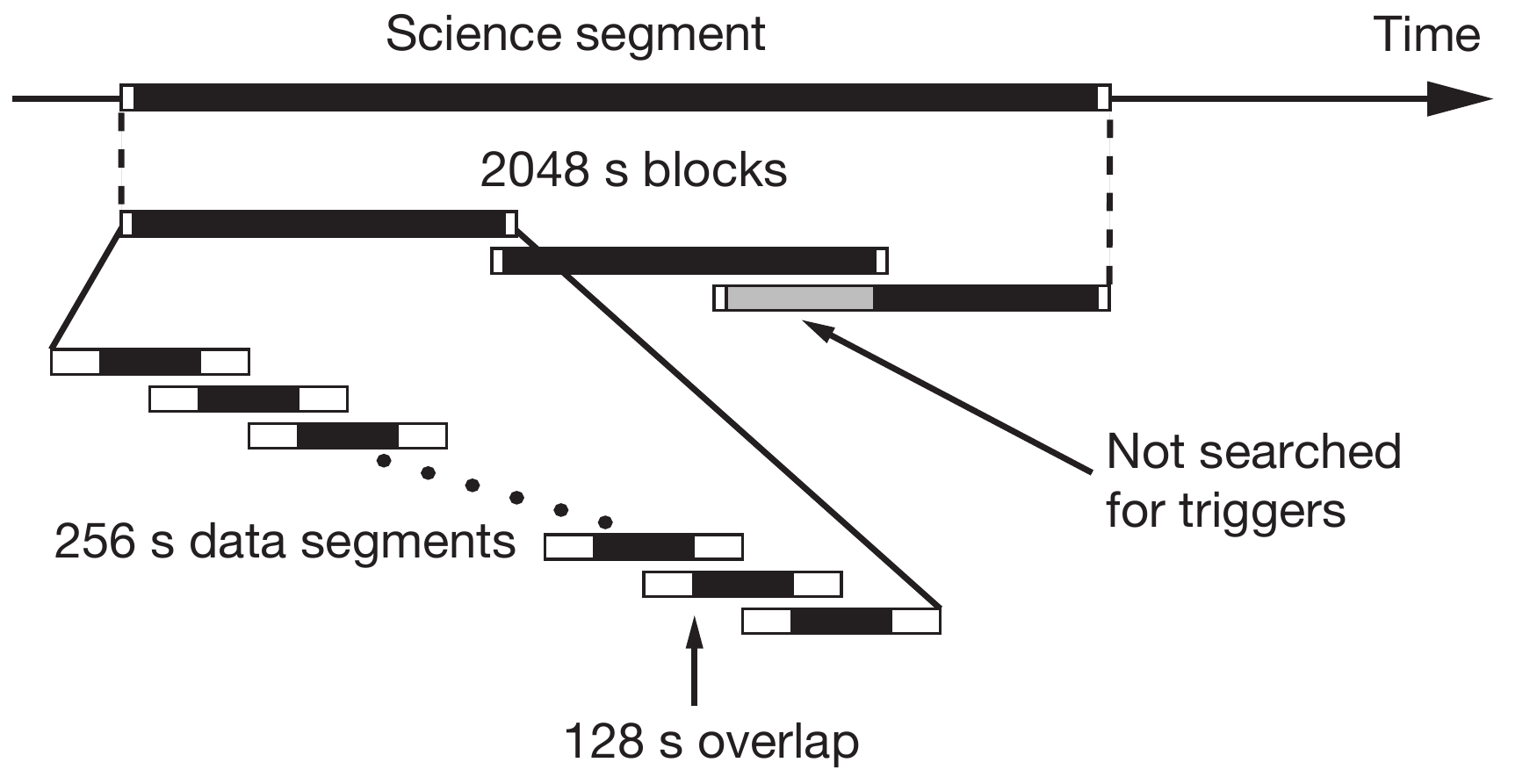}
\caption{
Figure showing the subdivision of a science segment. 
This figure, originally produced by Duncan Brown, 
was reproduced from B. Abbot et al., Phys. Rev. D {\bf 72}, 082001 (2005) 
\cite{LIGOS2iul}, with permission from the authors. 
}
\label{fig:sciencesegment}
\end{center}
\end{figure}

\subsection{Data selection}
\label{sec:dataselection}
The matched-filter is found to be the optimal filter to find a
known signal in stationary and Gaussian noise.
In reality, we find that our detector data is neither Gaussian or stationary.

\begin{figure}
\begin{center}
\includegraphics[angle=0, width=0.45\textwidth]{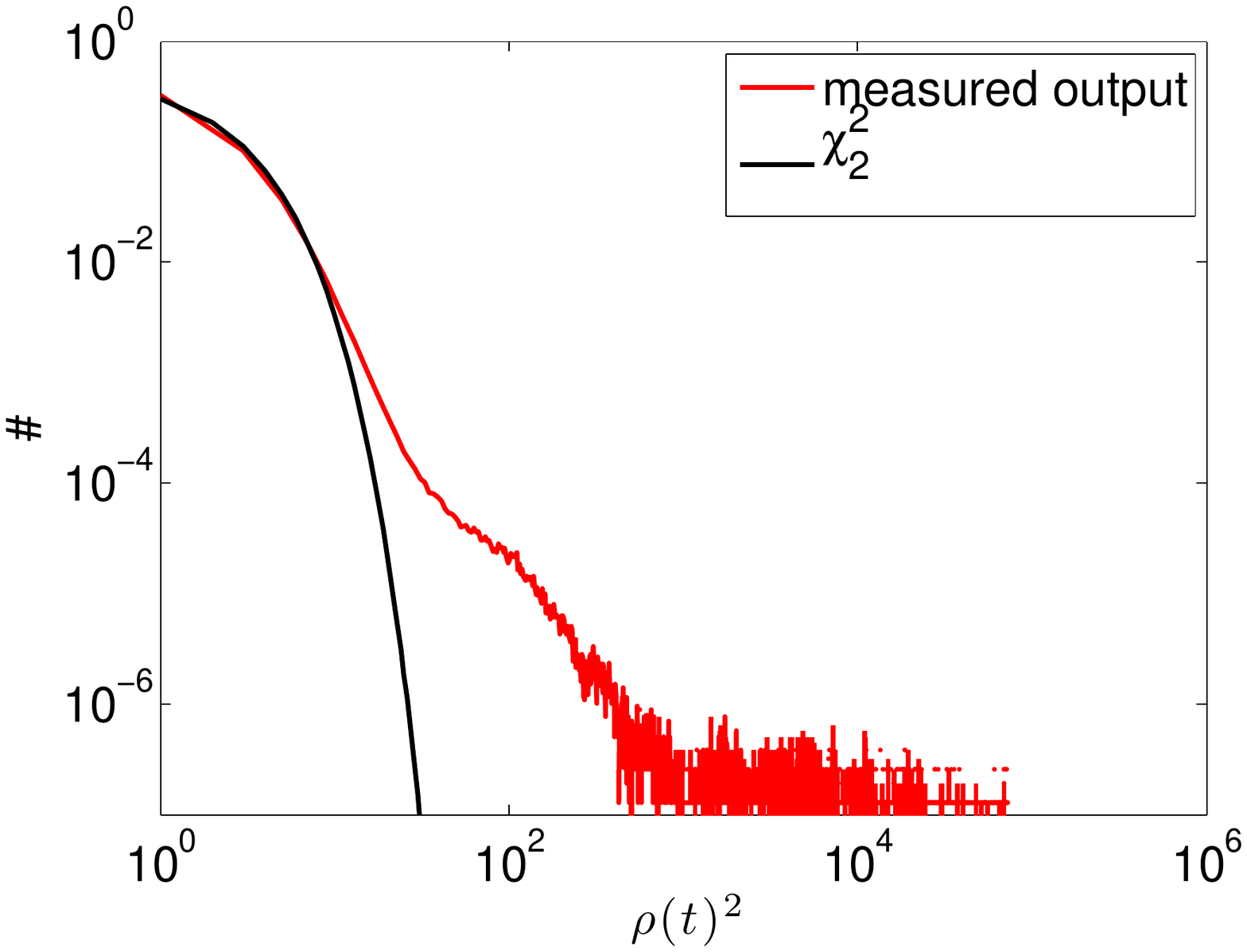}
\includegraphics[angle=0, width=0.45\textwidth]{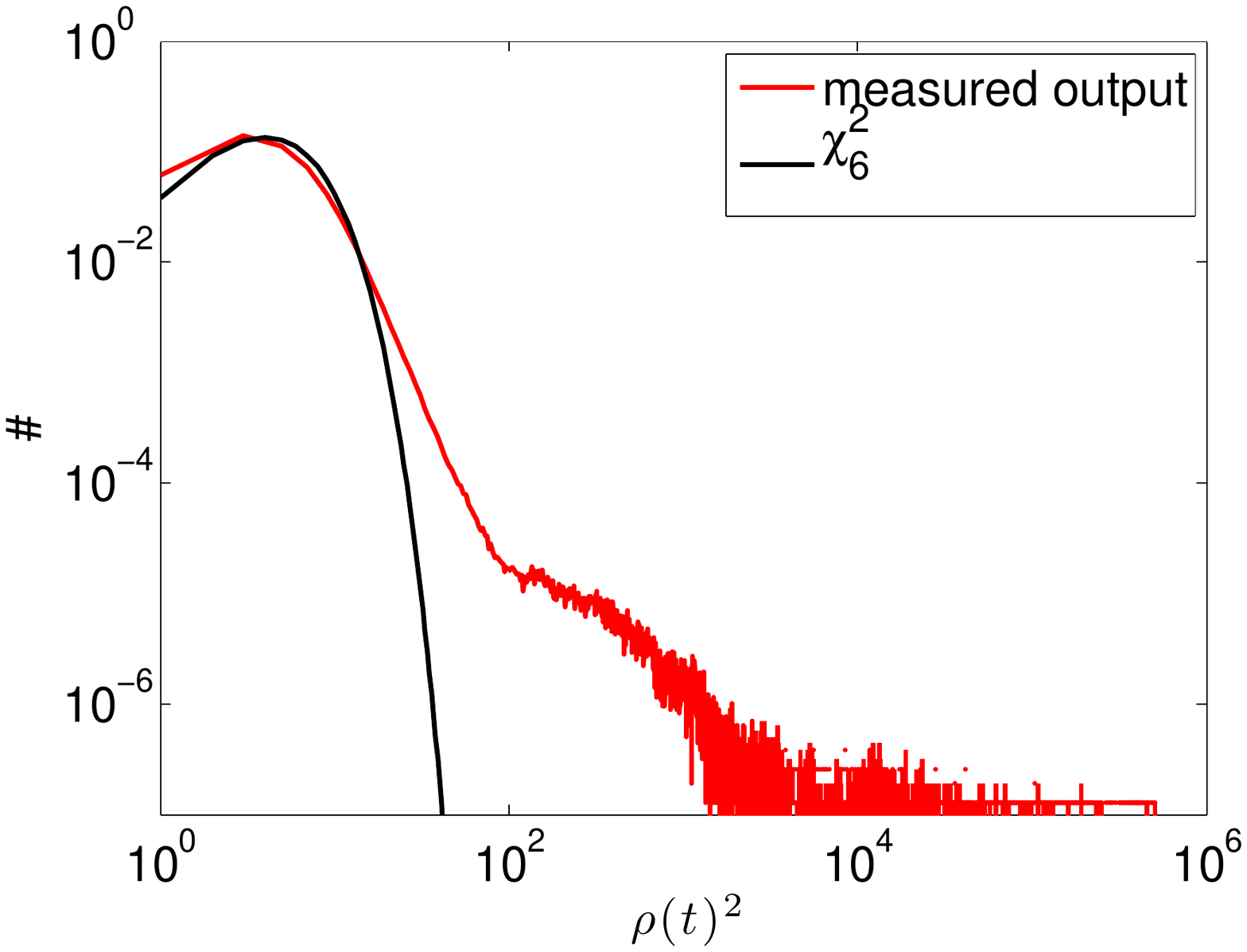}
\caption{Histograms of $\rho^{2}$ measured when matched-filtering real LIGO L1 data taken
during S3 with BCV2 templates with i) $\beta = 0$ (left hand plot) and ii) $\beta \neq 0$ (right hand plot). 
In Fig.~\ref{fig:rhosq_hist} we confirmed that $\rho^{2}$ will be distributed as a 
$\chi_{n}^{2}$ when the filtered data is Gaussian.
On the plots above we see an excess in the distribution of our measured output at
high values of $\rho^{2}$ indicating that our real detector data is non-Gaussian.
These high SNR events are caused by transients in the detector data and in Sec.~\ref{sec:dataselection}
we describe how data taken during times when the detector was performing poorly or when
there was a known disturbance is excluded from our analysis.
}
\label{fig:rhosq_hist_real}
\end{center}
\end{figure}

For Gaussian data we would expect the square of the SNR, $\rho^2$ obtained using the 
matched-filter templates of our DTF to follow a $\chi^{2}$ distribution
with 2 degrees of freedom when $\beta = 0$ and
6 degrees of freedom when $\beta \neq 0$ (see Sec.~\ref{testingDTF}).
Figure \ref{fig:rhosq_hist_real} shows a comparison between the distribution of $\rho^2$
we measured using real LIGO L1 data and the $\chi^{2}$ distribution we would expect
to observe if the data was truly Gaussian.
We observe a tail of high SNR triggers when matched-filtering the real data indicating
that the data is non-Gaussian.
%
Figure \ref{fig:H2PSDplot} shows the amplitude spectra for H2 estimated at two different
times during S3. 
We observe a flattening of the spectra as S3 progresses showing that the data is not stationary.
Also, in Gonzalez (2003) \cite{gonzalez-RayleighMonitor} (also see \cite{webRayleighMonitor})
the authors introduce the Rayleigh monitor which assesses how Gaussian and stationary
an interval of data is. A variant of this monitor is being used in more recent analysis
of LIGO data.

Indeed, transients in the data due to problems with the detector or 
environmental disturbances can lead to a huge number of {\it false alarm}
triggers, i.e., triggers caused by something other than a true gravitational 
wave signal.
Stretches of data during which the detector had poor performance
or when there was an environmental disturbance will be excluded from
analysis. 
The study of the detector's behaviour is called detector characterisation.
The detector characterisation carried out on S3 LIGO data is detailed in
Christensen (for the LIGO Scientific Collaboration, 2005) \cite{S3Vetoes}. 

We categorise times when we know the detector had poor operation
with data quality (DQ) flags.
Below we list and briefly describe the DQ flags used in this search 
(which were also used in all searches for inspiral signals in S3 data).
The numbers in brackets following the name of the DQ flag indicate
the percentage of data associated with that flag.

\begin{itemize}
\item {\tt NO\_DATA} ($0.01\%$): 
Some fault meant that the detector was not collecting data during these times.
\item {\tt NO\_RDS} ($0.02\%$): 
Under normal operation, we create a number of reduced data sets (RDS) which
contain a down-sampled time series of the gravitational wave channel as well as 
a selection of the auxiliary channels. This flag indicates that there was
some error meaning that the reduced data set was not created.
\item {\tt UNLOCKED} ($0.03\%$): \
When the detector is working correctly, 
the test mass mirrors of the interferometer will be ``locked'' into place so
that the laser beam will resonate in the optical cavity in each arm 
(see Sec.~\ref{subsecLIGO}). This flag indicates that the detector has
become unlocked.
\item {\tt INVALID\_TIMING} ($2.3\%$):
This flag indicates that the time-stamping of the data taken by the detector
is not valid. Knowledge of the exact time that data was taken is
crucial if we are to associate an event measured in one detector with an
event occurring in another detector (see Sec.~\ref{sub:coincAnalysis}).
Also, the accuracy of timing directly affects the accuracy to which we can
determine the sky location of a gravitational source by triangulation. 
\item {\tt CALIB\_LINE\_V03} ($2.0\%$):
Monochromatic sinusoidal oscillations are applied to the interferometer's test 
mass mirrors with known frequency and amplitude. 
These oscillations appear as spikes, known as calibration lines, in the amplitude
spectrum of the gravitational wave channel data.
By measuring the amplitude and frequency at which these lines appear in the data
it is possible to calibrate the amplitude of the gravitational wave channel strain data
\cite{S3Calib}.
This flag indicates that there is some problem with the oscillation 
of the test mass mirrors or the measurement of the calibration lines.
\item {\tt OUTSIDE\_S3} ($0.4\%$):
The database recording the state of the detector may include details of
data taken beyond the end of the S3 run. 
We will exclude data taken outside of the S3 run.
\end{itemize}

We construct a list of times for which each detector is 
operating well in what is known as {\it science mode} excluding times
associated with various DQ flags. 
As well as the use of DQ flags, later in the pipeline various short
stretches of data will be discarded or {\it vetoed}. There are two types of vetoes: 
signal-based vetoes and detector-based vetoes and these will be discussed 
in Sec.~\ref{sub:vetoes}.

Contiguous stretches of data taken when a detector is in science mode are
called {\it science segments}. 
The science segments are divided up into 2048 second {\it blocks}. 
Each block is divided into 15 overlapping 256 second 
{\it data segments}. 
Figure \ref{fig:sciencesegment} shows a single science segment and how it is
divided up for analysis.
Each data segment has 64 seconds overlap
with the preceding and subsequent data segment (except for the
first and fifteenth data segment which only have one adjacent
data segment).
The power spectrum of each 256 second data segment is estimated using
Welch's method with a Hann window 
(see Allen et al. (2005) \cite{allen:2005} for further details).
We then estimate the power spectrum of each 2048 second block as the
median average of the power spectra of its 15 data segments.
To be clear, we will measure the power contained within a particular 
frequency bin $f_{i} + \Delta f$ for each of the 15 data segments. 
We then take the median average of these powers and use that as
the power for that frequency bin in the block's power spectrum.
The median average is used rather than the mean to avoid 
biasing of the average power spectrum by short-duration non-stationary
noise events in the data.

There are two reasons we need to overlap data segments (and similarly
why we need to overlap 2048 second blocks).
This is discussed in detail in Allen et al. (2005) \cite{allen:2005}
and is briefly summarised here.
Firstly, the Fast Fourier Transform (FFT) we use in matched-filtering 
(see Sec. \ref{app:matchedfilter}) treats each data segment as
if it is periodic. The subsequent wrap-around effect causes a 
stretch of data the length of the template $T_{\rm{Template}}$ 
to be invalid at the start of each data segment.
Secondly, narrow lines (``spikes'') in the inverse noise power spectrum 
($S_{n}(f)^{-1}$) used in the inner product 
(see e.g., Eq.~(\ref{inner_real_BCV})) will cause corruption of data
throughout each data segment.
The inverse noise power spectrum is truncated in the time domain in order
to limit the corruption to stretches of data length $T_{\rm{Spec}}$ at the
start and end of each data segment. Note that we cannot choose 
$T_{\rm{Spec}}$ to be arbitrarily small since we would then lose
important information about the (inverse) noise power spectrum in the
frequency domain.
These two effects (wrap-around and ``spikes'') lead to stretches of corrupted 
data of duration $T_{\rm{Template}} + T_{\rm{Spec}}$ at the beginning
and $T_{\rm{Spec}}$ at the end of each data segment.
In order to avoid corrupted data we
only analyse and record triggers from the central 128 seconds 
of each 256 second data segment and ignore the first and last 
64 seconds of each data segment. 
We then overlap each data segment with the preceding and subsequent
data segment by 128 seconds to ensure that all the data will be analysed.
We do not analyse the first or last 64 seconds of each 2048 second 
block and we therefore overlap each block with the preceding and
subsequent block by 128 seconds 
(except for the first and last block in the 
science segment which only have only have one adjacent block). 

At the end of a science segment it might be necessary to overlap the final
two 2048 second blocks by more than 128 seconds to ensure that we analyse as
much of the remaining data as possible. Care is taken to ensure that the
same stretch of data is not analysed twice (i.e., the region marked
``Not searched for triggers'' in Fig.~\ref{fig:sciencesegment}).
Since it is not possible to overlap a 2048 second blocks at the very beginning
or end of a science segment, the first and last 64 seconds of  
each science segment will not be analysed.
Science segments shorter than 2048 seconds in duration will also not 
be analysed.

\subsection{Playground data}
We specify a subset of our data to be {\it playground data} which
we use to tune various parameters (e.g., SNR thresholds, coincidence windows).
The use of playground data allows the data analyst to tune parameters while
remaining {\it blind} to the remainder of the data set thus avoiding  
statistical bias. 
The set of playground times is defined formally as
\footnote{This formula uses set notation. In words, playground time consists
of intervals of time 600 seconds long that occur every 6370 seconds from GPS 
time 729273613.}
\begin{eqnarray} 
\mathcal{T} = 
\left\{ 
t \in \left[ t_{n}, t_{n} + 600 \right) : 
t_{n} = 729273613 + 6370 n, 
n \in \mathbb{Z}
\right\} 
\end{eqnarray} 
where $729273613$ is the GPS time of the beginning of LIGO's second
science run (from internal LIGO technical documents T030256 and T030020).
According to this definition playground times will account for, on average, 
$9.4\%$ of any given stretch of time.
Once parameters have been tuned using the playground data, the full data 
set (including playground) will be searched for gravitational wave
events. To avoid statistical bias, the values of the parameters chosen
after the analysis of playground data will remain fixed throughout the
subsequent analysis of the full data set.

Although it is possible to make a detection of a gravitational wave
during playground times, playground times are excluded from any 
upper limit calculations which are performed when no gravitational waves
have been detected.

\subsection{The S3 data set}
\label{subsec:S3data}
The third LIGO science run (S3) took place between
October 31st 2003 ($16:00:00$ UTC, $751651213$ GPS)
and
January 9th 2004 ($16:00:00$ UTC, $757699213$ GPS).
Data collected by the LIGO Hanford Observatory (LHO) detectors 
(i.e., H1 and H2) was only analysed when both detectors 
were in science mode. This was due to concerns that since both of these 
detectors share the same vacuum system, the laser beam of a detector in 
anything but science mode might interfere with the other detector
(see Sec.~\ref{sec:gwdetectors} for a description of the LIGO detectors).

We denote periods of time when all three detectors are in science
mode as H1-H2-L1 times and periods when only the LHO detectors are on as
H1-H2 times. A coincident trigger consisting of a trigger in the H1 detector 
and the L1 detector will be referred to as an H1-L1 coincident trigger and 
similarly for other combinations of detectors.
In this search we analysed 184 hours of H1-H2-L1 data and
604 hours of H1-H2 data (see Table \ref{tab:analysedtimes}).

\subsubsection{Lower frequency cutoff}
\label{subsub:lowfreqcut}
We must also choose the lower frequency cutoff $f_{\rm{low}}$ which
will be the lower limit of any integrals we perform in the frequency domain,
e.g., the computation of moments to calculate the template placement metric
or the inner product used to calculate the SNR. 
Note that the upper frequency cutoff will depend on the particular template 
used and the total mass of the source it represents 
(see Sec.~\ref{sec:fcutoff}).

There are competing factors that influence the choice of $f_{\rm{low}}$.
Binaries with larger masses will plunge and coalesce at lower frequencies
(see Eq.~(\ref{fLSO})).
Taking lower values of $f_{\rm{low}}$ means that we will be sensitive to 
binaries with larger total mass and will also observe more orbital cycles
of inspiraling binaries in general. 
The observation of more cycles allows for greater SNR's to be achieved but 
would require longer duration templates (and simulated waveforms) to be produced 
which would increase 
the computational cost of the search.
This increase in computational cost is far less for searches employing
templates which are generated in the frequency (rather than time) 
domain, such as the BCV2 DTF used in this analysis.
Note also, that although increasing the mass range of a search to include heavier 
binaries will increase the number of templates in the bank (for a given 
minimal match), the number of templates required to cover the higher-mass
region of parameter space is far smaller than the number of templates
required to cover the low-mass region (see, for example Fig.~5 of
Babak et al. (2006) \cite{babak:2006}).
However, at lower frequencies seismic activity causes the sensitivity 
of ground-based detectors to become worse and the spectrum to be non-stationary 
(see Sec.~\ref{sec:gwdetectors}). 

In practice we
take the lowest value of $f_{\rm{low}}$ for which the noise spectrum 
is approximately stationary. For searches of S3 LIGO data 
a value of $f_{\rm{low}} = 70 \rm{Hz}$ 
was chosen. 
As the detectors achieve better sensitivity and greater stationarity of noise,
the values of $f_{\rm{low}}$ we use have decreased allowing higher mass binaries to
be searched for. 
In searches of S4 and S5 LIGO data, lower cutoff frequencies as low as 
$f_{\rm{low}} = 40 \rm{Hz}$ have been used. 

\begin{table}[htdp]
\caption{Summary of the amount of data analysed in our various data sets. 
In S3 we only analyse data from the LHO detectors when both H1 and H2 are 
in science mode. Around $9.4\%$ of the data is classified as 
{\it playground data} and is used to tune the parameters of the search. 
Playground data is not included in the upper limit calculation but is 
still searched for possible detections.}
\begin{center}
\begin{tabular}{c|cc}
\hline
\hline
Data type & Total analysed (hours) & Non-playground (hours) \\
\hline
H1-H2    & 604   & 548 \\
H1-H2-L1 & 184   & 167 \\ 
\hline
\end{tabular}
\end{center}
\label{tab:analysedtimes}
\end{table}

\subsection{Template bank generation}
\label{subsec:tmpltbank}
Using the estimated PSD we will calculate the template placement metric
Eq.~(\ref{tmpltbankmetric}) and create a template bank for each
2048s blocks of data for each detector (H1, H2 and L1).
The metric calculation and template placement scheme used in this
search is described in Sec.~\ref{sec:templatebank}.

\begin{figure}
\begin{center}
\includegraphics[angle=0, width=0.9\textwidth]{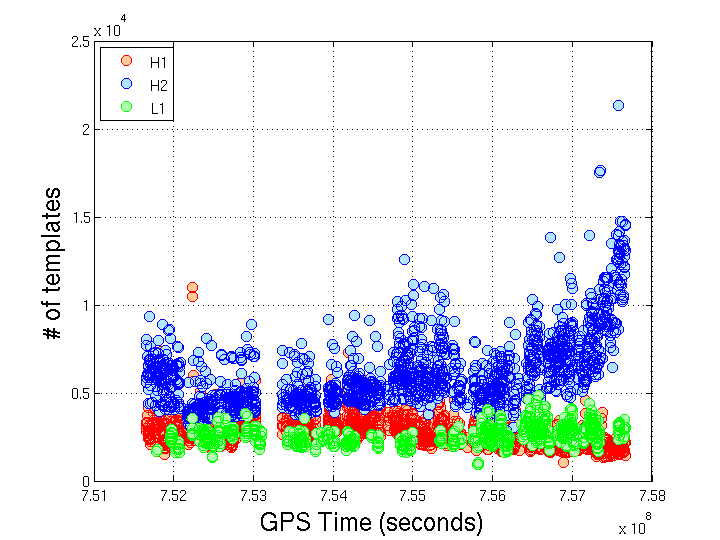}
\caption{Number of templates in each template bank. 
We generate a separate template bank for each 2048s block of data for each detector. 
We then matched-filter each template in the bank against the block of data and generate
a list of triggers which have SNR exceeding our threshold.
See Secs.~\ref{subsec:tmpltbank} and \ref{mfdetdata} for further details.
The large increase (a factor of $\sim 6$) in the size of the H2 template banks was caused
by a flattening of the its amplitude (and therefore power) spectral density profile as S3 progressed.
Figure \ref{fig:H2PSDplot} compares the amplitude spectra of H2 at two different times
during S3. 
}
\label{fig:FOMSize_VersusGPS_tmpltbank}
\end{center}
\end{figure}

\begin{figure}
\begin{center}
\includegraphics[angle=0, width=0.9\textwidth]{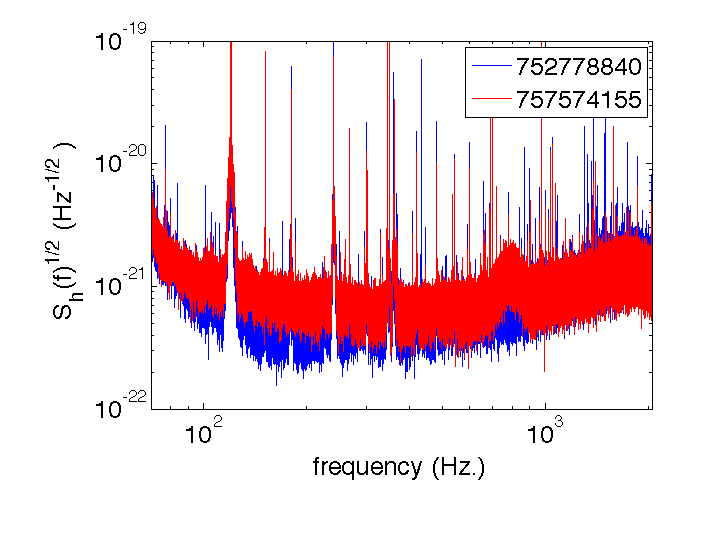}
\caption{Amplitude spectral density curves for H2 estimated at two different
times during S3. 
The GPS times in the legend of this plot indicate the start time of the 2048s block
of data that was used to estimate the spectrum.
As S3 progresses the amplitude (and therefore power) spectra of H2 become flatter
which leads to the increase in the number of templates required for H2 as shown in
Fig.~\ref{fig:FOMSize_VersusGPS_tmpltbank}.
}
\label{fig:H2PSDplot}
\end{center}
\end{figure}

In this search we use a minimal match of $0.95$.
Figure \ref{fig:FOMSize_VersusGPS_tmpltbank} shows the number of templates
in each template bank against GPS time throughout S3.
We see a large increase in the size of the H2 template banks which was caused
by a flattening of the its power spectral density profile as S3 progressed 
(see Fig.~\ref{fig:H2PSDplot}).
The output of template bank generation will be a list of the 
$\psi_{0}$, $\psi_{3}$ and $\beta$ values that we are required to search over 
for each 2048 second block for each detector.

\subsection{Matched-filtering of detector data}
\label{mfdetdata}
We matched-filter every 2048s block of data using each template in the associated 
template bank.
If the SNR measured by a particular template exceeds the SNR threshold,
we record a trigger which contains details of the template and the time
at which the SNR threshold was exceeded.
In practice we take the FFT of each 256 second data segment and matched-filter
each separately.
We do this because the power spectrum we use in the 
matched-filter has the frequency resolution of
the FFT of a 256 second data segment due to the way we estimate it 
(see Sec.~\ref{sec:dataselection}).
 
For Gaussian white noise, $\rho^2$ will, in general, have a $\chi^2$ distribution 
with 6 degrees of freedom. 
In the case where the spin parameter $\beta = 0$ we find that 
$\mathcal{\widehat{A}}_2$ and $\mathcal{\widehat{A}}_3$
both vanish and that $\rho^2$ is described by a $\chi^2$ distribution with 
2 degrees of freedom.
To reflect the increased freedom we choose a higher SNR threshold, 
$\rho_{*} = 12$ when $\beta \neq 0$ and a lower value of
$\rho_{*} \approx 11.2$ when $\beta = 0$.
These values were chosen to give approximately the same number of triggers when 
analysing Gaussian white noise and to ensure that the number of triggers produced 
during the real search was manageable.

We perform two stages of clustering in order to reduce the number of triggers
we are required to store.
The first stage involves identifying all triggers with SNR greater than our thresholds 
that were generated from one particular template.
We record only the trigger with the highest SNR over a stretch of data and discard
any other triggers generated by the same template within 16 seconds of it.
The second stage involves recording only the trigger with the highest SNR in 
each $100$ms stretch of data regardless of which template was used to generate it.

Due to the huge number of templates required by H2 during the end of S3 
we expected that the number of triggers generated might cause the search to 
become computationally unfeasible. 
However, through use of the clustering methods described here the number of
triggers recorded by the matched-filtering of H2 data was reduced to such a 
level ($\sim 5 \times 10^4$ triggers per 2048 second block)  that we decided 
to analyse data from all three LIGO detectors.
The output of the matched-filter stage is the list of triggers
with SNR exceeding the SNR threshold that survive clustering for each detector. 
Each trigger will contain information including the time $t$ it was recorded, 
the values of the intrinsic parameters of the template that was used 
to generate it (i.e., $\psi_{0}, \psi_{3}, \beta, f_{\rm{cut}}$),
its SNR $\rho$ and the values of the other extrinsic parameters
$\alpha_{1 \dots 6}$ that were used to obtain the (maximised) SNR. 

We know that the amplitude of the gravitational wave emitted by an inspiraling 
binary increases throughout the inspiral stage of its evolution before its 
components plunge and coalesce.
We expect that the time at which we measure the maximum SNR for a true 
gravitational signal from a  binary will correspond to the end of the inspiral 
stage of its evolution and, therefore, approximately to the (retarded) time of 
its coalescence.
The time recorded in our trigger is called the {\it coalescence time}.

\subsection{Coincidence analysis}
\label{sub:coincAnalysis}
To minimize the false alarm probability and to increase the
significance of a true detection we demand that a gravitational wave
signal be observed by two or more detectors with similar parameters.
In order to determine whether a trigger measured by one particular detector 
should be considered as coincident with a trigger in another detector we 
define a set of coincidence windows.

Suppose we measure a parameter $P$ to have a value $P_{1}$ at the 
first detector and $P_{2}$ at the second.
For the triggers to count as coincident we would demand 
\begin{eqnarray}
|P_{1} - P_{2}| < \Delta P_{1} + \Delta P_{2}
\end{eqnarray}
where $\Delta P_{1}$ and $\Delta P_{2}$ are our coincidence windows.
We have two choices to make: first we must decide from which of the
measured parameters we demand consistency and then we must choose or
{\it tune} the size of our coincident windows.

\subsubsection{Injection of simulated signals}
In order to choose and tune these coincidence windows
we perform software injections of simulated gravitational wave signals
into the data stream of each detector. 
Each injection will accurately mimic the detectors' 
(gravitational wave channel) output for a gravitational wave signal 
emitted by a particular simulated inspiraling binary source. 
The orientation, distance and direction of the simulated source 
relative to each detector is taken into account to ensure that the 
signal we inject into each detector is consistent with what we would expect
from a true source with the same parameters as the simulated source.
We use the target model described in Sec.~\ref{Sec:TargetModel} to generate
the waveforms we will inject.

We perform a large number of software injections ($\sim 8000$) choosing the 
parameters of each binary in our population at random within chosen ranges.
The spins are randomised such that
i) the spin magnitude of each of the compact objects is distributed uniformly
in the range $0 < \chi < 1$ and
ii) the direction of compact object's spin is uniformly distributed on the
surface of a sphere (that has radius $\chi$).
The physical distances $D$ of the simulated sources are chosen uniformly on a
logarithmic scale between $50$kpc and $50$Mpc.
The sky-positions and initial polarization and inclination angles of the
simulated sources are all chosen randomly
such that the direction of the initial orbital momenta will be uniformly
distributed on the surface of a sphere.
We simulated binaries with component masses distributed uniformly in the range
$1.0~M_{\odot} < m_{1}, m_{2} < 20.0~M_{\odot}$.

Having made the injections into the detectors data we measure the accuracy with
which we can measure the parameters of the simulated source.
This is made complicated because we describe our simulated source in terms of
physical parameters ($m_{1}$, $m_{2}$, $\chi_{1}$, $\chi_{2}$ \dots) and 
record the {\it non}-physical parameters ($\psi_{0}$, $\psi_{3}$, $\beta$ \dots)
of the detection template family in our triggers (see Sec.~\ref{mfdetdata}).
Although we have approximate relations between the physical and non-physical
parameters it is not clear how accurate these are.
In practice we choose to demand consistency between the 
coalescence time $t$, $\psi_{0}$ and $\psi_{3}$ since similar values of these
parameters are measured for nearby, high-SNR signals. 

In this search we demand that for triggers from different detectors to be
considered as coincident they must satisfy the following conditions:
\begin{eqnarray}
|t_{1} - t_{2}| & < & \Delta t_{1} + \Delta t_{2} + T_{1,2}, \\
|\psi_{0,1} - \psi_{0,2}| & < & \Delta \psi_{0,1} + \Delta \psi_{0,2} \and \\
|\psi_{3,1} - \psi_{3,2}| & < & \Delta \psi_{3,1} + \Delta \psi_{3,2}
\end{eqnarray}
where $t_{i}$, $\psi_{0,i}$ and $\psi_{3,i}$ are the time of
coalescence and phenomenological mass parameters measured using our
template bank in detector $i$;
$\Delta t_{i}$, $\Delta \psi_{0,i}$ and $\Delta \psi_{3,i}$
are our coincidence windows in detector $i$ and $T_{i,j}$ is the light
travel time between detector locations $i$ and $j$.
The light travel time between LHO and LLO is $\sim 10$ ms.
We must take the light travel time into account or with sufficiently
small values of $\Delta t_{i}$ we would risk missing coincidences
between the Hanford detectors and L1.

We tune our coincidence windows on the playground data in order to
recover as many of our simulated signals as possible whilst trying to 
minimize the false alarm rate caused by our non-gravitational wave background. 
The use of playground data allows us to tune our search parameters
without biasing the results of our full analysis.

\subsubsection{Background estimation}
\label{sub:background}
We estimate the rate of accidental coincidences, otherwise known as the
background or false alarm rate, for this search through analysis of
time-shifted data. 
We time-shift the triggers obtained from each detector relative to each other 
and then repeat our analysis, searching for triggers that occur in coincidence 
between 2 or more of the detectors. 
By choosing our time-shifts to be suitably large 
($\gg 10$ ms light travel time between LHO and LLO)
we ensure that none of the coincident triggers identified
in our time-shift analysis could be caused by a true gravitational wave signal
and can therefore be used as an estimate of the rate of accidental coincidences.
In practice we leave H1 data unshifted and time-shift H2 and L1 by increments
of $10$ and $5$ seconds respectively.
In this search, we analysed 100 sets of time-shifted data 
(50 forward shifts and 50 backward shifts).
For clarity we will use the term {\it in-time} to mean triggers which have not
been time-shifted.

\begin{figure}
\begin{center}
\includegraphics[width=0.9\textwidth]{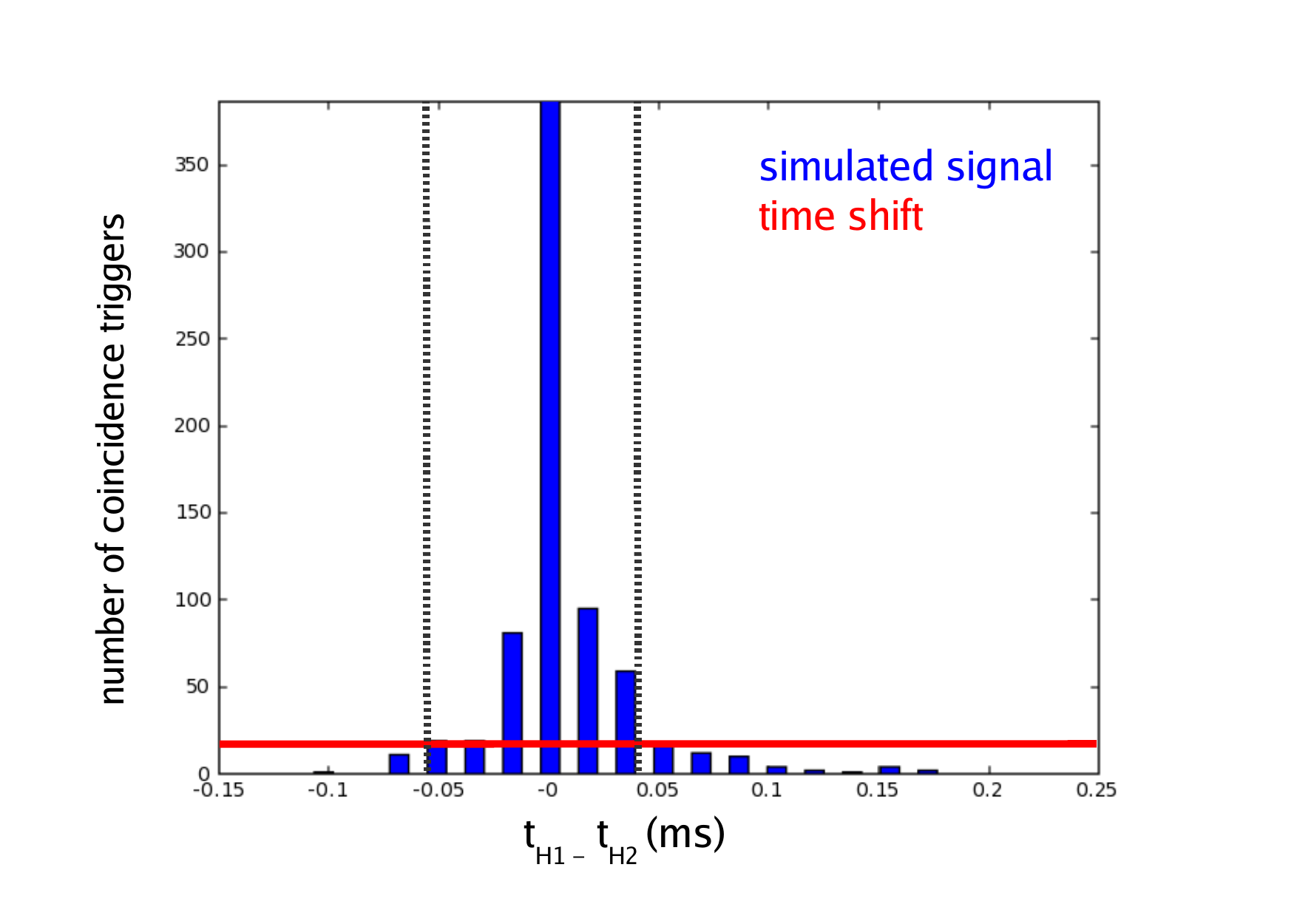}
\caption{A histogram of the number of triggers against the difference in 
coalescence time $t_{\rm{H1}} - t_{\rm{H2}}$ between H1 and H2.
The blue bars represent triggers caused by the software injection of
simulated signals and indicate where we might expect to observe triggers
caused by true gravitational wave signals (foreground).
The red line represent triggers found during analysis of time-shifted data
and are used to estimate the non-gravitational wave background.
We choose the coincidence windows (vertical dotted lines) so that we will
find all the simulated signals that lay above the background in coincidence.
Note that the plot shown here only uses nearby injections corresponding 
to simulated sources with physical distances $50 \rm{kpc} < D < 500 \rm{kpc}$. 
In order to find simulated sources at larger distances we extended our 
windows to $\Delta t = 100$ms.
}
\label{fig:H1H2_H1H2L1_plotcoincwindow_endtime}
\end{center}
\end{figure}

Figure ~\ref{fig:H1H2_H1H2L1_plotcoincwindow_endtime} shows a histogram
of the number of triggers against the difference in coalescence time 
$t_{\rm{H1}} - t_{\rm{H2}}$ between H1 and H2.
We choose the smallest possible values for our coincidence windows
that mean that all simulated signals that can be distinguished from
our background would be found in coincidence.

Using this tuning method we find our coincidence windows for each
detector to have values $\Delta t = 100$ ms,
$\Delta \psi_{0} = 40,000~ \rm{Hz}^{5/3} $ and
$\Delta \psi_{3} = 600~ \rm{Hz}^{2/3}$ (we rounded and symmetrized these
values for simplicity).
The value of $\Delta t$ used in this search is four times larger than the
$25$ ms value used in the S3 search for non-spinning binary black holes
\cite{LIGOS3S4all} indicating that the estimation of arrival time of a
gravitational waveform is less well determined in this search than in the
non-spinning search.

\subsection{Combined SNR}
\label{sub:combSNR}
\begin{figure}
\begin{center}
\includegraphics[width=0.9\textwidth]{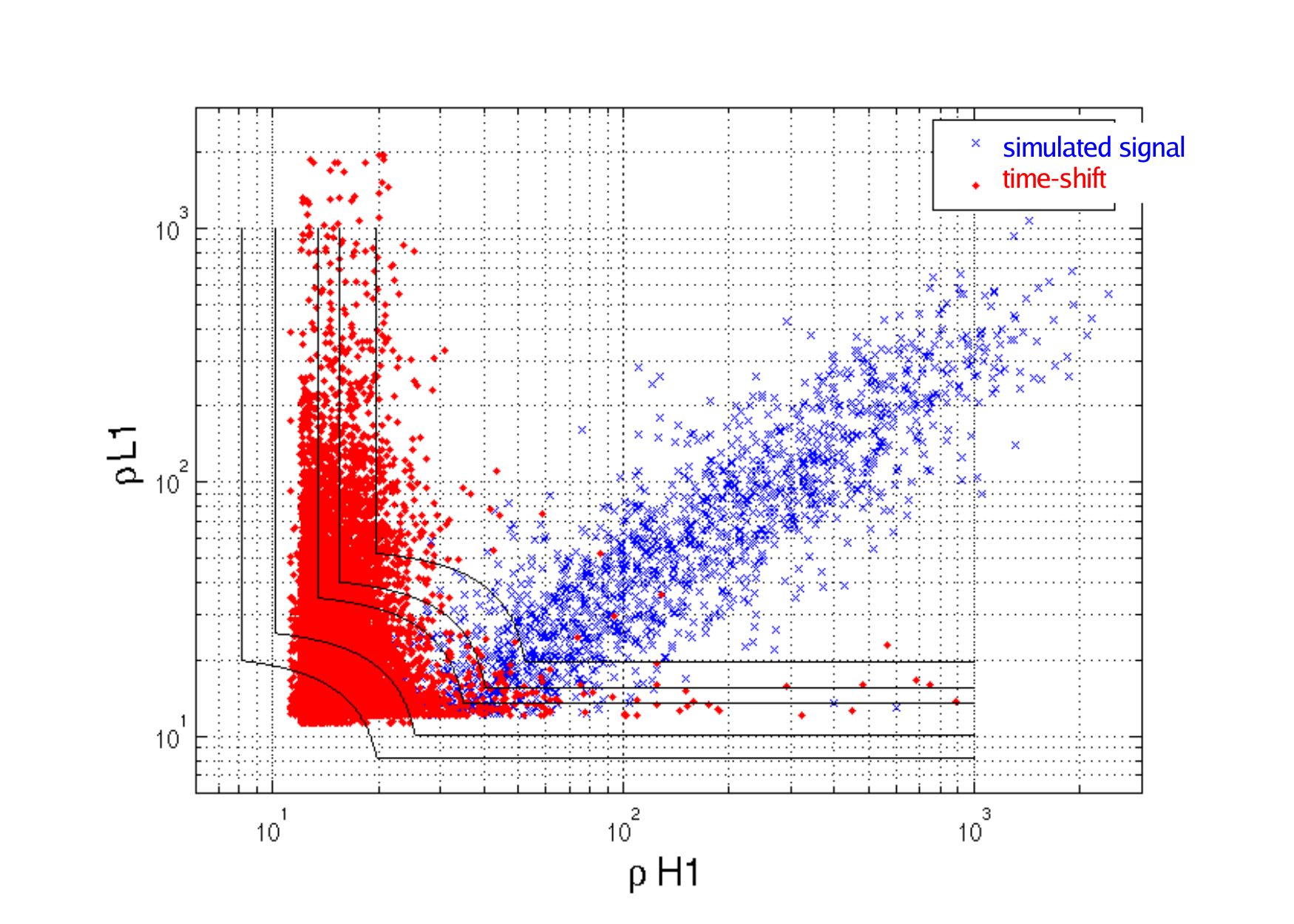}
\caption{Scatter plot of SNR measured in H1 and L1 for H1-L1 coincident triggers 
occurring in H1-H2-L1 times
(i.e., times when all three LIGO detectors were taking science quality data, 
see Sec.~\ref{subsec:S3data}).
The blue crosses represent triggers caused by the software injection of
simulated signals and indicate where we might expect to observe triggers
caused by true gravitational wave signals (foreground).
The red dots represent triggers found during analysis of time-shifted data
and are used to estimate the non-gravitational wave background.
The black curves show contours of constant combined SNR $\rho_{c}$ assigned
using Eq.~(\ref{combSNR_2ifo}).
Higher values of combined SNR are assigned to coincident triggers caused by
simulated signals allowing us to separate these from our estimated background.
}
\label{fig:H1L1_combinedSNR_inj_shift}
\end{center}
\end{figure}

We expect rough consistency between the SNR of triggers measured in 
different LIGO detectors if they originate from the same true gravitational 
wave signal (or software injected simulated signal)
\footnote{
The orientation of the Hanford and Livingston sites was chosen 
so that the detectors would be as closely as aligned as possible
(modulo $90^{\circ}$ that we can ignore
due to the quadrupolar nature of gravitational waves) 
in order to maximise their common response to
a signal, Abbott et al. (2004) \cite{abbott-2004-517}.
For misaligned detectors with poor overlap between their
antenna response patterns we would not be able to make this
assumption.
}.
Conversely, we would not necessarily expect any consistency between
the triggers measured in different detectors that are caused by spurious noise
events 
(however, we will see later that seismic activity at the Hanford site can cause 
triggers in H1 and H2 that are consistent with each other).
We assign a {\it combined signal-to-noise ratio} $\rho_{c}$ to our coincident 
triggers based upon the individual signal-to-noise ratios $\rho_{i}$
measured by each detector.
For triggers found in coincidence between two detectors we use
\begin{equation}
\label{combSNR_2ifo}
\rho_{c}^{2} = {\rm min} \left\{ \sum_{i}^{2} \rho_{\rm i}^2,
(a\rho_{\rm i}-b)^2 \right\}
\end{equation}
and for coincident triggers found in all three LIGO
detectors we use
\begin{equation}
\label{combSNR_3ifo}
\rho_{c}^{2} =  \sum_{i}^{3} \rho_{\rm i}^2.
\end{equation}
Equation~(\ref{combSNR_2ifo}) assigns higher combined SNR $\rho_{c}$ to 
coincident triggers with similar SNRs measured in both detectors
$\rho_{1} \sim \rho_{2}$ than those consisting of a very loud trigger
in one detector and a relatively quiet trigger in the other detector
$\rho_{1} \gg \rho_{2}$.
In practice the parameters $a$ and $b$ are tuned so that the contours
of false alarm generated using Eq.~(\ref{combSNR_2ifo}) separate triggers generated
by software injection of simulated signals and background triggers
as cleanly as possible \cite{LIGOS3S4Tuning}(see Sec.~\ref{sub:background} for 
details of how we estimate the background). 
In this search we used values $a = b = 3$ for all detectors. 
Figure \ref{fig:H1L1_combinedSNR_inj_shift} shows a scatter 
plot of the SNR measured in H1 and L1 with lines of constant $\rho_{c}$
(as assigned using Eq.~(\ref{combSNR_2ifo})).
We see that the combined SNR allows us to differentiate between foreground
(simulated signals) and background (estimated using time-shifts).

In some cases the presence of a weak (typically H2) trigger would cause
the combined SNR of a triple coincidence trigger (using Eq.~(\ref{combSNR_2ifo})) 
to be lower than the combined SNR of a double coincidence trigger where the 
weakest trigger of the triple coincidence has been neglected.
This is undesirable since triple coincident triggers are less likely to be
caused by noise events than double coincident triggers and we would like
to assign triple coincident triggers a higher value of combined SNR to
reflect their increased significance. 
By using Eq.~(\ref{combSNR_3ifo}) to assign the combined SNRs
of triple coincident candidates we ensure that the combined SNR 
of a triple coincident trigger will always be greater than the combined SNR of 
any two of its constituent triggers would be assigned from 
Eq.~(\ref{combSNR_2ifo}).

\subsection{Vetoes}

\label{sub:vetoes}

\subsubsection{Instrument-based vetoes}
We are able to veto some background triggers by observing correlation between 
the gravitational wave channel (AS\_Q) of a particular detector and one or 
more of its auxiliary channels which monitor the local physical environment. 
Since we would not expect a true gravitational wave signal to excite the 
auxiliary channels, we will treat as suspicious any excitation in the 
gravitational wave channel that is coincident in time with excitations in the 
auxiliary channels. 
A list of auxiliary channels found to effectively veto spurious 
(non-gravitational wave coincident triggers) were identified and used for all 
S3 searches \cite{S3Vetoes}.
Additional vetoes based upon other auxiliary channels were considered but were 
subsequently abandoned because the total amount of data these channels would 
have discounted, known as the {\it dead-time}, was unacceptably large.

\subsubsection{Signal-based vetoes}
\begin{figure}
\begin{center}
\includegraphics[width=0.9\textwidth]{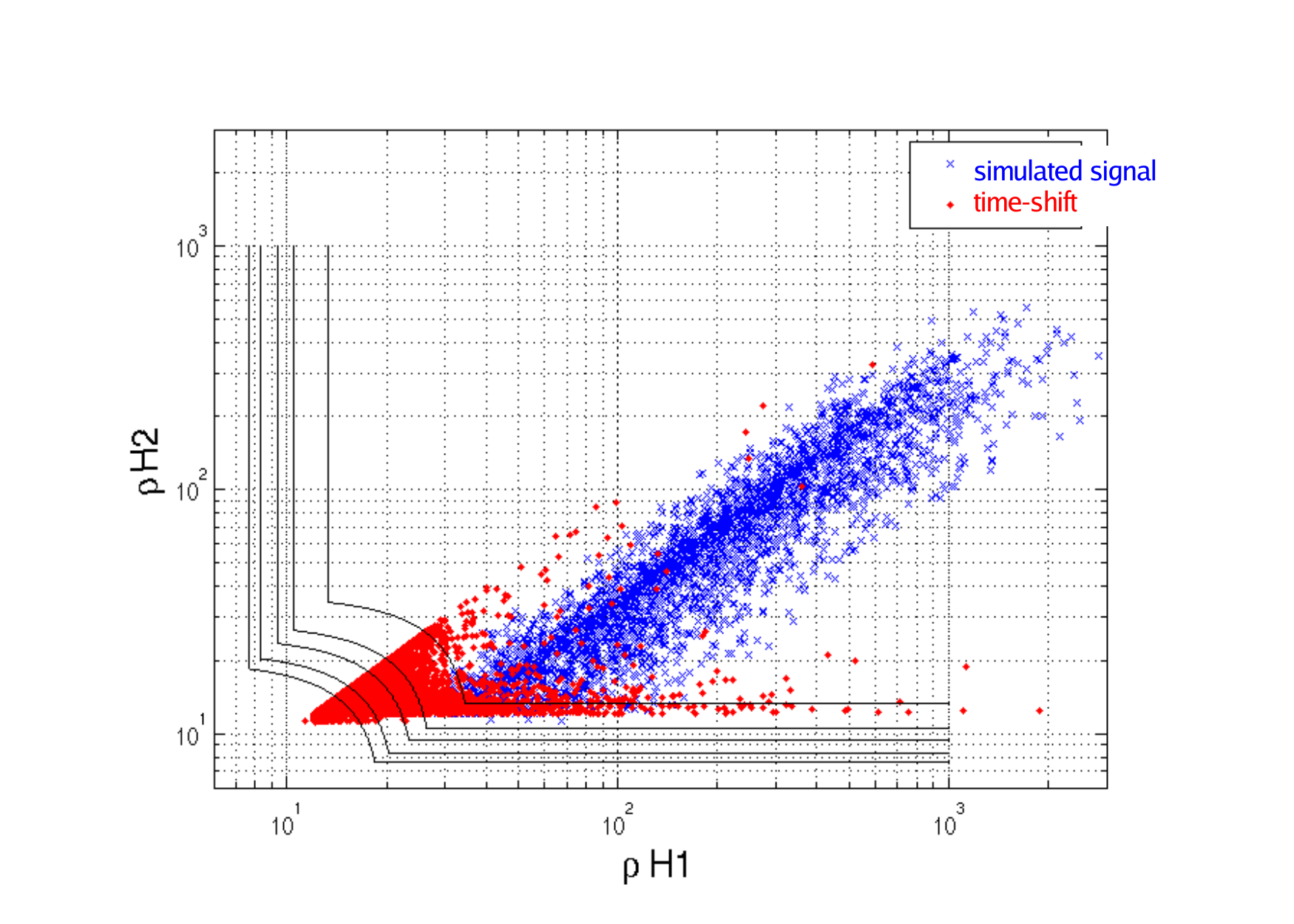}
\caption{Scatter plot of SNR for H1-H2 coincident triggers in H1-H2 times 
(see caption of Fig.~\ref{fig:H1L1_combinedSNR_inj_shift}).
We have removed coincident triggers that were measured to have a larger SNR in
H2 than in H1.
We find that applying this veto vastly reduces the number of background 
triggers but does not affect the number of simulated signals that were 
observed.
}
\label{fig:H1H2_combinedSNR_inj_shift}
\end{center}
\end{figure}

We can use the fact that the Hanford detectors are co-located to veto 
coincident triggers whose measured amplitude is not consistent between 
H1 and H2.
We check for consistency between the SNR values measured using H1 and H2 data 
for triggers found in coincidence. Since H1 is
the more sensitive instrument we simply required that the SNR measured in H1 
be greater than that measured in H2 for an event to survive this veto.
Figure \ref{fig:H1H2_combinedSNR_inj_shift} shows a scatter
plot of the SNR measured in H1 and H2 for triggers caused by simulated signals 
as well as those measured during time-shift analysis
with this veto applied. 
We find that the application of this veto will vastly reduce the number of
background triggers but does not affect the number of simulated signals that 
were observed.
Since H1 and H2 were only operated when both were in science mode during S3, 
this veto means that there will be no H2-L1 coincident triggers since this 
would indicate that H2 had detected a trigger which H1 was unable to detect.

The $\chi^{2}$ veto used for the primordial black hole and binary neutron star
searches~\cite{LIGOS3S4all} has not not been investigated for use in searches 
using detection template families (i.e., this search and the S2-S4 searches for
non-spinning binary black holes \cite{LIGOS2bbh, LIGOS3S4all}).


\section{Results and follow-up analysis}
\label{sub:results}
In the search of the S3 LIGO data described in this paper, no triple-coincident
event candidates (exceeding our pre-determined SNR threshold and satisfying the
coincidence requirements described in Sec.~\ref{sub:coincAnalysis}) were found 
in triple-time (H1-H2-L1) data. 
Many double-coincident event candidates were found in both triple-time and 
double-time (H1-H2) data.

A cumulative histogram of combined SNR for in-time and background coincident 
triggers is shown in Fig.~\ref{fig:cum_hist}. 
We see that, at the SNR threshold (i.e., the leftmost points on this figure), 
the number of in-time double-coincident triggers is consistent with the number
of coincident triggers yielded by the time-shift analysis.
The small excess in the number of in-time H1-H2 coincident triggers at higher 
SNRs indicates that there is some correlation between the LHO detectors. 
The coincident triggers contributing to this excess have been investigated and 
are not believed to be caused by gravitational waves. 
Seismic activity at the Hanford site has been recorded throughout S3 and can 
cause data to become noisy simultaneously in H1 and H2.  
Coincident triggers caused by seismic noise will predominantly cause only 
in-time coincidences (although time-shift coincidences caused by two seismic 
events separated in time but shifted together can occur) leading to an excess 
of in-time coincident triggers as we have observed in Fig.~\ref{fig:cum_hist}.
As mentioned previously, there were no coincident triggers observed by all three 
detectors.
A scatter plot of the SNRs measured for coincident triggers in H1-H2 times
is shown in Fig.~\ref{fig:snr_snr}. The distribution of our in-time triggers 
is consistent with our estimation of the background. This is also true for
the double-coincident triggers measured in H1-H2-L1 times.

\begin{figure}
\begin{center}
\includegraphics[width=0.9\textwidth]{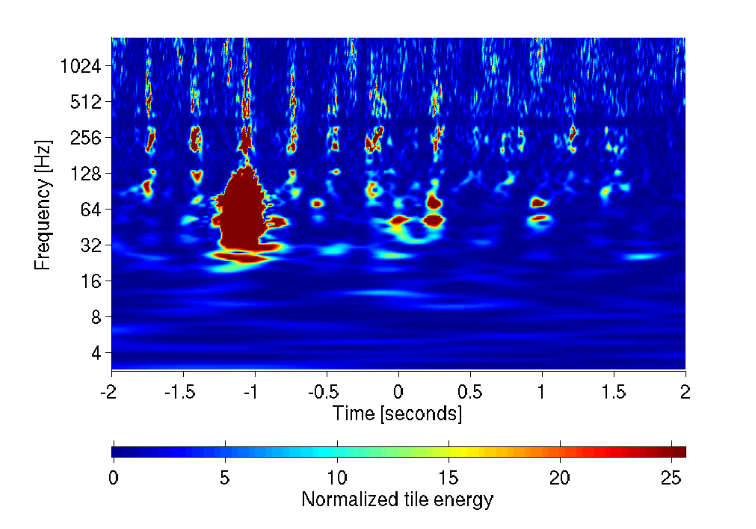}
\caption{Time-frequency image of the gravitational wave channel data taken by H1 
about the time of the loudest event candidate, an H1-H2 coincident trigger occurring
when only the Hanford detectors were in science mode.
A gravitational wave signal would occur at 0 seconds on the time scale of this figure.
This figure shows that the H1 gravitational wave channel is noisy at the 
time of this event and consequently does not improve the likelihood that this
candidate was caused by a true gravitational wave signal (see Sec.~ref{sub:results}). 
The H2 gravitational wave channel is also noisy at this time.
It is useful to compare this figure with Fig.~\ref{fig:spin_spec_thesis}
which shows time-frequency maps of the gravitational wave signals observed from
inspiraling binaries without the effects of detector noise.
This figure was produced using Q Scan \cite{Qscan, ShourovsThesis}.
}
\label{fig:QscanLoudestEvent}
\end{center}
\end{figure}

\begin{figure}
\begin{center}
\includegraphics[width=0.9\textwidth]{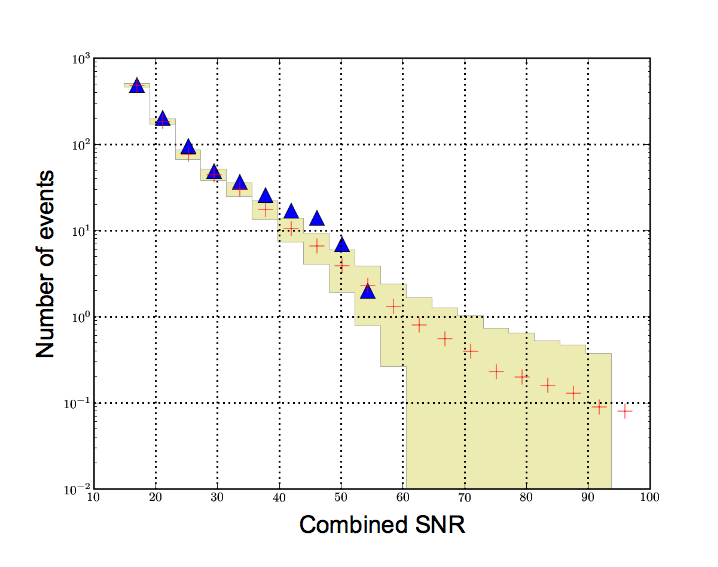}
\caption{
Cumulative histograms of the combined SNR, $\rho_{c}$ for in-time coincident triggers
(triangles) and our background (crosses with one-sigma deviation shown)
for all H1-H2 and H1-H2-L1 times within S3.
We see a small excess in the number of in-time coincident triggers with combined SNR
$\sim 45$. This excess was investigated and was caused by an excess of H1-H2 coincident
triggers. Since H1 and H2 are co-located, both detectors are affected by the same
local disturbances (e.g., seismic activity) which contributes to the number of in-time
coincidences but which is under-represented in time-shift estimates of the background.
}
\label{fig:cum_hist}
\end{center}
\end{figure}

\begin{figure}
\begin{center}
\includegraphics[width=0.9\textwidth]{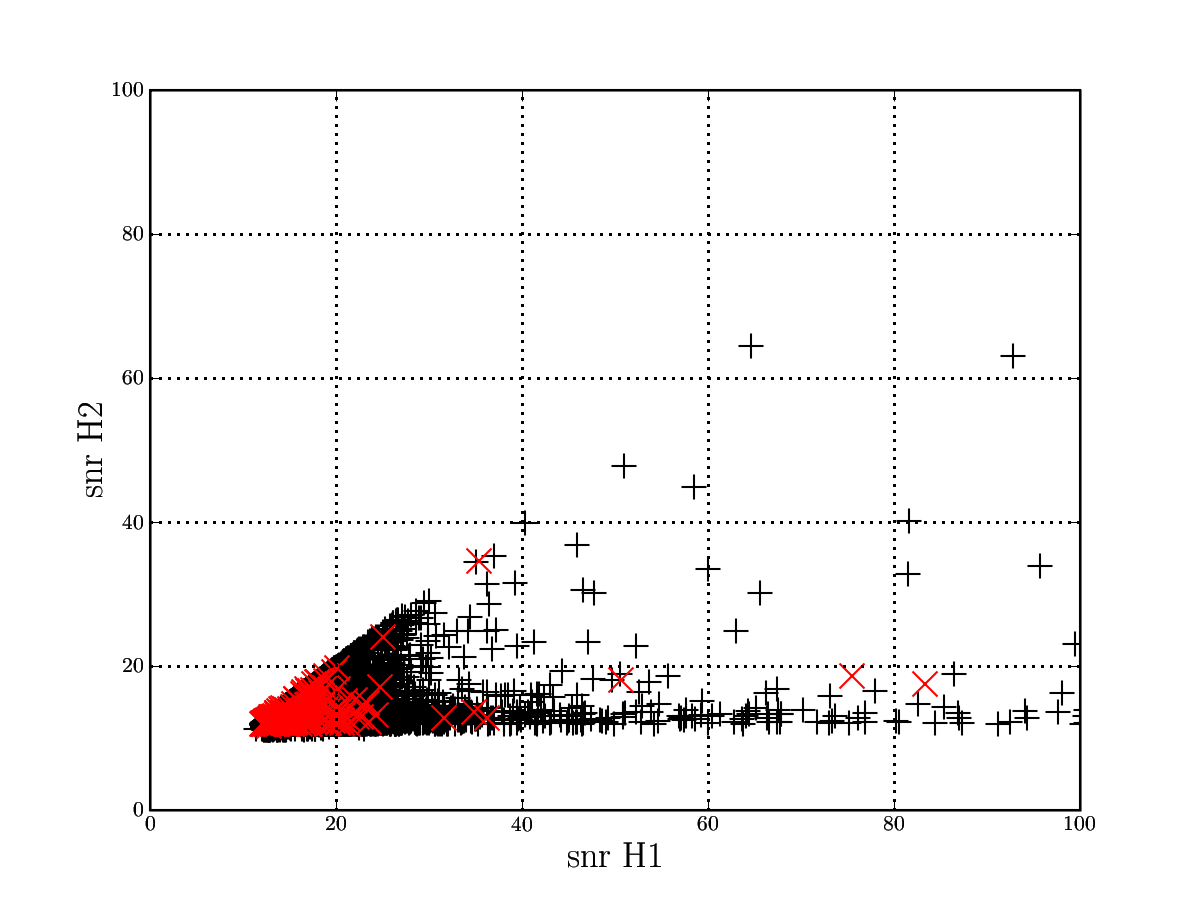}
\caption{Scatter plot of SNR for coincident triggers in H1-H2 times. 
The light coloured crosses represent in-time coincident triggers and the 
black pluses represent 
time-shift coincident triggers that we use to estimate the background.
Note that we observe more background triggers than in-time triggers
since we perform 100 time-shift analyses to estimate the background but
can perform only a single in-time analysis to search for true gravitational
wave signals (see Sec.~\ref{sub:coincAnalysis} for further details on
background estimation). 
Note that due to our signal-based veto on H1/H2 SNR
we see no coincident triggers with $\rho_{\mathrm{H1}} < \rho_{\mathrm{H2}}$.}
\label{fig:snr_snr}
\end{center}
\end{figure}

The loudest in-time coincident trigger was observed in H1-H2 when only the 
Hanford detectors were in science mode.
This event candidate is measured to have SNRs of 
$119.3$ in H1, 
$20.4$ in H2 
and a combined SNR of $58.3$.
The loudest coincident triggers are subjected to systematic follow-up 
investigations in which a variety of information (e.g., data quality at time 
of triggers, correlation between the detector's auxiliary channels and the 
gravitational wave channel) is used to assess whether the coincident triggers 
could be confidently claimed as detection of gravitational wave events.
This event is found at a time flagged for ``conditional'' vetoing. 
This means that during these times some of the detectors auxiliary channels 
exhibited correlation with the gravitational wave channel (AS\_Q ) and that 
we should be careful in how we treat event candidates found in these times.
For this particular coincident trigger an auxiliary channel indicated an
increased numbers of dust particles passing through the dark port beam
of the interferometer~\cite{S3Vetoes}.
Upon further investigation it was found that this coincident trigger occurred
during a period of seismic activity at the Hanford site and we
subsequently discounted this candidate as a potential gravitational wave event.
Time-frequency images of the gravitational wave channel around the time of this
candidate (see Fig.~\ref{fig:QscanLoudestEvent}) 
were inconsistent with expectations of what an inspiral signal should 
look like further reducing the plausibility of this candidate being a true 
gravitational wave event.
It is interesting, but unsurprising, to note that during the search for 
non-spinning binary black holes that also used S3 LIGO data, high-SNR triggers 
associated with this seismic activity were also detected \cite{LIGOS3S4all}.
Furthermore, the 20 next loudest event candidates were also investigated and 
none were found to be plausible gravitational wave event candidates.
Work is in progress to automate the follow-up investigative procedure and to 
include new techniques including null-stream and Markov chain Monte Carlo 
analysis for assessing the plausibility of coincident triggers as 
gravitational wave events. 

\section{Upper limit on the rate of binary coalescences}
\label{sub:upperlimit}

Given the absence of plausible detection candidates within the search 
described above, we calculate an upper limit on the rate of
spinning compact object coalescence in the universe. 
We quote the upper limit rate in units of 
coalescences per year per $\rm{L}_{10}$ where 
$\rm{L}_{10} = 10^{10} \, \rm{L}_{\odot,B}$ is $10^{10}$ times the blue 
light luminosity of the Sun.

We assume that binary coalescences only occur in galaxies.
The absorption-corrected blue light luminosity of a galaxy infers its massive 
star formation rate, and therefore supernova rate, which we assume scale with the rate of 
compact binary coalescence within it \cite{Phinney:1991ei}.
This assumption is well justified when the population of galaxies reached
by the detector 
(i.e., those galaxies which are close enough that it would be possible to 
detect a stellar mass binary inspiral from within them) 
is dominated by spiral galaxies with ongoing star formation
(e.g., the Milky Way).
Papers reporting on S1 and S2 ~\cite{LIGOS1iul, LIGOS2iul, LIGOS2bbh} 
have quoted the upper limit in units of Milky Way Equivalent Galaxy (MWEG) 
which is equivalent to about $1.7 \, \rm{L}_{10}$.
Upper limits on the rate of coalescences calculated during other searches
using S3 and S4 LIGO are given in units of $\rm{L}_{10}$ \cite{LIGOS3S4all}.

Our primary result will be an upper limit on the rate of coalescence of 
precessing neutron star - black hole binaries with masses 
$m_{\rm{NS}} \sim 1.35 M_{\odot}$ and $m_{\rm{BH}} \sim 5 M_{\odot}$.
These mass values correspond to NS-BH binaries with component masses similar 
to those used to assess the NS-NS and BH-BH upper limits in \cite{LIGOS3S4all}.
We will now detail the calculation of the upper limit on the rate of
binary coalescence before applying it to our search of S3 data for systems
with spinning components.

The setting of upper limits on rates is discussed in the 
following publications which were used by the author
in writing this Section:
Biswas et al. (2007) \cite{ul},
Brady and Fairhurst (2007) \cite{systematics}
Brady et al. (2004) \cite{loudestGWDAW03}.

\subsection{Calculating the upper limit}
We will treat arrival of a gravitational wave at our detectors
as a rare event which can be described by a Poissonian distribution.
The probability of detecting no gravitational waves
(emitted during binary coalescence) with combined SNR greater than
some value $\rho_{c}$ is given by
\begin{eqnarray}
P_{F}(\rho_{c}) = e^{- \nu(\rho_{c})} 
\end{eqnarray}
where $\nu(\rho_{c})$ is the mean number of gravitational wave events
detected with combined SNR greater than $\rho_{c}$ during the
course of a search (e.g., a science run).
We can write $\nu$ more explicitly as the product of 
i) the rate of binary coalescence (per year per $\rm{L}_{10}$),
ii) the total (cumulative) luminosity $C_{L}(\rho_{c})$ (in $\rm{L}_{10}$) 
that the detectors were sensitive to with combined SNR greater than $\rho_{c}$ 
and 
iii) the total observation time $T$ (in years).
We can therefore write
\begin{eqnarray}
P_{F}(\rho_{c}|R,T) = e^{-R \, T \, C_{L}(\rho_{c})}.
\end{eqnarray}
The subscript $F$ stands for foreground and is used to distinguish
this probability from the probability of measuring a background event
with combined SNR greater than $\rho_{c}$ which we shall call
$P_{B}(\rho_{c})$.

The probability of measuring no event candidates 
(true gravitational wave foreground or noise-induced background) 
with combined SNR greater than that of our loudest observed event candidate
$\rho_{c,\rm{max}}$ is given by
\begin{eqnarray}
P(\rho_{c,\rm{max}}|B,R,T) &=& P_{B}(\rho_{c,\rm{max}}) P_{F}(\rho_{c,\rm{max}}) 
\nonumber\\
&=& P_{B}(\rho_{c,\rm{max}}) e^{-R \, T \, C_{L}(\rho_{c,\rm{max}})}.
\end{eqnarray}
We can calculate the probability density as
\begin{eqnarray}
p(\rho_{c,\rm{max}}|B,R,T) &=& \frac{d}{d \rho_{c}} P(\rho_{c,\rm{max}}|B,R,T) \\
&=&
P_{B}^{\prime}(\rho_{c,\rm{max}}) e^{-R \, T \, C_{L}(\rho_{c,\rm{max}})  } 
\times \nonumber \\
& &
\left[
1 + R \, T \, C_{L}(\rho_{c,\rm{max}}) \, \Lambda    
\right]
\end{eqnarray}
where we have defined 
\begin{equation}
\label{likelihood}
  \Lambda =
  \frac{|\mathcal{C}_{L}^{\prime}(\rho_{c,\rm{max}})|}
       {P_{B}^{\prime}(\rho_{c,\rm{max}})}
  \left[
  \frac{\mathcal{C}_{L}(\rho_{c,\rm{max}})}{P_{B}(\rho_{c,\rm{max}})}
  \right]^{-1}
  \, ,
\end{equation}
where the derivatives are with respect to combined SNR $\rho_{c}$.
$\Lambda$ is a measure of the likelihood that the loudest event measured 
during a search is consistent with being a true gravitational wave signal 
(foreground) rather than being caused by noise (background).
We know by definition that the cumulative luminosity $C_{L}(\rho_{c})$ 
our detectors were sensitive to with combined SNR greater than $\rho_{c}$ 
will decrease as $\rho_{c}$ increases and therefore that 
$C_{L}^{\prime}(\rho_{c})$ will always be negative.

Using Bayes' theorem we can find the posterior probability 
distribution of the rate $p(R|\rho_{c,\rm{max}}, T, B)$ using our 
prior knowledge or guess of its distribution $p(R)$ and our
probability distribution $p(\rho_{c,\rm{max}}|B,R,T)$ for the number of 
events exceeding the combined SNR of the loudest measured event:
\begin{eqnarray}
\label{Bayes}
p(R|\rho_{c,\rm{max}}, T, B) = 
\frac{p(R) \, p(\rho_{c,\rm{max}}|B,R,T) }
{\int  p(R) \, p(\rho_{c,\rm{max}}|B,R,T) \, dR  } .
\end{eqnarray}

Since this is the first dedicated search for gravitational waves emitted 
by binaries with spinning component bodies we have no prior knowledge about 
the rate.
To reflect this, we use a uniform prior, $p(R) = \rm{constant}$.
In upper limit calculations for future searches we will be able to use
the posterior probability distribution calculated in this search
as the prior.
Integrating the denominator of Eq.~(\ref{Bayes}) by parts yields 
\begin{eqnarray}
&\int& p(R) p(\rho_{c,\rm{max}}|B,R,T) dR
\nonumber \\ 
&=& P_{B}^{\prime}(\rho_{c,\rm{max}}) p(R)
\int
e^{-R \, T \, C_{L}(\rho_{c,\rm{max}}) }
\left[
1 + R \, T \, C_{L}(\rho_{c,\rm{max}}) \, \Lambda
\right]
dR 
\end{eqnarray}
where we can take the prior outside the integral since it is
a constant and $P_{B}^{\prime}$ outside the integral since it
will clearly not depend on the rate of true gravitational wave events.
Evaluating the integrand over all possible rates 
(from $R=0$ to $R= \infty$) we find
\begin{eqnarray}
\int_{0}^{\infty} p(R) p(\rho_{c,\rm{max}}|B,R,T) dR
&=& 
P_{B}^{\prime} p(R) \frac{1 + \Lambda}{T \, C_{L}}. 
\end{eqnarray}
In practice we can use a finite upper bound on this integral by choosing
a large value for the rate $R$ with a suitably low probability of occurring. 

Substituting this result back into Bayes' theorem Eq.~(\ref{Bayes})
we find the posterior distribution to be:
\begin{eqnarray}
p(R|\rho_{c,\rm{max}}, T, B) 
&=&
e^{-R \, T \, C_{L}} \frac{T \, C_{L}} {1 + \Lambda} 
\left[
1 + R \, T \, C_{L} \, \Lambda
\right]
\end{eqnarray}

To find the upper limit $R_{\alpha}$ on the rate of coalescences 
with confidence $\alpha$ we evaluate
\begin{eqnarray}
\alpha = \int_{0}^{R_{\alpha}} p(R|\rho_{c,\rm{max}}, T, B) \, dR. 
\end{eqnarray}
Integrating by parts yields
\begin{eqnarray}
\label{bayesianprob}
1 - \alpha = e^{-R_{\alpha}\, T \, \mathcal{C}_{L}(\rho_{c,\rm{max}})}
\left[ 1 + \left( \frac{\Lambda}{1+\Lambda} \right) 
\, R_{\alpha}\, T \, \mathcal{C}_{L}(\rho_{c,\rm{max}}) \right].
\end{eqnarray}
The corresponding statistical statement would be that we have 
$\alpha \times 100\%$ confidence that the rate of binary coalescences 
is less than $R_{\alpha}$.


We evaluate the cumulative luminosity $\mathcal{C}_{L}$ at the combined SNR
of the loudest coincident trigger seen in this search, 
$\rho_{c,\rm{max}} = 58.3$ (see Sec.~\ref{sub:results} for discussion of 
this coincident trigger). 

\subsection{Observation time}
We only use data that was taken during {\it non}-playground times in the
calculation of the upper limit.
The (in-time) non-playground dataset is {\it blinded} in the sense that 
all analysis parameters are tuned and fixed prior to its analysis
in order to avoid statistical bias 
(as described in Sec.~\ref{sec:pipeline})
The observation time $T$ is taken from Table~\ref{tab:analysedtimes},
where we use the {\it non}-playground analysed times.

\subsection{Calculating the cumulative luminosity}
The cumulative luminosity $\mathcal{C}_{L}(\rho_{c})$ 
to which our search was sensitive to is a function of the detection
efficiency of our search $\mathcal{E}$ and the predicted luminosity $L$ of 
the local universe.

\subsubsection{Effective distance and inverse expected SNR}
\label{effDistInvExpSNR}
In searches for systems consisting of non-spinning bodies 
detection efficiency $\mathcal{E}$ is found as a function of its chirp mass 
$\mathcal{M} = M \eta^{3/5}$ and effective distance which are combined to 
construct a quantity called the ``chirp distance'' which describes how 
detectable a given source is \cite{systematics, LIGOS3S4all}.
For low values of chirp distance we would expect high detection efficiency
and vice versa. 
We find that the effective distance is not well defined for a source
consisting of spinning bodies and we find an alternative.

For a binary source located at a distance $D$ from a detector, 
the effective distance $D_{\rm eff}$ is the distance at
which it would produce the same SNR if it was positioned directly overhead 
the detector and with optimal orientation 
(i.e., face on to the detector, $\iota=0$) \cite{allen:2005}.
For a system consisting of {\it non}-spinning bodies effective distance can be
calculated using
\begin{eqnarray}
D_{\rm eff} = \frac{D}
{\sqrt{F_+^2 (1 + \cos^{2}\iota)^{2}/4 + F_\times^{2} ( \cos\iota)^2}}
\end{eqnarray}
where $D$ is the distance between the binary and the observer,
$\iota$ is the inclination angle of the binary with respect to the
observer and $F_{+}$ and $F_{\times}$ are the antenna patterns of the detector
(see Eq.~(\ref{Fpluscross})).
The effective distance $D_{\rm eff}$ of a binary will always be
equal or greater than the physical distance $D$.

For a system consisting of spinning bodies, its inclination $\iota$
with respect to a detector will evolve during the course of the inspiral making
the calculation of effective distance complicated (it would in fact be time
dependent if we used the formula above).
Instead, in this search we find efficiency and predicted source luminosity as a
function of the inverse of the {\it expected SNR} of a source.
The expected SNR is defined as the SNR that would be obtained for a given
simulated source assuming we use a template that perfectly matches the emitted
gravitational waveform (i.e., fitting factor $= 1$)
and a detector whose noise power spectrum we can estimate accurately.
We therefore define $\rho_{\rm{expected}} = \left< s,h \right>$ where
$h$ is our template and $s = A h$ is our signal. 
We can calculate the {\it combined} expected SNR using the formulae in 
Sec.~\ref{sub:combSNR}.

By taking our distance measure $D_{\rho}$ as the inverse of the expected SNR 
we obtain a quantity which behaves similarly to the chirp distance by taking 
larger values for signals which are detectable with a high SNR and by taking 
smaller values as the signals become less detectable.
Since a binary system will have slightly different orientations with
respect to the two LIGO observatories, detectors at different sites will
measure slightly different expected SNRs and therefore slightly
different $D_{\rho}$.
We will denote $D_{\rho, \rm{H}}$ as the inverse expected SNR that
would be measured at the Hanford site and $D_{\rho, \rm{L}}$ as the
inverse expected SNR that would be measured at the Livingston site.

We will find the detection efficiency $\mathcal{E}$ and the luminosity 
of the nearby universe $L$ both as functions of $D_{\rho, \rm{H}}$
and $D_{\rho, \rm{L}}$ and we need to perform a two-dimensional
integration in order to obtain $C_{L}$:
\begin{eqnarray}
  \mathcal{C}_{L}(\rho)=
  \int_0^{\infty}
  \int_0^{\infty}
  \mathcal{E}(D_{\rho,\rm{H}},D_{\rho,\rm{L}},\rho) \,
  L(D_{\rho,\rm{H}}, dD_{\rho,\rm{L}}) \,
  dD_{\rho,\rm{L}} \, dD_{\rho,\rm{H}}.
\end{eqnarray}
As mentioned earlier, we evaluate $\mathcal{C}_{L}$ at the combined SNR of our
loudest event candidate $\rho_{c,\rm{max}} = 58.3$.

\subsubsection{Detection efficiency}
We define detection efficiency as
\begin{eqnarray} 
\label{detEff}
\mathcal{E}(\rho_{c}) = 
\frac{\#_{\rm{found}}(\rho_{c})}
{\#_{\rm{found}}(\rho_{c}) + \#_{\rm{missed}}(\rho_{c})}
\end{eqnarray}
where $\#_{\rm{found}}(\rho_{c})$ is the number of simulated signals with
combined SNR greater than some $\rho_{c}$ that were detected (found) during
the search and similarly $\#_{\rm{missed}}(\rho_{c})$ is the number of simulated
signals with combined SNR greater than $\rho_{c}$ that were not detected 
(missed).

We use software injection of a population of simulated signals 
(the target waveforms described in Sec.~\ref{Sec:TargetModel})
to evaluate the detection efficiency $\mathcal{E}$ for observing events
with combined SNR greater than $\rho_{c}$, as a function of the source's 
inverse expected SNR $D_{\rho}$.
In order to sample the parameter space of the binary as thoroughly as possible
and to obtain a good estimate of the detection efficiency $\mathcal{E}$ we 
perform thousands of software injections. 
We choose the parameters of each binary in our population at random as we 
described in Sec.~\ref{sub:coincAnalysis} when
discussing software injections for the tuning of coincidence windows.

\begin{figure}
\begin{center}
\includegraphics[width=0.9\textwidth]{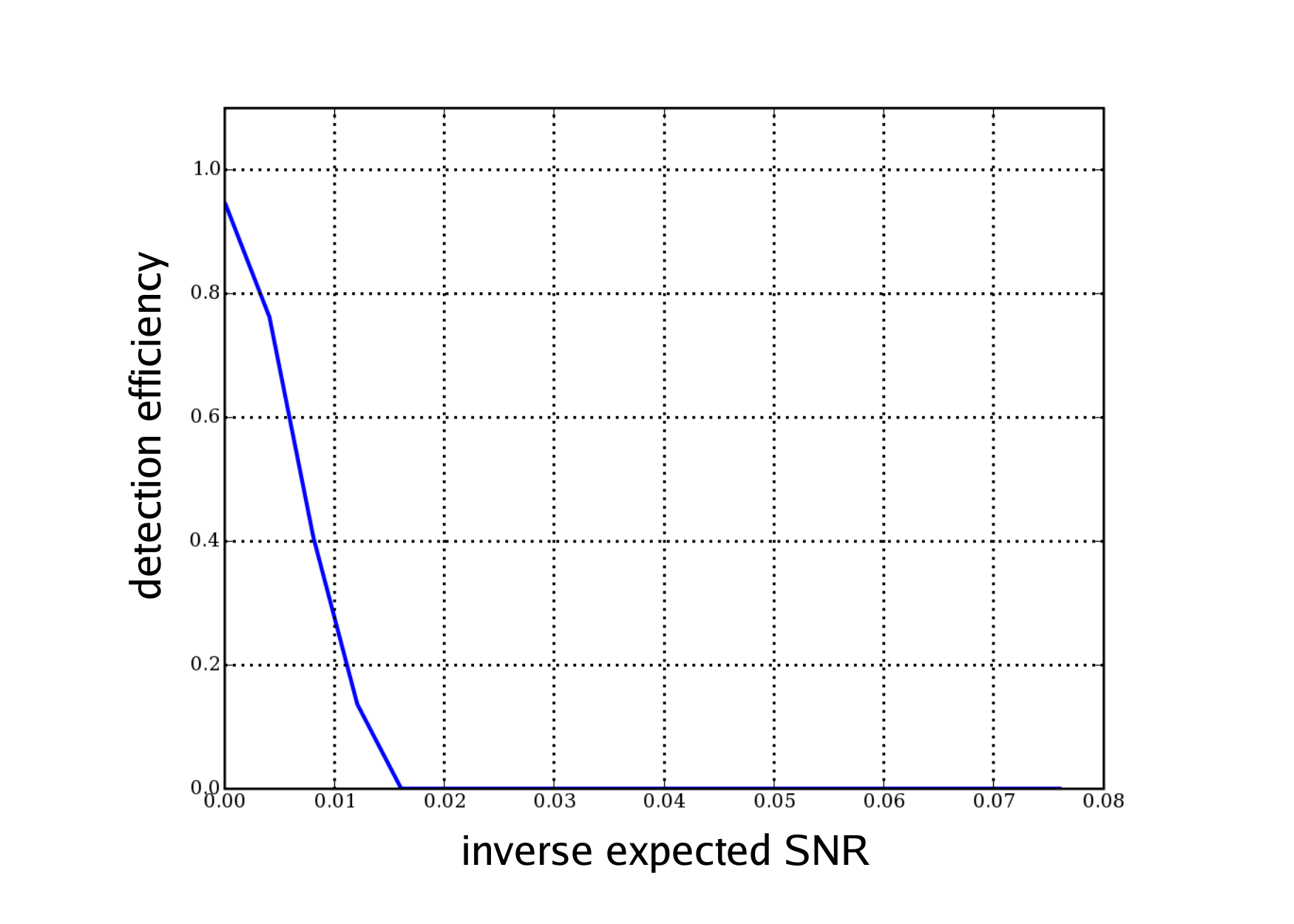}
\caption{
Detection efficiency for recovering software injected simulated signals 
measured against the inverse of the sources expected SNR.
This figure contains results for recovering injections in coincidence between H1
and H2 in H1-H2 times only.
The reason we only achieve $\sim 95\%$ efficiency at low inverse expected
SNR values is because we veto around $5\%$ of H1-H2 times and therefore
veto around $5\%$ of our injections which are then subsequently classified
as ``missed'' (see Eq.~(\ref{detEff})).
}
\label{fig:H1H2_H1H2_invexpSNR_efficiency}
\end{center}
\end{figure}

We evaluated the detection efficiency of this search for 
binaries with component masses distributed uniformly in the range
$1.0~M_{\odot} < m_{1}, m_{2} < 20.0~M_{\odot}$.
During S3, LIGO's efficiency to binaries in this range was dominated by
sources within the Milky Way for which detection efficiency was high
across the entire mass range investigated due to the proximity of
these sources to Earth.
Figure \ref{fig:H1H2_H1H2_invexpSNR_efficiency} shows the detection efficiency
measured for recovering software injections in coincidence between H1
and H2 in H1-H2 times against the simulated sources inverse expect SNR.

To calculate the upper limit on the rate of coalescence of NS-BH binaries
we will use a Gaussian distribution to generate the component masses of 
each binary. 
For the neutron star mass we assume a mean $\mu_{\rm{NS}} = 1.35 M_{\odot}$ and
standard deviation $\sigma_{\rm{NS}} = 0.04 M_{\odot}$.
This choice is motivated by the mass measurements of radio pulsars by 
Thorsett and Chakrabarty (1999) \cite{Thorsett:1998uc}.
Drawing upon analysis of stellar mass black hole observations
(Orosz (2002) \cite{orosz-2002}) and theoretical black hole population 
studies (Belczynski et al. (2002) \cite{Belczynski:2002}), 
O'Shaughnessy and Kalogera \cite{BBHMasses}
 recommend that upper limits on the rate of
binary coalescence assume a black hole mass distribution with
mean $\mu_{\rm{BH}} = 5 M_{\odot}$ and
standard deviation $\sigma_{\rm{BH}} = 1 M_{\odot}$.
This choice, although slightly ad hoc, corresponds to likely values of BH mass 
predicted by the population studies in Ref.~\cite{BBHMasses}.
Also, by assuming relatively low mass black holes that will appear less luminous
to our detector the upper limit we calculate will be correspondingly conservative.

\subsubsection{Luminosity of the nearby universe}
As well as the detection efficiency $\mathcal{E}$ we will also require an 
estimate of the expected distribution of coalescing binary sources in the 
nearby universe in order to evaluate the cumulative luminosity $C_{L}$
that our search was sensitive to.

\begin{figure}
\begin{center}
\includegraphics[width=0.9\textwidth]{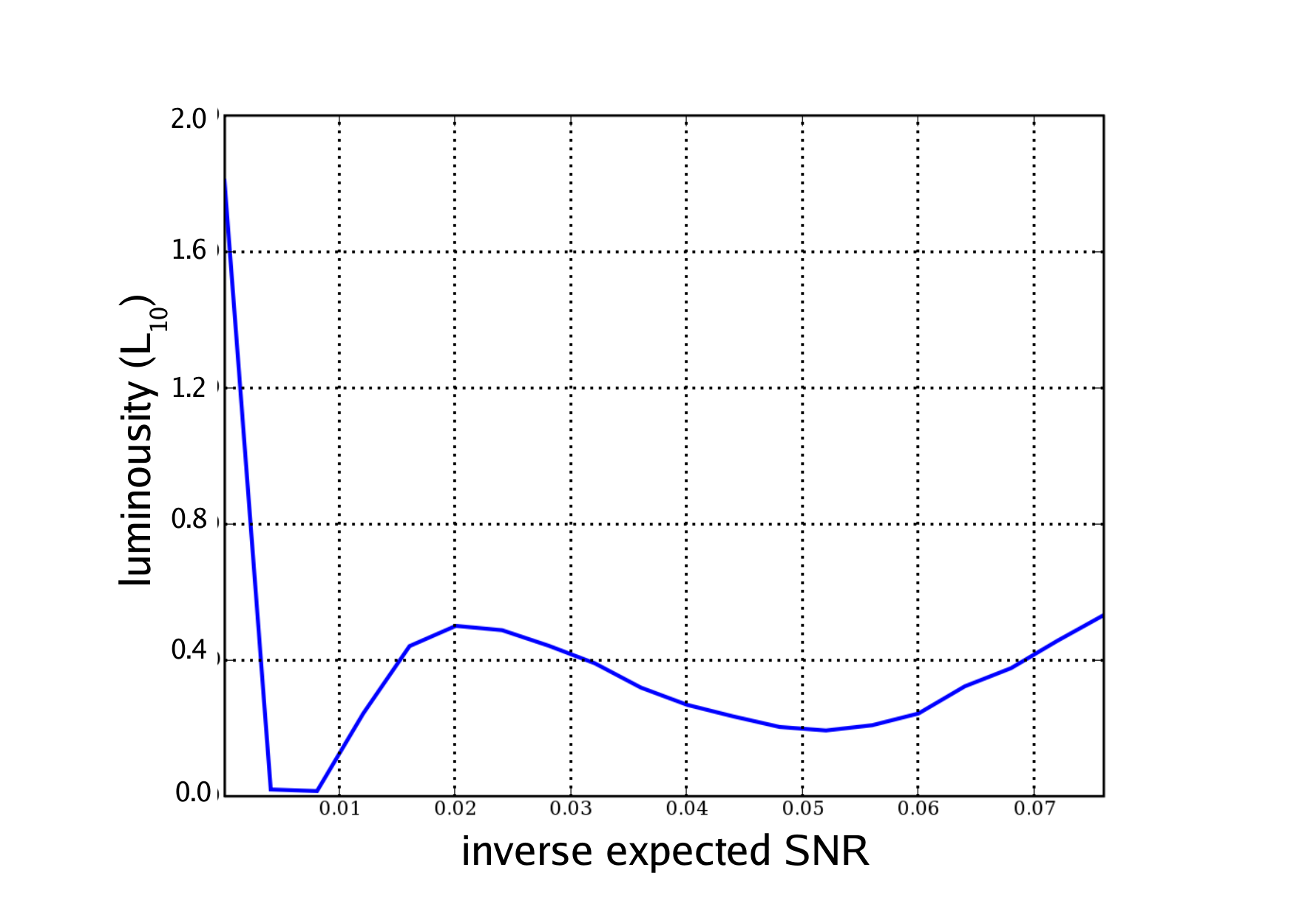}
\caption{
Estimated luminosity of the nearby universe against the inverse expected SNR
of our simulated sources.
Comparing this plot with a similar plot made for the search for black hole
binaries with non-spinning components (which had effective distance along the
$x$-axis, see Sec.~\ref{effDistInvExpSNR}) we are able to find an approximate 
conversion between inverse expected SNR $D_{\rho,H}$ and 
effective distance $D_{\rm{eff}}$. 
We find that $D_{\rm{eff}} \rm{(Mpc)}  \simeq 63 D_{\rho,H}$.
We identify the left most peak on this plot to be caused by the Milky Way,
the peak at $D_{\rho,H} \simeq 0.02$ to correspond to Andromeda (M31,NGC0224)
and the peak at $D_{\rho,H} \simeq 0.07$ to correspond to Centaurus A (NGC5128).
}
\label{fig:H1H2_H1H2_invexpSNR_luminousity}
\end{center}
\end{figure}

We calculate the luminosity of binary inspirals in the nearby universe
by generating a population of simulated signals using information on the 
observed distribution of sources from standard astronomy catalogues.
We use a model based on the work of 
Kopparapu et al. (2007) \cite{LIGOS3S4Galaxies} for the distribution
of blue light luminosity throughout the nearby Universe which we assume 
is proportional to the rate of binary coalescences (see start of this Section).
We will use the same distribution of spins and mass for this population
of binaries as we did when assessing the detection efficiency.
For each simulated signal we calculate the expected SNR as we did when
assessing the efficiency of the search.
Figure \ref{fig:H1H2_H1H2_invexpSNR_luminousity} shows the luminosity
distribution of the nearby universe.

From our search of S3 data we measure the cumulative luminosity 
$C_{L}(\rho_{c,\rm{max}})$
and
$C_{L}^{\prime}(\rho_{c,\rm{max}})$ 
 for H1-H2 times and H1-H2-L1 times separately.
We find
\begin{eqnarray} 
C_{L}(\rho_{c,\rm{max,H1-H2}}) &=& 1.76 \, \rm{L}_{10}, \nonumber\\
|C_{L}^{\prime}(\rho_{c,\rm{max,H1-H2}})| &=& 9.2 \times 10^{-3} \, 
\rm{L}_{10} \, \rho_{c}^{\phantom{c} -1}, \nonumber\\
C_{L}(\rho_{c,\rm{max,H1-H2-L1}}) &=& 2.23 \, \rm{L}_{10},  \nonumber\\
|C_{L}^{\prime}(\rho_{c,\rm{max,H1-H2-L1}})| &=& 1.5 \times 10^{-2} \, 
\rm{L}_{10} \, \rho_{c}^{\phantom{c} -1}.
\end{eqnarray}
The averaged values of $C_{L}(\rho_{c,\rm{max}})$ and $C_{L}^{\prime}(\rho_{c,\rm{max}})$ 
used to calculate the upper limit on the rate (Eq.~(\ref{bayesianprob}))
are simply
\begin{eqnarray}
C_{L}(\rho_{c,\rm{max}}) &=& 
\frac{  
T_{\rm{H1-H2}}\, C_{L}(\rho_{c,\rm{max,H1-H2}}) +
T_{\rm{H1-H2-L1}}\, C_{L}(\rho_{c,\rm{max,H1-H2-L1}})
}
{ T_{\rm{H1-H2}} + T_{\rm{H1-H2-L1}} }, \nonumber \\
C_{L}^{\prime}(\rho_{c,\rm{max}}) &=&
\frac{  
T_{\rm{H1-H2}}\, C_{L}^{\prime}(\rho_{c,\rm{max,H1-H2}}) +
T_{\rm{H1-H2-L1}}\, C_{L}^{\prime}(\rho_{c,\rm{max,H1-H2-L1}})
}
{ T_{\rm{H1-H2}} + T_{\rm{H1-H2-L1}} } \nonumber
\end{eqnarray}
where $T_{\rm{H1-H2}}$ and $T_{\rm{H1-H2-L1}}$
are the non-playground times listed in 
Table \ref{tab:analysedtimes}.

~\subsection{Background probability}
We estimate the background using time-shifts, see Sec.~\ref{sub:background}.
We estimate the probability $P_{B}(\rho_{c,\rm{max}})$ 
of there being no background events with combined SNR greater than that of the
loudest event as the fraction of time-shift events 
with combined SNR {\it less} than $\rho_{c,\rm{max}}$.
Our estimate of the probability density $p_{B}(\rho_{c,\rm{max}})$ is 
the gradient of $P_{B}(\rho_{c})$ with respect to $\rho_{c}$ at the combined
SNR of the loudest event $\rho_{c,\rm{max}}$.
For our search of S3 data we estimate $P_{B}(\rho_{c,\rm{max}}) = 0.23$
and $P_{B}^{\prime}(\rho_{c,\rm{max}}) = 0.026$.

Combining our results for the background probability $P_{B}$ and cumulative
luminosity $C_{L}$ we can find the likelihood (Eq.~(\ref{likelihood})) that 
the loudest event observed in this S3 was a true gravitational wave event to be 
~$\Lambda = 0.05$ (i.e., 20 times more likely to be caused by noise than a 
gravitational wave). 

We are now in a position to calculate the upper limit on the rate of 
coalescences of NS-BH binaries. Substituting the values we have calculated
for the observation time $T$, the cumulative luminosity $C_{L}$ and the
likelihood $\Lambda$ into Eq.~(\ref{bayesianprob}) we obtain the $90\%$ 
confidence upper limit on the rate to be
$R_{90\%} = 15.8 \,\mathrm{yr}^{-1}\,\mathrm{L_{10}}^{-1}$. 

\subsection{Marginalization of errors}
There are a number of systematic uncertainties in this calculation of
the upper limit arising from astrophysical and instrumental uncertainties as
well as the assumptions we have made during the calculation itself.
Systematic errors in the calculated upper limit rate can arise from
\begin{itemize}
\item uncertainties in the distances and luminosities of nearby galaxies,
\item uncertainties related to the calibration of data recorded by the 
detectors,
\item uncertainties due to the distribution of mass and spins assumed for
the population of binaries we use to assess the detection efficiency of our
search and the luminosity of the nearby universe,
\item uncertainties due to the limited number of software injections we 
performed in order to assess the detection efficiency and luminosity of the
nearby universe.
\end{itemize}

Note that for searches using different families of matched-filter templates 
that rely directly upon the modelling of the binary inspiral and the 
post-Newtonian approximation there is also an uncertainty associated with how 
well the templates match true gravitational wave signals.
However, since we use a {\it detection template family} designed to capture a 
broad range of signals based upon their wave shape 
(see discussion of detection template families in Sec.~\ref{sec:dtf}) 
we ignore this uncertainty.

In order to obtain the most accurate upper limit possible we will
marginalize over these uncertainties.
This involves specifying a prior distribution that describes how we expect
the uncertain parameter to behave.
For instance, suppose that our posterior on the rate 
$p(R|\rho_{c,\rm{max}},T,B,\Lambda)$
depended not only on
the combined SNR of the loudest event $\rho_{c,\rm{max}}$, observation time
$T$ and our background $B$ but also on some uncertainty in the likelihood 
$\Lambda$ (due to some uncertainty in $P_{B}$ or $C_{L}$). 
By assuming a prior distribution $p(\Lambda)$ of the likelihood we would be
able to marginalize over this uncertainty using
\begin{eqnarray} 
p(R|\rho_{c,\rm{max}},T,B) = 
\int p(\Lambda) p(R|\rho_{c,\rm{max}},T,B,\Lambda) d \Lambda.
\end{eqnarray} 

\begin{figure}
\begin{center}
\includegraphics[width=0.9\textwidth]{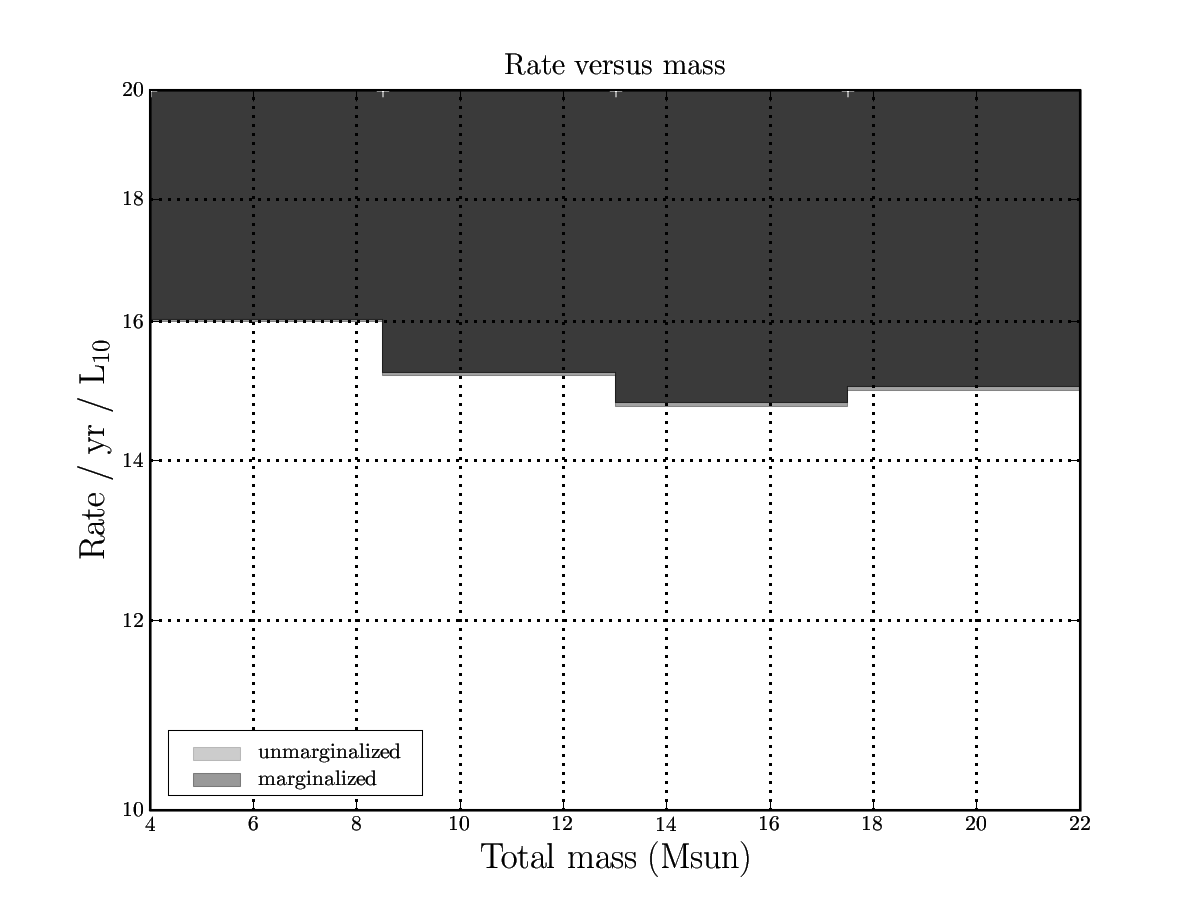}
\caption{Upper limits on the spinning binary coalescence rate per
$\mathrm{L}_{10}$ as a function of the total mass of the binary.
For this calculation, we have evaluated the efficiency of the search
using a population of binary systems with $m_1 = 1.35 M_{\odot}$
and $m_2$ uniformly distributed between $2$ and $20 M_{\odot}$.
The darker area on the plot shows the region excluded after marginalization
over the estimated systematic errors whereas the lighter region shows the
region excluded if these systematic errors are ignored.
The effect of marginalization is typically small ($<1\%$).
The initial decrease in the upper limit corresponds to the increasing
amplitude of the signals as total mass increases.
The subsequent increase in upper limit is due to the counter effect that as
total mass increases the signals become shorter and have fewer cycles in LIGO's
frequency band of good sensitivity.}
\label{fig:upperlimit}
\end{center}
\end{figure}

The process of marginalization is described further in 
Biswas et al. (2007) \cite{ul} and its application to searches for gravitational
waves emitted by binary systems is detailed in 
Brady and Fairhurst (2007) \cite{systematics}.
After marginalization over these errors we obtain an upper limit of
$R_{90\%} = 15.9 \,\mathrm{yr}^{-1}\,\mathrm{L_{10}}^{-1}$.
We also calculate upper limits for a range of binary systems with 
$m_1 = 1.35 M_{\odot}$ and $m_2$ uniformly distributed between $2$ and 
$20 M_{\odot}$.
These upper limits, both before and after marginalization are shown in 
Fig.~\ref{fig:upperlimit}. 
These upper limit results are around 7 orders of magnitude larger than the 
expected rates discussed
in Sec.~\ref{sec:mergerrates} so do not allow us to constrain the uncertainties
in them.

\subsubsection{Upper limits on the rate of binaries with non-spinning components}
There was no detection of gravitational waves in the S3 and S4 LIGO searches for 
binaries with non-spinning components and the upper limits on the rate of
their coalescence were calculated.
The S3 and S4 searches for binaries with non-spinning components are described
in Abbott et al. (2007) \cite{LIGOS3S4all}.
We briefly summarise the results of these searches and compare the upper limits
on the rates of coalescence calculated.
 
The S3 search for binary black holes with non-spinning components targeted systems
with component masses in the range 
$3.0 M_{\odot} < m_{1}, m_{2} < 40.0 M_{\odot}$.
The loudest event candidate observed was in H1 and H2 in H1-H2 times with combined
SNR $\rho_{c} = 106.5$.
We find that this event was the second loudest event observed in the search for 
binaries with spinning components where it was observed with $\rho_{c} = 53.2$ 
(we identified these events by the GPS time in which their peak SNR was measured).
This event was found by the search for binaries with non-spinning components 
with optimal $\psi_{0}$ and $\psi_{3}$ values well outside the region covered by 
the search for binaries with spinning components thus explaining
the higher SNR it achieved in the non-spinning search.
It was noticed that another two of the five loudest event candidates observed in 
H1-H2 times during the search for asymmetric binaries with spinning components
were among the loudest (four) event candidates observed in the search for binary 
black holes with non-spinning components.
In cases where both searches have triggers lying in (or very near) the range of 
$\psi_{0}$ and $\psi_{3}$ covered by the search for binaries with spinning
components we would expect it to yield higher SNR since the BCV2 detection 
template family used to capture spin-modulated signals has more degrees
of freedom (i.e., 6 when $\beta \neq 0$) than the BCV1 detection template
family used to capture signals from binaries with non-spinning components.  

The upper limit on the rate of coalescence of (approximately symmetric) 
binary bank holes consisting of non-spinning components with masses distributed 
with means $m_{\rm{BH}} = 5 M_{\odot}$
and standard deviations $\sigma_{\rm{BH}} = 1 M_{\odot}$ was calculated to be
$R_{90\%} = 23.6 \,\mathrm{yr}^{-1}\,\mathrm{L_{10}}^{-1}$.
The lack of a $\chi^{2}$ test and the large mass region the search covered
lead to the high combined SNR of the loudest event candidate which in turn
lead to a comparably high upper limit on the rate of coalescences.

The S3 search for binary neutron stars with non-spinning components targeted
systems with component masses in the range
$1.0 M_{\odot} < m_{1}, m_{2} < 3.0 M_{\odot}$.
The upper limit on the rate of coalescence for binary neutrons consisting of 
non-spinning components with masses distributed
with means $m_{\rm{NS}} = 1.35 M_{\odot}$
and standard deviations $\sigma_{\rm{NS}} = 0.04 M_{\odot}$ was calculated 
to be $R_{90\%} = 7.97 \,\mathrm{yr}^{-1}\,\mathrm{L_{10}}^{-1}$.
Again the value of this loudest event can be understood, at least partially,
in terms of the loudest event candidate observed in the search which
had combined SNR $\rho_{c} \sim 12$. 
This search utilised a $\chi^{2}$ test and actually used the effective SNR 
(higher for events that have good $\chi^{2}$ fit to the matched-filter template) 
to measure the loudness of the events. 

The expectation is that we will obtain more interesting (i.e., lower) values
for the upper limits on the rates of coalescences in future searches 
as the sensitivity of the detectors improves leading to larger 
detection efficiency (and therefore cumulative luminosities $C_{L}$) 
and improved detector stability leads to longer observation times $T$.
This is borne out by the results of the searches of S4 LIGO data for 
binaries with non-spinning components 
(B. Abbott et al. (2007) \cite{LIGOS3S4all}) which yielded
$R_{90\%} = 0.5 \,\mathrm{yr}^{-1}\,\mathrm{L_{10}}^{-1}$ 
for binary black holes and
$R_{90\%} = 1.2 \,\mathrm{yr}^{-1}\,\mathrm{L_{10}}^{-1}$
for binary neutron stars.

\chapter{Searching for Extreme Mass Ratio Inspirals using time-frequency algorithms}
\label{ch:emri}

In this Chapter will turn our attention to the development of data analysis 
techniques to enable the detection of inspiral events using the planned
LISA detector (described in Sec.~\ref{sec:gwdetectors}).
We will first describe the various sources LISA is expected to be sensitive to 
in Sec.~\ref{sec:LISAsources} before reviewing existing data analysis 
techniques in Sec.~\ref{sec:LISADA}.
Having identified the requirement for a computationally cheap method to
provide initial detection and rough parameter estimation of LISA sources,
in Sec.~\ref{sec:HACR} we will detail a time-frequency technique for this 
purpose, the Hierarchical Algorithm for Clusters and Ridges (HACR).
We will then go on to use HACR on a simulated LISA data set and assess its
ability to detect gravitational waves from our expected sources.

The analysis described in this Chapter was performed by the author 
(Gareth Jones) in collaboration with Dr. Jonathan Gair 
(Institute of Astronomy, University of Cambridge) and has been previously 
published in Gair and Jones (2007) \cite{HACR_GJ} and 
Gair and Jones (2007) \cite{MG11_GJ}.

\section{LISA sources}
\label{sec:LISAsources}

\subsection{Extreme Mass Ratio Inspirals}
Astronomical observations indicate that many galaxies contain 
a supermassive black hole (SMBH) in its nuclei 
\cite{Schodel2002, Padovani2004}
with masses in the range $\sim 10^{5} - 10^{10} M_{\odot}$ 
\cite{JaffeBacker2003, greene-2007}.
Encounters between bodies in the surrounding star cluster can perturb
the orbit of one of those bodies so that its periapse becomes close
to the SMBH. 
The body will radiate energy in the form of gravitational
waves and will become bound to the central SMBH. 
If that body happens to be a compact object such as a white dwarf, 
neutron star or black hole, it will withstand the tidal forces exerted upon 
it and will inspiral into the central SMBH. 
These events are called {\it extreme mass ratio inspirals} (EMRIs).
The inspiral of compact objects into a SMBH of mass 
$\sim 10^{5} - 10^{7} M_{\odot}$
will emit gravitational waves that we expect to observe with LISA
during the final few years before plunge.
For a discussion of EMRIs with regard to data analysis, tests of General Relativity
and astrophysics see Amaro-Seoane et al. (2007) \cite{amaroseoane-2007}.

The rate at which these extreme mass ratio inspiral (EMRI) events occur in the 
Universe is highly uncertain, but is likely to be at most only a few times per 
year in each cubic Gpc of space, see Freitag (2001) \cite{freitag01} and 
Gair et al. (2004) \cite{jon04} (see particularly Table 2). 
LISA EMRI events are thus unlikely to be closer than $~1$Gpc, at which distance 
the typical instantaneous strain amplitude is $h\sim 10^{-22}$ (from Eq.~(1) of
Wen and Gair (2005) \cite{wengair05} which is similar to Eq.~(\ref{strainvalue}) 
of this thesis with the inclusion of the reduced mass $\mu$  in order to take 
into account the extreme mass asymmetry of these systems). 
Comparing this value to the characteristic noise strain in the LISA detector of 
$\sim 5 \times 10^{-21}$ at the floor of the noise curve near 5 mHz 
(see Cutler (1998) \cite{curt98}, Barack and Cutler (2004) \cite{leor04}
and Fig.~\ref{PSDcurves_all}) we can see that the {\it instantaneous} 
(rather than accumulated coherent) SNR will be no more than $\sim 0.1$.   

\subsection{Merger of supermassive black holes}
LISA should also detect $\sim$ 1-10 signals per year (Sesana et al. (2005) 
\cite{sesana05}, see particularly Fig.~5) 
from the inspiral and merger of supermassive black hole binaries (SMBHBs) 
of appropriate mass ($\sim10^5 M_{\odot}$ - $10^7 M_{\odot}$).
These events will occur during the merger of the host galaxies of the
supermassive black holes and will be visible out to very high redshifts 
appearing in the LISA data stream with very high signal-to-noise ratio.

\subsection{Inspiral of white dwarfs}
We expect to detect gravitational waves from many millions of compact binaries 
(composed of white dwarfs (WDs) or neutron stars (NSs)) in the nearby Universe.
The orbital shrinkage of these binaries is slow and they generate essentially 
monochromatic gravitational wave signals (modulo modulation caused by the motion 
of LISA).
At low frequencies the huge number of these binaries will form a confusion 
foreground, but at higher frequencies 
we hope to individually resolve several thousands of these binaries 
(Danzmann et al. (1998) \cite{LISAppa}, see Fig 1.3 and discussion). 

\section{Data analysis and detection schemes}
\label{sec:LISADA}
In this Section we will briefly summarise different methods for the analysis
of LISA data in order to detect EMRI signals. 
We shall see that due to the complexity and duration of these sources
matched-filter based analysis will be very computationally expensive (in some
cases unfeasible) and we will suggest and subsequently develop a time-frequency
based approach.

\subsection{Matched-filtering for EMRIs}
EMRI waveforms will be detectable for several years before plunge, which makes
detection possible by building up the signal-to-noise ratio over many waveform 
cycles using matched filtering as discussed for the inspiral of stellar mass 
compact objects previously.
An EMRI waveform depends on $17$ parameters (although several of these are not 
important for determining the waveform phasing
and we can neglect the spin of the lower mass component) 
and LISA will detect up to 
$\sim10^5$ cycles of the waveform prior to plunge. 
Estimating that one template might be required per cycle in each parameter, 
and $\sim 6$ important (intrinsic) parameters, gives an estimate of 
$10^{30}$ templates required for the simplest case of a search for a single EMRI 
embedded in pure Gaussian noise, this is far more templates than the few
$\times 10^{3} - 10^{4} $ required in the search for stellar mass spinning
systems, see Fig.~\ref{fig:FOMSize_VersusGPS_tmpltbank}.
This is far more than can be searched in a reasonable computational time
Gair et al. \cite{jon04}.

\subsubsection{Markov Chain Monte Carlo techniques}
As well as the large number of parameters required to describe an EMRI, the
analysis of LISA data is further complicated by the fact that it is signal 
dominated, i.e., at any moment the data stream includes not only instrumental 
noise but thousands of signals of different types which overlap in time and 
frequency. 
The optimal matched-filter template should, therefore, be a superposition of all 
the signals that are present. 
Techniques exist to construct such a global matched filter iteratively, 
such as Markov Chain Monte Carlo (MCMC) methods, and are currently being 
investigated in the context of LISA 
(Cornish and Crowder (2005) \cite{cornish05}, 
Umst\"{a}tter et al. (2005) \cite{umstatter05},
Wickham et al. (2006) \cite{wickham06},
Cornish and Porter (2007) \cite{cornishporter07}),
including for characterisation of LISA EMRIs \cite{stroeer06emri}. 

However, even when performed efficiently the MCMC approach still requires the
matched-filtering of $\sim 10^7$ templates which need to be either generated on 
the fly or looked up in a template bank.
For EMRIs, the computational cost of either approach may be prohibitively high, 
unless some advance estimate has been made of the parameters of the signals 
present in the data. 
To devise such parameter estimation techniques, it is reasonable to first 
consider the problem of detecting a single source in noisy data, before using 
and adapting the methods to the case of multiple sources. 
It is this second problem, searching for a single source while ignoring source 
confusion, that work on EMRI searches has concentrated on so far.

\subsubsection{Semi-coherent matched-filtering for EMRIs}
Another possibility for the detection of EMRIs in LISA data is a 
semi-coherent approach. 
Rather than search for the full waveform (which may last the majority of
LISAs run) we first perform a coherent matched-filter search for $\sim2 - 3$ 
week segments of EMRI waveforms. 
Subsequently the power in each segment is (incoherently) summed 
(see e.g., Gair et al. (2004) \cite{jon04} Sec.~3 for a description of this
technique).
Assuming reasonable computational resources, this technique could detect 
individual EMRI events out to a redshift $z \approx 1$ 
(Gair et al. (2004) \cite{jon04}), which would mean as many as several 
hundred EMRI detections over the duration of the LISA mission, although this
result is clearly dependent on the intrinsic astrophysical rate of EMRI events. 
The semi-coherent method, although computationally feasible, makes heavy use 
of computing resources. 

However, the high potential event rate suggests that it might be possible to 
detect the loudest several EMRI events using much simpler, template-free 
time-frequency techniques, at a tiny fraction of the computational cost.

\section{Time-frequency techniques}
A promising technique for the detection of EMRIs, and other types of LISA 
sources, is a time-frequency analysis.
We divide the full LISA data set into segments of shorter duration
($\sim2 - 3$ weeks) and construct a Fourier spectrum of each, hence creating a 
time-frequency spectrogram. 
We then search this time-frequency map for features. 
The simplest possible time-frequency algorithm is an Excess Power search, 
where we search the time-frequency map of our data for unusually bright pixels.
While Excess Power searches perform poorly when applied to the basic 
time-frequency map, if the pixels of the time-frequency map are binned first
the Excess Power method is able to detect typical EMRI events at distances of 
up to $\sim 2.25$Gpc (Wen and Gair (2005) \cite{wengair05} and Gair and Wen 
(2005) \cite{gairwen05}) which is about half the distance of the semi-coherent 
search (Gair et al. (2004) \cite{jon04}). 
The disadvantage of the Excess Power method is that it does not by itself
provide much information about the source parameters, but merely indicates 
that a source is present in the data. 
A follow up analysis must therefore be used to extract information about 
events identified by the Excess Power search (Wen et al. (2006) \cite{wen06}).

In this analysis we consider a somewhat more sophisticated time-frequency 
algorithm, the Hierarchical Algorithm for Clusters and Ridges (HACR) 
(Heng et al. (2004) \cite{heng04}). 
This method involves first identifying unusually bright pixels in the 
time-frequency map, then constructing a cluster of bright pixels around it, 
before finally using the number of pixels in the cluster as a threshold to 
distinguish signals from noise events. 
The properties of the HACR clusters encode information about the source, 
and thus in a single analysis HACR allows both detection and parameter 
estimation.

The HACR search encompasses the Excess Power search as a subset (with the pixel 
threshold set to 1), which will allow us to compare HACR's performance to the 
performance of the Excess Power algorithm in this analysis.
We have found that when HACR is applied to the unbinned spectrogram, it 
performs poorly, but if the spectrogram is first binned via the same 
technique used for the Excess Power search (Wen and Gair (2005) 
\cite{wengair05} and Gair and Wen (2005) \cite{gairwen05}), 
we find that HACR outperforms the Excess Power search, as we would expect. 
HACR is able to detect typical EMRI events at distances of $\sim 2.6$Gpc, 
which is a little further than the Excess Power technique. 
However, the HACR clusters associated with detection events tend to have 
several hundred pixels, and thus encode a significant amount of information 
about the source. 
The HACR search can be tuned to be sensitive to a specific source at a specific 
distance, or to a specific source at an unknown distance, or to an unknown 
source at an unknown distance. 
While the detection performance for a specific source does depend on how the 
HACR thresholds are tuned, we find that the variation of detection rate is not 
huge and so a single HACR search could be used to detect multiple types of 
events in a search of the LISA data. 

\section{The LISA data set}
The LISA detector was described in Sec.~\ref{subsec:LISA}.
The main source of noise in LISA is random variations in the frequency of the
laser it uses to measure the change in (proper) distance between the spacecraft.
However, this laser frequency noise can be suppressed without eradicating the 
gravitational wave signal through use of Time Delay Interferometry (TDI, see 
for instance Vallisneri (2005) \cite{vallis05} and references therein).
At high frequencies, there are three independent TDI channels in which the 
noise is uncorrelated which are typically denoted $A$, $E$ and $T$. 
At low frequencies, there are essentially only two independent data channels
since LISA can be regarded as a superposition of two static  
Michelson ($90^{\circ}$) interferometers at $45^{\circ}$ to each other 
over the relevant timescales. 
These two low-frequency response functions, denoted $h_{I}$ and $h_{II}$, 
are defined in Cutler (1998) \cite{curt98}. 
In this analysis we treat the LISA data stream as consisting of only these
two channels, since our sources are at comparatively low frequencies, 
and the responses of the two Michelson interferometers are quick and easy to compute. 
While not a totally accurate representation of LISA, this approach 
incorporates the modulations due to the detector motion in a reasonable way 
and so is sufficient for the qualitative nature of the current analysis.

\subsection{LISA's noise spectral density}
\label{sec:LISAPSD}
To characterise the search, we need to include the effects of detector noise. 
To do this, we use the noise model from Barack and Cutler (2004) \cite{leor04}
\footnote{NB The published version of this paper contains an error in the 
expression for $S_h$, which has been corrected in the preprint gr-qc/0310125. 
We use the corrected expressions here.}, 
which includes both instrumental noise and ``confusion noise'' from the 
unresolvable white dwarf binary foreground. 
The noise power spectral density is given by
\begin{eqnarray}
\label{LISAsh}
&& S_h(f) = {\rm min} \left\{S_h^{\rm{inst}}(f) \, 
\exp\left(\kappa T^{-1}_{\rm{mission}}\, dN/ df\right), 
S_h^{\rm{inst}}(f) + S_h^{\rm{gal}}(f) \right\}
\nonumber\\
& & \hspace{15 mm}  + S_h^{\rm{ex. gal}}(f) \nonumber\\
\mbox{where } \, && S_h^{\rm{inst}}(f) = 9.18\times10^{-52} \,\, 
f^{-4} + 1.59 \times 10^{-41} + 9.18 \times 10^{-38} \,\,f^2 {\rm Hz}^{-1} \nonumber\\
\mbox{and }&& S_h^{\rm{gal}}(f) = 50 \, S_h^{\rm{ex. gal}}(f) = 
2.1 \times 10^{-45} \left(\frac{f}{1\,{\rm Hz}} \right)^{-7/3} \,\, {\rm Hz}^{-1}.
\end{eqnarray}
In this, the parameter $\kappa\,T^{-1}_{\rm{mission}}$ measures how well white 
dwarfs of similar frequency can be distinguished, and we take 
$\kappa\,T^{-1}_{\rm{mission}} = 1.5/{\rm yr}$ as in Barack and Cutler (2004) 
\cite{leor04}. 
In practice, rather than adding coloured noise to the injected signal, we first 
whiten the signal using this theoretical noise prescription and then add it 
to white Gaussian noise. 
These procedures are equivalent under the assumption that the LISA data stream 
can be regarded as stationary and supposing that the noise spectral density is 
known or can be determined. 
This is likely to be a poor assumption, but a more accurate analysis is 
difficult and beyond the scope of this project.

\section{A typical EMRI}
\label{sourcemodel}
In this analysis we concentrate on the issue of detection of EMRI events and to 
do so we must consider a typical EMRI signal. 
Work by Gair et al. (2004) \cite{jon04} using a the semi-coherent search
suggested that the LISA EMRI event rate would be dominated by the inspiral 
of black holes of mass $m \sim 10M_{\odot}$ into SMBHs with mass 
$M\sim 10^6 M_{\odot}$. 
An EMRI will be detectable for the last several years of the inspiral, and 
hence could last for a significant fraction of the 
LISA mission duration ($\sim 3 - 5$ years). 
Moreover, since the stellar mass black hole will typically be captured with 
very high eccentricity and random inclination with respect to the equatorial 
plane of the SMBH, its orbit as it inspirals is likely to have some residual 
eccentricity and inclination at plunge. 

Theoretical models (Volonteri et al. (2005) \cite{volon05}) and some 
observational evidence indicate that most astrophysical black holes will 
have significant spins 
(see Table~\ref{tab:xrb} for a summary of the measured spins of stellar 
mass black hole binaries).
A super massive black hole that has accreted substantial mass via accretion of
material with constant angular momentum axis (e.g. a non-precessing disk)
would spin up to near the maximal value allowed (i.e. 0.998).
A super massive black hole formed by the merger of two objects of comparable
mass is expected to have substantial spin whereas 
a super massive black hole that has
accreted mass via capture of low-mass objects from random directions would
not accumulate significant spin (see Rees and Volonteri (2007) \cite{rees-2007}
Sec II and references therein).
Since our analysis, Brenneman and Reynolds (2006) \cite{brenneman06} 
analysed XMM-Newton (X-ray) observations of the supermassive
black hole
(M = $3-6 \times 10^{6} M_{\odot}$ from McHardy et al. (2005) 
\cite{mchardy-2005-359}) in the centre of galaxy MCG-06-30-15.
Their analysis (X-ray spectroscopy) strongly ruled out that the
black hole is non-spinning and instead infers a value 
$\chi \simeq 0.989$.

Bearing all this in mind, we choose as a ``typical'' EMRI event 
(which we shall refer to as source ``A'') the inspiral of a $10 M_{\odot}$ 
black hole into a $10^6 M_{\odot}$ SMBH with spin $a=0.8 M$. 
We assume conservatively that the LISA mission will last only three years 
($3 \times 2^{25}$s) and that the EMRI event is observed for the whole of 
the LISA mission, but plunges shortly after the end of the observation. 
This sets the initial orbital pericentre to be at $r_p \approx 11M$. 
We take the eccentricity and orbital inclination at the start of the 
observation to be $e = 0.4$ and $\iota = 45^{\circ}$ and fix the sky 
position in ecliptic coordinates to be $\cos\theta_S = 0.5$, $\phi_S = 1$. 
The orientation of the SMBH spin is chosen such that if the SMBH was at the 
Solar System Barycentre, the spin would point towards ecliptic coordinates 
$\cos(\theta_K)=-0.5$, $\phi_K=4$. 
These latter orientation angles were chosen arbitrarily, but are non-special. 

We generate the EMRI waveform using the approximate, ``kludge'', approach 
described in Babak et al. (2007) \cite{kludgepaper} and Gair and 
Glampedakis (2006) \cite{GG06}. 
These kludge waveforms are much quicker to generate than accurate 
perturbative waveforms, but capture all the main features of true EMRI 
waveforms and show remarkable faithfulness when compared to more accurate 
waveforms. 

\begin{figure}
\begin{center}
\includegraphics[angle=0, width=0.7\textwidth]{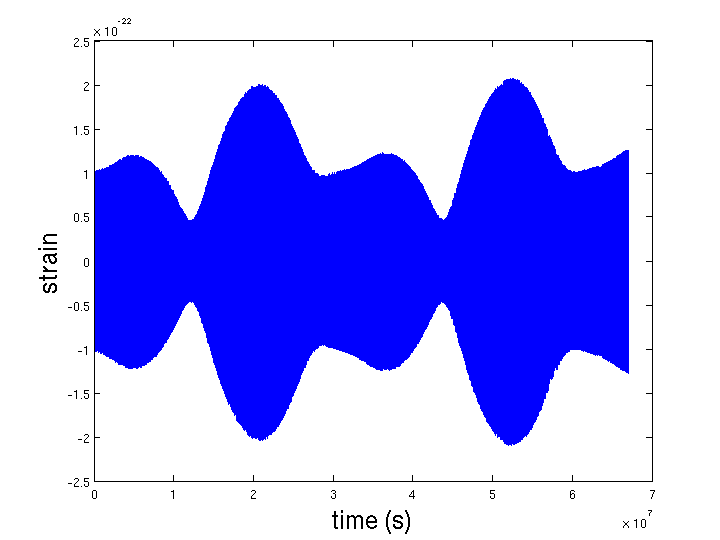}\\
\includegraphics[angle=0, width=0.7\textwidth]{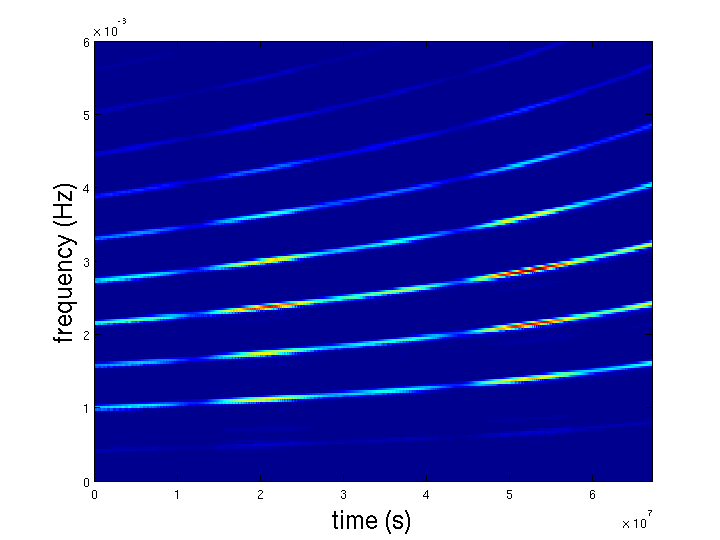}
\caption{
Time-series (upper panel) and spectrogram (lower panel) plots of EMRI ``A'' as 
it would be observed by LISA.
The amplitude modulation of the observed signal due to LISA's orbit about the sun
is clearly visible in these plots.
In the spectrogram various harmonics of the fundamental gravitational wave frequency are
observed (see caption of Fig.~\ref{EMRIKN_plots} for further details).
Note the ``chirping'' nature of the individual tracks on the spectrogram showing the
increase of gravitational wave amplitude and frequency as the system evolves.
}
\label{EMRIA_plots}
\end{center}
\end{figure}

\begin{figure}
\begin{center}
\includegraphics[angle=0, width=0.7\textwidth]{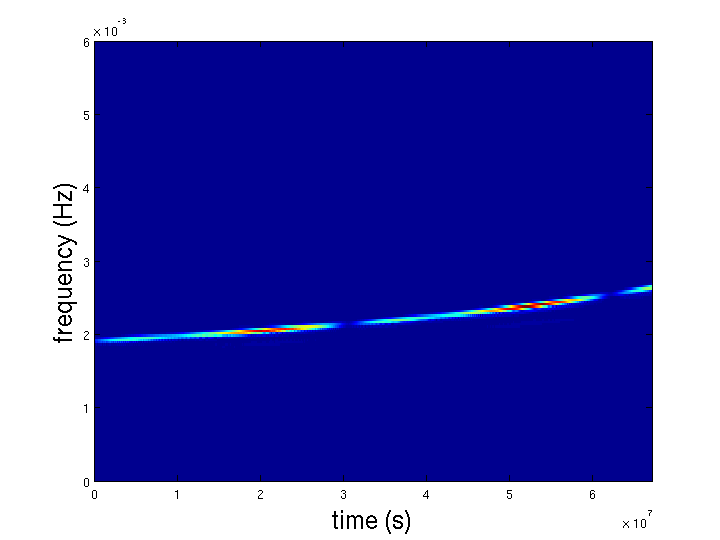}\\
\includegraphics[angle=0, width=0.7\textwidth]{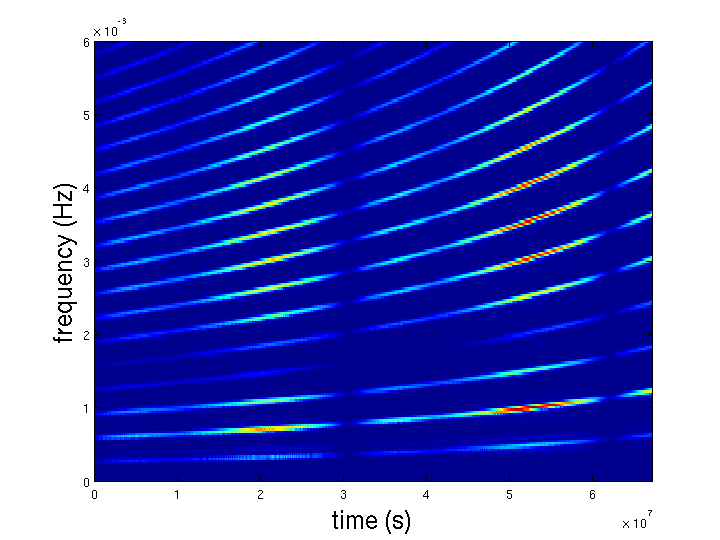}
\caption{
Spectrograms of EMRI ``K'' (upper panel) and ``N'' (lower panel) as they would be observed by LISA.
The splitting of signal power into the different harmonics of the fundamental gravitational
wave frequency is a function of the EMRI's orbital eccentricity $e$.
For eccentric orbits, like those of sources ``A'' and ``N'' most of the gravitational
radiation is emitted at the periapse of the orbit.
For more eccentric orbits, these peaks in the emission of gravitational radiation become 
more concentrated in time than for less eccentric orbits and higher harmonics in the 
frequency domain are observed (see Sec.~III of Peters and Matthews (1964) \cite{PetersMatthews}).
Indeed, more harmonics are observed for EMRI ``N'' ($e = 0.7$) than EMRI ``A'' 
($e = 0.4$, see Fig.~\ref{EMRIA_plots}) or EMRI ``K'' ($e = 0$).
Estimation of an EMRI's parameters using time-frequency representations of an observed signal is
described in Wen et al. (2006) \cite{wen06} and Gair et al. (2007) \cite{Gair:2007bz} 
(see Sec.~\ref{paramest}).
For example, the separation between the time-frequency tracks corresponding to
different harmonics can be used to estimate the system's orbital frequency near peripase.
Precession of the system's orbital plane, discussed in the previous Chapter for
stellar mass binaries, will cause splitting of each of the tracks into different sidebands.
The separation of these sidebands can be used to estimate the rate of precession of the
orbital plane and the orientation of the SMBH's spin \cite{Gair:2007bz}.    
}
\label{EMRIKN_plots}
\end{center}
\end{figure}

In addition to source ``A'', we will consider two other EMRI injections. 
These have the same parameters as ``A'', except for the initial orbital 
eccentricity, which is taken to be $e=0$ for source ``K'' and $e=0.7$ for 
source ``N''. 
The waveforms and waveform labels used are the same as those examined in the 
context of the Excess Power search in
Gair and Wen (2005) \cite{gairwen05} (see Table 1) to facilitate comparison.
Figures \ref{EMRIA_plots} and \ref{EMRIKN_plots} show how these signals can
expect to be observed by LISA.

In Section~\ref{otherperformance}, we will examine the performance of HACR in 
detecting other LISA sources, namely white dwarf binaries and SMBH mergers. 
For both of these sources, we take the waveform model, including detector 
modulations, from Cutler (1998) \cite{curt98}. 
Although more sophisticated SMBH merger models are available, the prescription 
in Cutler (1998) \cite{curt98} is sufficiently accurate for our purposes. 
The waveform model for a non-evolving white dwarf binary is very simple and has 
been well understood for many years and is summarised in 
Cutler (1998) \cite{curt98}.

\section{The Hierarchical Algorithm for Clusters and Ridges}
\label{sec:HACR}

\begin{figure}
\begin{center}
\includegraphics[angle=0, width=0.9\textwidth]{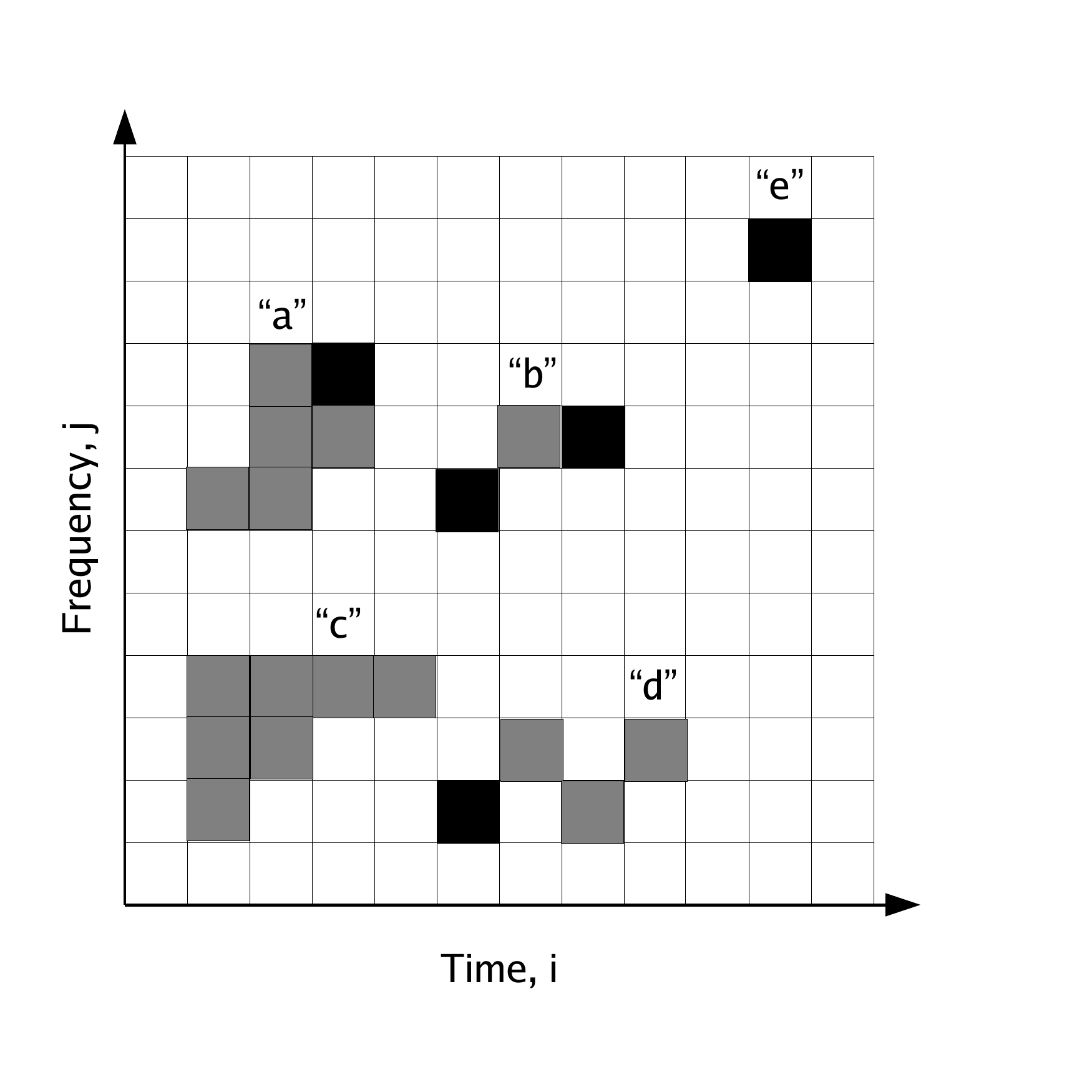}
\caption{A simple time-frequency map illustrating properties of the HACR
algorithm.
Pixels with power $P_{i,j} > \eta_{\rm{up}}$ are classified as {\em black pixels}.
Surrounding pixels with $P_{i,j} > \eta_{\rm{low}}$ are then classified as
{\em grey pixels} building a cluster around the black pixel. The
cluster is classified as an event candidate if the number of pixels it
contains, $N_{c}$, exceeds the threshold $N_{p}.$
Assuming $N_{p} < 6$ we would classify cluster ``a'' as an event candidate.
Clusters (``b'') may contain more than more black pixel (or even consist
solely of black pixels) but still require $N_{c} > N_{p}$ to be classified
an event candidate.
Clusters of any size (``c'') require at least one black pixel to be classified
an event candidate.
Pixels connected by their corners (``d'')  only still count as connected.
In the limit $N_{p} =1$ HACR will mimic a simple Excess Power search identifying
all black pixels as event candidates (e.g., ``e'').  
}
\label{fig:TFMap_simple}
\end{center}
\end{figure}

The HACR algorithm identifies clusters of pixels containing excess power in a 
time-frequency map (not necessarily a spectrogram) and represents a variation 
of the TFClusters algorithm described in Sylvestre (2002) \cite{jsylvestre}. 
In a given time-frequency map, we denote the power in a pixel as $P_{i,j}$ 
where $i$ and $j$ are the time and frequency co-ordinates of the pixel. 
HACR employs two power thresholds, $\eta_{\rm{up}} > \eta_{\rm{low}}$ and a 
threshold on the number of pixels above the power thresholds, $N_p$. 
At the first stage, the algorithm identifies all {\it black pixels} 
with $P_{i,j} > \eta_{\rm{up}}$ and all {\it grey pixels} with 
$P_{i,j} > \eta_{\rm{low}}$. 
At the second stage, HACR takes each black pixel in turn and counts all 
the grey pixels that are connected to that black pixel through a path
of touching grey pixels. 
Touching is defined as sharing an edge or corner. 
This process is repeated for each black pixel. 
To be classified as an {\it event candidate} a cluster of pixels must have
$N_{c} > N_{p}$ where $N_{c}$ is the number of pixels contained
in a particular cluster.
The algorithm is illustrated in Fig.~\ref{fig:TFMap_simple}.

There is some degeneracy between the thresholds, particularly $\eta_{\rm{low}}$
and $N_p$. 
Choosing a low value of $\eta_{\rm{low}}$ tends to make clusters larger but we
can limit the number of these clusters which become an event candidate by using a 
larger value for the pixel threshold, $N_p$. 
In our preliminary analysis, we fixed the value of $\eta_{\rm{low}}$ and tuned 
only $N_p$. 
However, tuning $\eta_{\rm{low}}$ as well can enhance the detection rate by 
$10-15\%$. 
The results shown in this analysis use tuning over both thresholds. 
The thresholds affect not only the detection rate, but also parameter 
extraction. 
Reducing $\eta_{\rm{low}}$ in order to make clusters larger 
might increase the detection rate, but it will also increase the number of 
noise pixels in each event candidate which will hamper parameter extraction. 
The optimal thresholds for the final search will ultimately come from a 
compromise between greater reach and more accurate parameter extraction. 
In a future paper, when we explore parameter estimation, we will examine 
this issue more carefully. 
In the current analysis, we look only at maximizing the detection rate.

\subsection{Investigating binning of the time-frequency maps}
It is possible to improve the performance of the search by ``binning''
the time-frequency maps. This binning procedure was the key stage in
the simple Excess Power search discussed in 
Wen and Gair (2005) \cite{wengair05} and Gair and Wen (2005) \cite{gairwen05}.

This binning procedure involves constructing an average power map using boxes 
of a particular size. The average power contained within a box is defined by
\begin{equation}
P_{i,j}^{n,l}= \frac{1}{m}\,\sum^{n-1}_{a=0} \sum^{l-1}_{b=0} P_{i+a,j+b}
\end{equation}
where $n$, $l$ are the lengths of the box edges in the time and frequency 
dimensions respectively, and $m=n\times l$ is the number of data points
enclosed. 
This average power is computed for a box aligned on {\it each} pixel in the 
(original unbinned) time-frequency map. 
Adjacent pixels in the binned time-frequency map are therefore not independent. 
In practice, for ease of computation we choose the alignment so that the pixel 
we use to label each box is in the top left hand corner of that particular box.
As in Wen and Gair (2005) \cite{wengair05}, we use only box sizes 
$(n,l)=(2^{n_t},2^{n_f})$ for all possible integer values of $n_t$ and $n_f$. 
We denote the total number of different box shapes used as $N_{\rm{box}}$.

For a given source, the box size that will do best for detection will be large 
enough to include much of the signal power but small enough to avoid too much 
contribution from noise. 
This optimum will be source specific due to the wide variation in waveforms. 
The inspiral of a $0.6 M_{\odot}$ white dwarf occurs much more slowly than that of 
a $10 M_{\odot}$ black hole, so in the first case, the optimal box size is likely 
to be longer in the time dimension. 
Gravitational waves from an inspiral of a compact object into a rapidly spinning 
black hole or from a highly eccentric inspiral orbit are characterised by many 
(often closely packed) frequency harmonics. 
In this case, a box that is wider in the frequency dimension may perform well. 
In designing a search, a balance must therefore be struck between having 
sensitivity to a range of sources and increasing the reach of the search for a 
specific source. 
We will consider this more carefully in Sec.~\ref{targsearch}.

\subsubsection{Efficient ``binning'' method}
\label{effbinning}
\begin{figure}
\begin{center}
\includegraphics[angle=0, width=0.9\textwidth]{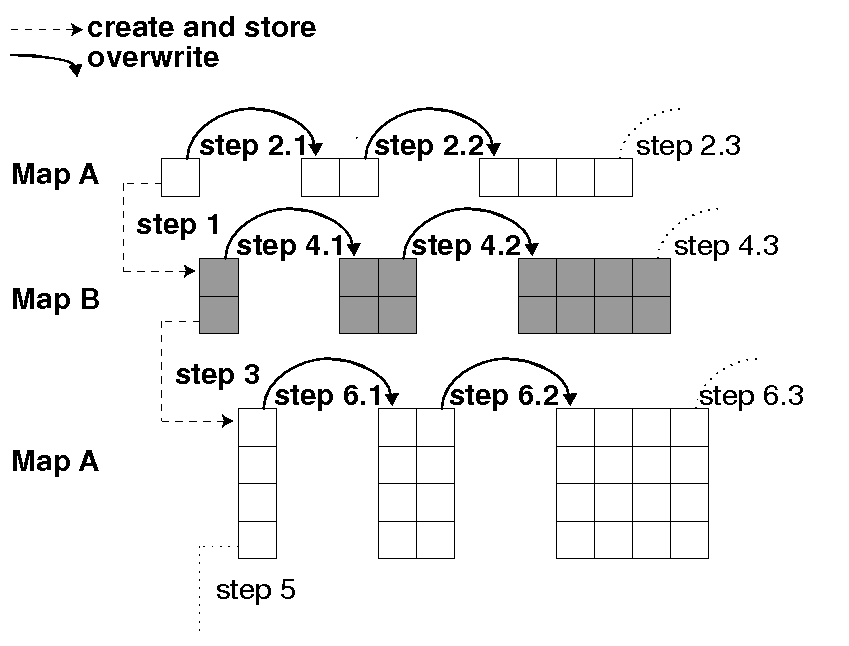}
\caption{
A schematic showing how we bin the pixels of our time-frequency map in an 
efficient manner following the algorithm described in Sec.~\ref{effbinning}.
This Figure was designed by Gair and Jones for \cite{HACR_GJ} and was drawn by
Gair.
}
\label{fig:binning}
\end{center}
\end{figure}

The binned spectrograms for each box size can be generated in a particular 
sequence that improves the efficiency and speed of the search as shown in
Figure~\ref{fig:binning}.
We first construct the unbinned $(n=1,l=1)$ map of the data and store it as 
map A.
Before analysing map A we construct the $(n=1,l=2)$ map by summing
the powers in vertically adjacent pixels and storing this as map B 
(step 1).
We then search map A using HACR before summing the power of pixels in 
horizontally adjacent pixels to construct the $(n=2,l=1)$ map, and overwrite 
map A (step 2). 
Repeating this procedure on this new map A, we construct and search all the box
sizes $(n=2^{n_t},l=1)$. 
Before analysing the $(n=1,l=2)$ map stored as map B we construct
the $(n=1,l=4)$ map and store this as map A (step 3).
Using and overwriting map B, we construct and search all the box sizes 
$(n=2^{n_t},l=2)$ (step 4).
We repeat this procedure until we have searched all possible box sizes 
up to the limit imposed by the size of our time-frequency map.

This efficient binning method requires the storage of only two time-frequency 
maps at any given time and reduces the number of floating point operations 
needed through careful recycling of maps. 
It is therefore very computationally efficient.

We set the HACR thresholds separately for each binned time-frequency map and 
label them according to the dimensions of the box they are tuned for, i.e., 
$\eta_{\rm{low}}^{n,l}$,
$\eta_{\rm{up}}^{n,l}$ and $N_{p}^{n,l}$.
A HACR detection occurs if there is an event candidate (i.e., a cluster
satisfying our thresholds) in {\it at least one} 
binned time-frequency map.

To characterise the entire search (over all box sizes) we define an overall 
false alarm probability ($OFAP$). 
This is defined as {\it the fraction of LISA missions} in which HACR would make 
{\it at least one false detection} in {\it at least one of the binned 
time-frequency maps} in the absence of any gravitational wave signals. 
Each box size that we use to analyse the data could be allowed to contribute
a different amount to $OFAP$, but to avoid prejudicing our results, we choose 
to assign an equal false alarm probability to each box size. 
We call this quantity the additional false alarm probability ($AFAP$).
To be clear, $AFAP$ is the probability of a false alarm in a time-frequency map 
with a particular box size, 
i.e., the fraction of LISA missions in which {\it that particular box size} would 
yield a false detection. 
The way in which the thresholds are computed ensures that time-frequency maps with 
each box size adds $AFAP$ to the overall false alarm rate (hence ``additional''), 
despite the fact that the binned time-frequency maps are not all independent. 
This will be described in Sec.~\ref{HACRtune}, and ensures that in 
practice $OFAP=N_{\rm{box}} \times AFAP$.

It is important to note that in the case $N_{p} =1$ then the HACR algorithm
is equivalent to the Excess Power method described in earlier papers 
Wen and Gair (2005) \cite{wengair05} and Gair and Wen (2005) \cite{gairwen05}.
A comparison between these two algorithms will be made in subsequent Sections 
of this thesis.

\subsection{Constructing spectrograms}
\label{sec:constructingspecs}
\begin{figure}
\begin{center}
\includegraphics[angle=0, width=0.9\textwidth]{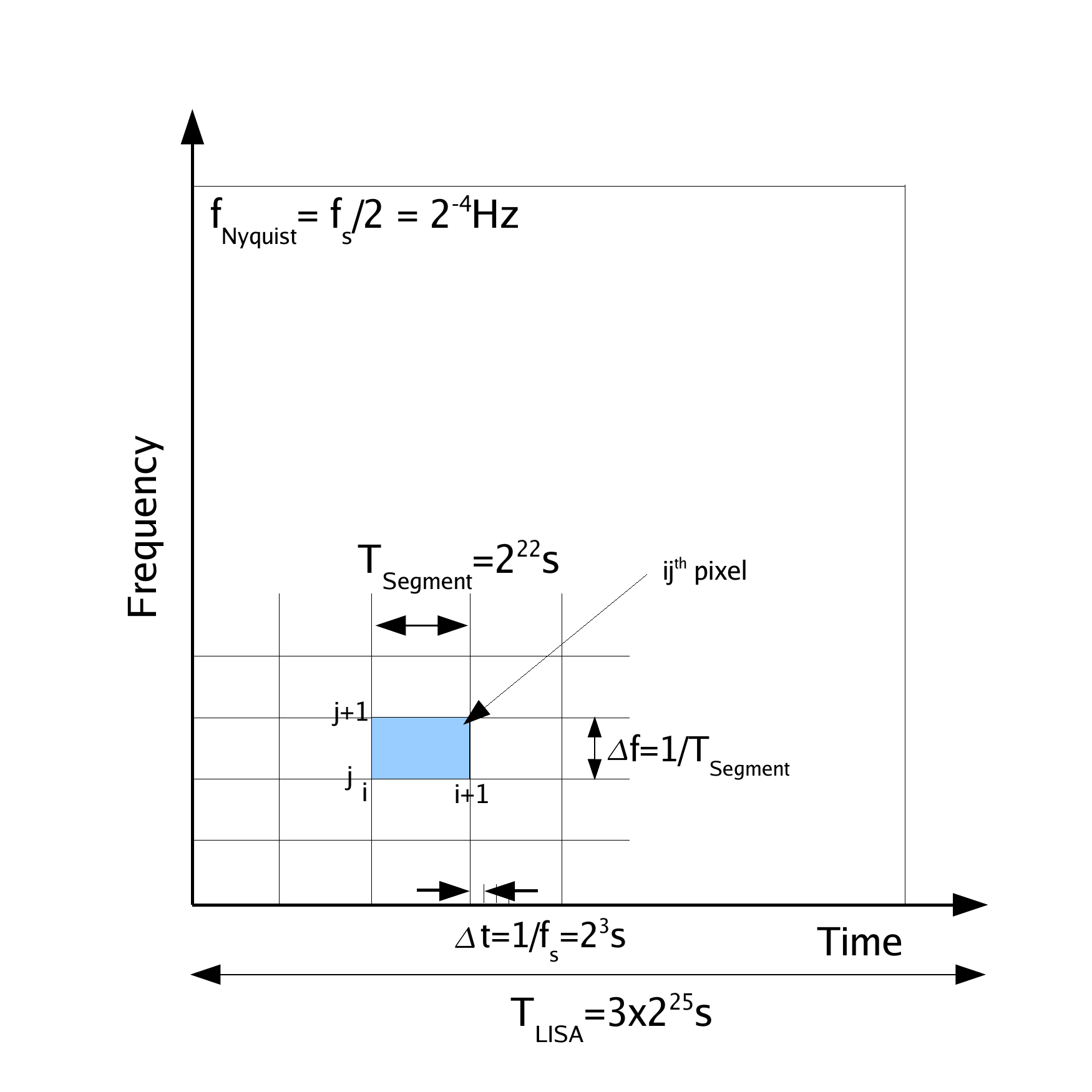}
\caption{Schematic diagram describing the construction of a spectrogram.
The LISA data set, length $T_{\rm{LISA}}$ is divided into $N_{\rm{segment}}$ 
time segments of length $T_{\rm{segment}}$.
The data stream is sampled at $f_{s} = 0.125$ Hz which corresponds to
a cadence of $\Delta t = 8$ seconds.
Each segment will contain $N_{\rm{sample}}$ samples, i.e., 
$T_{\rm{segment}} = N_{\rm{sample}} \Delta t$.
We have frequency resolution $\Delta f = 1 / T_{\rm{segment}}$ the maximum
frequency we can sample without aliasing is $f_{\rm{Nyquist}} = 0.5 f_{s}$.
We will therefore have $N_{\rm{samples}}$ frequency bins in our spectrogram.
}
\label{fig:construct_spec}
\end{center}
\end{figure}

We consider a three year LISA mission, and 
used $T_{\rm{LISA}} = 3 \times 2^{25}$ seconds of simulated LISA data sampled 
at $f_{s} = 0.125$ Hz (a cadence of $\Delta t = 8$ seconds). 
To construct the time-frequency map, we divided our time series data into
$T_{\rm{segment}} =  2^{20}$ seconds ($\sim 2$ week) time segments and an FFT was 
performed on each segment.
The frequency resolution of the spectrogram is 
$\Delta f = 1 / T_{\rm{segment}}$. 
The highest frequency we can sample without suffering low frequency aliasing
is determined by the frequency at which we sample our continuous stream of data, $f_s$.
The Shannon-Nyquist sampling theorem states that we can exactly reconstruct the original
continuous stream of data from our sampled data set as long as the original data stream is
band limited to contain only frequencies less than half the sampling frequency. This
critical frequency is known as the Nyquist frequency, $f_{\rm{Nyquist}} = 0.5 f_{s}$.
Figure \ref{fig:construct_spec} shows how the various quantities described above
define the resulting spectrogram.
The resulting time-frequency spectrograms consist of $96$ points in time and 
$65536$ points in frequency giving us $N_{\rm{box}} = 7 \times 17 = 119$ possible
box sizes of the form $n=2^{n_t}, l=2^{n_f}$ 
where $n_{t}= 0 \dots 6$ and $n_{f} = 0 \dots 16$.
Note that we do not use box size $(n=2^{6}, l=2^{16})$ since in this case our 
entire spectrogram will be represented by only a single box.

A power spectrogram was constructed separately for both LISA low-frequency 
channels, $h_{I}$ and $h_{II}$ and these were summed pixel by pixel to produce 
the time-frequency map searched by the HACR algorithm.
The power in the $ij^{th}$ pixel of the time-frequency map searched by HACR is given
by
\begin{eqnarray}
P_{i,j} = \sum_{\alpha = I,II} 
\left[ 
\frac
{2 | \tilde{h}_{\alpha, i, j} + \tilde{n}_{\alpha, i, j} |^{2}}
{\sigma^{2}_{\alpha, j}}
\right],
\end{eqnarray}
where we have written our data in the Fourier domain as 
$\tilde{x} = \tilde{h} + \tilde{n}$ 
(where $h$ is a signal and $n$ is noise)
and $\sigma^{2}$ is the expected variance of the noise component $n_{j}$.
This is given by 
\begin{eqnarray}
\sigma^{2}_{j} = 
\frac{S_{h}(f)}{2 (\Delta t)^{2} \Delta f}.
\end{eqnarray} 
In practice, the noise in the two LISA channels was taken to be Gaussian 
and white and the injected signals were whitened using the theoretical LISA 
noise curve $S_{h}(f)$ described in Sec.~\ref{sec:LISAPSD}. 
In this approach, in the absence 
of a signal the power, $P_{i,j}$, in each pixel of the unbinned spectrogram 
will be distributed as a $\chi^2$ with 4 degrees of freedom.

The division into $\sim 2$ week segments was chosen to facilitate comparison 
with the Excess Power search 
(Wen and Gair (2005) \cite{wengair05} and Gair and Wen \cite{gairwen05}), 
and it is a fairly reasonable choice for EMRIs. 
We would ideally like to ensure that the power measured in the spectrogram
corresponding to a particular gravitational wave signal (e.g., a given
harmonic of an inspiral chirp) is not split into too many small clusters.
The maximum segment length that ensures a source whose frequency is changing 
at $df/dt$ does not move by more than one frequency bin  
over the duration of the segment (i.e., $df/dt = \Delta f / T_{\rm{segment}}$) 
is $T_{\rm{segment}} = 1/\sqrt{df/dt}$. 
In the extreme mass ratio limit, the leading order post-Newtonian rate of 
change of frequency is 
\begin{eqnarray}
\label{dfdtcirc}
\frac{df}{dt} = \frac{192}{5} \frac{m}{M^{3}} \left(\frac{M}{r}\right)^{11/2} 
\end{eqnarray}
for a circular 
orbit of radius $r$ (in units with $c=G=1$) 
(use Eq.~(5.6) of Peters (1964) \cite{Peters:1964} for $dr/dt$ with Kepler's 
third law). 
For the inspiral of a $m=10M_{\odot}$ object into a $M=10^6 M_{\odot}$ this gives 
a maximum segment length of $\sim 1$ day when the orbital radius is $r=10M$. 
At that radius, the frequency would change by $\sim 10$ bins during one $2$ 
week time segment. 
This change is less rapid earlier in the inspiral and more rapid later in the 
inspiral. 
If we choose time segments that are too short, the spectrogram will be 
dominated by short timescale fluctuations in the detector noise, and the 
frequency bins will be broad, so we lose resolution in the time frequency map. 
Time segments with length $\sim 1$ week seem like a reasonable compromise. 
In the future, we plan to experiment with shorter and longer time segments. 
However, the choice of time segment length should not significantly affect our 
results thanks to the binning part of the search algorithm.

\subsection{Computational cost of a HACR search}
\label{HACRCompcost}
The computational cost of running the HACR search is very low. 
We divide the LISA data stream ($T_{\rm{LISA}}$) into $N_{\rm{segments}}$ 
time segments of length $T_{\rm{segment}}$. 
Each time segment contains $N_{\rm{samples}}$ time samples.
To FFT one time segment we perform $\sim N_{\rm{samples}} \log N_{\rm{samples}}$ 
floating point operations.
Therefore, to construct spectrograms for the full LISA data stream 
($N_{\rm{segments}}$) for both channels, $h_{I}$ and $h_{II}$, 
must perform $\sim 2 N_{\rm{segments}} N_{\rm{samples}} \log N_{\rm{samples}}$
operations.

The efficient binning algorithm ensures that only two floating point 
operations are needed to evaluate the average power for a given pixel in any 
one of the binned spectrograms (as opposed to $n\times l + 1$ operations if 
the binned spectrogram was computed directly from the unbinned map). 
Our unbinned spectrogram will have $N_{\rm{segment}}$ points in the time
and $N_{\rm{sample}}$ points in frequency.
The number of operations required to construct all the binned spectrograms is 
therefore less than 
$N_{\rm{segments}} N_{\rm{samples}} 
\log_2 N_{\rm{segments}} \log_2 N_{\rm{samples}}$ 
(less since the average power is not defined for pixels in the last $n-1$ 
columns and $l-1$ rows when using the $n \times l$ box size). 

The HACR algorithm first identifies pixels as black, grey or neither 
($N_{\rm{segments}} N_{\rm{samples}}$ operations) and then counts the number 
of pixels in each cluster. 
For a given cluster, counting the size involves $9$ operations per pixel 
($8$ comparisons to see if the 8 possible neighbours are also in the cluster 
and one addition to increment pixel count $N_p$). 
If HACR has identified $N_c$ clusters, and cluster $c_j$ has $N_p(c_j)$ pixels, 
this makes $N_c (9N_p(c_j) + 1)$ operations in total, assuming no overlap 
between the clusters. 
We do not know in advance how many clusters HACR will identify, nor how many 
pixels will be in each one.
However, we know that we cannot have more clusters than pixels 
$N_c < N_{\rm{segments}}\,N_{\rm{samples}}$. 
We also limit the number of pixels in a cluster to $N_p < 50000$ by choosing 
the minimal lower power threshold $\eta_{\rm{low}}$ 
(this will be described in the Sec.~\ref{searchchar}).

In practice, to run the HACR search with a single set of tuned thresholds on a 
spectrogram containing a single source, and with LISA and box size parameters 
as described in Sec.~\ref{sec:constructingspecs}, takes about $1$ minute on a 
$3.5$GHz workstation. 
If more sources were present, this time would be larger since more clusters 
would be identified, but $10$ minutes would be an upper limit. 
This should be compared to the cost of the semi-coherent search which 
requires $\sim 3$ years on a $50$Tflops cluster 
(Gair et al. (2004) \cite{jon04}). 
It should be noted that noise characterisation and tuning of HACR is more 
expensive, since it involves using low thresholds (thus increasing the 
number of HACR clusters identified), and repeating over many noise 
realisations. 
However, to complete $1000$ tuning runs using $40$ nodes of a typical computer 
cluster still takes only a few hours.

\section{Searching characterisation}
\label{searchchar}
Tuning HACR is a two step process. 
Firstly, simulated noisy data is analysed in order to identify triplets of our
thresholds $\eta^{n,l}_{low}$, $\eta^{n,l}_{up}$ and $N^{n,l}_{p}$ which yields
a specified false alarm probability $AFAP$ for each box size $n\times l$. 
Secondly, a stretch of simulated data containing both noise and a signal is 
analysed using these threshold triplets and the detection rate 
(or detection probability) is measured. 
For each value of false alarm probability considered we can then choose the 
threshold triplet that gives the largest detection rate. 
In this way, we obtain the optimal Receiver Operating Characteristic (ROC) 
curve for the detection of a particular source.

We will use the terms detection {\it rate} and false alarm {\it probability} 
in order to make a distinction between event candidates caused by a signal or 
by noise. 
What we are really computing as the detection {\it rate} is the detection 
probability, i.e., the fraction of LISA missions in which a particular source 
would be detected if we had many realisations of LISA. 
A more relevant observational quantity is the fraction of sources of a given 
type in the Universe that would be detected in a {\it single} LISA observation. 
The population of LISA events will have random sky positions, plunge times and 
plunge frequencies. 
They therefore sample different parts of the time-frequency spectrogram, which 
to some extent mimics averaging over noise realisations. 
The detection rate can thus be taken as a guide to the fraction of sources 
similar to the given one that would be detected in the LISA mission. 
A more accurate assessment of the fraction of sources detected requires 
injection of multiple identical sources simultaneously, but we do not consider 
that problem here.

To characterise the noise properties of the search we used $10000$ noise 
realisations and analysed them for twenty choices of $\eta_{\rm{low}}^{n,l}$ 
and with the threshold $\eta_{\rm{up}}^{n,l}$ set as low as is sensibly 
possible, recording the peak power, $P_{\rm{max}}$, and size, $N_{c}$, of 
every cluster detected. 
With such a list of clusters, it is possible during post-processing 
(discussed in Sec.~\ref{sec:postproc}) to obtain 
the number of false alarm detections that would be made using any of the 
twenty lower thresholds, $\eta_{\rm{low}}$, any value of 
$\eta_{\rm{up}} > (\eta_{\rm{up}}^{n,l})_{\rm{min}}$ and any value of $N_{p}$. 
The value of $(\eta^{n,l}_{\rm{up}})_{\rm{min}}$ has to be chosen carefully. 
If it is too low, many clusters will be found in every noise realisation, 
increasing the computational cost. 
If it is too high, too few clusters will be identified to give reasonable 
statistics. 
We used values of $(\eta^{n,l}_{\rm{up}})_{\rm{min}}$ chosen to ensure that 
a few clusters were found for each box size in each noise realisation.  
The lower threshold has to be less than or equal to 
$(\eta^{n,l}_{\rm{up}})_{\rm{min}}$. 
If it is set too low, large portions of the time-frequency map can be 
identified as a single cluster. 
Therefore, we choose the minimum value of $\eta_{\rm{low}}^{n,l}$ to 
ensure that all clusters are of reasonable size, 
which we define to be less than $50000$ pixels. 
By examining cluster properties in a few thousand noise realisations, we 
found suitable empirical choices to be
\begin{eqnarray}
\alpha^{n,l} & = & 4 + \frac{10 \sqrt{2}}{(n l)^{5/9} } \\
(\eta_{\rm{low}}^{n,l})_{\rm{min}} & = & 4 + 4 \sqrt{ \frac{10}{50000 + n l} } \\
(\eta_{\rm{up}}^{n,l})_{\rm{min}}  & = & 
\rm{max}[ \alpha^{n,l} , (\eta_{\rm{low}}^{n,l})_{\rm{min}} ].
\end{eqnarray}

We note that for large box sizes, 
$\alpha^{n,l} < (\eta^{n,l}_{\rm{low}})_{\rm{min}}$ and so we set 
$\eta_{\rm{up}}^{n,l} = \eta_{\rm{low}}^{n,l}$. 
Above this point, we no longer ensure that at least one cluster is found for 
each box size, as this is inconsistent with the more important requirement 
that no cluster exceeds $50000$ pixels. 
We emphasise that our search is robust to the somewhat arbitrary choice of 
these minimal values. 
For box sizes where $(\eta_{\rm{low}})_{\rm{min}} < (\eta_{\rm{up}})_{\rm{min}}$, 
we use 20 values of $\eta_{\rm{low}}$ spaced evenly between 
$(\eta_{\rm{low}})_{\rm{min}}$ and $(\eta_{\rm{up}})_{\rm{min}}$. 
Where $(\eta_{\rm{low}})_{\rm{min}}=(\eta_{\rm{up}})_{\rm{min}}$ we use only 
one lower threshold $\eta_{\rm{low}}=(\eta_{\rm{up}})_{\rm{min}}$. 
This comparatively small number of lower threshold choices is sufficient to 
find the maximum detection rate thanks to the degeneracy between $N_p$ 
and $\eta_{\rm{low}}$ mentioned earlier.

\subsection{Post-processing}
\label{sec:postproc}
For each box size and each lower threshold value we can consider values of 
$\eta_{\rm{up}}^{n,l}$ between $(\eta_{\rm{up}}^{n,l})_{\rm{min}}$ and the 
maximum power measured $(\eta_{\rm{up}}^{n,l})_{\rm{max}}$, 
and construct a list of all clusters with peak power 
$P_{\rm{max}} > \eta_{\rm{up}}^{n,l}$, ordered by increasing number of
pixels $N_{c}$. 

\begin{figure}
\begin{center}
\includegraphics[angle=0, width=0.7\textwidth]{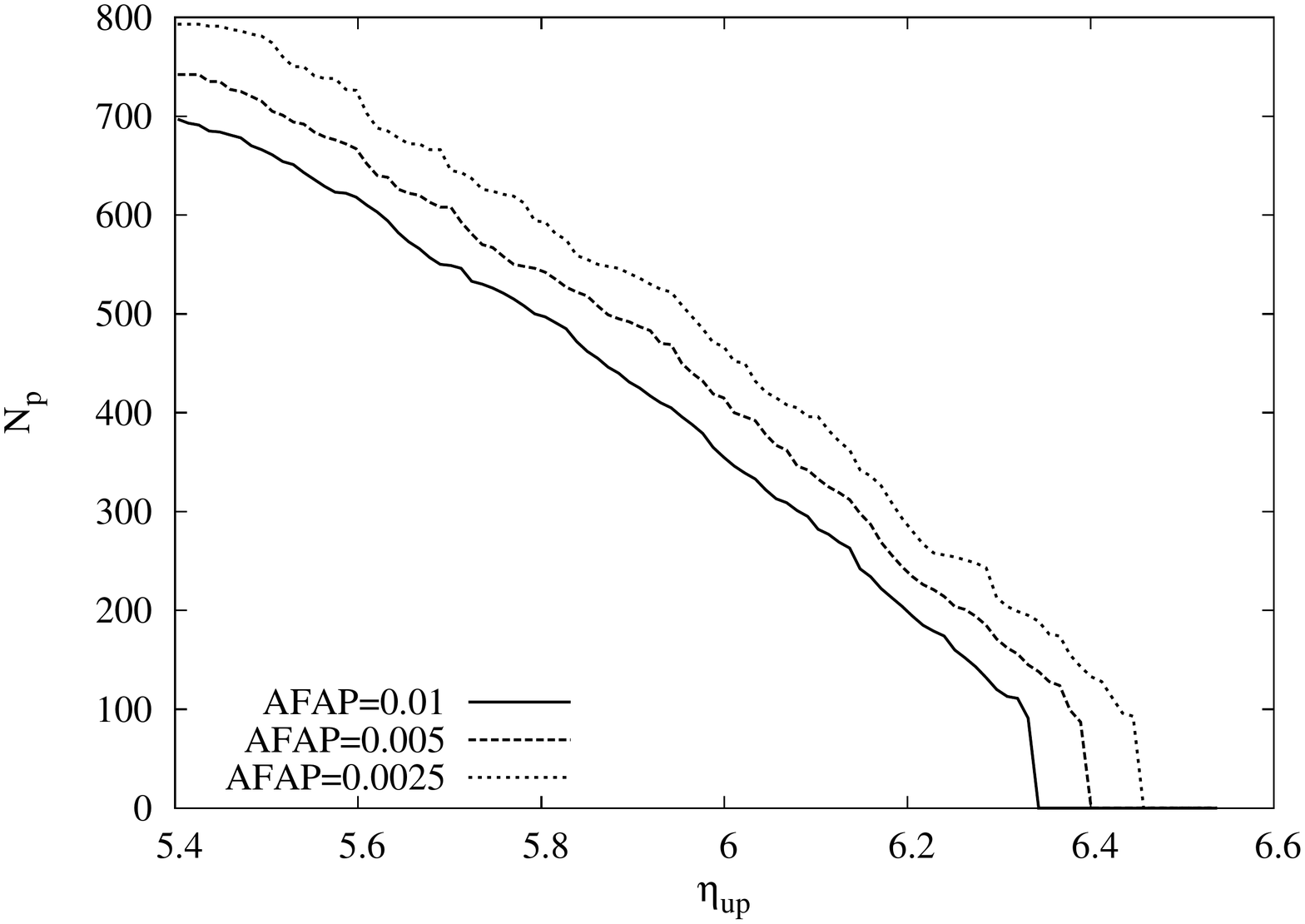}\\
\includegraphics[angle=0, width=0.7\textwidth]{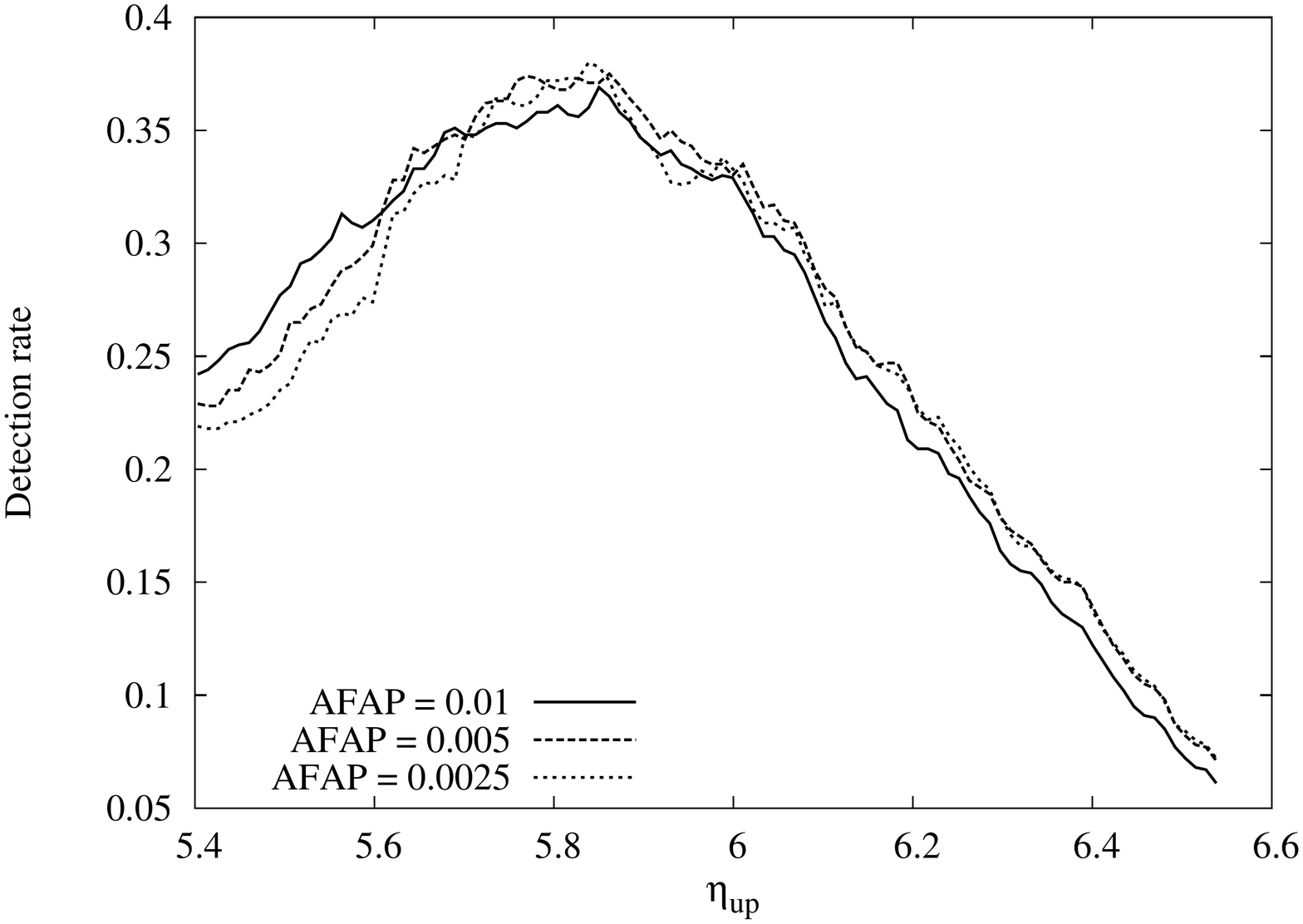}
\caption{
Upper panel: Contours of constant (additional) false alarm probability $AFAP$ 
for the box size $n=1,l=64$ and one particular lower threshold value. 
Pairs of thresholds $\eta_{\rm{up}}$ and $N_{p}$ which lie on a contour yield 
the same additional false alarm probability. 
Lower panel: Rate of signal detection plotted against choice of threshold, 
again for fixed lower threshold. 
Each point on the $x$-axis represents a set of thresholds which yield a 
particular value of $AFAP$. 
By choosing the threshold set which yields the largest value of 
detection rate, plotted on the $y$-axis, we can maximise the rate of 
signal detection for a given false alarm probability.
}
\label{threshrel}
\end{center}
\end{figure}

If we have set the false alarm rate of each box size to be $AFAP$,
we expect to see $N_{\rm{rls}} \times AFAP$  false alarms in $N_{\rm{rls}}$ 
noise realisations. 
By looking at the list of clusters, we can identify a value of the threshold 
$N_{p}^{n,l}$ with each pair of values for $\eta_{\rm{up}}^{n,l}$ and 
$\eta^{n,l}_{\rm{low}}$ that would give the correct number of false alarms in 
the realisations considered. 
If HACR was used to analyse pure noise with those three thresholds and only 
one box size $(n,l)$, it would yield a detection rate $AFAP$. 
A typical relationship between $\eta_{\rm{up}}^{n,l}$ and $N_{p}^{n,l}$ for a 
fixed choice of $\eta_{\rm{low}}^{n,l}$ is shown in the upper panel of 
Figure~\ref{threshrel}. 
This was generated for a box size $n=1$, $l=64$, the $6$th lower threshold value 
of the $20$ examined, and for three choices of $AFAP = 0.01, 0.005$ and $0.0025$.

To determine which combination of thresholds is optimal, we subsequently 
analyse spectrograms containing both noise and an injected signal. 
As mentioned earlier, since we are using white noise to generate the noise 
realisations, the signal is first whitened using the noise model described 
in Sec.~\ref{sec:LISAPSD} before injection. 
For each box size we may then select the triplet of thresholds which yields 
the largest detection rate. 
The lower panel of Fig.~\ref{threshrel} shows detection rate plotted against 
upper threshold value for EMRI source ``A'' at a distance of 2Gpc using the 
box size $n=1$, $l=64$ with $AFAP = 0.01, 0.005$ and $0.0025$ and for a fixed 
lower threshold value (the $6$th of the $20$ values used). 
Although only the $\eta_{\rm{up}}$ threshold value is shown a corresponding 
value of $N_{p}$ is inferred, determined by the assigned $AFAP$. 
This stage of the analysis will be discussed further in the next Section.

The full search uses multiple box sizes, searched in a particular order. 
We want the thresholds in a given box size to contribute an {\it additional} 
false alarm rate of $AFAP$. 
When determining the threshold triplets we therefore need to ignore 
realisations in which false alarms have already been found. 
The procedure above is thus slightly modified when considering more than one 
box size. 
If we are using $N_{\rm{rls}}$ noise realisations to determine the thresholds, 
each box size should give $N_{\rm{rls}}\times AFAP$ false alarms. 
The necessary threshold triplet can be determined for the first box size as 
described above. 
It is then possible to identify the realisations in which the false alarms 
were found for the first box size. 
This set of realisations will be somewhat different for each of the 
triplets of thresholds that give the desired $AFAP$. 
So, in practice we must do this in conjunction with the source tuning described 
in the next Section. 
This allows us to identify an {\it optimal threshold triplet} and we can find 
the noise realisations in which {\it that} threshold triplet gave false alarms. 
We then repeat the procedure described above, but now considering only clusters 
identified in the {\it remaining} realisations. 
This process is repeated for each box size in turn, ignoring in each subsequent 
box size any realisations in which false alarms have been identified in earlier 
box sizes. 
This means that the order in which the different box sizes are searched affects 
the thresholds. 
However, our results show that it does not matter in which order the box sizes 
are searched, provided the order is the same for tuning and the actual search. 
This will be discussed further in Sec.~\ref{compPerformance}.

\section{Performance of HACR in EMRI detection}

\subsection{Tuning HACR for a single specific EMRI source}
\label{HACRtune}
The fact that HACR has three thresholds allows the search to be tuned to
optimally detect a specific source at a specific distance. For a given choice
of false alarm probability, $AFAP$, we can choose triplet of thresholds for
each box size $\eta_{\rm{low}}^{n,l}$, $\eta_{\rm{up}}^{n,l}$ and $N_{p}^{n,l}$
that maximises the detection rate.
For this optimal threshold triplet, a Receiver Operating Characteristic (ROC) 
curve can be plotted for the HACR search tuned for that source. 
The ROC curve shows the detection rate as a function of the overall false 
alarm probability, $OFAP$, of the search using all box sizes.

In practice, the ROC is determined by generating a sequence of noise
realisations, injecting the whitened signal into each one, and then
constructing and searching the binned spectrograms.
A detection is defined as any realisation in which all thresholds are exceeded
in at least one box size.
The box sizes are searched in the order they were constructed (see
Fig.~\ref{fig:binning}). 
As discussed in the previous Section, if a detection has been made for one 
box size, we want to ignore that realisation when we search with subsequent 
box sizes. 
This ensures that we always choose the threshold triplet for a box size 
that provides the maximum number of {\it additional} detections. 
In practice, we achieve this goal using the following algorithm
\begin{itemize}
\item Search all realisations using the first box size, for threshold triplets 
(typically $\sim 100$ upper thresholds and $20$ lower thresholds) that all yield 
the assigned $AFAP$ (obtained through tuning of the pixel threshold).
\item Choose the threshold triplet that yields the highest detection rate. 
Identify every realisation in which {\it this optimal threshold triplet} 
gives a detection.
\item Move onto the second box size and repeat this procedure, but only
search realisations in which the optimal threshold triplet for the first
box size did {\it not} yield any detection.
\item Repeat for all other box sizes in order.
\end{itemize}
Once the optimal threshold triplets have been determined in this way, the 
detection rate must be measured by using these optimal thresholds to search 
a {\it separate} set of signal injections, to avoid biasing the rates. 
We experimented with using different numbers of injections and concluded 
that using 1000 signal injections to determine the thresholds and another 
1000 signal injections to measure the rate gave reliable answers. 
We estimate the error in the resulting ROC curve due to noise fluctuations to 
be less than $3\%$. 
All the results in this paper are computed in this way. 
To characterise the noise, we use the same set of 10000 pure noise realisations 
in all calculations.

\begin{figure}
\begin{center}
\includegraphics[angle=0, width=0.9\textwidth]{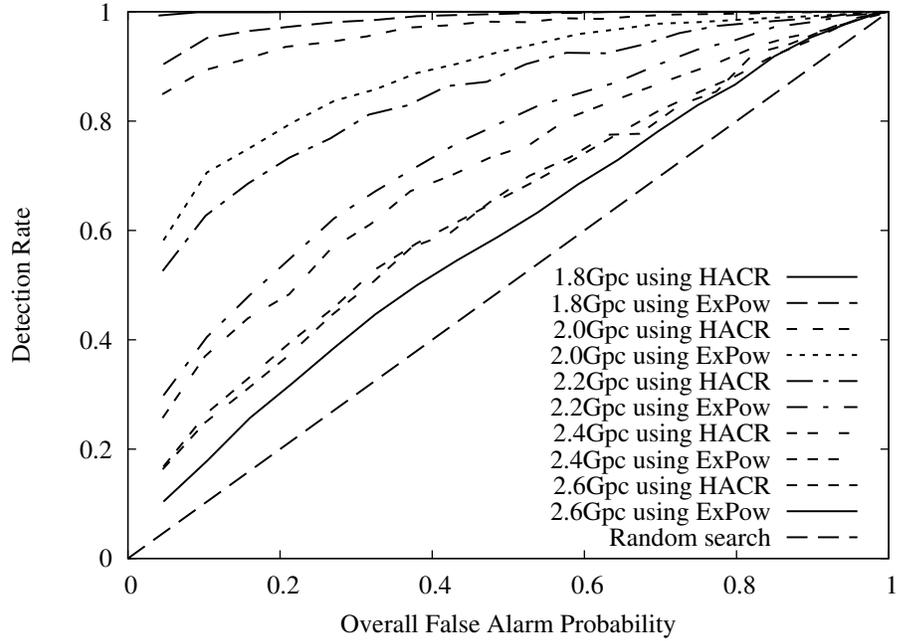}
\caption{
Receiver Operating Characteristic (ROC) curves for detection of an EMRI 
(source ``A'') at a range of distances from Earth. 
For each distance we show ROCs for HACR and the Excess Power search.
As expected HACR's performance either matches or exceeds that of the 
Excess Power search.
To aid interpretation of the ROC curve plots in this analysis, the ordering 
of the labels in the legend reflects the performance of the corresponding 
ROC curves, i.e. the second label from the top corresponds to the ROC curve
with the second best performance.
}
\label{ExPowHACRComparison}
\end{center}
\end{figure}

In Fig.~\ref{ExPowHACRComparison} we show the ROC curves for detection of
source ``A'' at a range of distances using HACR.
The random search line on this Figure represents a search for which the
detection rate and false alarm rate are equal.
This is the ``random limit'' since it is equivalent to tossing a coin and
saying that if it is heads the data stream
contains a signal and if it is tails it does not.
A search that yields a ROC curve equal to this random line is essentially
insensitive to signals.
In Fig.~\ref{ExPowHACRComparison}, we see that the source has very nearly $100\%$
detection rate for all $OFAP$s explored out to a distance of $\sim1.8$Gpc.
An overall false alarm probability of 10\% is probably quite a conservative
value,
since this is the probability that {\it in a given LISA mission} the entire
HACR search would yield just a single false alarm.
At a distance of 2Gpc, with the overall false alarm probability set to $10\%$,
HACR achieves a detection rate of $\sim 90\%$.
As the distance increases further, the detection rate further degrades, and the
source becomes undetectable at a distance of $\sim 3$Gpc.
The rate of EMRI events is somewhat uncertain, but the range for a
$10M_{\odot}$ black hole falling into a $10^6M_{\odot}$ black hole
is between $10^{-7}$ and $10^{-5}$ events per Milky Way equivalent galaxy per
year (Freitag (2001) \cite{freitag01} and Gair et al. (2004) \cite{jon04}.
Using the same extrapolation as in  Gair et al. (2001) \cite{jon04}, 
this gives $0.1 - 10$ events Gpc$^{-3}$ yr$^{-1}$.
Assuming a 3 year LISA mission, and that the detection rates quoted here are a
good approximation to the fraction of EMRI events that LISA would detect in a 
single realisation of the mission, these rates translate to a detection of
$\sim 15 - 1500$ events using this method (using a Euclidean volume-distance
relation). 
We note, however, that at the high end of this range, source confusion will 
be a significant problem and it has been ignored in the current work.

\subsubsection{Comparing the performance of HACR and the Excess Power method}
\label{compPerformance}
In Fig.~\ref{ExPowHACRComparison} we also show ROC curves for using the Excess 
Power search to detect source ``A'' at a range of distances. 
Since HACR effectively performs the Excess Power search when 
$N_{p}^{n,l} = 1$ we expect that HACR will always do
at least as well as the Excess Power search.
Due to the extra levels of tuning allowed by the HACR algorithm we find that it
can obtain a somewhat higher detection rate for a given false alarm probability.
The increase is in the range of $5-20\%$ for an $OFAP$ of $10\%$, but this 
translates to a significantly enhanced event rate.
With the source at a distance of 1.8Gpc both methods achieve very high detection 
rates; both have detection rates $>95\%$ with an $OFAP$ of $10\%$.
At intermediate distances (e.g. $\sim$2.2Gpc) HACR outperforms Excess Power 
considerably, but once the source is at 2.6Gpc, there is very little difference 
in the performance of the two searches.
However, as illustrated in Fig.~\ref{threshrel}, the optimal HACR pixel
threshold tends to be significantly greater than $1$.
Thus, HACR identifies clusters containing significant numbers of pixels, while 
the Excess Power search at the first stage identifies only individual pixels. 
Parameter extraction from the Excess Power method requires an additional track
identification stage. 
Such algorithms are currently being investigated (Wen et al. (2006)
\cite{wen06}), but HACR is more efficient, combining both stages into one.
The information contained in the structure of HACR clusters should allow
parameter estimation which can be used as input for later stages in a
hierarchical search. 
This will be discussed in more detail in Sec.~\ref{paramest}.

\section{Targeted searches}
\label{targsearch}
\begin{figure}
\begin{center}
\includegraphics[angle=0, width=0.9\textwidth]{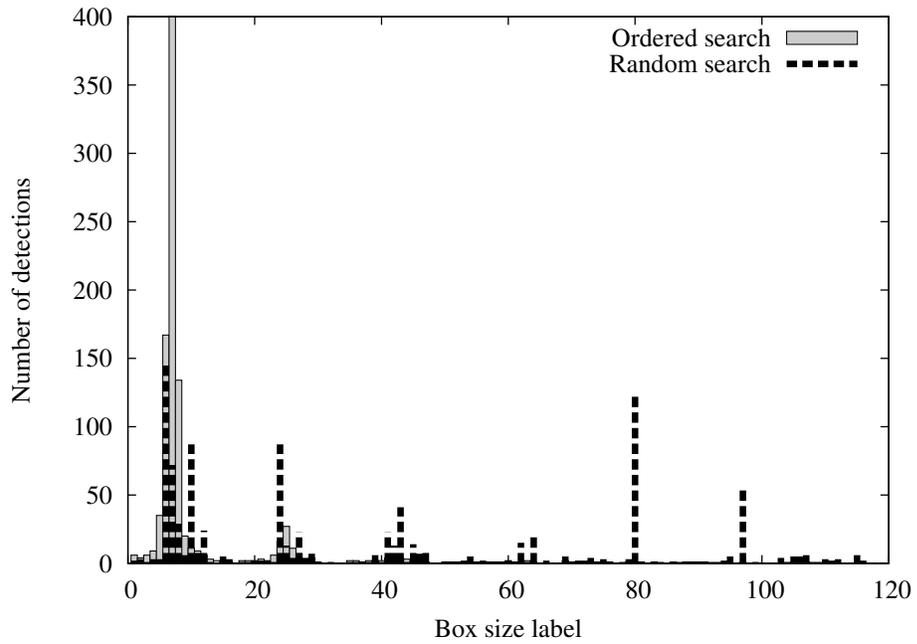}
\caption{
Number of detections as a function of box size when searching 1000 
realisations of source ``A'' at 2Gpc. 
Results are shown when using the ordered search, and when the box 
size search order is randomized. 
The $x$-axis is the box size label, which corresponds to the order in 
which the boxes are analysed in the {\it ordered} search.
}
\label{RateVsBinLab}
\end{center}
\end{figure}

In Fig.~\ref{RateVsBinLab} we show how the detection rate depends on the box
size.
This Figure shows the number of detections made for each box size over the 1000
realisations used in the
determination of the ROC curve for source ``A'' at 2Gpc.
It is clear that there is not only one single box size that makes detections, 
but several box sizes are important.
This is because random noise fluctuations will sometimes make one box size
better than another.
However, it is also clear that many of the box sizes do not make any detections
and are apparently not very useful for the detection of this particular source.
This is partially due to the box size search order.

As mentioned earlier, the fact that realisations in which detections are made
are omitted for the search of subsequent box sizes treats the earlier box 
sizes preferentially.
Fig.~\ref{RateVsBinLab} also shows the detection rate as a function of the
box size label when the search order was randomized.
Although the distribution is qualitatively similar, the box sizes that make
the detections are different in this case. 
It is clear that there are several box sizes that are equally good at detecting 
this source (these have approximately the same dimension in frequency, but
different dimensions in time).
Whichever of these equivalent box sizes is used first will make the detection.
However, we find that the overall search performance is independent of 
the box size search order and we recommend using the order specified by the 
efficient binning algorithm described in Sec.~\ref{effbinning} because of the 
computational savings.

Given that we have specified thresholds so that each box size contributes
equally to the overall false alarm probability we might expect the search 
to perform better if we restrict it to use only those few box sizes 
responsible for most of the detections of the injected signal. 
By eliminating box sizes that make few detections, we expect to reduce the 
overall false alarm probability while keeping the overall detection rate 
approximately constant, thereby improving the overall ROC performance. 

This can be investigated by re-analysing the data using only a small subset 
(i.e., $20$) of the 119 box sizes originally considered, choosing the box sizes
that were responsible for the most detections of EMRI source ``A''. 
Having performed the search using only $20$ box sizes, we can eliminate the 
box size which has the worst performance (i.e., the least
number of detections) {\it in the $20$ box search} and then repeat the
search with the remaining $19$ box sizes. 
This process can be repeated, eliminating one box size each time, until only 
one box size remains. 
The box size that contributes the fewest detections depends to a limited extent
on the (additional) false alarm probability assigned to each box size.
We used the additional false alarm probability that gave an overall search
false alarm probability of $\sim 10\%$ since, as argued earlier, this would
be a reasonable value to use in the final LISA search.

The results of this targeting procedure are summarised in 
Table~\ref{TargSearchROC}.
When the number of box sizes is reduced from $119$ to $20$, the ROC performance
does improve as the overall FAP reduces, while the detection rate remains 
largely unchanged. 
This improvement is of the order of $5\%$ in detection rate.
As the number of box sizes used is reduced further, the ROC performance remains
roughly constant until only $4$ box sizes are being used. 
Using fewer than $4$ box sizes leads to performance that degrades and is always 
worse than the full search.
This is in keeping with the understanding that several box sizes are needed for
efficient detection of a source due to the effect of noise fluctuations.
We also computed results for the Excess Power search (full and targeted), 
and these are also summarised in the same Table. 
The trend as box sizes are removed is the same and HACR always outperforms the 
Excess Power search.

We conclude that it is possible to improve the performance of the search for a
specific source by selecting fewer box sizes.
However, the improvement is not hugely significant.
This is consistent with what was found for the Excess Power search
(Gair and Wen (2005) \cite{gairwen05}).
Since the box sizes that are efficient for the detection of one particular
source will almost certainly not be the same as those
that are efficient for other sources, the best approach is to include all the
box sizes in the search.
However, since there are certain box shapes that are good for detecting certain
types of source, the box size for which a given
detection is made provides a diagnostic of the source system.
This will be discussed further in Sec.~\ref{multtune}.

\begin{table}
\caption{Detection rates for various overall false alarm probabilities when 
using the HACR or Excess Power searches with a restricted number of box sizes. }
\label{TargSearchROC}
\begin{center}
\begin{tabular}{c|c|c|c|c}
\hline
\hline
Search&\multicolumn{4}{c}{Detection rate at}\\ 
\hline
&OFAP=$5\%$&OFAP=$10\%$&OFAP=$30\%$&OFAP=$60\%$ \\ 
\hline
HACR, All bins & 84.9\% & 89.3\% & 95.5\% & 98.7\% \\ 
HACR, 20 bins  & 90.2\% & 92.9\% & 98.2\% & 99.7\% \\ 
HACR, 10 bins  & 90.5\% & 93.4\% & 98.4\% & 99.6\% \\ 
HACR, 7 bins   & 92.0\% & 94.7\% & 98.4\% & 99.4\% \\ 
HACR, 4 bins   & 92.7\% & 95.0\% & 98.5\% & 99.4\% \\ 
HACR, 1 bin    & 81.7\% & 87.5\% & 95.2\% & 99.0\% \\ 
\hline
Excess Power, All bins  & 63.8\% & 71.5\% &87.1\% & 95.4\% \\ 
Excess Power, 10 bins   & 72.6\% & 81.4\% &94.0\% & 98.2\% \\ 
Excess Power, 7 bins    & 66.0\% & 76.0\% &91.0\% & 98.1\% \\ 
Excess Power, 4 bins    & 68.7\% & 78.5\% &91.3\% & 98.4\% \\ 
Excess Power, 1 bin     & 47.8\% & 59.1\% &79.7\% & 93.8\% \\ 
\hline
\end{tabular}
\end{center}
\end{table}

\subsection{Detection of other EMRI sources}
\begin{figure}
\begin{center}
\includegraphics[angle=0, width=0.9\textwidth]{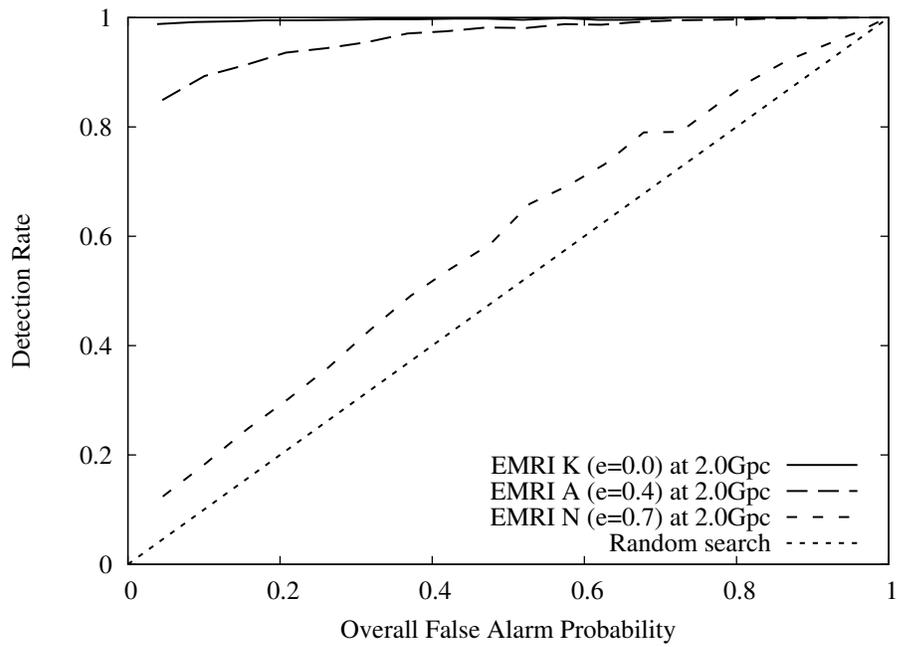}
\caption{
ROC curves for detection of EMRI sources ``A'', ``K'' and ``N'' at a distance 
of 2Gpc using HACR. 
These sources all have the same parameters except for their eccentricity.
}
\label{SigAKN_2GpcROC}
\end{center}
\end{figure}

The results described in the preceding Sections have focused on the detection
of one particular EMRI, source ``A''.
We have also explored the performance of HACR in detecting some of the other 
EMRI sources used for the investigation of the Excess Power search 
(Gair and Wen (2005) \cite{gairwen05}, see Table 1 for a summary).
Specifically we used the sources ``K'' and ``N'', which have the same
parameters as source ``A'' except for eccentricity.
The source ``K'' is initially circular, while source ``N'' has eccentricity
of $0.7$, compared to $e=0.4$ for source ``A''.
We placed these sources at a range of distances between 1.8Gpc and 2.6Gpc, and
injected them into noise realisations.
We were thus able to determine ROC curves for detection of these sources via 
the method described in Sec.~\ref{HACRtune}.
In Fig.~\ref{SigAKN_2GpcROC} we compare the ROC curves for detection of these
sources with HACR when they are at a distance of 2Gpc.
We see that our ability to detect a system at a given distance is better for
binaries in circular orbits (source ``K'')
than for systems with eccentric orbits (sources ``A'' and ``N''). 
This is consistent with what was found for the
Excess Power search in Gair and Wen (2005) \cite{gairwen05}.
The predominant effect of orbital eccentricity is to split the gravitational wave radiation
power into multiple harmonics. 
As the eccentricity increases, the frequencies of these harmonics become 
increasingly separated. 
As a consequence, a given box in the time-frequency map contains a smaller 
ratio of signal power to noise power. 
The detectability of EMRI sources therefore decreases as the eccentricity 
is increased.

The overall detectability of these new sources (``K'' and ``N'') with HACR 
follows the same pattern as the Excess Power search.
HACR has a slightly greater detection rate than Excess Power when the source is
nearby, but as the source is put further away, the performance of HACR and 
Excess Power become comparable before the random limit is reached.
However, in all cases, the HACR detection is made with a smaller upper threshold 
($\eta_{\rm{up}}^{n,l}$) than Excess Power, compensated by a larger pixel 
threshold ($N_p^{n,l}$).
Thus, HACR detections identify clusters with significant numbers of pixels, the
properties of which will be invaluable for subsequent parameter estimation.
This will be discussed in Sec.~\ref{paramest}.

\subsection{Tuning HACR for multiple EMRI sources}
\label{multEMRItune}
In the preceding Sections, we have focused on detection of a single EMRI source 
at a fixed distance.
However, in the actual analysis of LISA data, we will not know a priori what
sources will be in the data stream, and so the HACR thresholds need to be tuned 
as generally as possible.
Even in the case of a single EMRI source, the optimal threshold combination
depends to some extent on the distance at which the source is placed.
This is in contrast to the Excess Power search, where there is only one
threshold that is uniquely determined by the choice of false alarm
probability.
There are two possible approaches to constructing a general HACR search: 
1) have several separate HACR searches, targeting different sources and using 
different sets of thresholds or 
2) have a single HACR search with a set of thresholds chosen to be sensitive 
to as many LISA sources as possible.
We have focussed on the latter approach, since our results have shown that it 
is possible to do almost as well with a single set of ``generic'' thresholds 
as with source specific thresholds.

As a first step, we took the thresholds designed to optimally detect source 
``A'' at 2Gpc and used those thresholds to search for sources ``K'' and ``N''. 
We found that there was some degradation of performance, but that this was 
negligible. 
At an $OFAP$ of $10\%$, the detection rate for source ``K'' changed 
from $99.3\%$ to $99.7\%$, and that of source ``N'' changed from 
$18.4\%$ to $17.9\%$. 
This is a promising result and suggests that certain threshold combinations 
do well at detecting all the EMRI events. 
It is also possible to tune the thresholds to be generally sensitive to many 
different sources. 
This is not really necessary for the case of EMRI detection, but we will 
describe the procedure here as it will be needed when other types of sources
are included in the search (this is discussed in Sec.~\ref{otherperformance}).

We want to tune the search to maximize the total LISA event rate
(i.e., the number of events observed).
If we knew in advance which sources would be present in the LISA data, we could
tune the search by considering multiple noise realisations with that family of 
sources injected and choosing the threshold combination that gives the maximum 
total detection rate for given $OFAP$. 
Since we do not know what the actual sources in the LISA data will be, we can 
instead tune the thresholds to be as sensitive as possible to a single event 
of unknown type, using prior knowledge to weight the relative likelihood of
different types of events.
This procedure ignores issues of source confusion, but should ensure that the
loudest events are detected, no matter of what type or at what distance they 
might be.

In practice, tuning for multiple sources is done as follows:
\begin{itemize}
\item Generate realisations of noise with injected signals for each of the 
sources $s$  we want to include in the tuning.
\item For the first box size, determine the rate of detections, $R_s({\bf t_i})$, 
of each of the signals when using HACR with each threshold triplet, 
${\bf t_i}$, that yields a pre-chosen $AFAP$.
\item Construct a sum over these rates for each threshold triplet, 
$\sum w_s R_s({\bf t_i})$, using an appropriate weighting factor, 
$w_s$, for each source.
\item Choose the threshold triplet that maximizes this weighted sum. 
For each signal, identify the realisations in which that optimal threshold 
triplet gave a detection.
\item Move onto the next box size, but for each signal search only realisations 
in which the optimal thresholds for the previous box size(s) did not yield any 
detections.
\item Repeat for all box sizes.
\end{itemize}
One question is what to use for the weighting factors.
If we knew that only one type of source existed in the Universe, but it was
equally likely to be at any point in space,
we want a volume weighted average.
This is done by taking our set of sources to be a single given source placed at a
sequence of distances, $d_i$.
The source at distance $d_i$ can then be regarded to be representative of all
sources in the range $d_{i-1} < d < d_i$, and
should be weighted by the (Euclidean) volume of space in that range, $w_{i}
\propto 4\pi(d_i^3 - d_{i-1}^3)/3$.
We carried out this procedure using source ``A'' at distances of $1.8$Gpc,
$2.0$Gpc, ..., $2.6$Gpc, with
weightings $1.8^3 = 5.832$, $2.0^3 - 1.8^3 = 2.168$, $2.2^3 - 2.0^3 = 2.648$
... $2.6^3 - 2.4^3 = 3.752$
(we have neglected common factors of $4 \pi / 3$). 
We took the closest source to be at $1.8$Gpc since up to that distance, the 
detection rate is always $100\%$.
This appears to give artificial weight to the $1.8$Gpc source, but in practice 
this does not happen since virtually every threshold combination gives a $100\%$ 
detection rate for that source, and the variation in rate is determined 
primarily by the other injections.
We used distance weighted thresholds to search for source ``A'' at various 
distances. 
The thresholds did change to some extent, but these changes were small since 
the optimal thresholds are almost independent of distance, and the overall ROC 
performance was largely unaffected. 
We deduce that it is possible to detect a given EMRI source at any distance with 
a single set of thresholds.

LISA will see more than one type of source, and we can fold in prior information
about the relative abundance of different
events by adjusting the weighting factors.
We repeated the above, tuning for sources ``A'', ``K'' and ``N'' at a single distance
of $2$Gpc, and given equal weighting.
In that case too, we found that the ROC performance was not significantly
changed when tuning for these multiple sources.
We also tuned for all three sources, placed at all the
distances, $1.8$Gpc, ..., $2.6$Gpc, with the volume
weightings listed previously.
Once again, the ROC performance was not significantly altered.
Thus, there is a single set of HACR thresholds that can detect all three EMRI
sources at any distance.

These results may not be truly generic, since the three EMRI sources are quite
similar, differing only in eccentricity.
It is therefore perhaps unsurprising that a single set of thresholds can detect
all three sources almost optimally.
However, we will see in Sec.~\ref{multtune} that this result carries over to the 
case when the sources have quite different characteristics. 
This is not totally surprising, since we know that HACR includes the Excess Power 
search as the pixel threshold $N_p=1$ limit. 
The Excess Power search thresholds are independent of the tuning source at fixed 
assigned FAP. 
Thus, a HACR search tuned for a collection of sources can do no worse than the 
Excess Power search for each of those sources.
Since the HACR search does not seem to hugely outperform the Excess Power
search, we would not anticipate that this combined tuning procedure would lead 
to a serious degradation of performance even when considering very different 
classes of sources.

\section{Performance of HACR in detection of other LISA sources}
\label{otherperformance}
We have shown that HACR may be successfully tuned in order to detect multiple 
EMRI sources with different parameters. 
In this Section we investigate HACR's ability to detect other classes of 
signals, specifically white dwarf (WD) binaries and supermassive black hole 
(SMBH) binary mergers. 
We expect these other classes of signals to have quite different structures in a 
time-frequency map. 
A typical EMRI signal consists of several frequency components (due to the 
eccentricity of the orbit), which ``chirp'' slowly over the course of the 
observation, i.e., the frequency and amplitude increase. 
By contrast, the gravitational wave emission from a WD binary is essentially 
monochromatic. 
A SMBH binary inspiral also gives a chirping signal, but the chirp occurs much 
more quickly than the EMRI due to the increased mass ratio 
(see Eq.~(\ref{dfdtcirc})), so it will be characterised by a signal that is 
broader in frequency. 
This difference in structure allows HACR to be tuned for all three types of 
source simultaneously.

\subsection{A typical SMBH binary source}
As a preliminary investigation, we repeated the tuning procedure described 
earlier, injecting a typical SMBH binary inspiral and a typical WD binary at 
various distances. 
The SMBH binary waveform represented the inspiral of two $10^6 M_{\odot}$ 
non-spinning black holes, placed at a random sky position, and with merger 
occurring $\sim3$ weeks before the end of the observation. 
As mentioned in Sec.~\ref{sourcemodel}, our SMBH injections use the waveform 
model given in Cutler (1998) \cite{curt98}. 
This is a restricted post-Newtonian waveform accurate to 1.5PN. 
More accurate waveforms are available in the literature, with post-Newtonian 
corrections up to 3.5PN. 
However, the simple model captures the main features of a SMBH merger signal 
and is accurate enough for the more qualitative nature of this preliminary study. 
The quoted masses are the intrinsic masses of the black holes, 
i.e., not redshifted. 
When the source was placed at higher redshift there are two effects --- 
an increase in the luminosity distance to the source, 
and a redshifting effect --- which pushes the signal into the less 
sensitive part of the LISA noise curve. 

In Figure~\ref{SMBHWDAllDist} we show the ROC curves for detection of this 
SMBH binary source at a range of redshifts. 
At each redshift the optimal thresholds were chosen using the tuning method 
described in Section~\ref{HACRtune}. 
We find that SMBH binary sources at redshifts $z \leq 3$ are detected with 
almost perfect efficiency using HACR, but we stop being able to resolve signals 
for redshifts $z > 3.5$. 
This is primarily because the (matched-filtering, coherent) SNR of the source 
decreases significantly due to the redshifting effect mentioned above.

\subsection{A typical white dwarf binary source}
\begin{figure}
\begin{center}
\includegraphics[angle=0, width=0.8\textwidth]{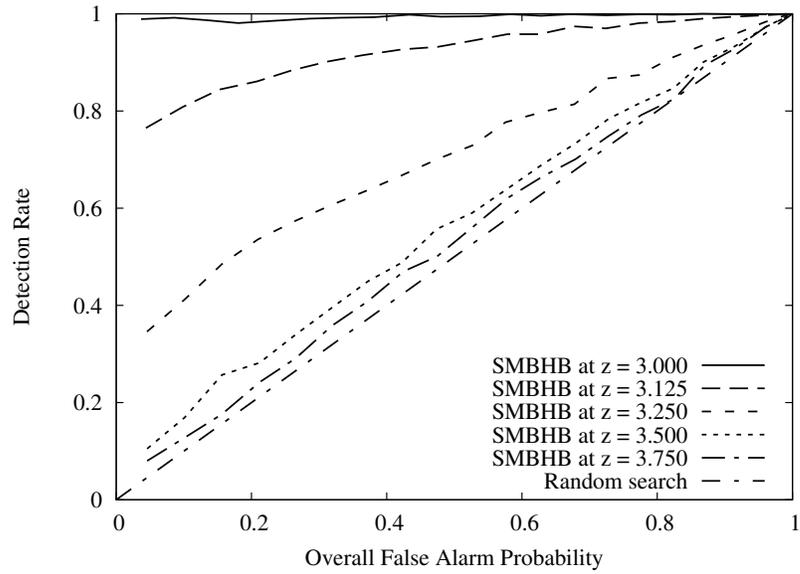}\\
\includegraphics[angle=0, width=0.8\textwidth]{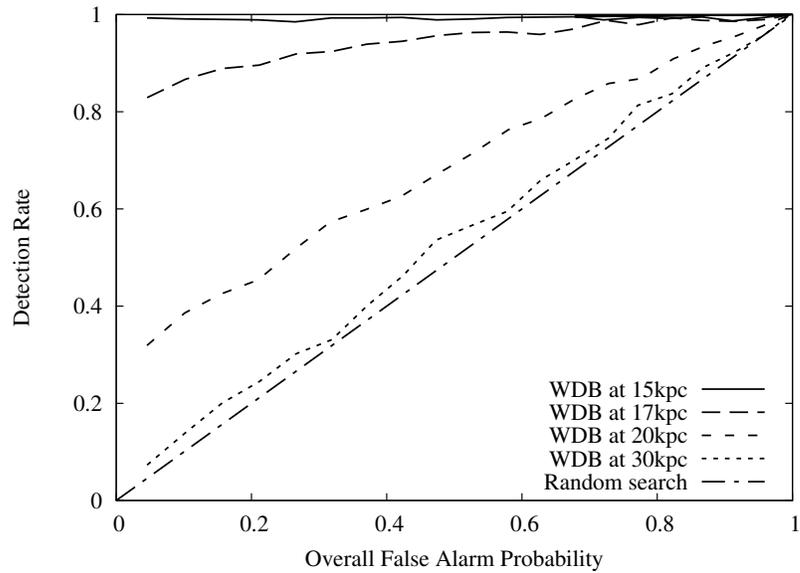}
\caption{
ROC curves for detection of a SMBH binary merger (upper panel) and a WD binary 
inspiral (lower panel) at various distances. 
The optimal thresholds for each distance were chosen using the tuning method 
described in Sec.~\ref{HACRtune}
}
\label{SMBHWDAllDist}
\end{center}
\end{figure}

The ``typical'' white dwarf binary was chosen to have the parameters of 
RXJ0806.3+1527 (one of LISA's ``verification binaries'' described in 
Stroeer and Vecchio (2006) \cite{stroeer06}), 
except for distance and sky position. 
The latter was chosen randomly, but this choice, and the noise model used 
meant the SNR of this source at a distance of $1$kpc was approximately a 
factor of $3$ greater than that quoted in Stroeer and Vecchio (2006) \cite{stroeer06}. 
This should be born in mind when considering the distances quoted in the 
following discussion. 
In Fig.~\ref{SMBHWDAllDist} we show the ROC curve for this WD source, 
injected at various distances. 
At distances $ \leq 15$kpc, we obtain near perfect detection using HACR. 
The sensitivity falls off rapidly for greater distances and the source becomes 
undetectable at greater than $\sim 20$kpc. 
Even allowing for the SNR discrepancy mentioned above, this source would be 
detectable at $\sim6$--$7$kpc, i.e., almost at the distance of the galactic 
center. 
Since this particular source is estimated to be at a distance of $300$--$1000$pc, 
it would be detectable via this method. 
We would expect to detect other similar white dwarfs at distances of $1$--$10$kpc 
depending on the source parameters. 
This does not allow for source confusion, as we have only injected single sources 
into the data stream, but the conclusion for RXJ0806 should be robust, since it 
radiates at $\sim 6$mHz, which is in the regime where WD binaries are well 
separated in frequency (this can be seen in the results of population synthesis 
models described in Nelemans et al. (2001) \cite{nelemans01} and is reflected in 
the LISA noise curve (Eq.~(\ref{LISAsh})) in which the contribution from WD binaries, 
accounting for resolvability of sources, is below the instrumental noise at $6$mHz).

\begin{figure}
\begin{center}
\includegraphics[angle=0, width=0.9\textwidth]{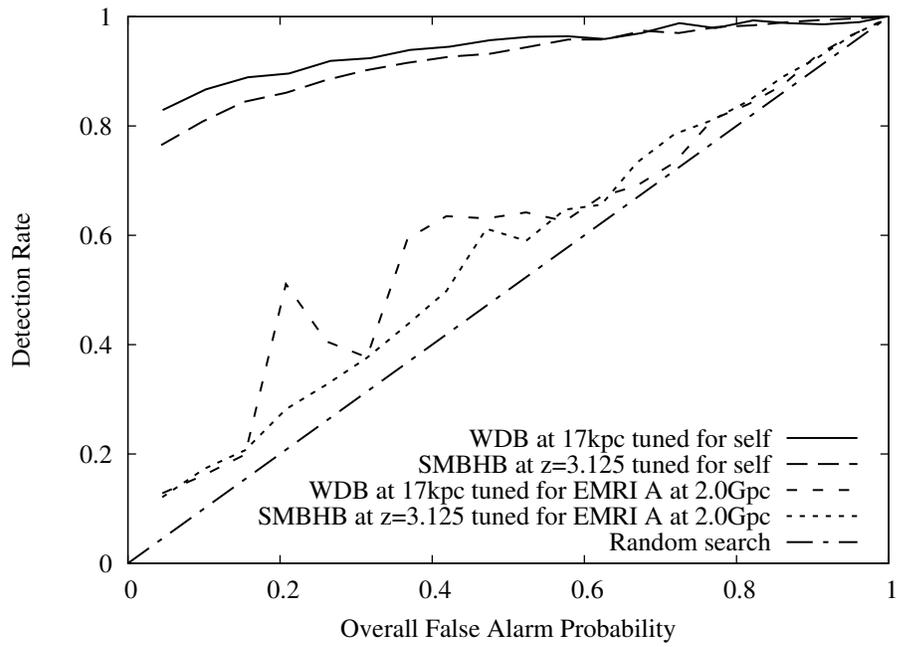}
\end{center}
\caption{ROC curves for detection of the SMBH and WD binary sources using 
thresholds tuned for EMRI source ``A'' at a distance of 2Gpc. 
For comparison, the ROC performance when the search is tuned for the 
source in question is also shown.}
\label{TestSigA2GpcThresh}
\end{figure}

In the preceding plots, the HACR thresholds have been tuned to detect the source 
in question (either an EMRI or a WD binary or SMBH merger), at a particular 
distance. 
If instead we imagined that we would use only one set of thresholds, tuned for 
EMRI source ``A'' at a distance of 2Gpc, then the ROC performance for detection 
of the SMBH binary and WD binary events is significantly degraded. 
This is shown in Fig.~\ref{TestSigA2GpcThresh}, which compares the ROC curve for 
detection of the SMBH binary at a redshift of $z = 3.125$ and the WD binary at a 
distance of $17$kpc when the EMRI thresholds are used, versus the result when 
the source specific tuned thresholds are used. 
We chose distances of $z=3.125$ and $17$kpc since in that case the sources are loud, 
but have less than a $100\%$ detection rate, so we will be able to see ROC variations. 
Figure~\ref{TestSigA2GpcThresh} shows that using the EMRI thresholds to detect other 
sources typically reduces the detection rate by a factor of $\sim 5$ at an $OFAP$ of 
$10\%$.

\subsection{Tuning HACR for multiple classes of sources}
\label{multtune}
One solution to this problem in a LISA search would be to run several 
independent searches focussed on different source families. 
However, it is also possible to tune a single set of HACR thresholds to be 
sensitive to all three types of sources simultaneously. 
This is done in the same way as the source and distance-averaged tuning 
described in Sec.~\ref{multEMRItune}, but now we inject not only EMRI 
signals, but also WD and SMBH signals.
When the thresholds are tuned using EMRI source ``A'' at 2Gpc, the WD binary 
at 17kpc and the SMBH binary at $z=3.125$ with equal weighting, the detection 
rate at an $OFAP$ of $10\%$ for the EMRI source ``A''  at 2Gpc is $87.0\%$ as 
opposed to $89.3\%$ using optimal tuning.

This difference is of the same order as the expected error in our ROC estimates 
(see Sec.~\ref{HACRtune}) and is therefore considered to be negligible.
For the SMBH binary at $z=3.125$ and the WD binary at 17kpc the change in 
detection rate when using the thresholds tuned for all three sources when compared 
to the detection rate obtained using the optimal thresholds is also negligible.
It is clear that when the thresholds are tuned for all three types of source, 
the performance of HACR is almost as high as the source specific searches, and 
still exceeds the performance of the Excess Power search. 

\begin{figure}
\begin{center}
\includegraphics[angle=0, width=0.9\textwidth]{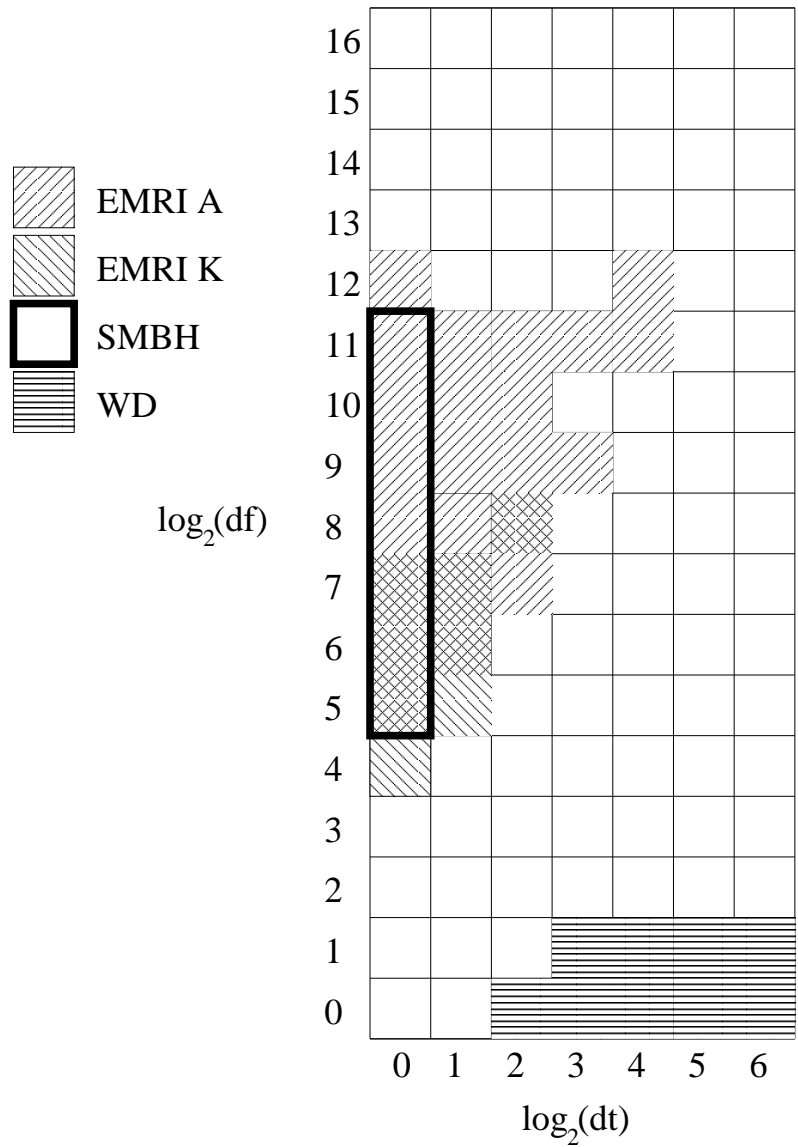}
\end{center}
\caption{
Box sizes in which the majority of detections are made for various sources. 
For each of four different sources --- EMRI ``A'' at 2Gpc, EMRI ``K'' at 2Gpc, 
the SMBH binary at z=3.125 and the WD binary at 17kpc --- we indicate all 
box sizes which were responsible for $>1\%$ of the detections of that source 
in 1000 realisations. 
The sources are identified by the patterns in the key. Box sizes that were 
good for several sources are indicated by multiple patterns, e.g. the box 
with co-ordinates (0,7).}
\label{BestBin}
\end{figure}

This is due to the different time-frequency properties of the three types of 
sources. 
The time-frequency properties of a source determine which box sizes are good 
for its detection. 
This is illustrated in Fig.~\ref{BestBin}, which shows schematically all box 
sizes that contribute more than $1\%$ of the detection rate for four different 
sources: EMRIs ``A'' and ``K'', the WD binary and the SMBH binary inspiral. 
Physically, we expect WD binary tracks to be virtually monochromatic, and of 
long duration. 
Therefore, we might expect to detect such sources in box sizes that are long 
in time but very narrow in frequency. 
The SMBH binary inspiral (at that redshift) is fairly short in duration, but 
sweeps through a reasonable range in frequency and is also quite
loud. Therefore, we might expect to see it in boxes that are narrow in time, and
broader in frequency. 

EMRIs are similar in structure to SMBH binary inspirals, but last longer in time 
and evolve more slowly. 
For a circular EMRI (e.g., source ``K''), one might expect to detect it in boxes 
that were long in time and quite narrow in frequency, although shorter in time 
and slightly broader in frequency than the WD binary 
(since the frequency changes as the source inspirals). 
However, an eccentric EMRI (e.g., source ``A'') will have multiple frequency 
harmonics, and one might expect to do better using a slightly broader box in 
frequency which then includes more of the frequency components. 

The distribution in Fig.~\ref{BestBin} fits precisely with this physical 
intuition. When tuning for multiple sources, the threshold in a given box size 
will be determined by the source that the box size is most suited to detecting. 
The fact that the various types of sources favour distinct groups of box sizes 
means the overall performance is comparable to the source specific performance. 
The box sizes in which HACR detections are made thus provide an additional way 
to classify the source type.

\section{Using HACR for parameter estimation}
\label{paramest}
We have emphasised throughout this paper that, although the HACR search does 
not provide a much greater detection rate than the Excess Power search, the 
clusters it identifies may be used to characterise the source. 
Parameters estimated from HACR clusters can then be used as input for other 
algorithms in subsequent stages of a hierarchical search of the LISA data. 

An Excess Power detection essentially contains only two pieces of information: 
the time and frequency at which the detection is made. 
Since we are using binning as part of the search, there is also some information 
contained in the box sizes used to bin the spectrograms in which the detections 
are made. 
To gain further information, a detection made by Excess Power must be followed 
by a track identification stage, and this is currently being investigated 
Wen et al. (2006) \cite{wen06}. 
In contrast, a cluster identified by HACR consists not of one but many pixels. 
Thus, in addition to the previous properties, the HACR cluster has shape 
information which is potentially a much more powerful diagnostic. 
The information that we can extract includes the size of the event in time 
and frequency and the shape and curvature of the boundary of the cluster. 
An event that is short in the time direction but broad in frequency might be an 
instrumental noise burst, whereas events long in time and narrow in frequency 
are probably inspiral events.

The difference in frequency between the latest and earliest pixels in the 
cluster divided by the difference in time provides an estimate of the 
rate of change of frequency (or chirp rate) of the event. 
In Wen et al. (2006) \cite{wen06} and
Gair et al. (2007) \cite{Gair:2007bz}
the authors show that by measuring the
evolution of the frequency $f_{n}(t)$ and its derivative $\dot{f}_{n}(t)$
we can estimate six of the EMRI's intrinsic parameters including both 
component masses, the spin of the SMBH, the orbital eccentricity and
its inclination with respect to the spin of the SMBH as well as the system's
orbital frequency. 
The power profile along an inspiral track would reveal the 
modulations associated with the motion of the detector and thus provide a 
method to find the sources sky position (although it turns out that opposite
points in the sky are degenerate). 
Figures \ref{EMRIA_plots} and \ref{EMRIKN_plots} show spectrograms of
our three EMRI signals.
The amplitude modulation caused by LISA's motion and the different harmonics 
caused by the orbit's eccentricity can be seen.

\subsubsection{Application of HACR for the Mock LISA Data Challenge}
In Gair et al. (2007) \cite{Gair:2007bz} the authors used 
HACR to detect EMRI's in simulated LISA data as part of the Mock LISA Data
Challenge (MLDC, see Arnaud et al. (2006) \cite{arnaud-2006} for an overview).  
In each of the five data sets provided (called $1.3.1$ - $1.3.5$) a single
EMRI signal was embedded in simulated LISA noise (i.e., no confusion between
different sources).
HACR performed well and identified four of the five EMRI's with clusters that
enabled the authors to estimate the sources parameters to reasonable accuracy
using the methods previously discussed.
(see Table 1 of Gair et al. (2007) \cite{Gair:2007bz}).
In the fifth case, an EMRI with relatively low SNR, only a small number of
bright pixels were identified and parameter estimation was not performed.

\subsubsection{Source confusion}
As mentioned previously, source confusion is a major issue for LISA, with many 
events likely to be overlapping in time and frequency in the data stream. 
A detection in the time-frequency plane could therefore either be a single source 
or several overlapping sources. 
An analysis of the cluster boundary should be able to distinguish these two 
cases in certain situations, i.e., distinguish a ``cross'' from a ``line''.

The shape parameters presented in Sahni et al. (1998) \cite{sahni98} may 
provide diagnostics which might be able to distinguish instrumental bursts from 
astrophysical bursts from long lived astrophysical events. 
A further use of a detected signal's power profile would be to distinguish crossing 
tracks (clusters) caused by different inspiral events. 
In a more sophisticated analysis, cluster properties would allow different 
clusters that are generated by the same event to be identified. 
An EMRI is characterised by several different frequency components and these 
might well appear as different clusters in a time-frequency analysis 
(see spectrograms in Wen and Gair (2005) \cite{wengair05}). 
However, these tracks remain almost parallel as they evolve, and so the 
rate of change of frequency provides a way to connect the tracks in a second 
stage analysis of the HACR clusters. 
If tracks can be identified like this, the properties, such as the track 
separation, encode information about the orbital eccentricity etc.

One complication in all of this is that the construction of the binned 
spectrograms makes use of bins that overlap in time and frequency. 
This has the effect of smearing out tracks from astrophysical sources and 
noise events in the data, which complicates cluster characterisation and 
parameter extraction. 
In analysing cluster properties, this effect must be accounted for, or methods 
developed to deconvolve the effect of binning once a source has been identified.

It is clear that HACR cluster properties are a potentially powerful tool both 
for vetoing, i.e., distinguishing astrophysical events from instrumental 
artifacts, and for parameter estimation. 
Work is currently underway to investigate which of these and other cluster 
properties are most powerful as diagnostics, and how the system's parameters 
may be estimated from them.

\chapter{Summary and Conclusions}
\label{ch:conclusions}

From Einstein's General Theory of Relativity we identify gravitational waves 
as perturbations to the curvature of spacetime caused by the acceleration of matter
and which propagate at the speed of light.
Gravitational waves cause a periodic strain (i.e., stretching and contraction) of the 
proper distance between points in spacetime as they pass and we describe how they can
be detected using laser-interferometers.

Binary systems will lose energy and angular momentum through the emission of gravitational waves
causing their orbits to shrink and leading to their eventual coalescence. In Chapter 2 we consider
the challenging prospect of detecting gravitational waves from the orbital decay or {\it inspiral}
of stellar mass binary systems with spinning components using the ground-based LIGO detectors.
Using approximations to the Einstein equations we are able to produce predictions of the gravitational
wave signals that would be observed from the inspiral of binaries consisting of compact objects, 
such as black holes and neutron stars.
We employ matched-filtering, a method which requires accurate predictions of the gravitational wave 
signals we expect to observe, in order to identify gravitational wave signals in the noise-dominated detector data.
The accurate predictions of the observed gravitational wave signal are our {\it templates}.

Interactions between the orbital angular momenta of the binary and the spin angular momenta of its
components will cause the binaries orbital plane to precess which in turn leads to modulation
of the amplitude and phase of the gravitational wave signal that will be observed.
Matched-filter searches using templates which do not include the effects of spin may miss the
gravitational wave signals emitted by binaries with spinning components.

Using post-Newtonian approximations to the Einstein equations we are able to produce templates for
spin-modulated gravitational wave signals that are functions of the 17 physical parameters used to describe 
a binary system with spinning components.
Unfortunately, using templates with this many parameters is very computationally expensive.
Instead, we use a {\it detection template family} (DTF) which captures the essential features of the true
gravitational wave signal but which is a function of fewer non-physical or {\it phenomenological}
parameters. 
We use the post-Newtonian approximated waveforms as a target model used to assess the
ability of our DTF to capture spin-modulated gravitational waves.

We describe the methods and results of the first dedicated search for gravitational waves emitted during
the inspiral of compact binaries with spinning component bodies.
Using the BCV2 DTF we performed a matched-filter search of $788$ hours of LIGO data collected during its
third science run (S3). 
Details of the implementation of the detection template family and calculation of the
signal to noise ratio are given in the Appendix.
No detection of gravitational wave signals was made, but by estimating our search pipeline's
sensitivity to gravitational wave signals we are able to set a Bayesian upper limit on the rate
of coalescence of stellar mass binaries.
The upper limit on the rate of coalescence for prototypical NS-BH binaries with spinning component
bodies was calculated to be $\mathcal{R}_{90\%} = 15.9 \,\mathrm{yr}^{-1}\,\mathrm{L_{10}}^{-1}$
once uncertainties had been marginalized over (see Sec.~\ref{sub:upperlimit}).
The upper limits on the rate of coalescence we calculate are around 7 orders of magnitude larger
than the rates predicted by population synthesis studies (see Sec.~\ref{sec:mergerrates}) 
and therefore do not allow us to
constrain uncertainties in these studies.  

Future searches for gravitational waves will benefit from improvements to the detectors used
to collect the data as well as the algorithms we use to analyse it with.
Data taken during LIGO's fifth science run (S5) is greatly improved in both sensitivity and
observation time (i.e., $\sim 1$ year of data with all three LIGO detectors simultaneously taking science quality
data) than previous data sets.
In 2007, during the final months of LIGO's S5 run, the French-Italian detector Virgo began taking its 
first science quality data.

Preparation for a search of LIGO S5 data for binaries with spinning components utilising
templates described by physical (rather than phenomenological) parameters \cite{Pan:2003qt, PBCV2}
is well underway.
This new template family, which we shall call the PBCV family, has two significant advantages over the BCV2 DTF.
In using physical parameters to describe the templates the PBCV family consists only of the physical
waveforms predicted by our target model and does not allow for any non-physical waveforms that can arise 
using the BCV2 DTF.
Therefore, in describing spin-modulated gravitational waves using fewer degrees of freedom than the BCV2 DTF, the
PBCV family will have a lower false alarm rate and will consequently be able to use lower detection (SNR) thresholds.
Also, since the PBCV templates are described using the physical parameters of the binary source they
are better suited to parameter estimation than the BCV2 templates. 

We found that the BCV2 DTF has good sensitivity to binary sources consisting of non-spinning, 
as well as spinning, components (see Fig.~\ref{fig:banksims}).
However, compared to dedicated searches for systems with non-spinning components \cite{LIGOS3S4all}, 
the BCV2 DTF requires a larger number of templates (see Table II of Abbott et al. \cite{LIGOS3S4all}) 
in order to capture the effects of spin and suffers from requiring a larger SNR threshold in order 
to reduce the number of triggers generated to a reasonable level.
Instead, searches of LIGO S5 data for systems with non-spinning components are likely to use 
post-Newtonian waveforms which will benefit from using templates described by physical parameters 
(see discussion above regarding the PBCV family).

In Chapter 3 we turn our attention to developing data analysis algorithms for the planned
space-based mission, LISA (Laser Interferometer Space Antenna).
LISA will be sensitive to extreme mass ratio inspirals (EMRIs) during which a stellar
mass compact object orbits and finally merges with a super massive black hole.
An EMRI waveform will depend on up to 17 parameters (similarly to the stellar mass binaries
we considered previously although in this case eccentricity cannot be neglected whereas the spin
of the smaller body can be) and will be observable throughout the duration of LISA's
operation ($\sim 3$ years).
Due to the long duration and complexity of the EMRI signals, matched-filter based searches
will be extremely computationally expensive.
We describe a less sensitive but computational cheap time-frequency based method
that can be used to quickly identify the loudest few EMRI events. 

The time-frequency method we describe combines and improves upon two previous algorithms.
The Excess Power algorithm \cite{wengair05, gairwen05} searches a time-frequency map 
(e.g., a spectrogram) for unusually bright pixels. 
This method works best when the power contained in the pixels of the time-frequency 
map are combined or {\it binned} so that a significant fraction of a gravitational wave signal's power is 
contained within a single {\it box}.
The Hierarchical Algorithm for Clusters and Ridges (HACR) \cite{heng04} is somewhat more 
sophisticated and works by identifying an unusually bright pixel and then building around it
a cluster of pixels whose power exceeds another (lower) threshold.

Our new algorithm, combines the binning stage of Excess Power with the cluster identification stage of HACR.
We call our new algorithm HACR since it is simply an extension to the existing algorithm 
of the same name.
The distance to which EMRI signals could be detected was similar for both HACR and for Excess Power. 
However, by associating a gravitational wave signal with a cluster of pixels 
rather than just one, we are able to extract more information about its source making HACR a 
potential first stage analysis in a hierarchical detection scheme. 
The estimation of parameters from time-frequency map events identified by HACR
could be used to perform targeted (and therefore less computationally expensive) matched-filter based
searches.

We are able to tune the thresholds involved in classifying a cluster of pixels as an 
event candidate in order to improve our sensitivity to particular EMRI source at a 
particular distance. 
We find that by setting different thresholds for the different boxes (created when we bin 
the power in the pixels of our time-frequency map) we are able to remain sensitive to a range
of EMRI signals whilst also being able to detect white dwarf binaries (WDBs) and the merger of
super massive black holes (SMBHs).
This is possible because EMRIs, WDBs and SMBHs occupy different
shaped regions of the time-frequency map and are therefore found by separate sets of boxes 
(see Fig.~\ref{BestBin}).

HACR was subsequently used to analyse data generated as part of the Mock LISA Data Challenge (MLDC)
and identified four of the five EMRI signals \cite{Gair:2007bz}. 
Future developments to HACR should include both refinement of the estimation of the source's parameters
and methods to deal with the issue of source confusion, the overlapping of signals in the time-frequency
plane discussed in Sec.~\ref{paramest}.

\chapter{Appendix}

\section{Miscellaneous Derivations}

\subsection{Proof of Eq.~(\ref{TijInt})}
\label{proofTijInt}

We will prove that 
\begin{eqnarray}
\int T^{ij} d^{3}x =
\frac{1}{2} \frac{d^{2}}{dt^{2}} \int x^{i} x^{j} T^{00} d^{3}x.
\end{eqnarray}
Using integration by parts:
\begin{eqnarray}
\int_{a}^{b} u(x) \frac{d v(x)}{dx} dx
= \left[ u(x) v(x) \right]_{a}^{b} - \int_{a}^{b} v(x) \frac{d u(x)}{dx} dx
\end{eqnarray}
we can write:
\begin{eqnarray}
\int
\frac{\partial}{\partial x^{k}}
\underbrace{\frac{\partial T^{kl}}{ \partial x^{l}} }_{v(x)}
\underbrace{x^{i} x^{j}}_{u(x)}
d^{3}x
&=&
\underbrace{\left[ x^{i} x^{j} \frac{\partial T^{kl}}{ \partial x^{l}} \right]}_{=0}
-
\int \frac{\partial T^{kl}}{ \partial x^{l}}
\frac{\partial}{\partial x^{k}} \left( x^{i} x^{j} \right) d^{3}x \nonumber \\
&=&
-
\int \frac{\partial T^{kl}}{ \partial x^{l}}
\left( \delta^{kj} x^{i} + \delta^{ki}  x^{j} \right) d^{3}x
\end{eqnarray}
where the integrand term in square brackets goes to zero since we
evaluate the integral over a surface far from the source
where $T^{kl} = 0$.
Integrating by parts again we can write:
\begin{eqnarray}
-
\int \frac{\partial}{\partial x^{l}} \underbrace{ T^{kl}}_{v(x)}
\underbrace{\left( \delta^{kj} x^{i} + \delta^{ki}  x^{j} \right)}_{u(x)}
d^{3}x
&=&
- \underbrace{\left[
\left( \delta^{kj} x^{i} + \delta^{ki} x^{j} \right) T^{kl}
\right]}_{=0} \nonumber \\
&+&
\int T^{kl}
\frac{\partial}{\partial x^{l}}
\left( \delta^{kj} x^{i} + \delta^{ki} x^{j} \right) d^{3}x \nonumber \\
&=&
\int T^{kl}
\left(
\delta^{li} \delta^{kj} + \delta^{lj} \delta^{ki}
\right) d^{3}x \nonumber \\
&=&
2 \int T^{ij} d^{3}x.
\end{eqnarray}

\subsection{Response of Gaussian random variables to linear transformations}
\label{gaussianresponse}

Consider a set of random variables $\xv$ where $\xv^{T}=[x_1, x_2 \dots x_N]$ 
is a row vector of random variables. We will use matrix notation for 
convenience and have used a superscript $T$ to denote the transpose.
We consider a set of random variables $\xv$  described by a multivariate 
Gaussian probability density function
\begin{eqnarray}
p_{x}(\xv) =
\frac{1}{ (2 \pi )^{N/2} |\Cv|^{1/2}}
\mathrm{exp} \left[
-\frac{1}{2} (\xv-\muv)^{T} \Cv^{-1} (\xv-\muv)
\right]
\end{eqnarray}
where $\muv$ are the means of $\xv$, i.e., 
$\muv^{T}=[\mu_1, \mu_2 \dots \mu_N]$.
$\Cv$ is the covariance matrix of the $\xv$ and $|\Cv|$ is the determinant 
of $\Cv$.  The covariance matrix $\Cv$ of $\xv$ is defined as
\begin{eqnarray}
C_{x}
=
\left( \begin{array}{cccc}
\sigma^{2}_{1}  &  \dots  &  \rho_{1,N} \sigma_{1} \sigma_{N}\\
\vdots          &  \ddots &  \vdots \\
\rho_{1,N} \sigma_{1} \sigma_{N}  & \dots  &  \sigma^2_{N} 
\end{array} \right).
\end{eqnarray}
where $\sigma^{2}_{1}$ is the variance of $x_{1}$ and $\rho_{1,2} \sigma_{1} \sigma_{2}$ 
is the covariance between $x_{1}$ and $x_{2}$.

Finally, we will be interested in the linear transform of a multivariate 
Gaussian where each random variable $x_{i}$ is described by an (independent)
Normal distribution with mean $\mu_{i} = 0$ and variance $\sigma_{i}^{2} = 1$.
 
We define a linear transform $\Lv$ such that
\begin{eqnarray}
\label{lintrans}
\yv = \Lv \xv
\end{eqnarray}
and where its inverse is given by
\begin{eqnarray}
\xv = \Gammav \yv
\end{eqnarray}
where we have $\Lv \Gammav = \Iv$ and $\Iv$ is the identity matrix.

We will now find the probability density function $p_{y}$ of the output $\xv$ of
the linear transformation Eq.~(\ref{lintrans}).
There will be a one-to-one mapping between the values of $x_{i}$ and $y_{i}$.
Following \cite{whalen:1971} (Eqs. (1.12) to (1.14)) we find that
\begin{eqnarray}
p_{y}(\yv) = |\Jv| p_{x}(\xv)
\end{eqnarray}
where $|\Jv|$ is the Jacobian determinant of the linear transformation $\Lv$.
The Jacobian of the transform $\Lv$ from $\xv$ to $\yv$ is defined as
\begin{eqnarray}
J_{xy}
= \frac
{\partial(x_1,x_2,\dots,x_N)}
{\partial(y_1,y_2,\dots,y_N)}
=
\left( \begin{array}{ccc}
\frac{\partial x_1}{\partial y_1 } & \dots  & \frac{\partial x_N}{\partial y_1 }  \\
\vdots                             & \ddots & \vdots \\
\frac{\partial x_1}{\partial y_N } & \dots  & \frac{\partial x_N}{\partial y_N }  \\
\end{array} \right).
\end{eqnarray}
The determinant of $\Jv$ is simply the determinant of the reverse transformation,
i.e., $|\Jv| = |\Gammav|$. From the standard relation $|A| = 1/|A^{-1}|$ we
find that $|\Jv| = 1/|\Lv|$.

Therefore we find the probability density function of $\Yv$ to be
\begin{eqnarray}
p_{y}(\yv) &=& |\Jv| p_{x}(\xv) \\
&=&
\frac{1}{ (2 \pi )^{N/2} |\Lv| \cdot |\Cv|^{1/2}}
\mathrm{exp} \left[
-\frac{1}{2} \xv^{T} \Cv^{-1} \xv
\right] \\
&=&
\frac{1}{ (2 \pi )^{N/2} |\Lv| |\Cv|^{1/2}}
\mathrm{exp} \left[
-\frac{1}{2} (\Gammav \yv)^{T} \Cv^{-1} (\Gammav \yv)
\right]
\end{eqnarray}
where we have used $\xv = \Gammav \yv$ to write the probability density
function in terms of $\yv$.
Following the derivation in \cite{whalen:1971} (Eqs. (4.26) to (4.28))
we define a new matrix $\Fv = \Lv \Cv \Lv^{T}$.
Using this definition and standard matrix identities we can see that
\begin{eqnarray}
(\Gammav \yv)^{T} \Cv^{-1} (\Gammav \yv) &=&
   \yv^{T} \Gammav^{T} \Cv^{-1} \Gammav \yv \\
&=&  \yv^{T} (\Lv^{T})^{-1} \Cv^{-1} \Lv^{-1} \yv \\
&=&  \yv^{T} (\Lv^{T} \Cv \Lv )^{-1} \yv \\
&=&  \yv^{T} \Fv^{-1} \yv.
\end{eqnarray}
Using $|\Av|^{T} = |\Av|$ we can write the determinant
$|\Fv| = |\Lv| |\Cv| |\Lv^{T}| = |\Lv|^{2} \cdot |\Cv|$.
Rewriting $p_{y}(\yv)$ we find that it has the form of a multivariate
Gaussian with covariance matrix $\Fv$,
\begin{eqnarray}
p_{Y}(\yv) =
\frac{1}{ (2 \pi )^{N/2} |\Fv|^{1/2}}
\mathrm{exp} \left[
-\frac{1}{2} \yv^{T} \Fv^{-1} \yv
\right].
\end{eqnarray}
We have therefore shown that the linear transformation (e.g., the matched-filtering)
of a multivariate Gaussian distribution is also a multivariate Gaussian
distribution.

\section{Construction of orthonormalised amplitude functions}
\label{app:orthonorm}

\subsection{Definitions}
The amplitude functions $\mathcal{A}_{k}(f_{\rm{cut}},\beta;f)$ to be 
orthonormalised are given below;
\begin{eqnarray}
\mathcal{A}_1(f_{\rm{cut}},\beta;f) & = & 
f^{-7/6} \theta (f_{\rm{cut}}-f), \nonumber\\
\mathcal{A}_2(f_{\rm{cut}},\beta;f) & = & 
f^{-7/6} \cos(\beta f^{-2/3}) \theta (f_{\rm{cut}}-f), \nonumber\\
\mathcal{A}_3(f_{\rm{cut}},\beta;f) & = & 
f^{-7/6} \sin(\beta f^{-2/3}) \theta (f_{\rm{cut}}-f). 
\end{eqnarray}
We shall denote the orthonormalised amplitude vectors as 
$\mathcal{\widehat{A}}_{k}(f_{\rm{cut}},\beta;f)$ 
and we shall use the Gram-Schmidt method to perform the orthonormalisation. 
The moments of the noise that will be used to abbreviate the expressions for 
the orthonormalised amplitude functions are given below;
\begin{eqnarray}
I &=& 4 \int_{f_{\rm{low}}}^{f_{\rm{cut}}} f^{-7/3} 
\frac{df} {S_{n} (f) }, \nonumber\\
J &=& 4 \int_{f_{\rm{low}}}^{f_{\rm{cut}}} f^{-7/3} 
\cos ( \beta f^{-2/3}) \frac{df} {S_{n} (f) }, \nonumber\\
K &=& 4 \int_{f_{\rm{low}}}^{f_{\rm{cut}}} f^{-7/3} 
\sin ( \beta f^{-2/3}) \frac{df} {S_{n} (f) }, \nonumber\\
L &=& 2 \int_{f_{\rm{low}}}^{f_{\rm{cut}}} f^{-7/3} 
\sin (2\beta f^{-2/3}) \frac{df} {S_{n} (f) }, \nonumber\\
M &=& 2 \int_{f_{\rm{low}}}^{f_{\rm{cut}}} f^{-7/3} 
\cos (2\beta f^{-2/3}) \frac{df} {S_{n} (f) },   
\label{app_moments}
\end{eqnarray}
where $S_{n}(f)$ is the one-sided noise power spectral density.
Throughout these derivations we shall use $||a(f)||$ to represent the inner 
product of a function $a(f)$ with itself:
\begin{equation}
||a(f)|| = \left < a(f), a(f) \right >
\end{equation}
and we shall also abbreviate our equations by writing 
$\mathcal{A}_{k}(f_{\rm{cut}},\beta;f)$ as $\mathcal{A}_{k}$ 
(and similarly for the orthonormalised functions) with no change in meaning.

\subsection{Finding $\mathcal{\widehat{A}}_1$}
To perform the transformation from $\mathcal{A}_{1}$ to 
$\mathcal{\widehat{A}}_{1}$ we use

\begin{equation} 
\label{A1hat}
\mathcal{A}_1 \to \mathcal{\widehat{A}}_1 = 
\frac {\mathcal{A}_1}{||\mathcal{A}_{1}||^{1/2}}.
\end{equation}

Finding $|| \mathcal{A}_{1}||$;
\begin{eqnarray}
||\mathcal{A}_{1}|| & = & \left < \mathcal{A}_1 , \mathcal{A}_1 \right > 
                          \nonumber\\
                    & = & 4 \int_{f_{\rm{low}}}^{f_{\rm{cut}}} \mathcal{A}_{1}^{*} (f) 
                          \mathcal{A}_{1} (f) \frac {df} {S_{n} (f)}.
\end{eqnarray}
Substituting in for $\mathcal{A}_1$, multiplying terms together 
and rewriting integrals in terms of moments of the noise;

\begin{equation} 
\label{A2hat}
|| \mathcal{A}_{1} ||
= 4 \int_{f_{\rm{low}}}^{f_{\rm{cut}}} f^{-7/3} \frac {df} {S_{n} (f)} = I.
\end{equation}
Substituting back into Eq.~(\ref{A1hat}) for $\mathcal{\widehat{A}}_1$;
\begin{equation}
\mathcal{\widehat{A}}_1 = \frac { f^{-7/6} }  { I^{1/2}}. 
\end{equation}

\subsection{Finding $\mathcal{\widehat{A}}_2$}
To perform the transformation from $\mathcal{A}_{2}$ to 
$\mathcal{\widehat{A}}_{2}$ we use;

\begin{eqnarray}
\mathcal{A}_2 \to \mathcal{\widehat{A}}_2 
& = & 
\frac {\mathcal{A}_2 - \left < \mathcal{A}_2,\mathcal{\widehat{A}}_1 \right >
      \mathcal{\widehat{A}}_1 }
      {|| \mathcal{A}_2 - 
      \left < \mathcal{A}_2,\mathcal{\widehat{A}}_1 \right >
      \mathcal{\widehat{A}}_1 || ^{1/2} }.
\end{eqnarray}

Finding $\left < \mathcal{A}_2 , \mathcal{\widehat{A}}_1 \right >$;

\begin{equation}
\left < \mathcal{A}_2 , \mathcal{\widehat{A}}_1 \right > 
=  4 \int_{f_{\rm{low}}}^{f_{\rm{cut}}} 
\mathcal{A}_{2}^{*}(f) \mathcal{\widehat{A}}_{1}(f) \frac {df} {S_{n} (f)}.
\end{equation}
Substituting in for $\mathcal{A}_2$ and $\mathcal{\widehat{A}}_1$, multiplying 
terms together and rewriting integrals in terms of moments of the noise;
\begin{eqnarray}
 \left < \mathcal{A}_2 , \mathcal{\widehat{A}}_1 \right > 
& = &  \frac{4} {I^{1/2}} \int_{f_{\rm{low}}}^{f_{\rm{cut}}} 
 f^{-7/3} \cos(\beta f^{-2/3})                                                                                  
\frac {df} {S_{n} (f)} \nonumber\\
& = & \frac{J} {I^{1/2}}.
\end{eqnarray}
Finding numerator of $\mathcal{\widehat{A}}_2$, $\mathcal{A}_2 - 
\left < \mathcal{A}_2 , \mathcal{\widehat{A}}_1 \right > 
\mathcal{\widehat{A}}_1$;
\begin{eqnarray}
\mathcal{A}_2 - \left < \mathcal{A}_2 , 
\mathcal{\widehat{A}}_1 \right > \mathcal{\widehat{A}}_1   
=  f^{-7/6} \cos(\beta f^{-2/3})                  
-  \frac{J}{I} f^{-7/6}.                                                                     \end{eqnarray}
Finding $||  \mathcal{A}_2 - \left < \mathcal{A}_2 , 
\mathcal{\widehat{A}}_1 \right > \mathcal{\widehat{A}}_1 ||$; 
\begin{eqnarray} \label{N2}
||  \mathcal{A}_2 - \left < \mathcal{A}_2 , 
\mathcal{\widehat{A}}_1 \right > \mathcal{\widehat{A}}_1 ||
& = & \left <  \mathcal{A}_2 - \left < \mathcal{A}_2 , 
\mathcal{\widehat{A}}_1 \right > \mathcal{\widehat{A}}_1 , 
 \mathcal{A}_2 - \left < \mathcal{A}_2 , \mathcal{\widehat{A}}_1 \right > 
\mathcal{\widehat{A}}_1 \right > \nonumber\\
& = &  4 \int_{f_{\rm{low}}}^{f_{\rm{cut}}} (\mathcal{A}_2 - 
\left < \mathcal{A}_2 , \mathcal{\widehat{A}}_1 \right > 
\mathcal{\widehat{A}}_1) ^{2} \frac {df} {S_n (f)}. 
\end{eqnarray}
Finding $(\mathcal{A}_2 - \left < \mathcal{A}_2 , 
\mathcal{\widehat{A}}_1 \right > \mathcal{\widehat{A}}_1) ^{2}$; 
\begin{eqnarray}
(\mathcal{A}_2 - \left < \mathcal{A}_2 , \mathcal{\widehat{A}}_1 \right > 
\mathcal{\widehat{A}}_1) ^{2}  
& = & \bigg[ f^{-7/6} \cos(\beta f^{-2/3})                  
-  f^{-7/6} \frac{J}{I} \bigg] ^{2}  \nonumber\\
& = & f^{-7/3} \cos^{2}(\beta f^{-2/3})    
+ \frac{J^{2}}{I^{2}}  f^{-7/3} \nonumber\\
& - & 2 \frac{J}{I} f^{-7/3}  \cos(\beta f^{-2/3}).
\end{eqnarray}
Substituting into Eq.~(\ref{N2}) for 
$|| \mathcal{A}_2 - \left < \mathcal{A}_2 , \mathcal{\widehat{A}}_1 \right > 
\mathcal{\widehat{A}}_1 ||$; 
\begin{eqnarray}
|| \mathcal{A}_2 &-& \left < \mathcal{A}_2 , \mathcal{\widehat{A}}_1 \right > 
\mathcal{\widehat{A}}_1 || \nonumber\\
& = &                              
4 \int_{f_{\rm{low}}}^{f_{\rm{cut}}}                            
\bigg[
 f^{-7/3} \cos^{2}(\beta f^{-2/3})    
+ \frac{J^{2}}{I^{2}}  f^{-7/3}
- 2 \frac{J}{I} f^{-7/3}  \cos(\beta f^{-2/3})
\bigg]
\frac {df} {S_n (f)}
\nonumber\\
& = & 
4 \int_{f_{\rm{low}}}^{f_{\rm{cut}}} 
f^{-7/3} \cos^{2}(\beta f^{-2/3}) \frac {df} {S_n (f)}      
+ 4 \frac{J^{2}}{I^{2}} \int_{f_{\rm{low}}}^{f_{\rm{cut}}} f^{-7/3} 
\frac {df} {S_n (f)} \nonumber\\ 
& - & 8 \frac{J}{I} \int_{f_{\rm{low}}}^{f_{\rm{cut}}}  f^{-7/3}  
\cos(\beta f^{-2/3}) \frac {df} {S_n (f)}.
\end{eqnarray}
Using $\cos^{2}(\beta f^{-2/3}) = \frac{1}{2} 
[1+\cos(2\beta f^{-2/3})]$ and rewriting integrals in terms of moments of 
the noise;
\begin{eqnarray}
|| \mathcal{A}_2 - \left < \mathcal{A}_2 , \mathcal{\widehat{A}}_1 \right > 
\mathcal{\widehat{A}}_1 || 
= M + \frac{I}{2} - \frac{J^{2}}{I}. 
\end{eqnarray}
Substituting back into Eq.~(\ref{A2hat}) for $\mathcal{\widehat{A}}_2$;
\begin{eqnarray}                                        
\mathcal{\widehat{A}}_2 = 
\frac{ f^{-7/6} \cos(\beta f^{-2/3}) - \frac{J}{I} f^{-7/6} } 
{ \bigg[ M + \frac{I}{2} - \frac{J^{2}}{I} \bigg] ^{1/2} } 
\end{eqnarray}
and simplifying;
\begin{eqnarray}
\mathcal{\widehat{A}}_2 = 
\frac{ f^{-7/6} \bigg[ \cos(\beta f^{-2/3}) - \frac{J}{I} \bigg] I^{1/2} } 
{ \bigg[ IM + \frac{I^{2}}{2} - J^{2} \bigg] ^{1/2} }. 
\label{app_A2hat}
\end{eqnarray}

\subsection{Finding $\mathcal{\widehat{A}}_3$}
To perform the transformation from $\mathcal{A}_{3}$ to 
$\mathcal{\widehat{A}}_{3}$ we use;
\begin{equation}\label{A3}
\mathcal{A}_3 \to \mathcal{\widehat{A}}_3 = 
\frac 
{\mathcal{A}_3 - \left < \mathcal{A}_3 , \mathcal{\widehat{A}}_1 \right > 
\mathcal{\widehat{A}}_1 - \left < \mathcal{A}_3 , 
\mathcal{\widehat{A}}_2 \right > \mathcal{\widehat{A}}_2 }
{|| \mathcal{A}_3 - \left < \mathcal{A}_3 , \mathcal{\widehat{A}}_1 \right > 
\mathcal{\widehat{A}}_1 - \left < \mathcal{A}_3 , 
\mathcal{\widehat{A}}_2 \right > \mathcal{\widehat{A}}_2 || }. 
\end{equation}
Finding $\left < \mathcal{A}_3 , \mathcal{\widehat{A}}_1 \right >$;
\begin{equation}
\left < \mathcal{A}_3 , \mathcal{\widehat{A}}_1 \right > =  
4 \int_{f_{\rm{low}}}^{f_{\rm{cut}}} \mathcal{A}_{3}^{*} (f) 
\mathcal{\widehat{A}}_{1} (f) \frac {df} {S_{n} (f)}
\end{equation}
Substituting in for $\mathcal{A}_3$ and $\mathcal{\widehat{A}}_1$, 
multiplying terms together and rewriting integrals in terms of moments 
of the noise;
\begin{eqnarray}
\left < \mathcal{A}_3 , \mathcal{\widehat{A}}_1 \right > 
& = & \frac{4}{I^{1/2}} \int_{f_{\rm{low}}}^{f_{\rm{cut}}} f^{-7/3} \sin(\beta f^{-2/3})
\frac{df}{S_n(f)} \\
& = & \frac{K}{I^{1/2}}.
\end{eqnarray}
Finding $\left < \mathcal{A}_3 , \mathcal{\widehat{A}}_2 \right >$;
\begin{equation}
\left < \mathcal{A}_3 , \mathcal{\widehat{A}}_2 \right > 
=  4 \int_{f_{\rm{low}}}^{f_{\rm{cut}}} \mathcal{A}_{3}^{*} (f) \mathcal{\widehat{A}}_{2} (f)
\frac {df} {S_{n} (f)}.
\end{equation}
Substituting in for $\mathcal{A}_3$ and $\mathcal{\widehat{A}}_2$ and 
multiplying terms together;
\begin{eqnarray}
\left < \mathcal{A}_3 , \mathcal{\widehat{A}}_2 \right > & = &
\frac{I^{1/2}}{ \bigg[ IM + \frac{I^{2}}{2} - J^{2} \bigg] ^{1/2} }
\nonumber \\
& \times & 
\left[
4 \int_{f_{\rm{low}}}^{f_{\rm{cut}}} f^{-7/3} \sin(\beta f^{-2/3}) \cos(\beta f^{-2/3}) 
\frac {df} {S_{n} (f)} \right. \\
& - & \left. \frac{4J}{I} \int_{f_{\rm{low}}}^{f_{\rm{cut}}} f^{-7/3} 
\sin(\beta f^{-2/3}) \frac{df}{S_{n} (f)}
\right].
\end{eqnarray} 
Using $2\sin(\beta f^{-2/3})\cos(\beta f^{-2/3})=\sin(2\beta f^{-2/3})$;
\begin{eqnarray}
\left < \mathcal{A}_3 , \mathcal{\widehat{A}}_2 \right > & = &
\frac{I^{1/2}}{ \bigg[ IM + \frac{I^{2}}{2} - J^{2} \bigg] ^{1/2} }
\nonumber\\
& \times &
\left[
2 \int_{f_{\rm{low}}}^{f_{\rm{cut}}} f^{-7/3} \sin(2 \beta f^{-2/3}) 
\frac {df} {S_{n} (f)} \right. \nonumber\\
& - & \left. \frac{4J}{I} \int_{f_{\rm{low}}}^{f_{\rm{cut}}} f^{-7/3} 
\sin(\beta f^{-2/3}) \frac{df}{S_{n} (f)}
\right].
\end{eqnarray} 
Rewriting in terms of moments of the noise;
\begin{eqnarray}
\left < \mathcal{A}_3 , \mathcal{\widehat{A}}_2 \right > & = &
\frac{I^{1/2}}{ \bigg[ IM + \frac{I^{2}}{2} - J^{2} \bigg] ^{1/2} }
\bigg[
L - \frac{JK}{I}
\bigg] \nonumber\\
& = & \frac{IL - JK}{I^{1/2} 
\bigg[ IM + \frac{I^{2}}{2} - J^{2} \bigg] ^{1/2} }.
\end{eqnarray} 
Finding numerator of $\mathcal{\widehat{A}}_3$, $\mathcal{A}_3 - 
\left < \mathcal{A}_3 , \mathcal{\widehat{A}}_1 \right > 
\mathcal{\widehat{A}}_1 - \left < \mathcal{A}_3 , 
\mathcal{\widehat{A}}_2 \right > \mathcal{\widehat{A}}_2$;
\begin{eqnarray} \label{A3num}
&&\mathcal{A}_3 - \left < \mathcal{A}_3 , \mathcal{\widehat{A}}_1 \right > 
\mathcal{\widehat{A}}_1 - \left < \mathcal{A}_3 , 
\mathcal{\widehat{A}}_2 \right > \mathcal{\widehat{A}}_2 
 = f^{-7/6} \nonumber\\ 
& \times &
\left[ \sin(\beta f^{-2/3}) - \frac{K}{I}
 -  \frac{IL - JK}{\bigg[ IM + \frac{I^{2}}{2} - J^{2} \bigg]} 
\left[\cos(\beta f^{-2/3}) - \frac{J}{I} \right] \right].
\end{eqnarray}
Finding $|| \mathcal{A}_3 - \left < \mathcal{A}_3 , 
\mathcal{\widehat{A}}_1 \right > \mathcal{\widehat{A}}_1 - 
\left < \mathcal{A}_3 , \mathcal{\widehat{A}}_2 \right > 
\mathcal{\widehat{A}}_2 ||$;

\begin{eqnarray} \label{N3}
|| \mathcal{A}_3 &-& \left < \mathcal{A}_3 , \mathcal{\widehat{A}}_1 \right > 
\mathcal{\widehat{A}}_1 - \left < \mathcal{A}_3 , 
\mathcal{\widehat{A}}_2 \right > \mathcal{\widehat{A}}_2 || \nonumber\\
& = & 
4 \int_{f_{\rm{low}}}^{f_{\rm{cut}}}
\left[ \mathcal{A}_3 - 
\left < \mathcal{A}_3 , \mathcal{\widehat{A}}_1 \right > 
\mathcal{\widehat{A}}_1 - 
\left < \mathcal{A}_3 , \mathcal{\widehat{A}}_2 \right > 
\mathcal{\widehat{A}}_2 \right] ^{2}
\frac {df} {S_n (f)}.
\end{eqnarray}
Finding $\left[ \mathcal{A}_3 - 
\left < \mathcal{A}_3 , \mathcal{\widehat{A}}_1 \right > 
\mathcal{\widehat{A}}_1 - 
\left < \mathcal{A}_3 , \mathcal{\widehat{A}}_2 \right > 
\mathcal{\widehat{A}}_2 \right] ^{2}$; 
\begin{eqnarray}
\left[ \mathcal{A}_3 \right.  & - & \left.
\left < \mathcal{A}_3 , \mathcal{\widehat{A}}_1 \right > 
\mathcal{\widehat{A}}_1 - 
\left < \mathcal{A}_3 , \mathcal{\widehat{A}}_2 \right > 
\mathcal{\widehat{A}}_2 \right] ^{2} \nonumber\\ 
& = &
f^{-7/3} \Bigg[
\sin^{2}(\beta f^{-2/3}) + \frac{K^{2}}{I^{2}} \nonumber\\ 
& + & \frac{(IL - JK)^{2}}{\bigg[ IM + \frac{I^{2}}{2} - J^{2} \bigg]^{2}} 
\bigg[ \cos^{2}(\beta f^{-2/3}) - \frac{2J}{I} \cos(\beta f^{-2/3}) + 
\frac{J^{2}}{I^{2}} \bigg] \nonumber\\
& - & \frac{2K}{I} \sin(\beta f^{-2/3}) 
  +  \frac{2(IL - JK)}{ IM + \frac{I^{2}}{2} - J^{2} } 
\bigg[ \frac{K}{I}  \sin(\beta f^{-2/3}) - \frac{JK}{I} \nonumber\\
& - & \sin(\beta f^{-2/3}) \cos(\beta f^{-2/3}) + 
\frac{J}{I} \sin(\beta f^{-2/3})\bigg]
\Bigg].
\end{eqnarray}
Using double angle formulas to simplify the equation;
\begin{eqnarray}
(\mathcal{A}_3 &-& \left < \mathcal{A}_3 , \mathcal{\widehat{A}}_1 \right > 
\mathcal{\widehat{A}}_1 - 
\left < \mathcal{A}_3 , \mathcal{\widehat{A}}_2 \right > 
\mathcal{\widehat{A}}_2) ^{2} \nonumber\\
& = &
f^{-7/3} \Bigg[
\frac{1}{2}-\frac{1}{2}\cos(2 \beta f^{-2/3}) + \frac{K^{2}}{I^{2}} \nonumber\\
& + & \frac{(IL - JK)^{2}}{\bigg[ IM + \frac{I^{2}}{2} - J^{2} \bigg]^{2}} 
\bigg[ \frac{1}{2} + \frac{1}{2}\cos(2 \beta f^{-2/3}) - \frac{2J}{I} 
\cos(\beta f^{-2/3}) + \frac{J^{2}}{I^{2}} \bigg] \nonumber\\
& - & \frac{2K}{I} \sin(\beta f^{-2/3}) 
 +  \frac{2(IL - JK)}{ IM + \frac{I^{2}}{2} - J^{2} } 
\bigg[ \frac{K}{I}  \cos(\beta f^{-2/3}) - \frac{JK}{I^{2}} \nonumber\\
& - & \frac{1}{2} \sin(2 \beta f^{-2/3}) + \frac{J}{I} 
\sin(\beta f^{-2/3})\bigg]
\Bigg].
\end{eqnarray}
Substituting into Eq.~(\ref{N3}) for $|| \mathcal{A}_3 - 
\left < \mathcal{A}_3 , \mathcal{\widehat{A}}_1 \right > 
\mathcal{\widehat{A}}_1 - 
\left < \mathcal{A}_3 , \mathcal{\widehat{A}}_2 \right > 
\mathcal{\widehat{A}}_2 ||$;
\begin{eqnarray}
|| \mathcal{A}_3 &-& 
\left < \mathcal{A}_3 , \mathcal{\widehat{A}}_1 \right > 
\mathcal{\widehat{A}}_1 - 
\left < \mathcal{A}_3 , \mathcal{\widehat{A}}_2 \right > 
\mathcal{\widehat{A}}_2 ||
\nonumber\\
& = &
2 \int_{f_{\rm{low}}}^{f_{\rm{cut}}} f^{-7/3} \frac {df} {S_n (f)}
- 2 \int_{f_{\rm{low}}}^{f_{\rm{cut}}} f^{-7/3} \cos(2 \beta f^{-2/3}) 
\frac {df} {S_n (f)}\nonumber\\
& + & \frac{4 K^{2}}{I^{2}} \int_{f_{\rm{low}}}^{f_{\rm{cut}}}  f^{-7/3} 
\frac {df} {S_n (f)} \nonumber\\
& + &  \frac{(IL - JK)^{2}}
{\bigg[ IM + \frac{I^{2}}{2} - J^{2} \bigg]^{2}}
\bigg[ 
2  \int_{f_{\rm{low}}}^{f_{\rm{cut}}} f^{-7/3} \frac {df} {S_n (f)} \nonumber\\
& + & 2 \int_{f_{\rm{low}}}^{f_{\rm{cut}}} f^{-7/3} \cos(2 \beta f^{-2/3}) 
\frac {df} {S_n (f)} \nonumber\\
& - & \frac{8 J}{I} \cos(\beta f^{-2/3}) \frac {df} {S_n (f)}
+ \frac{4 J^{2}}{I^{2}}  \int_{f_{\rm{low}}}^{f_{\rm{cut}}} f^{-7/3} 
\frac {df} {S_n (f)} 
\bigg] \nonumber\\
& - & \frac{8 K}{I}  \int_{f_{\rm{low}}}^{f_{\rm{cut}}} f^{-7/3} 
\sin(\beta f^{-2/3}) \frac {df} {S_n (f)} \nonumber\\
& + & \frac{(IL - JK)}{ IM + \frac{I^{2}}{2} - J^{2} } 
\bigg[
\frac{8 K}{I}  \int_{f_{\rm{low}}}^{f_{\rm{cut}}}  f^{-7/3} 
\cos(\beta f^{-2/3}) \frac {df} {S_n (f)}\nonumber\\
& - & \frac{8 JK}{I^{2}}  \int_{f_{\rm{low}}}^{f_{\rm{cut}}}  f^{-7/3} \frac {df} {S_n (f)}
 -  4 \int_{f_{\rm{low}}}^{f_{\rm{cut}}}  f^{-7/3}  \sin(2 \beta f^{-2/3}) 
\frac {df} {S_n (f)}\nonumber\\
& + & \frac{8 J}{I} \int_{f_{\rm{low}}}^{f_{\rm{cut}}}  f^{-7/3} 
\sin(\beta f^{-2/3}) \frac {df} {S_n (f)}
\bigg].
\end{eqnarray}
Rewriting integrals in terms of moments of the noise;
\begin{eqnarray} \label{A3den}
|| \mathcal{A}_3 &-& 
\left < \mathcal{A}_3 , \mathcal{\widehat{A}}_1 \right > 
\mathcal{\widehat{A}}_1 - 
\left < \mathcal{A}_3 , \mathcal{\widehat{A}}_2 \right > 
\mathcal{\widehat{A}}_2 ||
\nonumber\\
& = & \frac{1}{I} \bigg[
 \frac{I^{2}}{2} - IM - K^{2} 
 -  \frac{(IL - JK)^{2}}{ IM + \frac{I^{2}}{2} - J^{2} } \bigg]
\end{eqnarray}
Substituting in Eq.~(\ref{A3num}) and Eq.~(\ref{A3den}) into 
Eq.~(\ref{A3}) for $\mathcal{\widehat{A}}_{3}$;
\begin{eqnarray}
\mathcal{\widehat{A}}_3 & = &
\frac{f^{-7/6}
\Bigg[ \sin(\beta f^{-2/3}) - \frac{K}{I}
 -  \frac{IL - JK}{ IM + \frac{I^{2}}{2} - J^{2} } 
\bigg[\cos(\beta f^{-2/3}) - \frac{J}{I} \bigg] 
\Bigg] I^{1/2}  }
{
\bigg[
 \frac{I^{2}}{2} - IM - K^{2} 
 -  \frac{(IL - JK)^{2}}{ IM + \frac{I^{2}}{2} - J^{2} } \bigg] ^{1/2}
}.\nonumber\\ 
\label{app_A3hat}
\end{eqnarray}

\section{Documentation of the BCVSpin matched-filter engine}
\label{app:matchedfilter}

\subsection{Introduction}
This document describes the functions that have been written to perform 
matched-filtering of time-domain interferometric detector data using the 
detection templates developed by Buonanno, Chen and Vallisneri 
in BCV2 \cite{BCV2}. 
These functions have been written in C within the LSC Algorithm Library 
(Sec.~\ref{LAL}) and can be found in {\tt lal/packages/findchirp/src} 
in the following files:
\begin{center}
\begin{tabular}{l|l}
Function & Filename \\
\hline \\
LALFindChirpBCVSpinTemplate()      & FindChirpBCVSpinTemplate.c \\
LALFindChirpBCVSpinFilterSegment() & FindChirpBCVSpinFilter.c
\end{tabular}
\end{center}   

These functions draw heavily upon pre-existing LAL functionality.

\subsubsection{LSC Algorithm Library}
\label{LAL}
The LSC Algorithm Library (LAL) is a software package which has been developed 
by the LIGO Laboratory (LL) and the LIGO Scientific Collaboration for the 
purpose of analysing data from interferometric gravitational wave detectors. To enable as many 
contributors as possible LAL is written in C which was thought to be the 
language the majority of contributors would be most proficient in. A full 
specification of LAL as well as downloadable versions of the code and 
documentation is available at the LAL Home Page \cite{LAL}. 
Functions written in LAL 
(e.g., calibration of detector data, estimation of power spectral density, 
calculation of template bank metric, matched-filtering) 
are organised to 
perform higher level tasks 
(e.g., creating a template bank and measuring of triggers)
using the LALapps (LAL 
Applications) package.
LAL is freely available and distributed under the GNU General Public License.
The FindChirp package, included within LAL, which performs the matched-filter
routines used for inspiral searches is described in 
Allen et al. (2005) \cite{allen:2005}.


\subsection{Definitions}
Consider a time series $h(t)$ sampled at $N$ ({\tt numPoints}) consecutive 
points with sampling interval $\Delta t$
({\tt deltaT}), that is;
\begin{equation}
h_{j} \equiv h(t_{j}) \qquad t_{j} = j\Delta t 
\end{equation}
where the sampling frequency $f_{s}$ is given by
\begin{equation}
f_{s} = \frac{1}{\Delta t}
\end{equation}
and the sampling interval $\Delta f$ is given by
\begin{equation}
\Delta f = \frac{1}{N \Delta t}.
\end{equation} 
\subsubsection{Discrete Fourier Transforms}
The forward DFT used by LAL is
\begin{equation}
\tilde{h}_{k} = \sum_{j=0}^{N-1} h_{j} e^{-2\pi i j k / N}
\end{equation}
where $i = \sqrt{-1}$.
We can recover the frequency series using
\begin{equation}
\tilde{h}(f_{k}) = \Delta t \tilde{h_{k}}.
\end{equation}
The reverse DFT used by LAL is 
\begin{equation}
h(t_{j}) = \frac{1}{N} \sum_{k=0}^{N-1} \tilde{h}_{k} e^{2\pi i j k / N}.
\end{equation}

In practise we use the Fast Fourier Transform to
perform (forward and reverse) DFTs.
A DFT would typically require $\sim N^{2}$ arithmetic (floating point) 
operations. 
Using Fast Fourier Transforms (FFTs) 
(see e.g., Chapter 12 of Ref.~\cite{NumericalRecipesInC} for a 
description) these transformations can be performed
in only $\sim N \log N$ operations.

\subsubsection{Discrete inner product}
In practice the factor $1/N$ is omitted by the function performing the reverse 
DFT. The inner product of two time series $x(t_{j})$ and $y(t_{j})$ is defined 
as
\begin{equation} \label{BSEinner}
\left<x(t_{j}),y(t_{j})\right> = 
4\Delta f \Re \sum_{k=0}^{N/2} 
\frac{ \tilde{x}(f_{k})^{*} \tilde{y}(f_{k}) }
{ S_{n}( f_{k} ) } 
\end{equation}
which is equivalent to
\begin{equation} 
\left<x(t_{j}),y(t_{j})\right> = 
\frac{4 \Delta t}{N} \Re \sum_{k=0}^{N/2} 
\frac{ \tilde{x}_{k}^{*} \tilde{y}_{k} }
{ S_{n}( f_{k} ) } 
\end{equation}
where $S_{n}( f_{k} )$ is the one-sided noise power spectral density defined as
\begin{equation}
\overline{ \tilde{n}(f_{k})\tilde{n}^{*}(f'_{k}) } 
= \frac{1}{2} S_{n}( f_{k} ) \delta(f_{k} - f'_{k}) 
\end{equation}
and a superscript $*$ above a quantity indicates that its complex conjugate has 
been taken. We will define a normalised template (or waveform) $\hat{h}$ such 
that $\left<\hat{h},\hat{h}\right> = 1$. To normalise a template $h$ we
say that $\hat{h} = Ah$. Therefore
\begin{equation} 
\left<\hat{h},\hat{h}\right> 
= \frac{4 A^{2} \Delta t}{N} 
\Re \sum_{k=0}^{N/2} \frac{ \tilde{h}_{k}^{*} \tilde{h}_{k} }{ S_{n}( f_{k} ) }
= 1 
\end{equation}
It follows that
\begin{equation}
A = \left [ \frac{4 \Delta t}{N} 
\Re \sum_{k=0}^{N/2} 
\frac{ \tilde{h}_{k}^{*} \tilde{h}_{k} }{ S_{n}( f_{k} ) } \right]^{-1/2}  
\end{equation}
\subsubsection{BCVSpin detection templates}
Here we define a set of orthonormal templates ${\hat{h}}$ (in the frequency domain)
\begin{equation}
\label{BSErecon}
\hat{h} = \sum_{l=1}^{2n} \hat{\alpha}_{l} \hat{h}_{l} \qquad n = 3
\end{equation}
where 
\begin{eqnarray}
\hat{h}_{l}  (t_{j};f_{k}) & = &  
\widehat{\mathcal{A}}_{l}(f_{k})e^{i\psi_{\rm{NM}}(f_{k})}e^{ijk/N} \nonumber\\
\hat{h}_{l+n}(t_{j};f_{k}) & = & 
i\widehat{\mathcal{A}}_{l}(f_{k})e^{i\psi_{\rm{NM}}(f_{k})}e^{ijk/N}  
\end{eqnarray}
where $l=1,2,3$, $n=3$ and $\hat{\alpha}_{l}$ are values corresponding to the 
amplitudes of each basis template (vector) $\hat{h}_{l}$.
The vectors $\widehat{\mathcal{A}}_{l}$ are called the orthonormalised 
amplitude vectors and are given later by
Eq. (\ref{BSEamp}). 
 To ensure that the templates are normalised, 
\begin{equation}
\left< \hat{h}, \hat{h} \right> = 1
\end{equation}
then it must be true that
\begin{equation}
\sum_{l=1}^{2n} \hat{\alpha}_{l}^{2} = 1 \qquad n=3
\end{equation}

Using the relation $f_{k} = k \Delta f = k/(N \Delta t) $ we can find the 
discretized form of the various powers of frequency we use in the 
construction of the detection templates
\begin{eqnarray}
f_{k}^q = \left ( \frac{k}{N \Delta t} \right )^{q} 
\qquad k = 1 \dots \frac{N}{2}.
\end{eqnarray}
In practise we store arrays containing $(k/N)^{q}$ and then 
include the factor $\Delta t ^{-q}$ as required
\footnote{ 
It would also be possible to simply store $k^{q}$ and then re-include
factors of $(N \Delta t)^{-q}$.
}.

\subsection{BCVSpin matched-filter engine}
\subsubsection{Calibrating the strain data}
Take FFT of time series data $x(t_{j})$ 
\begin{equation}
\tilde{x_{k}} = \sum_{j=0}^{N-1} x_{j} e^{-2\pi i j k / N}
\end{equation}
Calculate the strain $\tilde{s_{k}}$\\
\begin{equation}
\tilde{s_{k}} = \tilde{x_{k}} \times R_{k} \qquad k = 0 \dots \frac{N}{2}  
\end{equation}
where 
\begin{equation}
R_{k} = {\tt response}_{k} \times {\tt dynRange}
\end{equation}
where {\tt response} is a complex vector ($k=0 \dots N/2$) and {\tt dynRange}
is a single user defined value used to artificially adjust the magnitude of 
the strain. This is useful since realistic strains caused by gravitational 
wave events will be of the order $\sim 10^{-22}$ and it is easier to deal 
with quantities with values of around unity. Henceforth we shall refer to 
the strain $s$ as the detector output.

\subsubsection{Calculation of inverse noise power spectrum}
Calculating the inverse noise power spectrum $\tilde{\omega}_{v}(f_{k})$
\begin{displaymath}
\tilde{\omega}_{v}(f_{k}) =\left\{
\begin{array}{cl}
 0 & k = 0 \dots k_{\rm{min}}-1 \\   
 \frac{1}{ S_{n}( f_{k} ) } &  k = k_{\rm{min}} \dots N/2
\end{array} \right.
\end{displaymath}
where
\begin{equation}
k_{\rm{min}} = \frac{f_{\rm{low}}}{\Delta f}
\end{equation}
Converting $\tilde{\omega}_{v}(k)$ to $\tilde{\omega}_{h}(k)$\\
\begin{equation}
\tilde{\omega}_{h}(f_{k}) = \frac{\tilde{\omega}_{v}(f_{k})}{R_{k}R^{*}_{k} } 
\end{equation}


\subsubsection{LALFindChirpBCVSpinTemplate()}
We calculate the non-modulational phase of template as
\begin{displaymath}
\psi_{\rm{NM}}(f_{k})= \left\{
\begin{array}{cl}
0                                       & 
k = 0             \dots k_{\rm{min}}-1\\
f_{k}^{-5/3}(\psi_{0} + f_{k} \psi_{3}) & 
k = k_{\rm{min}}  \dots k_{\rm{max}}-1\\
0                                       & 
k = k_{\rm{max}}  \dots N/2
\end{array} \right.
\end{displaymath}
where
\begin{equation}
k_{\rm{min}} = \frac{f_{\rm{low}}}{\Delta f} \qquad 
k_{\rm{max}} = \frac{f_{\rm{cut}}}{\Delta f}. 
\end{equation}
$f_{\rm{low}}$ is the detectors lower frequency cutoff
(see Sec.~\ref{subsub:lowfreqcut}) 
and $f_{\rm{cut}}$ will eventually be supplied by the template bank.

We now calculate the moments of the noise required to construct the 
orthonormalised amplitude functions $\widehat{\mathcal{A}}_l$ 
\begin{eqnarray}
I & = & 4 \Delta f \, \sum_{k_{\rm{min}}}^{k_{\rm{max}}} f_{k}^{-7/3} 
                      \tilde{\omega}_{h}(f_{k}) 
                      \nonumber\\
J & = & 4 \Delta f \, \sum_{k_{\rm{min}}}^{k_{\rm{max}}} f_{k}^{-7/3} 
                      \cos(\beta f^{-2/3}) \tilde{\omega}_{h}(f_{k}) 
                      \nonumber\\
K & = & 4 \Delta f \, \sum_{k_{\rm{min}}}^{k_{\rm{max}}} f_{k}^{-7/3} 
                      \sin(\beta f^{-2/3}) \tilde{\omega}_{h}(f_{k}) 
                      \nonumber\\
L & = & 2 \Delta f \, \sum_{k_{\rm{min}}}^{k_{\rm{max}}} f_{k}^{-7/3} 
                      \sin(2\beta f^{-2/3}) \tilde{\omega}_{h}(f_{k}) 
                      \nonumber\\
M & = & 2 \Delta f \, \sum_{k_{\rm{min}}}^{k_{\rm{max}}} f_{k}^{-7/3} 
                      \cos(2\beta f^{-2/3}) \tilde{\omega}_{h}(f_{k}) 
                      \nonumber\\
\end{eqnarray}
and
\begin{displaymath}
\left.
\begin{array}{c}
I \\ J \\ K \\ L \\ M 
\end{array} 
\right\}
= \left\{
\begin{array}{cl}
0  & k = 0 \dots k_{\rm{min}}-1\\
0  & k = k_{\rm{max}}  \dots N/2.
\end{array} \right.
\end{displaymath}
In practise we use omit prefactors of $\Delta t ^{7/3}$ when calculating these
moments.
We can construct the orthonormalised amplitude functions, in the range 
$k_{\rm{min}} \geq k < k_{\rm{max}}$
\begin{eqnarray}
\label{BSEamp}
\mathcal{\widehat{A}}_{1}(f_{k})   & = &  f_{k}^{-7/6} 
                                  \frac {1}  { I^{1/2}} \\
\mathcal{\widehat{A}}_{2}(f_{k})   & = & f_{k}^{-7/6}     
                                  \frac{ \bigg [ \cos(\beta f^{-2/3}) - \frac{J}{I} \bigg ] I^{1/2} } { \bigg[ IM + \frac{I^{2}}{2} 
                                     - J^{2} \bigg] ^{1/2} } \nonumber\\
\mathcal{\widehat{A}}_{3}(f_{k})   & = &  f_{k}^{-7/6}             
                              \frac{ \bigg [ 
                              \sin(\beta f^{-2/3}) 
                            - \frac{K}{I} 
                            - \frac{IL  - JK}{IM + \frac{I^{2}}{2} - J^{2}} \big[\cos(\beta f^{-2/3}) -\frac{J}{I} \big ]
                                              \bigg ] I^{1/2} }
                            { \bigg [ 
                              \frac{I^{2}}{2} - IM - K^{2} - \frac{ (IL - JK)^{2}}{IM + \frac{I^{2}}{2} - J^{2} }   
                              \bigg ] ^{1/2} }  \nonumber
\end{eqnarray}
and
\begin{displaymath}
\left.
\begin{array}{c}
\mathcal{\widehat{A}}_{1}(f_{k}) \\ 
\mathcal{\widehat{A}}_{2}(f_{k}) \\
\mathcal{\widehat{A}}_{3}(f_{k})  
\end{array}
\right\}
= \left\{
\begin{array}{cl}
0  & k = 0 \dots k_{\rm{min}}-1\\
0  & k = k_{\rm{max}}  \dots N/2.
\end{array} \right.
\end{displaymath}
In practise we use omit factors of $\Delta t ^{7/6}$ from the $f_{k}^{-7/6}$
prefactor terms in these functions.
We find that the factors of $\Delta t$ in these terms and in the calculation
of the noise moments cancel meaning that our amplitude functions are
correctly scaled.

We can calculate the cross products of the orthonormalised amplitude 
functions
\begin{eqnarray}
\left< \widehat{\mathcal{A}}_{l}(f_{k}),
\widehat{\mathcal{A}}_{m}(f_{k}) \right>
= 
4 \Delta f \, \sum_{k=0}^{N/2} 
\widehat{\mathcal{A}}_{l}(f_{k}) \widehat{\mathcal{A}}_{m}(f_{k})
\tilde{\omega}_{h}(f_{k}) \qquad l,m = 1,2,3  
\end{eqnarray}
These results can be used to check that the amplitude functions are 
truly orthonormal
\begin{equation}
\left < \widehat{\mathcal{A}}_{l}(f_{k}),
\widehat{\mathcal{A}}_{m}(f_{k}) \right> = \delta_{l,m}
\end{equation}
where 
\begin{displaymath}
\delta_{l,m} = \left\{
              \begin{array}{cl}
              1 & l = m\\
              0 & l \neq m.
              \end{array} \right.
\end{displaymath}
\subsubsection{LALFindChirpBCVSpinFilterSegment()}
We calculate the quantities $\tilde{q}_{n}(k)$ and use these to 
calculate the overlaps between the detector output $s$ and
the orthonormalised basis templates $\hat{h}_{j}$
\begin{eqnarray}
\tilde{q}_{l}(k) = \widehat{\mathcal{A}}_{l}(f_{k}) 
                   e^{i \psi_{\rm{NM}}} \tilde{s}^{*}_{k}  
                  \tilde{\omega}_{h}(f_{k}) 
\end{eqnarray}
where $l = 1,2,3$. The overlaps between the detector output $s$ 
and the 6 basis templates $\hat{h}_{l}$ can then be found at every time
$t_{j}$:
\begin{eqnarray}
\left< s,  \hat{h}_{l} (t_{j}) \right>
    & = &  \frac{4}{N}  \Re \sum_{k=0}^{N-1} \tilde{q}_{l}(k) 
           e^{2 \pi i j k /N} \\
\left< s, \hat{h}_{l+n} (t_{j}) \right>
    & = &  -\frac{4}{N} \Im \sum_{k=0}^{N-1} \tilde{q}_{l}(k) 
           e^{2 \pi i j k /N} 
\end{eqnarray}
where $l = 1,2,3$ and $ n=3$. The factor of $1/N$ arises from the need to 
include a factor $\Delta t$ to convert from $\tilde{s}_{k}$ and 
$\tilde{s}(f_{k})$. Multiplying the existing factor of $\Delta f$ used in 
the definition of the inner product (see Eq. (\ref{BSEinner})) and the factor 
$\Delta t$ gives $1/N$. Using the overlaps calculated above we can find the 
signal to noise ratio $\rho$ of the detector output $s$ with the normalised 
template $\hat{h}$ at every time $t_{j}$:
\begin{equation}
\rho(t_{j}) = \sqrt{ \sum_{l=1}^{6} \left< s, \tilde{h}_l (t_{j}) \right> ^{2} }.
\end{equation}
We note that we can calculate $\rho(t)$ for all the times we filter using the
(Fast) Fourier transform (see Sec.~\ref{sec:maximisation}). 
We can now find the individual $\hat{\alpha}_{l}$ values which correspond 
to the maxima in $\rho$, $\rho_{\rm{max}}$:
\begin{equation}
\hat{\alpha}_{l} = 
\frac{\left< s, \tilde{h}_l \right>_{\rm{max}}}{\rho_{\rm{max}}} 
\qquad l=1,2 \dots 6.
\end{equation}
We can then use these values to reconstruct the (normalised) waveform which 
caused the peak in $\rho$ using Eq. (\ref{BSErecon}). The reconstruction of 
waveforms can be used to test the code when performing injections
of known waveforms.

\bibliographystyle{plain}
\bibliography{thesis}

\end{document}